\documentclass[aps,prx,twocolumn,showpacs,superscriptaddress,groupedaddress,floatfix,10pt,longbibliography]{revtex4-2}
\pdfoutput=1
\usepackage[utf8]{inputenc}
\usepackage{graphicx}
\usepackage[english]{babel}
\usepackage{amsmath,amssymb}
\usepackage{bm,bbm}
\usepackage{mathrsfs}
\usepackage{graphicx}
\usepackage{color}
\usepackage{mathtools}
\usepackage{amsfonts}
\usepackage{booktabs}
\usepackage{amsthm}
\usepackage{url}
\usepackage{textcomp}
\usepackage{bbold}
\usepackage[bb=boondox]{mathalfa}
\usepackage{dsfont}

\newcommand{\e}[1]{\mathrm{e}^{#1}}
\newcommand{\tmu}{\tilde{\mu}}
\newcommand{\thh}{\tilde{h}}
\newcommand{\tj}{\tilde{J}}
\newcommand{\tf}{\tilde{f}}
\newcommand{\rtf}{\tilde{\rm f}}
\newcommand{\mH}{\mathcal{H}}
\newcommand{\sigall}{\{\sigma_i\}}
\newcommand{\Zqc}{Z^{\mathrm{BG}}}

\newcommand{\ovX}{\bar{X}}
\newcommand{\bz}{\bar{z}}
\newcommand{\vphi}{\varphi}

\DeclareMathOperator\arctanh{arctanh}

\begin{document}
\title{Criticality in Cell Adhesion}
\author{Kristian Blom}
\author{Alja\v{z} Godec}
\email{agodec@mpibpc.mpg.de}
\affiliation{Mathematical bioPhysics Group, Max Planck Institute for Biophysical Chemistry, G\"{o}ttingen 37077, Germany}
\begin{abstract}
We illuminate the many-body effects underlying the structure, formation, and dissolution
of cellular adhesion domains in the presence and absence of forces. We
consider mixed Glauber-Kawasaki dynamics of a two-dimensional model of
nearest-neighbor interacting adhesion bonds with intrinsic
binding-affinity under the action of a shared pulling or pushing
force. \textcolor{black}{We consider adhesion bonds that are immobile due to being anchored to the
      underlying cytoskeleton as well as adhesion molecules that are transiently
      diffusing.} Highly accurate analytical results are obtained on the pair-correlation
level of the Bethe-Guggenheim approximation for the complete thermodynamics and kinetics of adhesion clusters of any size,
including the thermodynamic limit. A new kind of dynamical phase
transition is uncovered --- the mean formation and
dissolution times per adhesion bond change discontinuously with
respect to the bond-coupling parameter. At the respective critical points cluster formation and dissolution are fastest, while the statistically dominant transition path undergoes
a qualitative change --- the entropic barrier to complete binding/unbinding is rate-limiting below, and the phase transition between dense and dilute phases above the dynamical critical point. In the context of the Ising model the dynamical phase
  transition reflects a first-order discontinuity in the
  magnetization-reversal time. Our results provide a potential explanation for the mechanical regulation of cell
adhesion, and suggest that the quasi-static and kinetic response to changes in the membrane stiffness or applied forces is largest near the statical and dynamical critical point, respectively.     
\end{abstract}
\maketitle
\section{Introduction}
Cell adhesion refers to the specific binding of cells to neighboring
cells or the extracellular matrix. It plays a major role in cell regulation \cite{regulation}, intercellular communication
\cite{commun}, immune response \cite{zarnitsyna_t_2012}, wound healing \cite{wound}, morphogenesis \cite{morpho},
cellular function \cite{hynes_cell_1999}, and tumorigenesis
\cite{cancer,andl_misregulation_2010}. 
Cellular adhesion domains form as a
result of the association of transmembrane cellular adhesion molecules
(CAMs) that interact with the actin cytoskeleton \cite{anchor} \textcolor{black}{and can translocate over the membrane \cite{DEMOND20083286}}. \textcolor{black}{There are four major superfamilies of CAMs; the immunoglobulins, integrins, cadherins, and selectins, and throughout we generically refer to them as CAMs.} Biological adhesion bonds are typically
non-covalent with binding energies on the order of a few $k_{\rm B}T$
corresponding to forces on the order of $\simeq 4$\,pN${}\cdot{}$nm at
$T\simeq{}$300\,K \cite{bell_models_1978, schwarz_physics_2013}. As a
result of thermal fluctuations these bonds have finite lifetimes --
they can break and re-associate depending on the receptor-ligand
distance, their respective conformations and local concentrations, and depending on internal and external mechanical forces
\cite{schwarz_physics_2013,evans_dynamic_1997}. While it was
originally thought that the strength of adhesion is determined by the
biochemistry of CAMs alone, more recently, cellular mechanics
\cite{Schmitz_2008} and adhesion bond interactions induced by thermal
undulations of the membrane \cite{Bruinsma,Speck,speck_specific_2010,bihr_nucleation_2012,farago_fluctuation-induced_2010} emerged as essential physical
regulators of cellular adhesion.

Diverse aspects of biological adhesion have been investigated
experimentally by contact-area fluorescence recovery after
photobleaching \cite{tolentino_measuring_2008}, Förster
resonance energy transfer \cite{huppa_tcrpeptidemhc_2010},
metal-induced energy transfer \cite{baronsky_cellsubstrate_2017},
reflection interference contrast microscopy
\cite{limozin_quantitative_2009}, optical tweezers
\cite{fallman_optical_2004}, flow-chamber methods
\cite{alon_lifetime_1995,juliano_adhesion_1977}, centrifugation
assays \cite{piper_determining_1998, marlin_purified_1987},
biomembrane force probe
\cite{chen_monitoring_2008,evans_sensitive_1995}, micropipette
techniques \cite{prechtel_dynamic_2002, lomakina_micromechanical_2004}, and
atomic force spectroscopy \cite{evans_dynamic_1997,heymann_dynamic_2000, sanyour_spontaneous_2018,rico_temperature_2010, sagvolden_cell_1999,Makarov,Makarov_2}. Experiments unraveled a collective behavior of clusters of
adhesion bonds that
cannot be explained as a sum of their individual behavior
\cite{williams_quantifying_2001,huppa_tcrpeptidemhc_2010,wu_transforming_2011,zarnitsyna_t_2012,fenz_membrane_2017}
that is meanwhile well understood (see
e.g.\ \cite{merkel_force_2001, gao_probing_2011}). More specifically, the opening/closing of adhesion bonds is profoundly affected by membrane fluctuations even if their amplitude becomes as small as 0.5 nm -- smaller than the thickness of the membrane itself \cite{hu_binding_2013, steinkuhler2019membrane}.

These observations imply many-body physics to be at play,
i.e.\ an interplay between
the coupling of nearby adhesion bonds through deformations of the fluctuating membrane and mechanical forces
acting on the membrane \cite{Bruinsma,Speck,speck_specific_2010,bihr_nucleation_2012,
  farago_fluctuation-induced_2010, fenz_membrane_2017,
  hu_binding_2013, hu_binding_2015, krobath_binding_2009,
  reister-gottfried_dynamics_2008, schmidt_coexistence_2012,
  xu_binding_2015,zarnitsyna_t_2012, steinkuhler2019membrane}.
Supporting the idea are 
experimental observations of cells changing the membrane flexibility and/or membrane fluctuations through ATP-driven activity \cite{monzel2016measuring, gov2005red, tuvia1997cell, biswas_mapping_2017}, decoupling the F-actin network \cite{simson1998membrane} or remodelling the actomyosin cytoskeleton \cite{biswas_mapping_2017}, and through acidosis \cite{steinkuhler2019membrane}, in
order to alter adhesion binding rates and strength \cite{Fungal,hong_vasoactive_2014, zhu_temporal_2012, fenz_membrane_2017, hong_coordination_2012, steinkuhler2019membrane} or to
become motile \cite{swaminathan_mechanical_2011}.
There is also a striking correspondence between
membrane stiffness and the metastatic potential of cancer cells --
the stiffness of cancer cells was found to determine their migration
and invasion potential \cite{swaminathan_mechanical_2011}.
The effect is not limited to cells; the elastic
  modulus was similarly found to significantly affect the specific adhesion of polymeric
  networks \cite{wang2017elastic}.
\begin{figure}
    \includegraphics[width = 0.48\textwidth]{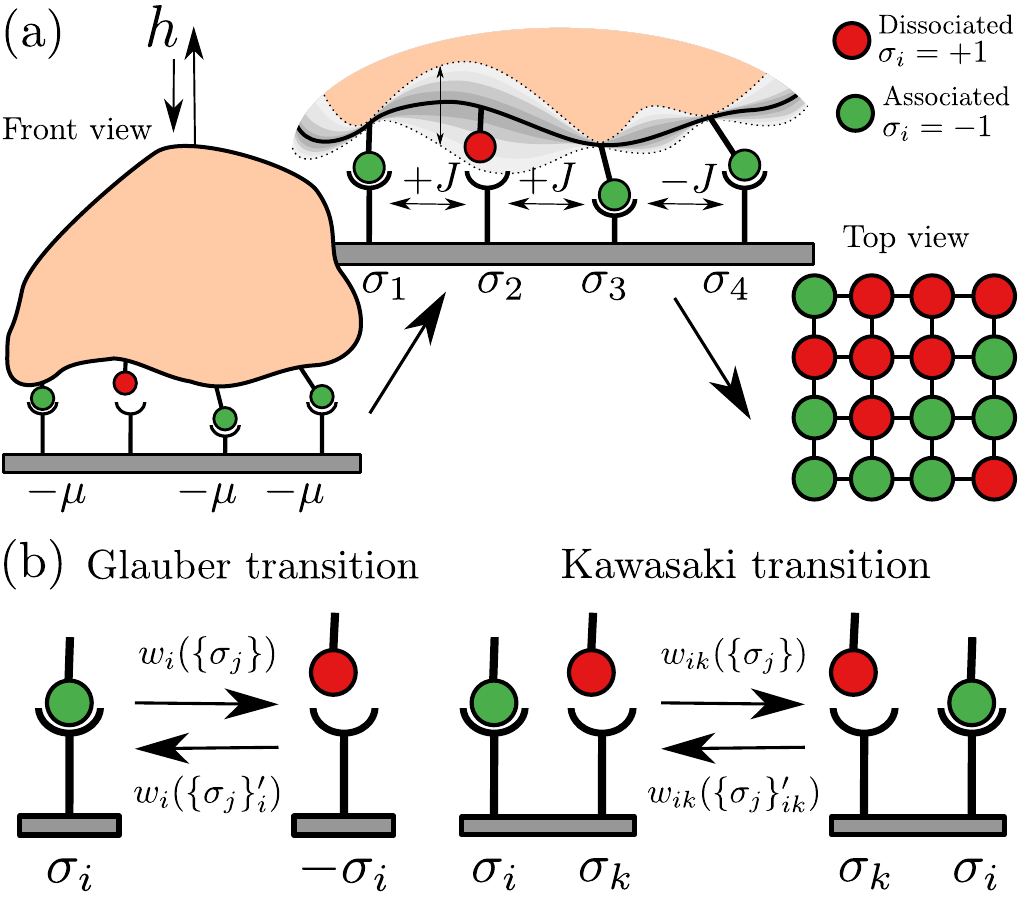}
    \caption{\textbf{A coarse-grained model for the cooperative association/dissociation of adhesion
    bonds.} (a) Schematic of the effective many-body model governed by Eq.~\eqref{Hamiltonian1} and \eqref{force}, depicting an
    adhesion domain on a cell-patch with
    16 CAMs anchored to stiff substrate. Adhesion bonds are arranged
    on a $4\times 4$ square lattice and can assume two
    states, $\sigma_{i}=\pm1$, where $+1$ corresponds to an open (red) and $-1$ to a closed bond (green).
    Nearest-neighbor bonds experience an effective interaction $J$
    induced by undulations of the anchoring
    membrane. An external force $h$ is
      pulling/pushing on the adhesion domain. Each adhesion bond
      has an intrinsic binding-affinity $\mu\geq0$ that favors a bound state. A small number of bonds is depicted for convenience only. 
    In the work we consider different system sizes including the thermodynamic limit. (b) Glauber and Kawasaki transition. A Glauber transition changes the binding state of a single adhesion bond to $\sigma_{i} \rightarrow -\sigma_{i}$ with transition rate $w_{i}(\{\sigma_{j}\})$ (see Eq.~\eqref{Glauber Transition rate}). A Kawasaki transition interchanges two nearest-neighbor adhesion bonds $\sigma_{i} \leftrightarrow \sigma_{k}$ with transition rate $w_{ik}(\{\sigma_{j}\})$ (see Eq.~\eqref{Kawasaki Transition rate}), corresponding to lateral diffusion.}
    \label{fig 1}
\end{figure} 

Most of our current understanding of the formation and stability of adhesion clusters
derives from the analysis of individual \cite{bell_models_1978} and non-interacting
adhesion bonds \cite{seifert_rupture_2000, erdmann_stochastic_2004,Dasanna},
and studies of collective effects in biomimetic vesicular model systems with floppy membranes
\cite{smith_progress_2009,reister-gottfried_dynamics_2008} and mobile
CAMs \cite{smith_force-induced_2008}. These results therefore do not
necessarily apply to cells, where membranes are stiffened by the
presence of, and receptors are anchored to, the stiff actin
cytoskeleton that can actively exert forces on the membrane
\cite{anchor}.

Notwithstanding all theoretical efforts \cite{Bruinsma,Speck,speck_specific_2010,gao_probing_2011,bihr_nucleation_2012,schmidt_coexistence_2012,farago_fluctuation-induced_2010,hu_binding_2013,
 hu_binding_2015,krobath_binding_2009}, a consistent 
and comprehensive physical picture of collective adhesion under the action of a
mechanical force that could explain the observations on live cellular systems  \cite{Fungal,hong_vasoactive_2014,zhu_temporal_2012,fenz_membrane_2017,zarnitsyna_t_2012,hong_coordination_2012,boudjemaa_direct_2019,swaminathan_mechanical_2011}
remains elusive. For example, whether the coupling of individual bonds causes
the collective association and dissociation rates to increase or decrease, respectively, was speculated
to depend on the intrinsic single-bond affinity
\cite{huppa_tcrpeptidemhc_2010, huang_kinetics_2010}, cell type
(i.e.\ surface corrugation) \cite{williams_quantifying_2001} and on the
state of the actin cytoskeleton \cite{huppa_tcrpeptidemhc_2010}.  
An understanding of cellular adhesion therefore must integrate
the complex interplay between the correlated, collective
(un)binding  \cite{bihr_nucleation_2012,fenz_membrane_2017,hu_binding_2015,krobath_binding_2009,reister-gottfried_dynamics_2008,schmidt_coexistence_2012,smith_progress_2009},  
the intrinsic affinity of anchored adhesion bonds
\cite{huppa_tcrpeptidemhc_2010, huang_kinetics_2010,affinity}, the cell
type and surface topology \cite{williams_quantifying_2001}, as well as
the integrity of, and forces generated by, the actin cytoskeleton
\cite{huppa_tcrpeptidemhc_2010,sanyour_spontaneous_2018,hong_vasoactive_2014,zhu_temporal_2012,hong_coordination_2012}
under physiological \cite{boudjemaa_direct_2019} or pathological
conditions \cite{buda_e-cadherin_2011,korb_integrity_2004,swaminathan_mechanical_2011,zeng_sept9_i1_2019}.

\textcolor{black}{While it is omnipresent in
  biological systems, cell adhesion displays subtle differences in the specific microsopic
  details. Here we aim to capture the essential general features
  of the physics of cell adhesion.} In order to arrive at a deeper understanding of the mechanical
regulation of cellular adhesion that would explain the collective
dynamics of adhesion bonds on the level of individual (un)binding
events we here consider mixed Glauber-Kawasaki dynamics of a generic, two-dimensional model of diffusing nearest-neighbor interacting adhesion bonds
with intrinsic affinity $\mu$ under the action of a shared force $h$ (see Fig.~\ref{fig 1}a).

Highly accurate analytical results on the
Bethe-Guggenheim level reveal the many-body (that is, beyond ``mean field'')
physics underlying biological adhesion. We consider in detail
cluster-sizes ranging from a few CAMs to the thermodynamic limit. 
In the thermodynamic limit we determine the equation of state and complete phase behavior that displays a phase separation and co-existence of dense and dilute
adhesion domains. The critical behavior is investigated in detail and striking differences are found between pulling- and
pushing-forces. Strikingly, we prove the existence of a seemingly new kind of
\emph{dynamical phase transition} -- the mean first passage time to
cluster formation/dissolution is proven to change discontinuously with
respect to
the coupling strength. This dynamical phase transition, and more generally the non-linear and
non-monotonic dependence on the membrane
flexibility, may explain the puzzling cooperative behavior of effective association
and dissociation rates measured experimentally.
\subsection{Outline of the work}
The paper is structured as follows. In Sec.~\ref{Model} we
present an effective mesoscopic model of adhesion clusters and provide
a practical roadmap to the diverse calculations and analyses. In
Sec.~\ref{Structure} we present explicit analytical 
results for the thermodynamic equation of state and complete phase behavior of
adhesion clusters, and in Sec.~\ref{Kinetics} we present
analytical results for the kinetics of cluster formation and
dissolution both in the presence and absence of forces. In Sec.~\ref{Implication} we discuss
the biological implications of our results and in particular the suggestive r\^ole of
criticality in the context of equilibrium adhesion strength and the
kinetic dissolution and formation rates, respectively. Finally, in
Sec.~\ref{2d Ising implication}
we  highlight the relevance of our results in the context of the Ising model.
We conclude in Sec.~\ref{Conc} with a summary and a
perspective on the importance and limitations of our
results, and mention possible extensions to be made in future studies. Details of calculations, explicit
asymptotic results and further technical information is presented in a
series of Appendices.
\section{\label{Model} Model of interacting adhesion bonds under
  shared force}
\subsection{Equilibrium}
We consider a two-dimensional patch of a cell surface with $N$ adhesion molecules embedded
in the cell membrane, their
lateral positions forming a lattice with coordination number $z$
(see Fig.~\ref{fig 1}). The results we derive hold for any lattice
but we focus the discussion mainly on the square lattice with free boundary conditions.
Opposing the patch is a stiff substrate
or a neighboring cell-patch with complementary adhesion
molecules occupying a commensurate lattice. The
state of individual bonds is denoted by $\sigma_{i}$, $i=1,2,...,N$,
where $\sigma_{i}=+1$ if bond $i$ is broken an $\sigma_{i}=-1$ if it
is closed.

In the presence of a timescale separation the opening/closing of nearest neighbor bonds is coupled via
membrane fluctuations. Following closely
the arguments of Ref.~\cite{speck_specific_2010} we can integrate out the
membrane degrees of freedom to obtain an effective Ising-like model
for the bonds within the patch with effective Hamiltonian
\begin{equation}
   \mathcal{H}(\{\sigma_{i}\})= -J\sum_{\langle ij \rangle} \sigma_{i}\sigma_{j} - \mu N_{c}(\{\sigma_i\}) +\mH_{\rm h}(\{\sigma_i\}),
    \label{Hamiltonian1}
\end{equation}
where $J\ge 0$ is the membrane-induced short-range coupling between the bonds, $\langle ij \rangle$
denotes all nearest-neighbor pairs, $\mu$ is the effective chemical potential (i.e.\ intrinsic
affinity) of individual bonds, and $\mH_{\rm h}(\{\sigma_i\})$ is the
Hamiltonian describing the effect of the mechanical force. The first term in Eq.~\eqref{Hamiltonian1} represents the effective
coupling between nearest neighbor bonds, and is isomorphic to the
interaction term in the Ising model \cite{ising_beitrag_1925}. It is an
effective measure of bond-cooperativity, i.e.\ it reflects that the (free) energy penalty
of closing/breaking a bond is smaller if neighboring bonds are
closed/open, respectively
\cite{speck_specific_2010}. Such an effective
  description in terms of bonds coupled via a
 short-range membrane-mediated interaction is feasible when bonds are
 flexible and/or the patch
 of the cell membrane is quite (but not completely) stiff and is thus rather pulled down as a whole instead of being
locally strongly deformed by the binding of individual bonds \cite{speck_specific_2010}.
In this limit the coupling strength is determined by the effective bending rigidity
of the cell membrane, $\kappa$, via $J \propto 1/\sqrt{\kappa}$ (see
\cite{speck_specific_2010} and Appendix~\ref{Appendix A}). That is,
in this regime
a
relatively floppier cell membrane with lower bending
rigidity induces a stronger 
cooperativity between neighboring bonds than a
relatively stiff
membrane. Notably, 
  a detailed comparison between the full model of specific adhesion
  (i.e.\ reversible adhesion bonds explicitly coupled to a dynamic
  fluctuating membrane) and the lattice model captured by the first
  term of Eq.~\eqref{Hamiltonian1} revealed a quantitative agreement
  (see e.g.\ Fig.~5 in \cite{speck_specific_2010}) in
  the range $0 \le J \lesssim 1.2\,k_{\rm B}T$ that lies entirely
  within the rather
  stiff limit
  \cite{speck_specific_2010}. This is the range of $J$ we are
  interested in and includes the values relevant for cell adhesion
  (see Sec.~\ref{Implication} below).

The second term in Eq.~\eqref{Hamiltonian1} reflects that each closed bond stabilizes the adhesion cluster by
an amount $-\mu$. Aside from the last term $\mH_{\rm h}(\{\sigma_i\})$
the Hamiltonian \eqref{Hamiltonian1} is isomorphic to the lattice gas
model developed in \cite{speck_specific_2010}, and a mapping between the two models is provided in Appendix \ref{Appendix A}.

The third term in Eq.~\eqref{Hamiltonian1}, $\mH_{\rm
  h}(\{\sigma_i\})$, accounts for the mechanical force $h$ acting on the membrane-embedded bonds  
that we assume to be equally shared between all $N_c$ \emph{closed} bonds of
a given configuration $\{\sigma_i\}$, i.e.\ $N_c(\{\sigma_i\})\equiv\sum_{i}\delta_{\sigma_{i},-1}$,
where $\delta_{ik}$ is Kronecker's delta.
More precisely, the force $h$ destabilizes the
bound state by introducing an elastic (free) energy penalty on all
closed bonds whereby broken bonds remain unaffected. If all bonds are
closed, $N_c=N$, this penalty is set to be $hx_0$, where $x_0$ is
a microscopic length-scale specific for a given CAM that merely sets
the energy scale associated with the elastic strain caused by $h$. Conversely, the
penalty must vanish in a completely dissolved configuration with
$N_c=0$, and is assumed to be a smooth and monotonic function of
$N_c$. A mathematically and physically consistent definition is
\begin{equation}
\mH_{\rm h}(\{\sigma_i\})=-2hx_0\left(\frac{1}{1+N_c(\{\sigma_i\})/N}-1\right).
\label{force}  
\end{equation}  
A 'pulling force', $h>0$, favors the dissociation of bonds while a
'pushing force', $h<0$, favors their association. We
are interested in strain energies on the order of the thermal energy per bond,
i.e.\ $|h|x_0/N=\mathcal{O}(k_{\rm B}T)$.
Note that the assumption of an equally shared force in Eq.~\eqref{force} is valid if either of the following conditions is satisfied: the
anchoring membrane has a large combined elastic modulus (i.e.\ stiff membranes
or membrane/substrate pairs), individual bonds are flexible, the
bond-density is low, or the membrane is prestressed by the actin
cytoskeleton
\cite{erdmann_stability_2004,gao_probing_2011,Qian_2008}. In the limit of a rather stiff membrane  both, a spin
  representation with effective coupling $J$ and a uniform
  force load are valid approximations to describe cell adhesion under
  force over a broad range of physically relevant parameters, as we
  detail below. The implications of a non-uniform force load are
  addressed in detail in Sec.~\ref{Conc} and Appendix~\ref{Appendix E.1.b}.
\subsection{Kinetics}
The breaking/closure and lateral diffusion of adhesion bonds are assumed to evolve as a discrete
time Markov chain with mixed single-bond-flip Glauber dynamics
\cite{glauber_timedependent_1963} and two-bond-exchange Kawasaki
dynamics \cite{PhysRev.145.224} (see Fig.~\ref{fig 1}b). For a single
jump in the Markov chain we define the probability to attempt a
Glauber transition as $p_{k}\in[0,1]$ which controls the diffusion
rate, and for the sake of generality is allowed to depend on the
number of closed bonds $k$. Similarly, the probability to attempt a
Kawasaki transition is given by $1-p_{k}\in [0,1]$.
\textcolor{black}{We consider two distinct scenarios, one in which
      adhesion bonds are immobile due to being anchored to the
      underlying cytoskeleton (i.e.\ $p_{k}=1 \ \forall k$), and the
      other in which adhesion molecules are allowed to transiently
      diffuse (i.e.\ $0<p_{k}<1
      \ \forall k$; see e.g. \cite{DEMOND20083286}). Conversely,
      permanently associated/dissociated 
      freely diffusing bonds (i.e.\ $p_{k}=0 \ \forall k$)
      will not be considered, since these are not relevant.}
Further details about the respective transition rates are given below.

\textit{Glauber transitions}: Let $\{\sigma_j\}'_i$ denote the
bond configuration obtained by flipping bond $i$ while keeping the configuration of
all other bonds fixed,
i.e.\ $\{\sigma_{j}\}'_i\equiv(-\sigma_i,\{\sigma_{j\ne
  i}\})$. Moreover, let
$w_i(\{\sigma_{j}\})$ denote the transition
rate from $\{\sigma_{j}\}$ to $\{\sigma_{j}\}'_i$
and $\Delta \mathcal{H}_i(\{\sigma_{j}\})\equiv
\mathcal{H}(\{\sigma_{j}\}_i') -\mathcal{H}(\{\sigma_{j}\})$ the energy
difference associated with the transition. These rates can be specified uniquely by limiting interactions to nearest-neighbors, imposing
isotropy in position space, and requiring that $w_i$ satisfies detailed
balance, i.e.\
$w_i(\{\sigma_{j}\})/w_i(\{\sigma_{j}\}_i')=\exp{(-\beta\Delta \mathcal{H}_i(\{\sigma_{j}\}))}$,
where $\beta=1/k_{\mathrm{B}}T$ is the inverse thermal energy. The general result reads
$w_i(\{\sigma_{j}\})=\alpha[1-\tanh(\beta\Delta \mathcal{H}_i(\{\sigma_{j}\})/2)]/2N$, where $\alpha$ is an intrinsic attempt-frequency that sets the fastest
timescale \cite{glauber_timedependent_1963}, and time will throughout be expressed in units of $\alpha^{-1}$. Introducing furthermore
the dimensionless quantities $\tj=\beta J$,
$\tilde{\mu}=\beta\mu$ and $\thh=\beta hx_0/N$
this leads to
\begin{equation}
    w_i(\{\sigma_{j}\}) =\frac{\alpha}{2N}\bigg\{1-\sigma_{i}\tanh\!\Big[\tilde{J} \sum_{\langle ij \rangle}\!\sigma_{j}-\frac{\tilde{\mu}}{2}\!+\!\Lambda^{\tilde{h}}_{\{\sigma_{j}\},i}\Big]\bigg\},
    \label{Glauber Transition rate}
\end{equation}
where we defined the auxiliary function
\begin{equation}
 \Lambda^{\tilde{h}}_{\{\sigma_{j}\},i}\equiv
\frac{\tilde{h}}{(1+N_c(\{\sigma_{j}\})/N)(1+N_c(\{\sigma_{j}\}'_i)/N)}.
\label{Lambdas}
\end{equation}

\textit{Kawasaki transitions}: Let $\{\sigma_j\}'_{ik}$ denote the
bond configuration upon interchanging the state of the nearest neighbor bonds $\sigma_{i}$ and $\sigma_{k}$ while keeping the configuration of
all other bonds fixed,
i.e.\ $\{\sigma_{j}\}'_{ik}\equiv(\sigma_i\leftrightarrow\sigma_{k},\{\sigma_{j\ne
  (i,k)}\})$. We denote the Kawasaki transition rate from $\{\sigma_{j}\}$ to $\{\sigma_{j}\}'_{ik}$ as 
$w_{ik}(\{\sigma_{j}\})$,
where $\Delta \mathcal{H}_{ik}(\{\sigma_{j}\})\equiv
\mathcal{H}(\{\sigma_{j}\}_{ik}') -\mathcal{H}(\{\sigma_{j}\})$ is the energy
difference associated with the transition. Imposing the same symmetry constraints as for the Glauber rates as well as detailed-balance yields the general expression \cite{PhysRev.145.224}
\begin{equation}
    w_{ik}(\{\sigma_{j}\}) =\frac{\alpha}{2N}\bigg\{1-\frac{\sigma_{i}-\sigma_{k}}{2}\tanh\!\Big[\tilde{J}(\sum_{\langle ij \rangle}\sigma_{j}-\sum_{\langle kl \rangle}\sigma_{l})\Big]\bigg\},
    \label{Kawasaki Transition rate}
\end{equation}
where we have used that $(\sigma_{i}-\sigma_{k})/2 \in \{-1,0,1\}$. 
As pointed out in \cite{PhysRev.145.224}, the transition is only
meaningful when $\sigma_{k}=-\sigma_{i}$, otherwise the transition
brings the system to an identical state, which is equivalent to no
transition. Note that the Kawasaki rates given by Eq.~\eqref{Kawasaki
  Transition rate} do not depend on the external force $\thh$ nor the
binding-affinity $\tmu$, since the Kawasaki transition conserves the
total number of open and closed adhesion bonds. However, if in
addition we introduce a position-dependent force/binding affinity the
Kawasaki rates also depend on $\thh$ and $\tmu$, which we analyze in Appendix~\ref{Appendix E.1.b}.
\subsection{Strategy roadmap}
\begin{figure}
    \includegraphics[width = 0.48\textwidth]{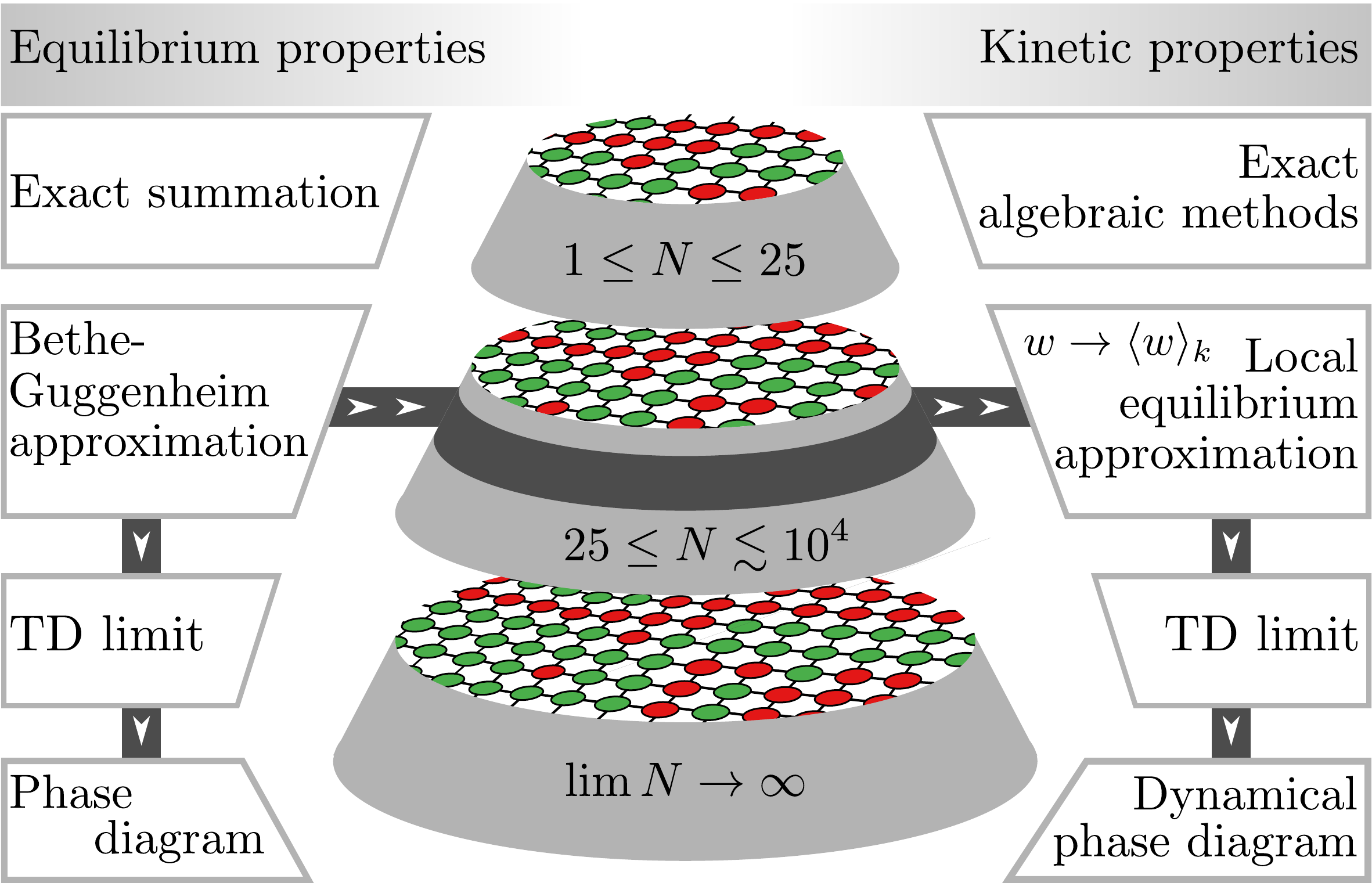}
    \caption{\textbf{Strategy roadmap.} Small system
      sizes $N\le 5\times 5$ are solved for exactly. The
      thermodynamics of larger systems
      is treated on the level of the highly accurate \emph{Bethe-Guggenheim}
      approximation and the kinetics by assuming \emph{local equilibrium}. Within the Bethe-Guggenheim approximation we take the thermodynamic (TD) limit
    $N\to\infty$ and determine the phase behavior, master scaling
    of dissolution/formation kinetics and analyze the statical and
    dynamical critical behavior.}
    \label{fig 2}
\end{figure}

We focus in detail on both, the equilibrium properties as well as the
kinetics of cluster formation and dissolution for all cluster
sizes. A roadmap to our extensive analysis is presented in
Fig.~\ref{fig 2}. 

For small to moderate cluster sizes, i.e.\ up to $50$ bonds for the
equilibrium properties and up to $25$ bonds in the case of formation/dissolution
kinetics, we obtain exact solutions using standard algebraic methods
\cite{iosifescu_finite_2014}. To circumvent the explosion of
combinatorial complexity for large system sizes we employ a
variational approach -- the so-called \emph{Bethe-Guggenheim}
approximation \cite{fowler_statistical_1939} --  to derive closed-form expressions for
the partition function, and finally carry out the thermodynamic limit to
derive explicit closed-form results for large adhesion
clusters. When considering the formation/dissolution
  kinetics of large clusters and in particular in the thermodynamic limit, we employ the \emph{local
equilibrium approximation}, where we assume that the growth and
dissolution evolves like a birth-death process on the free energy
landscape. 

We systematically test the accuracy of all approximations by comparing
them with exact results for system sizes that are amenable to
exact solutions. The results reveal a remarkable accuracy that improves further with the size of the system (e.g.\ see Fig.~\ref{fig A1}).  
\section{\label{Structure} Equilibrium behavior of adhesion clusters}
\subsection{Small and intermediate clusters}
In order to quantify the equilibrium stability of adhesion clusters we first analyze the equation of state for the average fraction of
closed bonds, $\langle \varphi\rangle\equiv\langle
N_{c}(\{\sigma_i\}) \rangle/N$ at given $\tmu,\tj$ and $\thh$. To this
end we require $Q_k$, the partition function constrained to the number of
closed bonds $N_c(\{\sigma_i\})=k$. We therefore write the
total canonical partition function $Q$ for a
system of $N$ adhesion bonds as $Q\equiv\sum_{\{\sigma_i\}}\e{-\beta\mH(\sigall)}\equiv\sum_{k=0}^NQ_k$, 
where
\begin{equation}
    Q_k\equiv\sum_{\{\sigma_i\}}\e{-\beta
    \mH(\sigall)}\delta_{N_c(\sigall),k}=\e{[\tmu+2\thh(k/N+1)^{-1}]k}Z_k,
    \label{QK}
\end{equation} 
and $Z_k\equiv\sum_{\{\sigma_i\}}\exp{(\tj\sum_{\langle
ij\rangle}\sigma_i\sigma_j)}\delta_{N_c(\sigall),k}$ is the partition function of the Ising model at
zero field conditioned to have a
magnetization $N/2-k$. The free energy density (per bond) in units of
thermal energy $k_{\rm B}T$ constrained to a given
fraction of closed bonds
$\varphi$, $\tf_N(\varphi)$, 
and the equation of state, $\langle\vphi(\tmu,\tj,\thh)\rangle$, are given by
\begin{equation}
\tf_N(\varphi)=-N^{-1}\ln Q_k,\quad \langle \varphi\rangle=N^{-1}\partial_{\tmu}\ln Q.
\label{ES_ex}  
\end{equation}
We note that $\e{-N\tf_N(\varphi)}/Q={\rm Prob}(N_c=N\varphi)$ in an
equilibrium ensemble of $N$ bonds.
The sum over constrained configurations in $Z_k$ contains
$\binom{N}{k}$ terms. Whereas it can be performed exactly for
$N\lesssim 50$ it explodes for larger system sizes. To
overcome the computational complexity we employ a variational approach --
the \emph{Bethe-Guggenheim} approximation
\cite{fowler_statistical_1939}, yielding
(see derivation in Appendix~\ref{Appendix B.1})
\begin{equation}
Z_k\approx\Zqc_k=\binom{N}{k}\frac{\psi_{\bz N}^{\bz
    k}(\bz X_k^*)}{\psi_{\bz N}^{\bz k}(\bz\ovX_k)}\e{-\bz\tj (2\ovX_k-N/2)} , 
\label{zqc}  
\end{equation}
where $\bz=\sum_{i=1}^Nz_i/N$ is the average coordination number in a
cluster with local coordination $z_i$ that accounts for finite-size
effects, $X_k^*\equiv k(N-k)/N$, we have defined 
\begin{equation}
\bar{X}_k \equiv\frac{2X_k^*}{[1+4X_k^*(\e{4\tj}-1)/N]^{1/2}+1},
\label{xx}  
\end{equation}  
and introduced the auxiliary function
\begin{equation}
\psi_a^b(x){\equiv}\Gamma([b-x]/2+1)\Gamma^{2}(x/2+1)\Gamma([a-b-x]/2+1),
\label{aux_QC}
\end{equation}
where $\Gamma(z)$ stands for the Gamma function. Note that by setting $\bar{X}_k=X_k^*$ in Eq.~\eqref{zqc} we recover the mean field result $Z_k^{\rm MF}$ (which happens automatically for $\tj=0$ or $k=0 \lor N$) which is discussed in Appendix~\ref{Appendix C.1}. 

Fig.~\ref{fig 3}(a-c) shows a
comparison of the
free energy density $\tf_N(\vphi)$ for a cluster of $40$ bonds for
various affinities $\tmu$ and external forces $\thh$, and confirms
the high accuracy of the Bethe-Guggenheim approximation on the one hand, and the
systematic failure of the mean field result on the other hand. This signifies that
correlations between adhesion bonds decisively affect cluster properties. Moreover, pairwise correlations captured by the Bethe-Guggenheim
approach are apparently dominant, whereas three-body and higher order correlations that
were ignored are apparently insignificant. 
\begin{figure*}[ht!]
    \includegraphics[width = 0.9\textwidth]{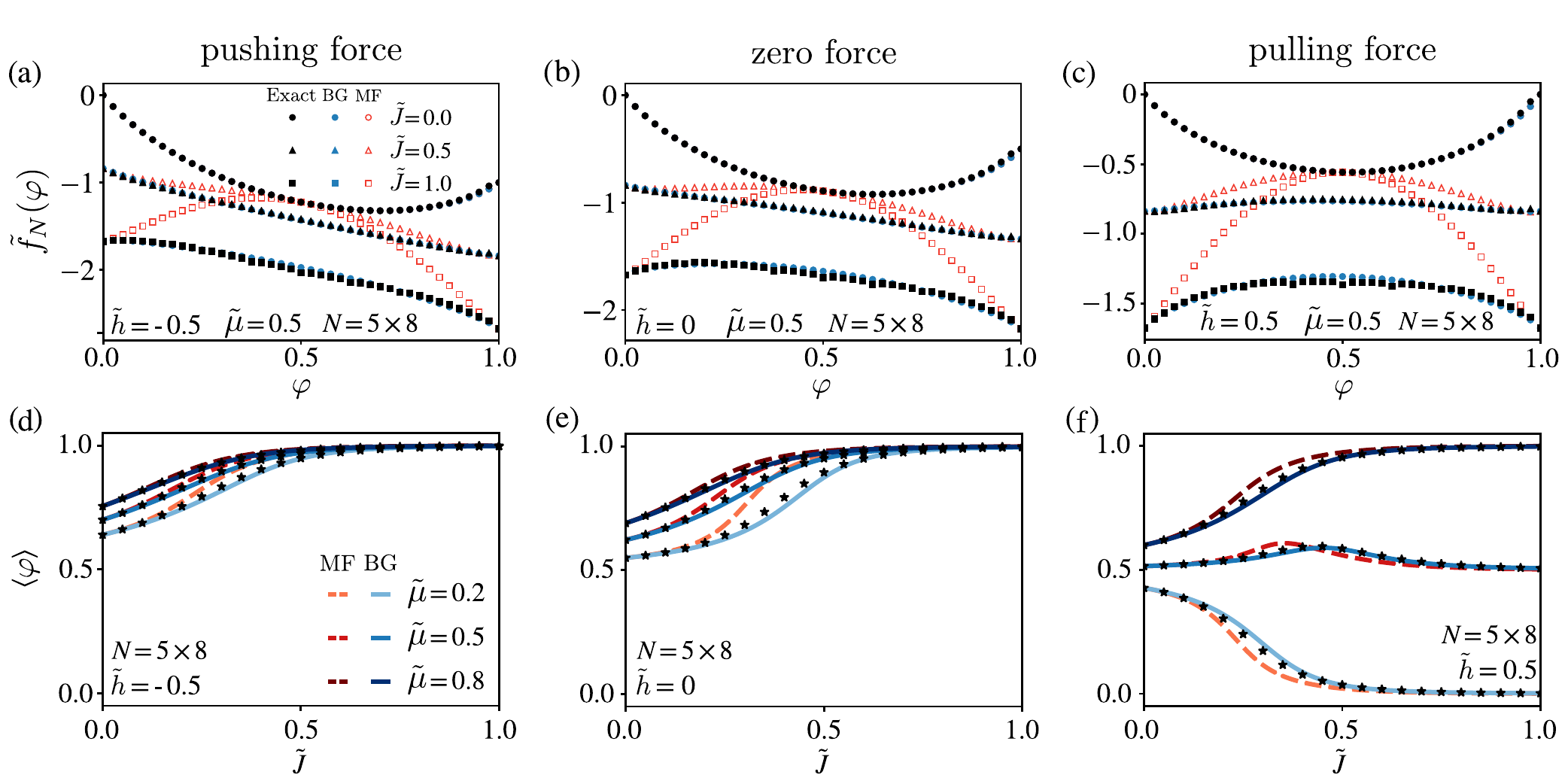}
    \caption{\textbf{Free energy landscape and equation of state for
        small clusters.} (a-c) Free energy density conditioned on $\varphi$,
      $\tf_N(\varphi)$ from Eq.~\eqref{ES_ex} for a system of $N=5\times
      8$ bonds on a square lattice for (a) a pushing
    force $\thh=-0.5$, (e) no force $\thh=0$, and (f) pulling force $\thh=0.5$;
    Black symbols depict exact results, blue symbols the
    Bethe-Guggenheim approximation, and red symbols the mean field
    result. (d-f) Equation of state, $\langle \varphi\rangle$, for a cluster
      of $5\times 8$ adhesion bonds on a square lattice as a function
      of the dimensionless coupling $\tj$ for (d) a pushing
    force $\thh=-0.5$, (e) no force and (f) a pulling force
    $\thh=0.5$. Symbols depict exact results, blue lines correspond to
    the Bethe-Guggenheim approximation and red to the mean field
    result.}
    \label{fig 3}
\end{figure*}

Similarly, in Fig.~\ref{fig 3}(d-f) we depict the equation of state for a cluster of 40 bonds. The Bethe-Guggenheim approximation (blue lines) is very accurate for
all values of $\tj$ whereas the mean field approximation (red lines) fails for intermediate values of the coupling. We observe striking differences in the dependence of $\langle
\varphi\rangle$ on the coupling $\tj$ (and hence membrane rigidity)
with respect to the intrinsic binding-affinity $\tmu$ in the presence of
a pulling force (see Fig.~\ref{fig 3}f).  At strong coupling
between adhesion bonds $\langle
\varphi\rangle$ depends strongly on $\tmu$. In the presence of a
pulling force adhesion bonds with a weak
affinity are on average all broken, whereas they are all closed if
the affinity is large. Notably, the dependence of $\langle \varphi \rangle$ on the coupling $\tilde{J}$ at zero force (see Fig.~\ref{fig 3}e) agrees qualitatively well with experimental observations \cite{huppa_tcrpeptidemhc_2010,
  huang_kinetics_2010,williams_quantifying_2001,huppa_tcrpeptidemhc_2010}
and hints at some form of critical behavior underneath, which we
discuss in more detail in Sec.~\ref{Implication}.
\subsection{Thermodynamic limit}
To explore the phase diagram in detail and analyze the
critical behavior we consider the thermodynamic limit of the
Bethe-Guggenheim (BG) and mean field (MF) free energy density, i.e.\ the scaling limit
\begin{equation}
    \rtf^{\rm BG, MF}(\varphi)\equiv \lim_{\substack{N\to\infty,\\k/N=\varphi={\rm
    const.}}}\tf^{\rm BG, MF}_N(\varphi),
    \label{TDlim}
\end{equation}
which exists and is given by 
\begin{widetext}
    \begin{equation}
        \rtf^{\rm BG}(\varphi)=-\tmu\varphi+2\thh\frac{\varphi}{1+\varphi}+\frac{1}{2}\tj\bz[4\Omega_\varphi-1]+\frac{\bz}{2}[\Xi(\varphi-\Omega_\varphi)+\Xi(1-\varphi-\Omega_\varphi)+2\Xi(\Omega_\varphi)] +(1-\bz)[\Xi(\varphi)+\Xi(1-\varphi)],
        \label{TD}
    \end{equation}
\end{widetext}
where $X_{\varphi}^*/N=\varphi(1-\varphi)$, $\Omega_\varphi\equiv\bar{X}_\vphi/N$, and we have introduced the auxiliary function $\Xi(x)\equiv x\ln x$. The result for $\rtf^{\rm MF}(\varphi)$ is given in Appendix~\ref{Appendix C.1}. Somewhat surprisingly the free energy density of a finite system, $\tf_N^{\rm BG}(\varphi)$, converges to the thermodynamic limit $\rtf^{\rm BG}(\varphi)$ already for $N\gtrsim 100$. For convenience we henceforth drop the superscript BG when considering the Bethe-Guggenheim result, i.e.\ $\rtf^{\rm BG}(\varphi)\to \rtf(\varphi)$. 
  
The equation of state in the thermodynamic limit is determined by means of the saddle-point method (for derivation see Appendix~\ref{Appendix D}), yielding a weighted sum over $\varphi^0_i$, the $M$ \emph{global}
minima of $\rtf(\varphi)$:
\begin{equation}
    \langle\varphi\rangle_{\rm TD}=\lim_{N\to\infty}N^{-1}\partial_{\tmu}\ln
    Q^{\rm BG}\simeq \sum_{i=1}^Mc_i\varphi^0_i,
    \label{TEos}
\end{equation}
where $\rtf(\varphi_i^0)=\rtf_{\rm min},\forall i$, and $\simeq$ stands for asymptotic equality in the thermodynamic limit. In practice $M$ is either 1 (unique minimum) or 2 (two-fold degenerate minima). The minima have the universal form $\varphi^0_m=\xi_{\tmu,\tj,\thh}^{4}/(1+\xi_{\tmu,\tj,\thh}^{4})$ with the coefficients $\xi_{\tmu,\tj,\thh}$ and weights $c_i$ given explicitly in Appendix~\ref{Appendix D}.
\begin{figure*}[ht!]
    \includegraphics[width = 0.9\textwidth]{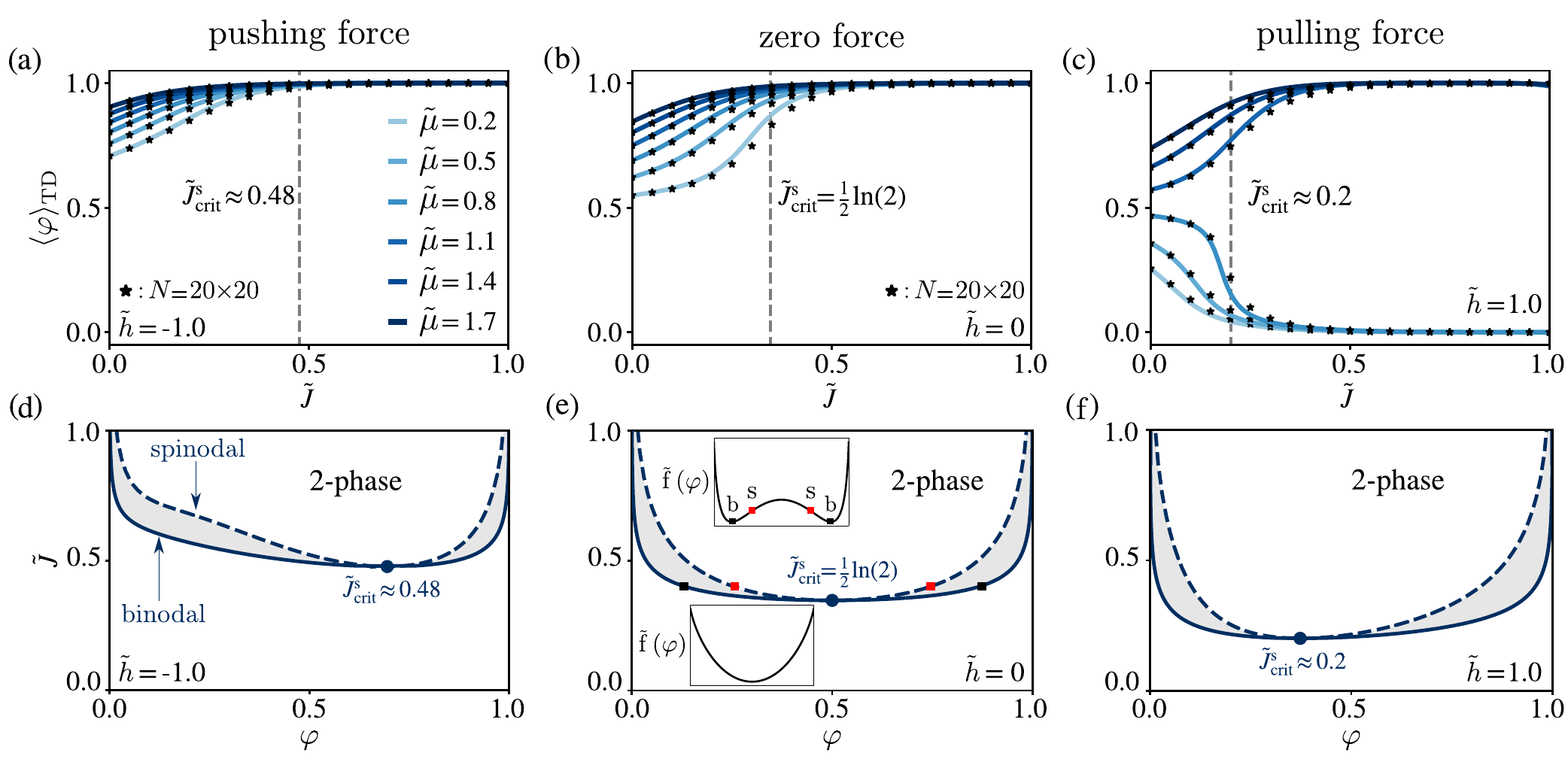}
    \caption{\textbf{Equation of state  and
        phase diagram in the thermodynamic limit.} (a-c) $\langle \varphi\rangle$ for a cluster
      of $20\times 20$ adhesion bonds on a square lattice (symbols)
      and the saddle-point asymptotic $\langle\varphi\rangle_{\rm TD}$
      from Eq.~\eqref{TEos} as a function
      of the dimensionless coupling $\tj$ for various affinities
      $\tmu$ and for (a) a pushing
    force $\thh=-1$, (b) no force, and (c) a pulling force
    $\thh=1$. The dashed vertical line denotes the (statical) \emph{critical}
    coupling strength $\tj^s_{\rm crit}$ whereupon the system
    phase-separates into dense and dilute phases of closed
    bonds. (d-e) Phase diagram for (d) a pushing
    force $\thh=-1$, (e) no force, and (f) a pulling force
    $\thh=1$; the full and dashed lines depict the binodal and
    spinodal line, respectively. The shaded area depicts the region where the
    system is metastable. The blue circle depicts the (statical) critical
    point $(\vphi^s_{\rm crit},\tj^s_{\rm crit})$. Inset in (e):
    Schematic of the free energy landscape $\rtf(\vphi)$ below the
    critical coupling $\tj<\tj^s_{\rm crit}$ (bottom) displaying a single minimum,
    and a bi-stable free energy landscape above the
    critical coupling $\tj>\tj^s_{\rm crit}$ (top), with the black and red
    symbols illustrating the meaning of phase compositions highlighted
    in the phase diagram.}
    \label{fig 4}
\end{figure*}
The equation of state $\langle\varphi\rangle$ for a finite cluster seems  to converge to the saddle-point asymptotic
$\langle\varphi\rangle_{\rm TD}$ already for $N\gtrsim 400$ for any value of the force $\thh$, bond affinity $\tmu$, and coupling $\tj$ (see Fig.~\ref{fig 4}(a-c)), and is qualitatively the same as for smaller clusters (compare Fig.~\ref{fig 4}(a-c) with Fig.~\ref{fig 3}(d-f)). However,
important differences emerge in the thermodynamic limit -- the system may undergo a phase transition and phase-separate into dense (``liquid'') and dilute (``gas'') phases of closed bonds with composition $\varphi_l$ and $\varphi_g$, respectively (see also \cite{schmidt_coexistence_2012}).
\subsection{Phase diagram and critical behavior}\label{Phase diagram and critical behavior}
To determine the phase diagram we require the binodal $\tj_b(\varphi)$ and spinodal $\tj_s(\varphi)$ line. The binodal line $\tj_b(\varphi)$ denotes the
\emph{onset} of phase separation and is determined by the ``common tangent'' construction, i.e.\ from the solution of the coupled equations
\begin{equation}
    \rtf'(\varphi_l)=\rtf'(\varphi_g),\quad
    \frac{\rtf(\varphi_l)-\rtf(\varphi_g)}{\varphi_l-\varphi_g}=\rtf'(\varphi_l),
    \label{binodal}
\end{equation}
where the prime denotes the derivative with respect to $\varphi$ at constant $\tj$. The spinodal line $\tj_s(\varphi)$, also known as the stability boundary, denotes the boundary between the metastable and unstable regimes and is determined by $\rtf''(\varphi)=0$. For a non-zero force, $\thh\ne 0$, we determine
$\tj_b(\varphi)$ numerically, whereas we obtain an exact result for a vanishing force $\thh=0$ that reads (see derivation in Appendix~\ref{Appendix B.2})
\begin{equation}
    \tj_b(\varphi,\thh)|_{\thh=0}=\frac{1}{2}\ln\left(\frac{1-\chi_\varphi}{\chi_\varphi^{1/\bz}-\chi_\varphi^{1-1/\bz}}\right),
    \label{bin}
\end{equation}
where we have introduced $\chi_\varphi=\varphi/(1-\varphi)$. The spinodal line for any force $\thh$ is in turn given exactly by
\begin{equation}
    \tj_s(\varphi,\thh)=\frac{1}{4}\ln\left\{\frac{[\vphi-\Phi(\vphi,\thh)][1-\vphi-\Phi(\vphi,\thh)]}{\Phi(\vphi,\thh)^2}\right\},
    \label{spin}
\end{equation}
with the auxiliary function
\begin{equation}
    \Phi(\vphi,\thh)\equiv2\vphi(1-\vphi)+\bz\left[\frac{1-\bz}{\vphi(1-\vphi)}-\frac{4\thh}{(1+\vphi)^3}\right]^{-1},
    \label{spinA}
\end{equation}
that is defined for $(2-\bz)/8\thh\le\vphi(1-\vphi)/(1+\vphi)^3\le(1-\bz)/4\thh$. Note that it follows from their respective definitions that neither $\tj_b(\varphi,\thh)$ nor $\tj_s(\varphi,\thh)$ depends on $\tmu$ (for a proof
see Appendix~\ref{Appendix B.2}). The phase diagram for a pushing, zero, and pulling force $\thh$ is shown in
Fig.~\ref{fig 4}(d-f) and displays, above the critical coupling strength
$\tj>\tj^s_{\rm crit}$, a phase separation into a
dense and dilute phase of closed bonds with compositions $\vphi_l$ and $\vphi_g$, respectively. A pushing force $\thh<0$ lifts the critical coupling and ``tilts'' the phase
diagram towards higher density, i.e.\ at a given coupling
$\tj>\tj^s_{\rm crit}$ the density of both phases
increases. Conversely, a pulling force $\thh>0$ lowers the critical coupling  and ``tilts'' the phase diagram towards lower density, i.e.\ at a given coupling $\tj>\tj^s_{\rm crit}$ the density of both phases decreases. The biological implications of these results will be discussed in Sec.~\ref{Implication}. The binodal and spinodal line in the mean field approximation are given in Appendix~\ref{Appendix C.2}.

\begin{figure}
    \includegraphics[width = 0.40\textwidth]{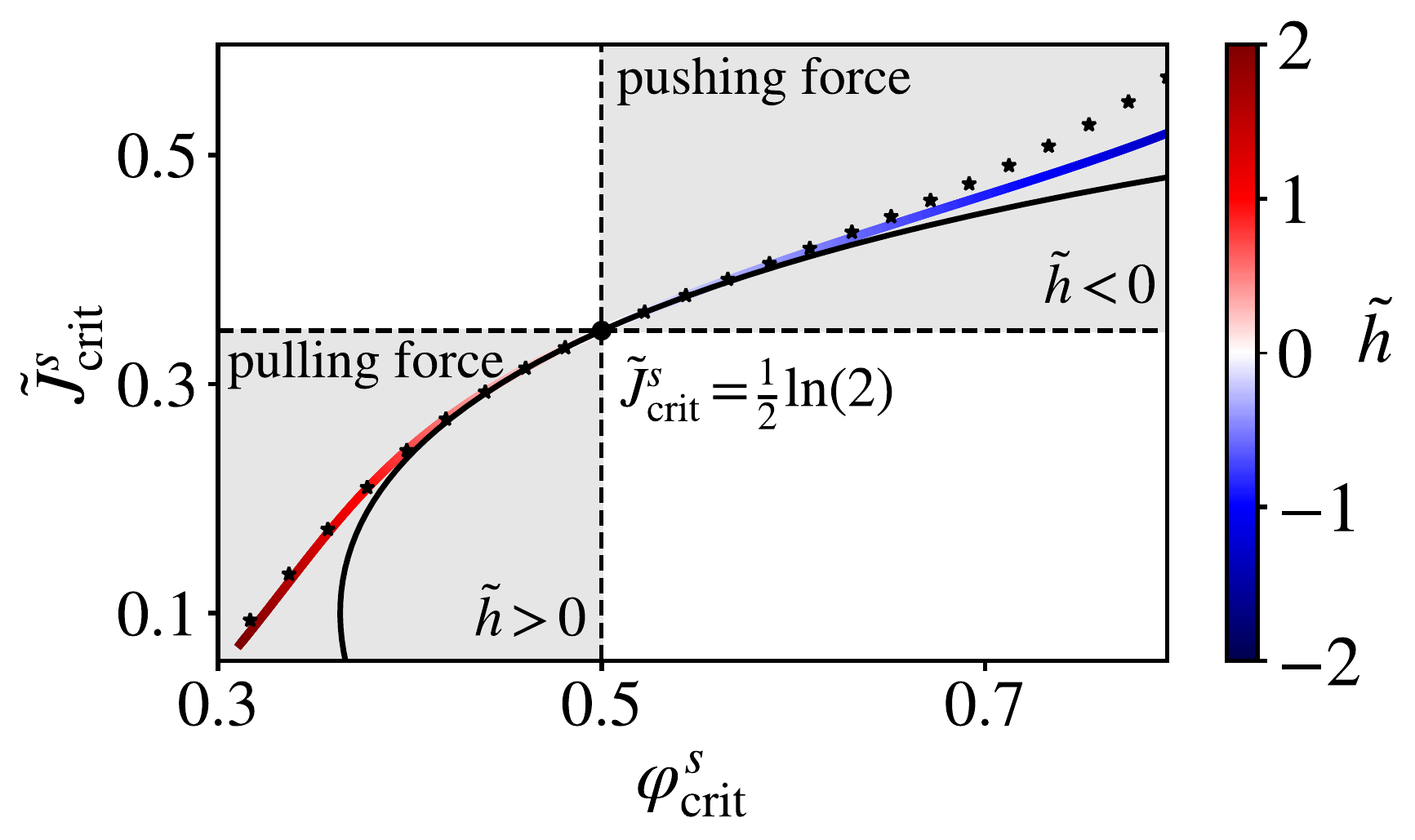}
    \caption{\textbf{Statical critical point} as a function of the force
    $\thh$. Symbols depict the exact solution using a converged
      Newton's series  and the gradient line depicts the two-term
      (so-called quadratic) approximation of the complete Newton's
      series, which is very accurate for any pulling- and up to a
      moderate pushing force, i.e.\ $\thh\ge -1$. Explicit expressions
      are given in Appendix~\ref{Appendix E}. The black line corresponds to the
      prediction of second order perturbation theory from
      Eqs.~\eqref{smallforce} and \eqref{smallforce1} that is valid for
      small forces.}
    \label{fig 5}
\end{figure}

We now address in detail the behavior of the statical critical point $(\varphi^s_{\rm crit},\tj^s_{\rm crit})$ -- the point where the binodal and spinodal merge, $\tj_b(\varphi^s_{\rm crit},\thh)=\tj_s(\varphi^s_{\rm crit},\thh)\equiv \tj^s_{\rm crit}(\varphi^s_{\rm crit},\thh)$. The critical point denotes the onset of phase separation and is the solution of $\rtf'''(\vphi)=0$,
which in absence of the force yields (for derivation see Appendix~\ref{Appendix B.2})
$(\varphi^{s,0}_{\rm crit},\tj^{s,0}_{\rm crit})\equiv\frac{1}{2}(1,\ln\frac{\bz}{\bz-2})$. In the presence of a force $\thh\ne 0$ we obtain the exact solution using a Newton's series approach \cite{Godec_2016,Hartich_2018,Hartich_2019} (for details regarding the Newton series, see 
Appendix~\ref{Appendix D.2}). The analytical result is non-trivial and is given explicitly in Appendix~\ref{Appendix B.2}. For small forces $|\thh|\ll
1$ we in addition derive a second order perturbation expansion $\tj^s_{\rm crit}= \tj^{s,0}_{\rm crit}-\delta \tj^s_{\rm crit}(\thh)+\mathcal{O}(\thh^3)$, where
\begin{equation}
    \delta \tj^s_{\rm crit}(\thh)= \frac{8}{27}\frac{1}{\bar{z}-2}\left(\thh+\frac{2}{27}\frac{\bar{z}+2}{\bar{z}-1}\thh^2\right),
    \label{smallforce} 
\end{equation}
and correspondingly $ \vphi^s_{\rm crit}=\vphi^{s,0}_{\rm crit}-\delta\vphi_{\rm crit}(\thh)+\mathcal{O}(\thh^3)$ with
\begin{equation}
    \delta\vphi_{\rm crit}(\thh)=\frac{2}{3}\frac{(\bar{z}/3)^2}{(\bar{z}-2)(\bar{z}-1)}\left[\thh+\frac{16}{9}\frac{(\bar{z}/3)^2-\bar{z}+1}{(\bar{z}-2)(\bar{z}-1)}\thh^2\right].
    \label{smallforce1}  
\end{equation}
The dependence of the
statical critical point on the external force is depicted in
Fig.~\ref{fig 5}. A pulling force (red) pulls the critical point towards lower $\tj$ and lower $\vphi$, whereas a pushing force (blue) effects the opposite and shifts the critical point towards larger coupling $\tj$ and higher density $\vphi$. The mean field statical critical point can be derived exactly as a function of the force $\thh$, and the result is given in Appendix~\ref{Appendix C.2}.
\section{\label{Kinetics} Kinetics of cluster formation and dissolution}
\subsection{Small and intermediate clusters}
We are interested in the kinetics of cluster formation from a completely unbound state, and cluster dissolution from a completely bound state. More general initial conditions are treated in Appendix~\ref{Appendix E}. We quantify the kinetics by means of the mean first passage time $\langle \tau_{d,f}\rangle$, where the subscripts $d$ and $f$ stand for dissolution and formation, respectively, and $\tau_{d,f}$ is the first passage time defined as
\begin{eqnarray}
    \tau_d&\equiv&
    \inf_{t}\left[\varphi(\{\sigma_i\}_t)=0|\varphi(\{\sigma_i\}_0)=1\right],\nonumber
    \\
    \tau_f&\equiv&
    \inf_{t}\left [\varphi(\{\sigma_i\}_t)=1|\varphi(\{\sigma_i\}_0)=0\right],
    \label{FPT}
\end{eqnarray}
where $\sigall_t$ denotes the instantaneous state at time $t$. A cluster with $N$ adhesion bonds has $2^N$ possible states $\sigall$. We enumerate them such that the first state corresponds to all bonds closed and the final state to all bonds broken. The transition matrix of the Markov chain describing mixed Glauber-Kawasaki dynamics on this state-space has dimension $2^N\times 2^N$, whereby we must impose absorbing boundary conditions on the fully dissolved and fully bound states, respectively. An exact algebraic result for $\langle \tau_{d,f}\rangle$ is given in Eq.~\eqref{MFPT} in Appendix~\ref{Appendix E.1} but requires the inversion of a $(2^N-1)\times(2^N-1)$ sparse matrix, followed by a sum over $2^N-1$ terms, which is feasible only for $N\lesssim 5\times 5$.

\begin{figure}
    \includegraphics[width = 0.41\textwidth]{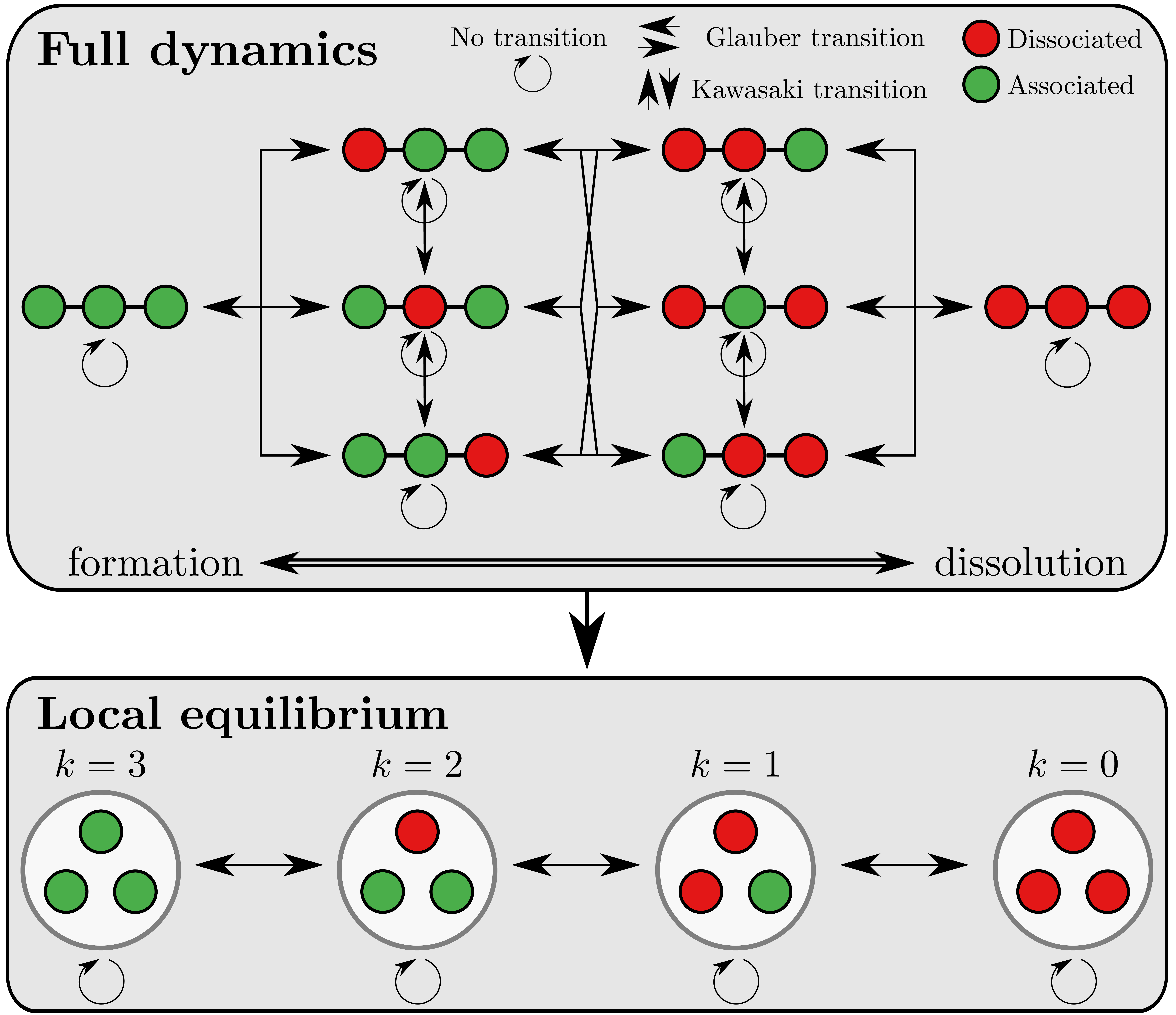}
    \caption{\textbf{Mapping the full dynamics onto a birth-death process.} For convenience, and without any loss of generality, we here show an example of a system composed of 3 adhesion bonds on a 1-dimensional lattice. The mapping holds for any lattice geometry. In the full dynamics each lattice configuration represents a different node, comprising a $2^{N}\times2^{N}$ transition matrix, whereas in the local equilibrium approximation we only need to distinguish between states with a different number of closed/open bonds, comprising a $(N+1)\times(N+1)$ transition matrix.}
    \label{fig 6}
\end{figure}

\begin{figure}[t!]
    \includegraphics[width = 0.35\textwidth]{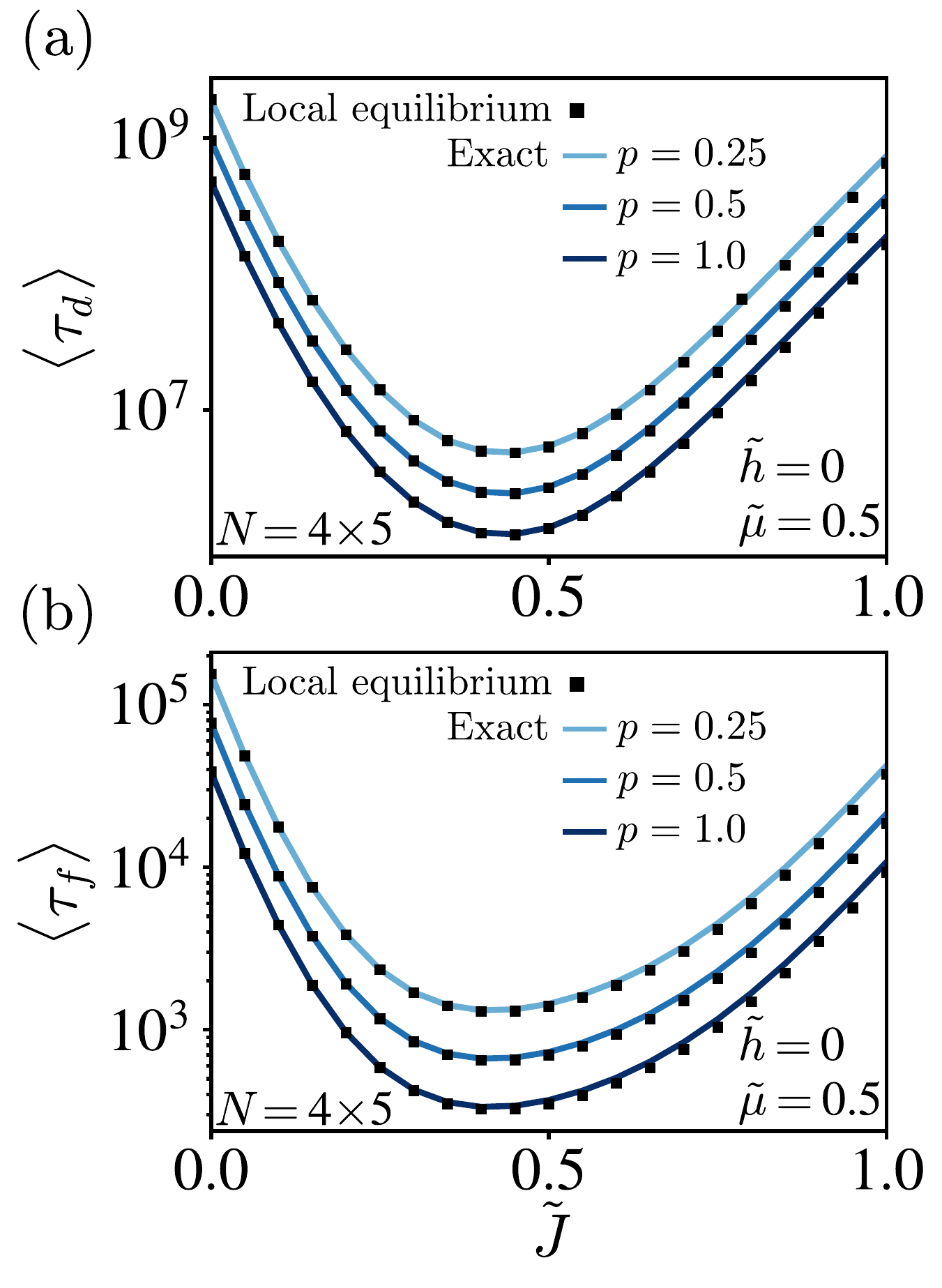}
    \caption{\textbf{Kinetics of dissolution and formation of small clusters.} Mean first passage time for cluster dissolution (a) and formation (b) as a function of the coupling $\tj$ for $N=4\times5$ adhesion bonds with intrinsic affinity $\tmu = 0.5$ in the absence of a force (for nonzero force values see Fig~\ref{fig A6}). Colored lines correspond to exact results obtained from Eq.~\eqref{MFPT} for various values of the Glauber attempt probability $p$, which we set to be constant $p_{k}\rightarrow p$, and symbols denote the local equilibrium approximation Eq.~\eqref{MFPT_le} evaluated with the exact $Q_{k}$ and $\bar{w}_{k\to k\pm1}$ from Eqs.~\eqref{QK} and \eqref{loceq} respectively.}
    \label{fig 7}
\end{figure}

As a result of the non-systematic cluster formation and dissolution at zero coupling $\tj = 0$, and motivated by the intuitive idea that the dynamics is
dominated by low energy (i.e.\ minimum action)  paths at large coupling $\tj\gg 1$, we make the so-called \emph{local equilibrium approximation} to treat large clusters. Thereby we map the dynamics of the $2^N \times 2^N$ state-space onto a one-dimensional birth-death process for the instantaneous number of closed bonds $k$ (see Fig.~\ref{fig 6}) with effective transition rates
\begin{equation}
    \bar{w}_{k\to k\pm1}\equiv \tilde{Q}^{-1}_{k}\sum_{\{\sigma_i\}}\e{-\beta
    \mH(\sigall)}w^{\pm}_{\rm exit}(\sigall)\delta_{N_c(\sigall),k},
    \label{loceq}  
\end{equation}
where we have defined the re-weighted canonical partition function $\tilde{Q}_{k}\equiv Q_{k}/p_{k}$ where $p_{k}$ is the Glauber attempt probability in state $k$, and we have introduced the exit rates from configuration $\sigall$ in the ``$+$'' (i.e.\ $N_c(\sigall'_j)=N_c(\sigall)+1$) and
``$-$'' (i.e.\ $N_c(\sigall'_j)=N_c(\sigall)-1$) direction, respectively, given by
\begin{equation}
    w^{\pm}_{\rm exit}(\sigall)\equiv \sum_{j=1}^Nw_j(\{\sigma_{i}\})\delta_{N_c(\sigall'_j),k\pm1}.
    \label{exit}  
\end{equation}
Note that only the Glauber transitions, given by Eq.~\eqref{Glauber Transition rate}, enter in Eq.~\eqref{exit}. The Kawasaki transitions given by Eq.~\eqref{Kawasaki Transition rate}, which conserve the total number of closed bonds, enter the dynamics through the diagonal of the transition matrix as the waiting rates $\bar{w}_{k\rightarrow k}=1-\bar{w}_{k\rightarrow k+1}-\bar{w}_{k \rightarrow k-1}$, where the right hand side follows from conservation of probability.
Within the local equilibrium approximation the mean first passage time for cluster dissolution and formation become, respectively
\begin{eqnarray}
    \langle \tau_d\rangle\approx \langle \tau_d^{\rm le}\rangle &=& \sum_{k=0}^{N-1}\frac{1}{\bar{w}_{k\to k+1}}\sum_{l=k+1}^{N}\frac{\tilde{Q}_l}{\tilde{Q}_k}\nonumber\\
    \langle \tau_f\rangle\approx \langle \tau_f^{\rm le}\rangle &=&
    \sum_{k=1}^{N}\frac{1}{\bar{w}_{k\to
    k-1}}\sum_{l=0}^{k-1}\frac{\tilde{Q}_l}{\tilde{Q}_{k}},
    \label{MFPT_le}  
\end{eqnarray}
where one can further use the detailed balance relation
$\tilde{Q}_{k}\bar{w}_{k\to k-1}=\tilde{Q}_{k-1}\bar{w}_{k-1\to k}$ (which we prove in Appendix \ref{Appendix E.2}) to interchange the backward and forward rate in the second line and change the summation according to $\sum_{k=1}^{N}\bar{w}_{k\to k-1}^{-1}\sum_{l=0}^{k-1}\tilde{Q}_l/\tilde{Q}_{k}\to \sum_{k=0}^{N-1}\bar{w}_{k\to
k+1}^{-1}\sum_{l=0}^{k}\tilde{Q}_l/\tilde{Q}_{k}$. In Appendix~\ref{Appendix E.3} we prove that Eq.~\eqref{MFPT_le} holds for any birth-death process where the transition rates obey detailed balance.
A comparison of the exact result given by Eq.~\eqref{MFPT} with the local equilibrium approximation in Eq.~\eqref{MFPT_le} shown in Fig.~\ref{fig 7} demonstrates the remarkable accuracy of the approximation already for $N\sim 20$ bonds, which increases further for larger $N$. The reason for the high accuracy can be found in the large entropic barrier to align bonds in an unbound/bound state, effecting a local
equilibration prior to complete formation/dissolution. Moreover, the local equilibrium approximation is expected to become asymptotically exact even for small clusters in the ideal, non-interacting limit $\tj\to0$ as well as for
$\tj\to\infty$ that is dominated by the minimum-action, ``instanton'' path. A further discussion of the local equilibrium approximation and an approximate closed form expression for Eq.~\eqref{MFPT_le} for larger systems is given in Appendices~\ref{Appendix E.4} and \ref{Appendix E.5}.

The mean first passage times for cluster dissolution/formation
shown in Fig.~\ref{fig 7} both display a strong
and non-monotonic dependence on the coupling parameter
$\tj$ with a pronounced minimum, hinting at some form of critical dynamics. As we prove below this minimum in the thermodynamic limit indeed corresponds to a \emph{dynamical critical coupling}.
\begin{figure*}[ht!]
    \includegraphics[width = 0.9\textwidth]{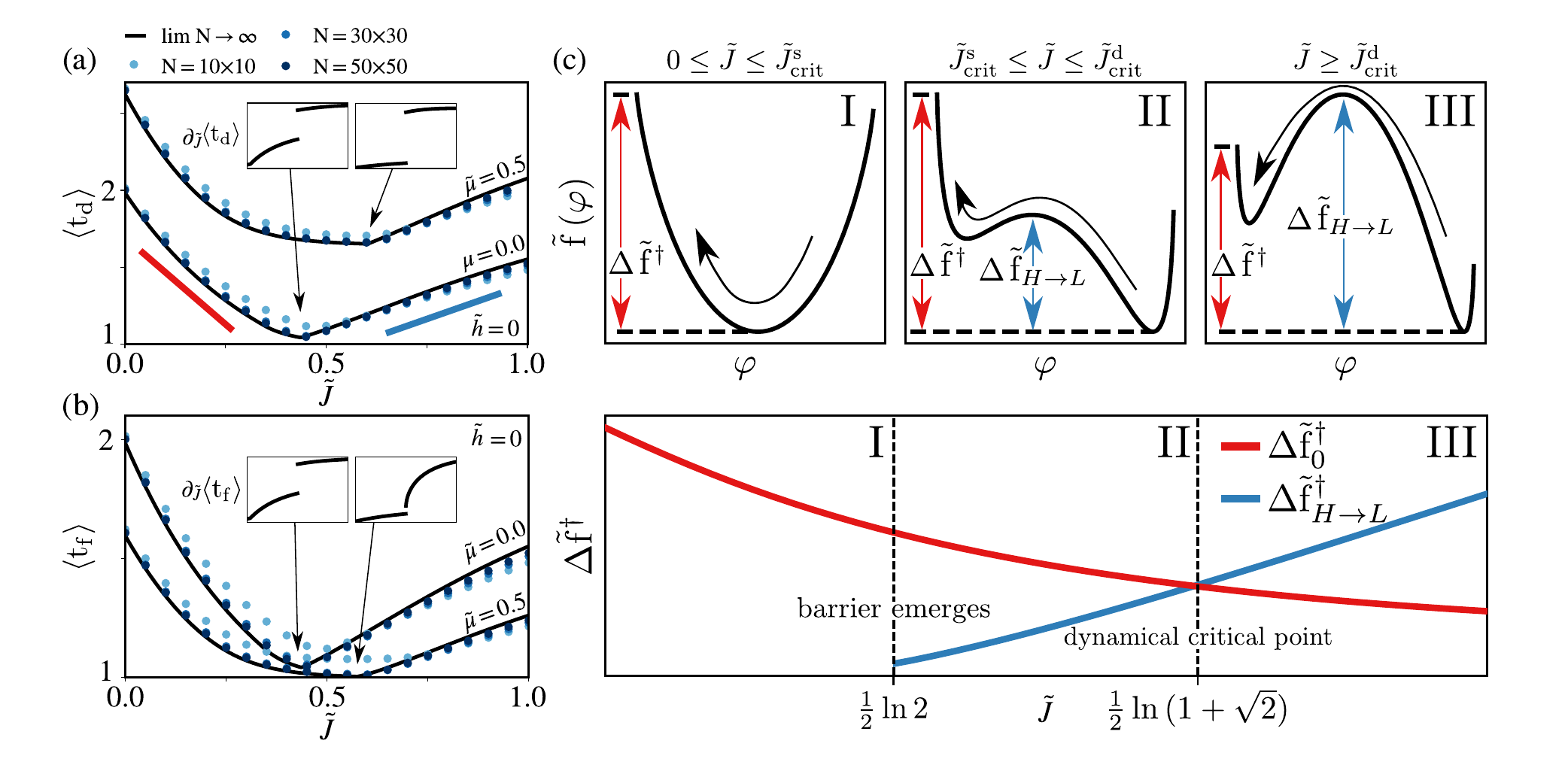}
    \caption{\textbf{Master scaling of mean dissolution and formation times per bond for finite clusters and in the thermodynamic limit, and the origin of the dynamical critical point}. $\langle {\rm
  t_{d,f}}\rangle$ for cluster dissolution (a) and formation (b) as a function of the
      coupling $\tj$ for a pair of intrinsic affinities $\tmu=0$ and
      $\tmu=0.5$ and various cluster sizes (symbols) as well as the
     thermodynamic limit (lines)
      in the absence of an external force; Symbols are evaluated with local equilibrium approximation Eqs.~\eqref{MFPT_le}
      using $Q_k^{\rm BG}$ (Eqs.~\eqref{QK} and \eqref{zqc}) and $\bar{w}_{k\to
  k+1}$ from Eq.~\eqref{loceq}. The discrepancy between the lines and
    symbols is due to finite-size effects. (c) In the thermodynamic limit and more generally for large clusters
      the mean dissolution/formation time $\langle{\rm t _{d,f}}\rangle$
      depends only on the \emph{largest} free energy barrier (see
      Eq.~\eqref{MDT}). For small coupling (regime I) the latter corresponds to
      the difference between the free energy minimum and the fully
      dissolved or bound configuration, $\Delta \rtf^\dagger=\Delta \rtf^\dagger_{0,1}$, respectively.  
      At the statical critical coupling value, $J^s_{\rm crit}$,
      (onset of regime II) a free energy barrier emerges
      separating the meta-stable from the stable phase, $\Delta \rtf^\dagger_{H \rightleftarrows L}$, but the
      largest free energy barrier is still $\Delta
      \rtf^\dagger=\Delta \rtf^\dagger_{0,1}$. At the \emph{dynamical
      critical coupling}, $\tj^d_{\rm crit}$, (onset of regime III) the free energy barrier
      separating the meta-stable from the stable phase becomes
      dominant, $\Delta \rtf^\dagger=\Delta \rtf^\dagger_{H
        \rightleftarrows L}$. The depicted free energy landscapes $\rtf(\vphi)$ correspond
    to Eq.~\eqref{TD} with $\tmu=0.05$ and $\thh=0$.}
    \label{fig 8}
\end{figure*}
\subsection{Thermodynamic limit}\label{Sec.TD}
We now consider dissolution and formation kinetics in very large clusters, i.e.\ in the limit $N\to\infty$. Note that while the mean first passage time formally diverges, i.e.\
$\lim_{N\to\infty}\langle\tau_{d,f}\rangle=\infty$, it is expected to
do so in a ``mathematically nice'', well-defined ``bulk
scaling''. In anticipation of an
  exponential scaling of relevant time-scales with the system size $N$
  we define the \emph{mean formation/dissolution time per bond} in the thermodynamic limit as $\langle{\rm t_{d,f}} \rangle\equiv\lim_{N\to\infty}\langle\tau_{d,f}\rangle^{1/N}$. Using the local equilibrium approximation for the mean first passage time given by Eq.~\eqref{MFPT_le}, and assuming that the Glauber attempt probabilities $p_{k}$ are strictly sub-exponential in $N$, we prove via a \emph{squeezing theorem} in Appendix~\ref{Appendix E.6} that the exact mean dissolution and formation time per bond in the thermodynamic limit reads
\begin{equation}
    \langle {\rm
    t_{d,f}}\rangle=\e{\rtf(\vphi^{d,f}_{\rm max})-\rtf(\vphi^{d,f}_{\rm min})}\equiv\e{\Delta\rtf^\dagger},
    \label{MDT}
\end{equation}
where
\begin{eqnarray}
    \vphi^d_{\rm max}&\equiv& \sup_{\vphi < 1}\rtf(\vphi),\quad
    \vphi^d_{\rm min}\equiv \inf_{\vphi > \vphi^d_{\rm
    max}}\rtf(\vphi),\nonumber\\
    \vphi^f_{\rm max}&\equiv& \sup_{\vphi > 0}\rtf(\vphi),\quad
    \vphi^f_{\rm min}\equiv \inf_{\vphi < \vphi^f_{\rm
    max}}\rtf(\vphi).
    \label{ext_TD}
\end{eqnarray}
Eq.~\eqref{MDT} shows that the mean first passage per bond in the thermodynamic limit is determined exactly by the largest left/right-approaching free energy barrier between the initial and final point, and is completely independent of the Glauber attempt probability $p_{k}$. We obtain analytical results for Eqs.~\eqref{MDT} and
\eqref{ext_TD} for arbitrary $\tj, \tmu$ and $\thh$. Since these results are somewhat complicated for $\tmu>0$ and $\thh \neq 0$ we present them in Appendix~\ref{Appendix E.7} and Fig~\ref{fig A7}. In the force-free case with zero intrinsic affinity, i.e.\ $\tmu=\thh= 0$, they turn out to be surprisingly compact and
given by
\begin{eqnarray}
    \langle {\rm
    t_{d,f}}\rangle=
    \left\{
    \begin{array}{lc}
    2\e{-2\tj}\cosh^2\tj, & 0 \le \tj\le \frac{1}{2}\ln{2} \\
    \displaystyle{4\frac{\sinh^2 2\tj}{\e{4\tj}-2}}, &\hspace{-3mm} \frac{1}{2}\ln{2}\le\!\tj\!\le\frac{1}{2}\ln{(1\!+\!\sqrt{2})}\\
    \displaystyle{8\e{2\tj}\frac{\sinh^2 \tj}{\e{4\tj}-2}}, & \tj\ge \frac{1}{2}\ln{(1\!+\!\sqrt{2})},
    \end{array}
    \right .
    \label{MDT_ff}  
\end{eqnarray}
such that for $\tj=0$ and $\tj\to\infty$ we have $\langle {\rm t_{d,f}}\rangle=2$ being the maximum, and the minimum occurs at $\tj = \ln{(1+\sqrt{2})}/2$ where $\langle {\rm t_{d,f}} \rangle =(4/7)(2\sqrt{2}-1)$. Fig.~\ref{fig 8}a,b shows a comparison of the prediction of Eq.~\eqref{MDT} with the results for finite system given by Eqs.~\eqref{MFPT_le} and \eqref{loceq} rescaled according to $\langle \tau_{d,f}\rangle^{1/N}$. Already for $N=900$ a nearly complete collapse to the thermodynamic limit \eqref{MDT} is observed for both, cluster formation as well as
dissolution. The mean field analogue of
  Eq.~\eqref{MDT_ff} is given by Eq.~\eqref{MFPT bare field MF
    explicit} for a general $\bar{z}$ and remarkably has a universal
  (i.e.\ $\bar{z}$-independent) minimum value of $\langle {\rm
    t_{d,f}} \rangle_{\rm MF} \approx 1.0785$ at the dynamical critical coupling
  $\tj = 2 \ln{(2)}/\bar{z}$ (see Eq.~\eqref{min time MF}). Moreover, $\langle {\rm
    t_{d,f}} \rangle_{\rm MF}$ displays an unphysical divergence in the limit $\tj \to \infty$
  (see Fig.~\ref{fig A8}).
\subsection{Dynamical phase transition and critical behavior}
Strikingly, the mean dissolution and formation time in
the thermodynamic limit \eqref{MDT} display a discontinuity as a function of the coupling $\tj$ (see jumps in $\partial_{\tj} \langle {\rm t_{d,f}}\rangle$ depicted in the insets in Fig.~\ref{fig 8}a,b). In particular, for zero affinity and external force we find from Eq.~\eqref{MDT_ff}
\begin{eqnarray*}
    \lim_{\tj \nearrow \frac{1}{2}\ln(1+\sqrt{2})}\partial_{\tj}\langle {\rm
    t_{d,f}}\rangle&=&-(4/7)^2(13\sqrt{2}-17)\\
    \lim_{\tj \searrow   \frac{1}{2}\ln(1+\sqrt{2})}\partial_{\tj}\langle {\rm
    t_{d,f}}\rangle&=&(8/7^2)(9\sqrt{2}-8).
\end{eqnarray*}
This implies the existence of a first order \emph{dynamical
phase transition} at the \emph{dynamical critical coupling}
$\tj^d_{\rm crit}$ and hence a qualitative change in the dominant dissolution/formation pathway. Coincidentally, the Bethe-Guggenheim \emph{dynamical critical point} for $\tmu=\thh=0$ coincides with the exact (Onsager's) statical
critical point for the two-dimensional zero-field Ising model \cite{PhysRev.65.117}. Similarly, the mean field dynamical critical point for $\tmu=\thh=0$ coincides with the Bethe-Guggenheim statical critical point (for a more detailed discussion see Appendices~\ref{Appendix E.7} and \ref{Appendix E.8}). Strikingly, the dynamic critical point always corresponds to the minimum of $\langle {\rm t_{d,f}}\rangle$. The explanation of the physics underneath the dynamical phase transition and the meaning of $\tj^d_{\rm crit}$ is given in Fig.~\ref{fig 8}c.

The qualitative behavior of $\langle {\rm t_{d,f}}\rangle$ has three distinct regimes. In regime I, where $0\le \tj<\tj^s_{\rm crit}$, the free energy landscape $\rtf(\vphi)$ has a single well and
according to Eq.~\eqref{MDT} $\langle {\rm t_{d,f}}\rangle$ is determined by  $\Delta \rtf^\dagger_{0,1}$ -- the
free energy difference between the minimum and the absorbing point (i.e.\ $\vphi=0$ for dissolution and $\vphi=1$ for formation, respectively). $\Delta \rtf^\dagger_{0,1}$ is a decreasing function of $\tj$.

At the statical critical coupling $\tj^s_{\rm crit}$, which marks the onset of regime II, a second free energy barrier emerges delimiting the phase-separated low ($L$) and a
high ($H$) density phase. We denote this free energy barrier by $\Delta \rtf^{\dagger}_{H \rightleftarrows L}$ where $\rightarrow$ and $\leftarrow$ stand for dissolution and
formation, respectively. $\Delta \rtf^{\dagger}_{H \rightleftarrows L}$ is an increasing function of $\tj$. In regime II, that is when $\tj^s_{\rm crit}\le \tj< \tj^d_{\rm
  crit}$, the dissolution and formation first
evolve through a (thermodynamic) phase transition and, finally, must also surmount the second, predominantly entropic barrier to the complete dissolved/bound state. In regime II, as in regime I, the largest free energy barrier
remains the free energy difference between the minimum and the absorbing point, i.e.\ $\Delta \rtf^\dagger_{0,1}>\Delta \rtf^{\dagger}_{H \rightleftarrows L}$. 

Exactly at the \emph{dynamical critical coupling} $\tj^d_{\rm
crit}$ the two barriers become identical, $\Delta
\rtf^\dagger_{0,1}=\Delta \rtf^{\dagger}_{H \rightleftarrows L}$ and for $\tj> \tj^d_{\rm crit}$ we always have $\Delta
\rtf^\dagger_{0,1}<\Delta \rtf^{\dagger}_{H \rightleftarrows L}$. Therefore, in regime III the rate-limiting event becomes the phase transition itself, whereas the fully dissolved/bound state is thereupon reached by typical density fluctuations. Since $\Delta\rtf^\dagger_{0,1}$ decreases with $\tj$ while $\Delta \rtf^{\dagger}_{H \rightleftarrows L}$ increases with $\tj$, the mean dissolution/formation time per bond at the dynamical critical coupling $\tj^d_{\rm crit}$ must be minimal.  This explains the dynamical phase transition completely. 

\textcolor{black}{Note that the dynamical phase transition is
  preserved under initial conditions that lie beyond the largest free
energy barrier (from the final/absorbing state). For example, we may consider $\varphi(\{\sigma_{i}\}_{0})=\varphi_{L,H}^{0}$ in
      Eq.~(\ref{FPT}), where $\varphi_{L,H}^{0}$ is the (meta)-stable minimum
      in the high and low density region for cluster
      dissolution and formation, respectively. In the thermodynamic limit the
      equilibration time from the initial condition
      $\varphi(\{\sigma_{i}\})=0\lor 1$ to the (meta)-stable minimum
      $\varphi_{L,H}^{0}$ becomes exponentially faster than the total
      transition time, which renders $\langle {\rm t}_{\rm d,f}
      \rangle$ unaffected.}
\begin{figure}[t!]
    \includegraphics[width = 0.35\textwidth]{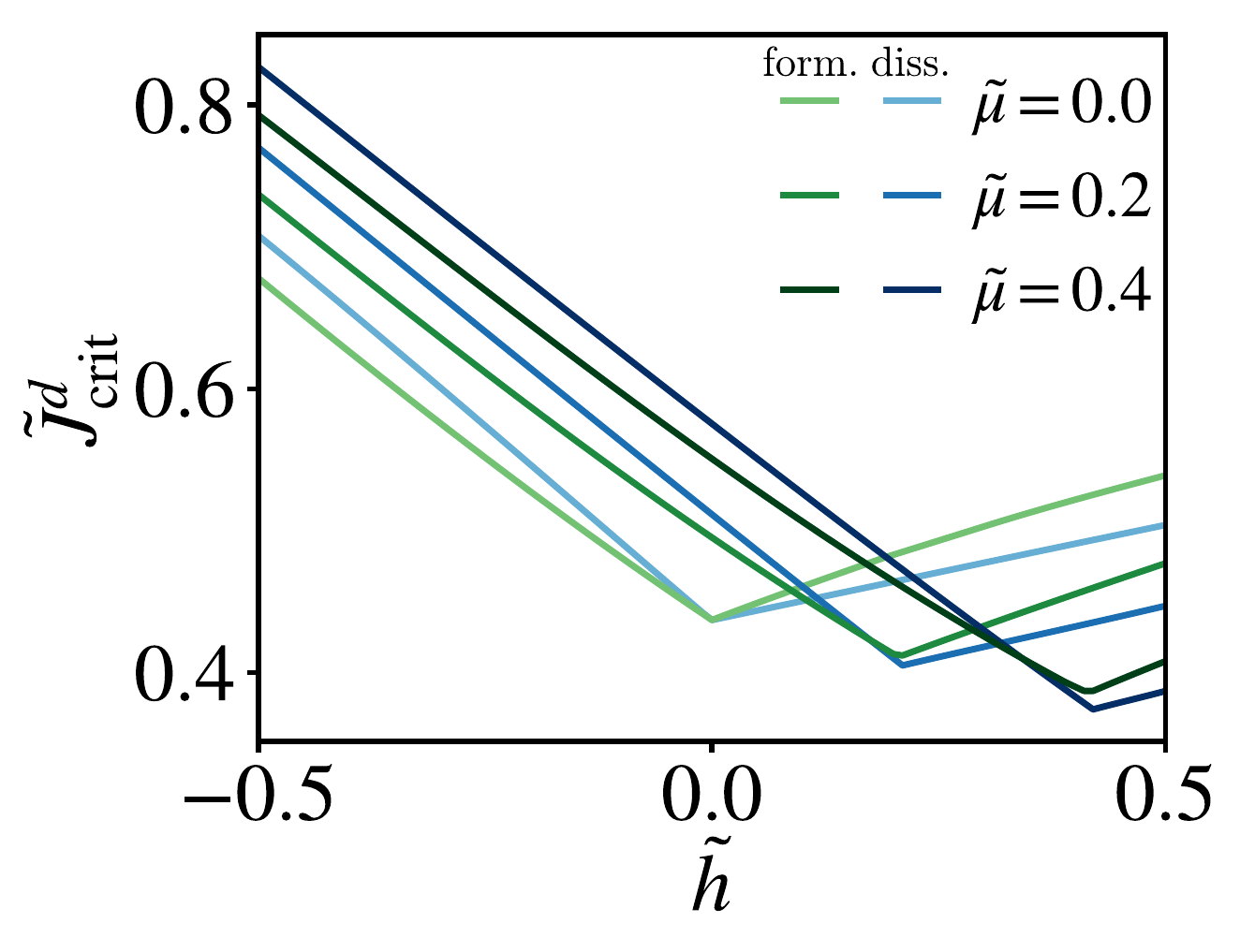}
    \caption{\textbf{Dynamical critical point.} Dynamical critical
      coupling $\tj^d_{\rm crit}$ as a function of the external force
      $\thh$ for several values of the intrinsic binding-affinity
      $\tmu$. Note that $\tj^d_{\rm crit}$ as a function of $\thh$ may be
      non-monotonic with
      a global minimum whose location depends on $\tmu$.
     }
    \label{fig 9}
\end{figure}

The dependence of $\tj^d_{\rm crit}$ on $\tmu$ and $\thh$ is
determined in the form of a Newton's series in Appendix ~\ref{Appendix E.7}, and is depicted
in Fig.~\ref{fig 9}. Depending on the intrinsic affinity
$\tmu$, the dependence of $\tj^d_{\rm crit}$ may be
non-monotonic. Note that in contrast to the statical critical coupling $\tj^{\rm s}_{\rm crit} $ that is independent of $\tmu$, the dynamical critical coupling $\tj^{\rm d}_{\rm crit}$ depends on the particular value of $\tmu$.
\section{Many-body physics in the mechanical
  regulation of adhesion}\label{Implication}
Our results tie the effective bending rigidity, $\kappa$, and in turn
interactions between neighboring adhesion bonds, $\tj\propto
\kappa^{-1/2}$ (see Appendix~\ref{Appendix A}), to the collective phase behavior of adhesion clusters at equilibrium, and
to distinct dynamical phases of cluster dissolution and
formation. Based on the quantitative relationship
  between the coupling strength $\tj$ and bending ridigity $\kappa$ given by Eq.~\eqref{Coupling-membrane-relation stiff membrane}, and an order-of-magnitude estimation of the relevant parameters listed in Table \ref{Table I} we find that the coupling strength in cellular systems lies within the range $0 \lesssim \tj \lesssim 2.5$. Notably, both the statical and dynamical critical point at moderate values of the external force values and/or intrinsic binding-affinity lie within said range (see Fig.~\ref{fig 5} and \ref{fig 9}). Yet, it remains to be explained why a near-critical coupling may be beneficial for cells, and how it may be regulated.

\begin{table}[t!]
        \caption{Estimated parameter values in cellular systems.}
        \centering
        \begin{tabular}{|p{3cm}||p{1.1cm}|p{2.3cm}|p{1.5cm}|}
            \hline
            \multicolumn{4}{|c|}{Estimated parameter values} \\
            \hline
            Parameter & Symbol & Estimated value / range & Source\\
            \hline
            Spring constant & $\beta k$ & $\sim 10^{-2} \ \ [{\rm nm}^{-2}]$ & \cite{paszek2009integrin, caputo2005effect, speck_specific_2010}\\
            Non-specific interaction strength & $\beta \gamma$ & $\sim 10^{-5} \ \ [{\rm nm}^{-4}]$ &
            \cite{speck_specific_2010}\\
            Bond separation distance & $h_{0}-l_{0}$ & $25 \sim 50 \ \ \ \ [{\rm nm}]$ &
            \cite{paszek2009integrin, nermut1988electron, pelta2000statistical}\\
            Bending rigidity & $\beta \kappa$ & $4 \sim 400$ &
            \cite{dimova2014recent, faizi2019bending, braig2015pharmacological, simson1998membrane}\\
            \hline
        \end{tabular}
        \label{Table I}
\end{table}

Our results provoke the hypothesis that the membrane rigidity (and hence the coupling strength) may lie close to the statical critical value for quasi-static, and near the dynamical critical value for transient processes. Mechanical regulation of the bending rigidity can be achieved through hypotonic swelling \cite{ayee_hypotonic_2018}, (de)polymerization of the F-actin
network \cite{callies_membrane_2011, sliogeryte_chondrocyte_2016}, by decoupling the F-actin network from the plasma membrane \cite{simson1998membrane}, through changes of the membrane composition
\cite{sanyour_membrane_2019,fowler_membrane_2016,dimova2014recent, faizi2019bending} or integral membrane proteins \cite{fowler_membrane_2016}, membrane-protein activity
\cite{faris_membrane_2009}, temperature modulation \cite{rico_temperature_2010, dimova2014recent, marlin_purified_1987}, and acidosis \cite{steinkuhler2019membrane}, to name but a few. Moreover, it has been shown experimentally that temperature modulations affects adhesion strength through changes in membrane fluidity \cite{marlin_purified_1987}, cell elasticity \cite{rico_temperature_2010}, or via a temperature cooperative process \cite{juliano_adhesion_1977}, albeit the denaturation of the binding proteins also provides a possible explanation \cite{sagvolden_cell_1999}.

Below we argue that the \emph{change} in the response of a cell to a perturbation, defined as a change in the equilibrium binding strength or association/dissociation rates, is largest near criticality. This results in either a very small or very large response, depending on the change of the underlying parameter. Here we follow the same kind of reasoning as rooted in the \emph{criticality hypothesis}, which states that systems undergoing an order--disorder phase transition achieve the highest trade-off between robustness an flexibility around criticality \cite{roli2018dynamical}.
\subsection{Criticality at equilibrium}
\begin{figure*}[ht!]
    \includegraphics[width = 1\textwidth]{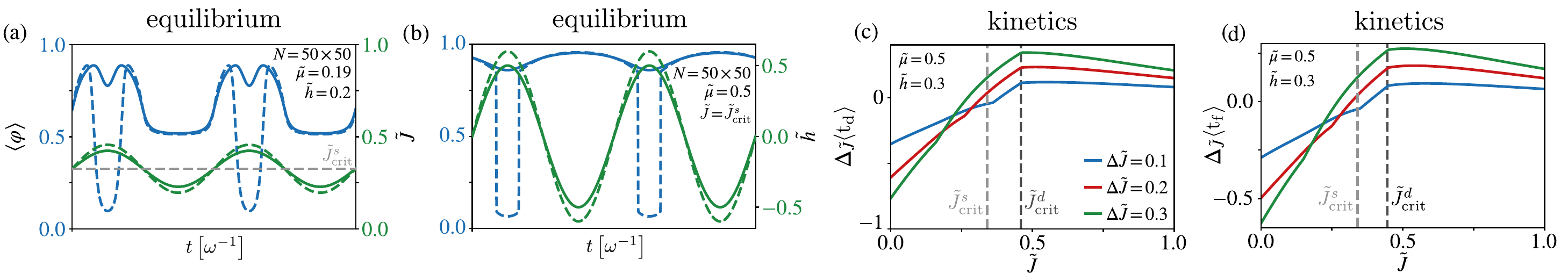}
    \caption{\textbf{Equilibrium and kinetic response to changes in
        cell stiffness, external force, or the binding-affinity.} Equilibrium response of the average
      fraction of closed bonds $\langle\vphi(t)\rangle$ to a slow (quasi-static),
      periodic modulation of (a) the coupling $\tj(t)=\tj+\delta\tj\sin(\omega
      t)$ (and hence membrane
      stiffness) and (b) the external force $\thh(t)=\thh+\delta\thh\sin(\omega
      t)$. Note the
      strong sensitivity of the response near the \emph{statical} critical coupling
      $\tj^s_{\rm crit}$. (c-d) Change in mean dissolution (c) and formation (d) time per bond, 
      $\Delta_{\tj}\langle{\rm t_{d,f}}\rangle\equiv\langle{\rm
        t_{d,f}}(\tj+\Delta\tj)\rangle-\langle{\rm
        t_{d,f}}(\tj)\rangle$, as a response to a change $\Delta\tj$
      of the coupling, as a function of $\tj$ for various $\Delta\tj$. The kinetic response is largest for $\tj$ near the \emph{dynamical}
      critical coupling.}
    \label{fig 10}
\end{figure*}

In Fig.~\ref{fig 10}a we depict how oscillations in the coupling strength (arising through oscillations in the bending rigidity $\kappa $) around the statical critical point affect the average fraction of closed bonds. Similar oscillatory patterns and their effect on the adhesion strength have been observed in vascular smooth muscle cells, where changes in the bending rigidity were concerted by the remodeling of the actin cytoskeleton
\cite{hong_vasoactive_2014, zhu_temporal_2012, sanyour_spontaneous_2018}. Minute changes in the amplitude, $\delta\tj$, can drive the systems's behavior from oscillations within a dense phase with $\langle \varphi(t) \rangle > 0.5$ to intermittent periods of nearly complete dissolution (compare full and dashed lines in Fig.~\ref{fig 10}a). Hence we find that the response (i.e. $\langle \varphi(t) \rangle$) is most sensitive to a change in the amplitude $\delta\tj$ when $\tj$ lies close to the statical critical point.

Similarly, in
Fig.~\ref{fig 10}b we show the response of $\langle \vphi(t)\rangle$ to a mechanical perturbation
oscillating quasi-statically between a pulling and a pushing force,
$\thh(t)=\thh_{\rm ref}+\delta\thh\sin(\omega t)$ (for practical
examples see e.g.\ \cite{Russel,Spatz_cyc}). Such mechanical perturbations can for example arise through changes in
active stresses generated within the cytoskeleton \cite{Parsons}. Here as well, a small change in the
force $\delta\thh$ acting on the cluster, can lead to stark differences
in the cluster stability $\langle \vphi(t)\rangle$. The sensitivity to a change in the
the force is most amplified near the statical critical coupling $\tj^s_{\rm
  crit}$ (compare full and dashed lines in
Fig.~\ref{fig 10}b), where a small change
in the amplitude, $\delta\thh$, can cause intermittent periods of essentially
complete cluster detachment.

Drastic changes in the average number of closed bonds
  have been observed experimentally in adhesion frequency assays and
  single-molecule microscopy \cite{huppa_tcrpeptidemhc_2010,
    huang_kinetics_2010}. There it was shown that binding affinities
  and binding dynamics for a T-cell receptor (TCR) interacting with
  the peptide-major histocompatbility complex (pMHC) are more than an
  order of magnitude smaller in solution (i.e.\ in 3D) as compared to
 when they are anchored to a cell membrane (i.e.\ in 2D). One
 possible contribution to the discrepancy between the 3D and 2D
 binding kinetics is the difference in the reduction of the entropy
 upon binding, which is larger in 3D than in 2D
 \cite{wu_transforming_2011}. However, it has been explicitly remarked
 that this contribution alone does \emph{not} explain the
 measured difference in the binding affinities
 \cite{wu_transforming_2011}. 
The authors of Ref.\ \cite{huppa_tcrpeptidemhc_2010,
  huang_kinetics_2010} rationalize these differences in binding in
terms of a cooperativity between neighboring TCRs due to the anchoring
membrane.  In particular, Fig.~11a in the Supplementary Material of
Ref.\ \cite{huang_kinetics_2010} shows the adhesion frequency
$P_{a}(t_{\rm c}) \in [0,1]$, defined as the fraction of observed
adhesion events between the TCR and pMHC as a function of the contact
time $t_{\rm c}$ between the anchoring membranes, derived from
Monte-Carlo simulations. Upon introducing a heuristic
neighbor-dependent amplification factor in the binding rates the
authors observe an amplification of the adhesion frequency $P_{a}$
(compare squares with diamonds), indicating an increase in binding
events in agreement with their experimental observations.

We may relate our results to the observations in
  Ref.\ \cite{huang_kinetics_2010} by recalling the relation between
  $P_{a}$ and $\langle N_{c} \rangle = N\langle \varphi \rangle$,
  i.e.\ $P^{\rm ss}_{a}\equiv \lim_{t_{\rm
      c}\rightarrow\infty}P_{a}(t_{\rm c})=1-\exp{(-N\langle \varphi
    \rangle)}$ (see \cite{chesla1998measuring} as well as Eqs.~(1) and
  (2) in \cite{huang_kinetics_2010}). In our model the aforementioned
  amplification factor arises naturally from a nonzero coupling
  strength $\tilde{J}$ due to the anchoring membrane. Indeed, in
  Fig.~\ref{fig 3}e  an increase in $\tilde{J}$ leads to an increase
  in $\langle \varphi \rangle$, which in turn causes an increase
  of the steady state adhesion frequency $P^{\rm ss}_{a}$. Hence we
  find that the amplification factor in \cite{huang_kinetics_2010} and
  coupling $\tilde{J}$ in our model have the same effect
  on the adhesion frequency.

A similar observation was made in
  \cite{williams_quantifying_2001} on the basis of a detailed analysis of
  the binding affinities of the adhesion receptor CD16b
  placed in three distinct environments: red blood cells (RBCs),
  detached Chinese hamster ovary (CHO) cells, and K562 cells. Based
  on Fig.~4a,b in \cite{williams_quantifying_2001} the adhesion
  frequency for RBCs is around a 15-fold larger than for CHO and
  K562 cells. In the discussion 
  the authors point towards the modulation of surface smoothness as an
  explanation  for the observed differences in adhesion
  frequency \cite{williams_quantifying_2001}. Since K562 cells are
  known to have a larger bending
  rigidity than RBCs \cite{zhelev1994role, sharma1993cellular} (we
  were unfortunately not able to find the corresponding information
  for  CHO cells in the existing literature), it is expected that the
  coupling  strength $\tilde{J}$ is generally higher in the latter
  (see Appendix~A),  which provides a potential explanation for the
  observed  difference in adhesion frequencies between RBCs and K562
  cells.
\subsection{Criticality in kinetics}
Many biological processes \cite{Detach,Yamada,De_Pascalis,Kirfel} and experiments \cite{Dwir,Erbeldinger,Zhou} involve adhesion under
transient, non-equilibrium conditions, where cells can become detached
completely from a substrate (for a particular realization with a
constant force see \cite{Dwir}). The duration of these transients may
be quantified by the mean dissolution and formation time, $\langle
{\rm t_{d,f}}\rangle$ (see Fig.~\ref{fig 8}). Imagine that the cell
can change the bending rigidity by an amount $\Delta\kappa$ that in
turn translates into a change in coupling, $\tj'=\tj+\Delta\tj\propto 1/\sqrt{\kappa+\Delta\kappa}$. If the
mechanical regulation is to be efficient, a small change of $\Delta\tj$
should effect a large change of $\langle{\rm
  t_{d,f}}\rangle$.

The efficiency of the regulation,
expressed as the change of mean dissolution/formation time in response to a change  $\Delta\tj$, $\Delta\langle{\rm t_{d,f}}\rangle\equiv\langle{\rm t_{d,f}}(\tj+\Delta\tj)\rangle-\langle{\rm
t_{d,f}}(\tj)\rangle$, is shown in Fig.~\ref{fig 10}(c-d). The results demonstrate that the regulation is most efficient, that is gives the largest change, when $\tj$ is poised near the \emph{dynamical} critical coupling, $\tj\simeq\tj^d_{\rm crit}$, regardless of the magnitude of the change $\Delta \tj$. Recall that the formation and dissolution rate, $1/\langle {\rm t_f}\rangle$ and $1/\langle {\rm t_d}\rangle$, respectively, are highest at the dynamical critical coupling (see Fig.~\ref{fig 8}). Therefore not only do we find the largest response to a change in $\tj$, but also the fastest formation and dissolution kinetics at the dynamical critical coupling $\tj\simeq\tj^d_{\rm crit}$. 

An example where fast kinetic (un)binding and a large sensitivity to the bending rigidity can be beneficial is found in tumor cells that undergo metastasis - the process through which tumor cells spread to secondary locations in the host's body. Recent studies suggest that cancer cells are mechanically more compliant than normal, healthy cells \cite{compliant}. Moreover, experiments with magnetic-tweezers have shown that membrane
stiffness of patient tumor cells and cancer cell-lines inversely
correlates with their migration and invasion potential
\cite{swaminathan_mechanical_2011}, and an increase of membrane rigidity alone is sufficient to inhibit invasiveness of cancer cells \cite{braig2015pharmacological}. Cells with the highest
invasive capacity were found to be five times less
stiff than cells with the lowest migration and invasion potential, but
the underlying mechanism behind this correlation remained elusive \cite{swaminathan_mechanical_2011}.

Based on our results a decrease in the bending rigidity, and hence the membrane stiffness, can alter both, the equilibrium strength of adhesion (see Fig.~\ref{fig 3}) as well as the kinetics of formation and dissolution of adhesion domains (see Fig.~\ref{fig 8}). This may provide a clue about the mechanical dysregulation of cell adhesion in metastasis in terms of a softening of the cell membrane.
\section{Criticality in the Ising model}
\label{2d Ising implication}
By setting $\thh=0$ and writing $\delta_{\sigma_{j},-1}=(1-\sigma_{j})/2$ we find that Eq.~\eqref{Hamiltonian1} is, up to a constant, identical to the Hamiltonian of the isotropic ferromagnetic Ising model in a uniform external magnetic field $M\equiv\mu/2$. Therefore, our findings, and in particular the uncovered dynamical phase transition, also provide new insight into equilibrium and kinetic properties of the Ising model in the presence of a uniform external magnetic field.

The equilibrium properties of the two-dimensional Ising model in the absence of a magnetic field, such as the total free energy per spin, statical critical point, and binodal line were obtained in the seminal work by Onsager \cite{PhysRev.65.117}. The effect of a uniform magnetic field has mostly been studied numerically \cite{PhysRevB.2.2660, PhysRevB.23.287}, e.g.\ by Monte-Carlo simulations \cite{PhysRevA.36.4439} and renormalization group theory \cite{PhysRevB.36.3697}, but hitherto no exact closed-form expression for the free energy per spin has been found. On the Bethe-Guggenheim level the free energy density, binodal line, spinodal line, and statical critical point were known \cite{de2013liquid}, but to our knowledge we are the first to provide an exact closed-form expression for the equation of state in the presence of a uniform  magnetic field (see Appendix~\ref{Appendix D}).

The kinetics of the two-dimensional Ising model have been studied in the context of magnetization-reversal times (i.e.\ the time required to reverse the magnetization) \cite{datta2018magnetisation, brendel2003magnetization, garcia1996nonhomogeneous}, nucleation times \cite{brendel2005nucleation, sear2006heterogeneous}, and critical slowing down \cite{RevModPhys.49.435, RevModPhys.58.801}. Here we report a new type of dynamical critical phenomenon related to a first-order discontinuity and a global minimum of the magnetization reversal time at the concurrent dynamical critical point (see Fig.~\ref{fig 8}), which is fundamentally different from the statical critical point. The dynamical phase transition reflects a qualitative change in the instanton path towards magnetization reversal, and has not been reported before.

In Table \ref{Table II} we summarize the values of the statical and dynamical critical points obtained by the mean field and Bethe-Guggenheim approximation in the absence of a magnetic field, and for a general coordination number $\bar{z}$ (for a derivation of the dynamical critical points see Appendices~\ref{Appendix E.7} and \ref{Appendix E.8}). We also state the exact statical critical point of the two-dimensional Ising model. Conversely, the exact dynamical critical point of the two-dimensional Ising model remains unknown as it requires the exact free energy density as a function of the fraction of down spins (see Eq.~\eqref{TD} for the result within the Bethe-Guggenheim approximation). A lower bound on the dynamical critical point is set by the statical critical point, as the latter denotes the onset of an interior local maximum that is required for the dynamical critical point (see Fig.~\ref{fig 8}). The exact dynamical critical point may provide further insight into the nature of the dynamical phase transition. Moreover, it also sets a lower bound on the magnetization reversal times per spin in ferromagnetic systems in the absence of an external force.

\begin{table}
        \caption{Statical and dynamical critical point as a function of the coordination number $\bar{z}$ obtained within the mean field and Bethe-Guggenheim approximation, alongside the exact statical critical point $\tj^{\rm s}_{\rm crit}$ for the two-dimensional Ising model at zero field and binding-affinity, $\tmu=\thh=0$. The exact dynamical critical point $\tj^{\rm d}_{\rm crit}$ for the two-dimensional Ising model remains unknown; a lower bound is given by the Onsager statical critical point.}
  \centering
    \begin{tabular}{|p{3.cm}||p{2.55cm}|p{2.55cm}|}
        \hline
        \multicolumn{3}{|c|}{Critical points} \\
        \hline
        Approximation & $\tilde{J}^{s}_{\rm crit}$ & $\tilde{J}^{d}_{\rm crit}$\\
        \hline
        Mean field & $\frac{1}{\bar{z}}$ & $\frac{2}{\bar{z}}\ln{2}$\\
        Bethe-Guggenheim & $\frac{1}{2}\ln{\left(\frac{\bar{z}}{\bar{z}-2}\right)}$  & $\frac{-1}{2}\ln{(2^{1-2/\bar{z}}-1)}$\\
        Exact 2D & $\frac{1}{2}\ln{\left(1+\sqrt{2}\right)}$  & $\geq \tilde{J}^{s}_{\rm crit}$\\
        \hline
    \end{tabular}
    \label{Table II}
\end{table}
\section{Concluding remarks}
\label{Conc}
The behavior of individual \cite{bell_models_1978} and non-interacting
\cite{seifert_rupture_2000,erdmann_stochastic_2004,erdmann_stability_2004,Dasanna} adhesion bonds under force, the
effect of the elastic properties of the substrate and pre-stresses in
the membrane \cite{gao_probing_2011,Qian_2008}, as well
as the physical origin of the interaction between opening and closing of
individual adhesion bonds due to the coupling with the fluctuating
cell membrane \cite{Bruinsma,Speck,speck_specific_2010,bihr_nucleation_2012,farago_fluctuation-induced_2010,krobath_line_2011,hu_binding_2013,
 hu_binding_2015,erdmann_impact_2007} are by now theoretically well
established. However, in order to understand the importance of these
interactions and their manifestation for the mechanical regulation of
cell adhesion in and out of equilibrium one must go deeper, and disentangle the response of
adhesion clusters of all sizes to external forces and how it becomes
altered by changes in membrane stiffness. This is paramount because
interactions strongly change the physical behavior of adhesion clusters under
force both, qualitatively as well as quantitatively.

Founded on firm background knowledge \cite{bell_models_1978,seifert_rupture_2000,erdmann_stochastic_2004,erdmann_stability_2004,Dasanna,gao_probing_2011,Qian_2008,Bruinsma,Speck,speck_specific_2010,bihr_nucleation_2012,farago_fluctuation-induced_2010,krobath_line_2011,hu_binding_2013,
  hu_binding_2015,erdmann_impact_2007} our explicit analytical results
provide deeper insight into cooperative
effects in cell-adhesion dynamics and integrate them into a
comprehensive physical picture of cell adhesion under force. We
considered the full range of CAM binding-affinities and forces, and
established the phase behavior of two-dimensional adhesion clusters at
equilibrium as well as the kinetics of their formation and dissolution.

We have obtained, to the best of our knowledge, the first theoretical
results on equilibrium behavior and dynamic stability of adhesion clusters in
the thermodynamic limit beyond the mean field-level (existing studies, even those addressing
non-interacting adhesion bonds \cite{erdmann_stochastic_2004,erdmann_stability_2004,Dasanna}, are limited to small
clusters sizes
\cite{gao_probing_2011,Qian_2008,Speck,speck_specific_2010,bihr_nucleation_2012}). We
explained the complete thermodynamic phase behavior, including the co-existence of dense and dilute
adhesion domains, and characterized in detail the corresponding critical
behavior. 

We demonstrated conclusively the existence of a seemingly new kind of dynamical phase transition in the kinetics of adhesion cluster formation and dissolution, which arises due to the interactions between the bonds and occurs at a critical coupling $\tilde{J}^{d}_{\rm crit}$, whose value depends on the external force $\tilde{h}$ and binding-affinity $\tilde{\mu}$. At
the dynamical critical coupling $\tj^d_{\rm crit}$, and in turn critical bending
rigidity $\kappa^d_{\rm crit}\propto (\tj^d_{\rm crit})^{-2}$, the dominant formation and dissolution pathways change
qualitatively. Below $\tj^d_{\rm crit}$ the rate-determining step for
cluster formation and dissolution is the surmounting of the (mostly)
entropic barrier to completely bound and unbound states,
respectively. Conversely, above $\tj^d_{\rm crit}$ the thermodynamic phase
transition between the dense and dilute phase for dissolution, and
between the dilute and dense phase for cluster formation, becomes
rate-limiting, whereas the  completely bound and unbound states, respectively, are
thereupon reached by typical density fluctuations.

\textcolor{black}{We expect the
      non-monotonic dependence of the mean first passage time to
      cluster dissolution/formation on the coupling strength $\tj$
      that is asymmetric around the minimum to be experimentally
      observable (though the notion of a fully bound state during
      cluster formation may be
      experimentally ambiguous). According to our theory the existence
      of such a minimum and its asymmetric shape would immediately
      imply a dynamical phase 
      transition in the thermodynamic limit.\\ 
\indent
      Measuring the mean dissolution/formation time
      (in the absence or presence of an external force) for an
      ensemble of cells adhering to a stiff substrate seems to be
      experimentally feasible. The effective membrane rigidity (and thus the
      coupling $\tj$) could in principle be controlled
      by varying the membrane composition (e.g.\ increasing the
      cholesterol concentration that in turn increases membrane
      rigidity \cite{dimova2014recent}), by tuning the osmotic pressure of the medium \cite{Ayee_2018}, or
      by the depolymerization of F-actin \cite{Bausch}. Testing for
      signatures of the theoretically predicted dynamical phase
      transition thus seems to be experimentally (at least conceptually)
      possible, and we hope that our results will motivate such investigations.}

We discussed the biological implications of our results in the
context of mechanical regulation of the bending rigidity around criticality. Based on our results we have suggested that the response of a cell to a change in the bending rigidity may be largest near the statical critical point for quasi-static processes, and near the dynamical critical point for transient processes. This observation agrees with the \emph{criticality hypothesis}, and might expand the list of biological processes hypothesized to be poised at criticality \cite{Munoz}. 

Finally, we discussed the implications of our result for the two-dimensional Ising model. The observed dynamical phase transition is related to a first-order discontinuity in the magnetization reversal time, and the exact dynamical critical point for the two-dimensional Ising model remains elusive (see Table~\ref{Table II}).

We now remark on the limitations of our results. The mapping onto
a lattice gas/Ising model (i.e.\ Eq.~\eqref{Hamiltonian1} and
Appendix~\ref{Appendix A}; see also
\cite{Speck,speck_specific_2010}) may not apply to
genuinely floppy membranes 
encountered in biomimetic vesicular systems 
\cite{smith_progress_2009,reister-gottfried_dynamics_2008}. Moreover, since we only allow for two possible states of the bonds, i.e.\ associated and dissociated, we neglect any internal degrees of freedom (e.g.\ orientations of the bonds) which may contribute to the entropy loss upon binding \cite{wu_transforming_2011}, thereby changing the free energy.

Likewise, the assumption of an equally
shared force is generally good for stiff membranes (stiffened by the
presence of, or anchoring to, the stiff actin
cytoskeleton \cite{anchor})
or stiff membrane/substrate pairs, flexible individual bonds, low
bond-densities, or the presence of pre-stresses exerted by the actin
cytoskeleton
\cite{erdmann_stability_2004,gao_probing_2011,Qian_2008}. In
  Appendix~\ref{Appendix E.1.b} we provide an analysis of the effect
  of a non-uniform force load. Based on this analysis we find that in
  the case of rather floppy membranes, corresponding to large values of the
  coupling strength $\tilde{J}=\mathcal{O}(1)$, the difference between a uniform and
  a non-uniform force load is negligible for a broad range of
  realizations of the non-uniform force distribution. Only under
  the extreme, non-physiological condition that the ratio of forces
  experienced by inner and outer bonds is larger than an order of
  magnitude, we observe significant differences. Therefore,  the
  dependencies of the statical and dynamical critical points on the
  external force (see Figs.~5 and 8, respectively) are expected to remain valid for a non-uniform force distribution over a large range of force magnitudes.

In their present form our
results may not apply to
conditions when cells actively contract  in response to a mechanical
force on a timescale comparable to cluster assembly or dissolution
\cite{constant_1}, as well as situations in which cells actively
counteract the effect of an
external pulling force and make adhesion clusters grow (see
results in terms of a change in membrane stiffness as well.

Finally, throughout we have considered clusters consisting of so-called ``slip-bonds'',
whereas cell adhesion may also involve ``catch-bonds'' that
dissociate slower in the presence of sufficiently large pulling forces
\cite{catch_rev}. The reason lies in a second, alternative dissociation
pathway that becomes dominant at large pulling forces
\cite{Hin,Hin1,Hin2,Hin3}. 
Our results therefore do not apply to focal
adhesions composed of catch-bonds and would require a generalization of the
Hamiltonian (\ref{Hamiltonian1}-\ref{force}) and rate \eqref{Glauber Transition rate}.
These open questions are beyond the scope of the present work and will be addressed in forthcoming publications. 
\section{Data availability}
The open source code for the evaluation of the equation of state and
mean first passage times to cluster dissolution and formation for
finite-size systems is available at \cite{Godec2021}.
\section{Acknowledgments}
\textcolor{black}{We thank the anonymous referees for their valuable
  comments.} The financial support from the ``Deutsche
Forschungsgemeinschaft'' (DFG) through the 
Emmy Noether Program "GO 2762/1-1" (to AG), and an IMPRS fellowship of
the Max-Planck-Society (to KB) are gratefully acknowledged. 
\appendix
\setcounter{equation}{0}
\setcounter{figure}{0}
\renewcommand{\thefigure}{A\arabic{figure}}
\renewcommand{\theequation}{A\arabic{equation}}
\section{Relation between membrane rigidity and coupling strength}\label{Appendix A}
Here we provide a quantitative relation between the
effective bending rigidity
$\kappa$ and the coupling strength $\tj$ based on the Results of Ref.~\cite{speck_specific_2010}.
Consider a set of adhesion bonds at fixed positions
$\{\mathbf{r}_{i}\}$ coupled to a fluctuating membrane. The effective bending rigidity quantifies the amount of energy needed to change the membrane curvature, and is supposed to depend on the membrane composition \cite{dimova2014recent, faizi2019bending}, state of the actin network \cite{simson1998membrane}, and other intrinsic factors that determine the mechanical stiffness of the cell. Let
$\{b_{i}\}$ describe the state of all bonds, where $b_{i}=1$ denotes a
closed and $b_{i}=0$ an open bond.
The bonds are represented by springs with constant $k$, resting length
$l_{0}$, and binding energy $\epsilon_{b}$.  Non-specific interactions
between the membrane and the opposing substrate are described by a
harmonic potential with strength $\gamma$, which arises from a Taylor
expansion around the optimal interaction distance $h_{0}$ between the
membrane and the substrate. Assuming a timescale separation between
the opening/closing of individual bonds and membrane fluctuations, the
following partition function for the state of bonds $\{b_{i}\}$ can be derived \cite{speck_specific_2010} 
\begin{equation}
    Z = \sum\limits_{\{b_{i}\}}\exp{\left(\sum\limits_{i\neq j}\tj_{ij}b_{i}b_{j}+\tmu \sum\limits_{i=1}^{N}b_{i}\right)}+\mathcal{O}\left(\frac{k^2}{\gamma \kappa }\right),
    \label{Partition function lattice gas}
\end{equation}
where $\tmu$ plays the role of an intrinsic binding-affinity and $\tj_{ij}$ is an effective interaction between the bonds given by
\begin{equation}
    \tj_{ij}\equiv\frac{\beta k^{2}(h_{0}-l_{0})^2}{16\sqrt{\gamma\kappa}} m\left(|\mathbf{r}_{i}-\mathbf{r}_{j}|\right),
    \label{effective interaction}
\end{equation}
with $m\left(r\right)=-\frac{4}{\pi}\rm
kei_{0}\left(r(\kappa/\gamma)^{1/4}\right)$, and $\rm kei_{0}(x)$ is a
Kelvin function defined as ${\rm kei_{0}}(x)\equiv {\rm Im}
K_{0}(x\e{3\pi i/4})$ where $K_{0}(z)$ is the zero-order modified
Bessel function of the second kind \cite{speck_specific_2010}.
A systematic comparison of the equation of state $\varphi$ of the
full/explicit model (i.e.\ reversible adhesion bonds coupled to a
dynamic, fluctuating membrane) and of the lattice gas governed by
  Eq.~\eqref{effective interaction} has been carried out
  in \cite{speck_specific_2010}. Using the following set of parameter
  values $\beta\kappa = 80$, $\beta\gamma = 10^{-5} \ \rm{nm}^{-4}$,
  $\beta k= 2.25\times10^{-2} \ \rm{nm}^{-2}$, $h_{0}-l_{0}= 45.9-50.3
  \ \rm{nm}$, and $m\left(1.5\right)=0.42194(6)$, corresponding to a
  coupling strength of $\tilde{J}\approx 1.0-1.2$, the authors found a
  \emph{quantitative} agreement between the full and lattice gas
  models (see Fig.~5 in Ref.~\cite{speck_specific_2010}). Note that
  the lattice gas model becomes exact in the limit $\kappa\to\infty$
  corresponding to $\tj=0$.

Our effective Hamiltonian, given by Eq.~\eqref{Hamiltonian1}, is directly derived from Eq.~\eqref{Partition function lattice gas} by considering the following arguments: First we note that the effective interaction $\tj_{ij}$ decays exponentially fast as a function of the lattice distance between the bonds, and therefore it suffices to only take into account nearest neighbor interactions \cite{speck_specific_2010}. Moreover, since we place the adhesion bonds on a lattice with equidistant vertices, the position dependence in Eq.~\eqref{effective interaction} drops out and we get $|\mathbf{r}_{i}-\mathbf{r}_{j}|=\Delta r$. Finally, upon introducing the variables $\sigma_{i}\equiv1-2b_{i}\in
[-1,1]$, and applying the transformations $\tmu \rightarrow \tmu - \bar{z}\tj$ and $\tj \rightarrow 4\tj$, we arrive at our effective Hamiltonian Eq.~\eqref{Hamiltonian1}.

\begin{equation}
    \lim\limits_{\substack{\kappa \rightarrow \infty \\ \gamma \neq 0, \ k < \infty, \ \Delta r < \infty}} \tj = \frac{\beta k^{2}\left(h_{0}-l_{0}\right)^{2}}{16\sqrt{\gamma\kappa}}+\mathcal{O}\left(\frac{\ln{\kappa}}{\kappa}\right).
    \label{Coupling-membrane-relation stiff membrane}
\end{equation}
Here we find the relation $\tilde{J}\propto1/\sqrt{\kappa}$, as
mentioned in the main text.
\setcounter{equation}{0}
\setcounter{figure}{0}
\renewcommand{\thefigure}{B\arabic{figure}}
\renewcommand{\theequation}{B\arabic{equation}}
\section{Bethe-Guggenheim approximation}\label{Appendix B}
\subsection{Partition function}\label{Appendix B.1}
Here we derive the partition function  of the spin-1/2
Ising model at zero field with fixed magnetization at $N/2-k$,
$Z_k\equiv\sum_{\{\sigma_i\}}\exp{(\tj\sum_{\langle
    ij\rangle}\sigma_i\sigma_j)}\delta_{N_c(\sigall),k}$, within the
Bethe-Guggenheim variational approximation. 
As a reminder we point out that $k$ denotes the number of closed bonds, and $N-k$ the number of open bonds. Since the sum in the exponent goes over nearest-neighbor terms, we can write
\begin{equation}
    \sum\limits_{\langle ij \rangle}\sigma_{i}\sigma_{j}= \left(N_{cc}+N_{oo}-N_{oc}\right),
    \label{Hamiltonian rewritten}
\end{equation}
where $N_{oo}, N_{oc}$ and $N_{cc}$ denote the total number of open-open, open-closed, and closed-closed adhesion pairs, respectively. Notice that every closed-closed adhesion pair consists of two closed adhesion bonds, and every open-closed adhesion pair consists of a single closed adhesion bond, hence $2N_{cc}+N_{oc} \approx \bar{z}k$. Similar reasoning applies to open-open adhesion pairs, resulting in the general relations
\begin{equation}
     N_{cc} \approx \frac{1}{2}\left(\bar{z}k-N_{oc}\right),\,\, 
    N_{oo} \approx \frac{1}{2}\left(\bar{z}\left(N-k\right)-N_{oc}\right).
    \label{open-open pairs}
\end{equation}        
Eqs.~\eqref{open-open pairs} become
exact for infinite lattices and lattices with periodic boundary
conditions. Instead of summing over all configurations
$\{\sigma_{i}\}$ with $N_{oc}$ open-closed pairs
we may formally sum over all distinct values of $N_{oc}$ and account
for their multiplicity by introducing a degeneracy factor
$\Psi_{\bar{z}N}^{\bar{z}k}(N_{oc})$ that counts the number of
configurations with a given $N_{oc}$ at fixed $k$, i.e.
\begin{equation}
\sum_{\{N_{oc}\}}\e{\tj\sum_{\langle ij\rangle}\sigma_{i}\sigma_{j}}\approx \sum_{N_{oc}}
\Psi_{\bar{z}N}^{\bar{z}k}(N_{oc})\e{\tilde{J}\bar{z}
  (N/2-2N_{oc}/\bar{z})},
\label{Canonical partition function with degeneracy factor}
\end{equation}
where we have used Eqs.~\eqref{open-open pairs}. The core idea is to approximate
$\Psi_{\bar{z}N}^{\bar{z}k}(N_{oc})$ by the variational, Bethe-Guggenheim approximation \cite{retter1987adsorption,
  guggenheim1935statistical, bethe1935statistical} (for an excellent
explanation of the method see \cite{fowler_statistical_1939,
  guggenheim1952mixtures, de2013liquid}).
For $\tj=0$ the degeneracy factor must obey
\begin{equation}
    Z_{k}|_{\tj=0}\approx\sum_{N_{oc}}\Psi_{\bar{z}N}^{\bar{z}k}(N_{oc})\stackrel{!}{=}\binom{N}{k}.
    \label{Constraint degeneracy factor}
\end{equation}
To implement this constraint it is convenient to normalize Eq.~\eqref{Canonical partition function with degeneracy factor} at zero coupling and write
\begin{equation}
    Z_{k}\approx\binom{N}{k}\frac{\sum\limits_{N_{oc}}\Psi_{\bar{z}N}^{\bar{z}k}(N_{oc})\e{\tilde{J}\bar{z} (N/2-2N_{oc}/\bar{z})}}{\sum\limits_{N_{oc}}\Psi_{\bar{z}N}^{\bar{z}k}(N_{oc})}\equiv\binom{N}{k}\frac{S_{1}}{S_{2}},
    \label{Canonical partition function with approximate degeneracy factor}
\end{equation}
where we will determine $S_{1,2}$ variationally.
Then we have for $\tilde{J}=0$ that $Z_{k}=\binom{N}{k}$. Let us now
consider placing pairs of adhesion bonds randomly onto the
lattice. The total number of unique lattice configurations for fixed
$N_{oo}, N_{oc}$ and $N_{cc}$ may be approximated by 
\begin{equation}
    \Psi_{\bar{z}N}^{\bar{z}k}(N_{oc}) \approx
    \frac{\left(N_{cc}+N_{oc}+N_{oo}\right)!}{\left(N_{cc}\right)!\left(N_{oc}/2\right)!^{2}\left(N_{oo}\right)!}.
    \label{degeneracy factor initial guess}
\end{equation}
Notice that for a two-dimensional square lattice that is either infinite
or has periodic boundary conditions, the number of open-closed pairs
is always even, and therefore the term $(N_{oc}/2)!$ is
well-defined. For a finite two-dimensional square lattice with
free boundary conditions the number of open-closed adhesion pairs can
be odd, which forces us to consider the generalized factorial (i.e. Gamma function). Replacing the factorial with the Gamma function we get 
\begin{equation}
    \Psi_{\bar{z}N}^{\bar{z}k}(N_{oc})\approx
    \frac{\Gamma\left(N_{cc}+N_{oc}+N_{oo}+1\right)}{\Gamma\left(N_{cc}+1\right)\Gamma\left(N_{oc}/2+1\right)^{2}\Gamma\left(N_{oo}+1\right)},
    \label{degeneracy factor with genereralized factorial}
\end{equation}
where $\Gamma\left(n\right)=\left(n-1\right)!$ for $n\in
\mathbb{Z}^+$. 
Substituting Eqs.~\eqref{open-open pairs} for $N_{cc}$ and $N_{oo}$
leaves $N_{oc}$ as the only free parameter in Eq.~\eqref{degeneracy
factor with genereralized factorial}. 

We approximate $S_{1,2}$ by an analytic continuation of the maximum
term method \cite{mcquarrie1965statistical, hill1986introduction} to
real numbers. First we analytical continue the summands over positive
real numbers using Eq.~\eqref{degeneracy factor with genereralized
  factorial}. We now approximate both sums in
Eq.~\eqref{Canonical partition function with approximate degeneracy factor}  by their respective largest term, i.e.\ by the solutions of
the pair of optimization problems
\begin{equation}
\sup_{x_{1}}\Psi_{\bar{z}N}^{\bar{z}k}(x_{1})\e{\tilde{J}\bar{z}(N/2-2x_{1}/\bar{z})},\quad
\sup_{x_{2}}\Psi_{\bar{z}N}^{\bar{z}k}(x_{2}),
\end{equation}
which yields 
\begin{equation}
    S_{1} \approx \Psi_{\bar{z}N}^{\bar{z}k}(\bar{z}\bar{X}_{k}) \e{\tilde{J}\bar{z} (N/2-2\bar{X}_{k})}, \ S_{2} \approx \Psi_{\bar{z}N}^{\bar{z}k}(\bar{z} X^{*}_{k}), 
    \label{S12 approximation}
\end{equation}
with $X^{*}_{k}\equiv k\left(N-k\right)/N$ and $\bar{X}_{k}$ defined in Eq.~\eqref{xx}. We used Stirling's approximation for the Gamma function to find the
local maxima,
i.e.\ $\Gamma\left(z\right)=\sqrt{2\pi/z}(z/\e{})^{z}[1+\mathcal{O}(1/z)]$,
and therefore expect the accuracy of Eq.~\eqref{S12 approximation} to
increase with increasing $N$. Using Eq.~\eqref{S12 approximation} and
introducing $\psi_{\bar{z}N}^{\bar{z}k}(x)=\Gamma(1+\bar{z}N/2)/\Psi_{\bar{z}N}^{\bar{z}k}(x)$ we
obtain Eq.~\eqref{zqc} in the main text. The
mean field solution is in turn recovered by setting $\bar{X}_{k}=X^{*}_{k}$, which happens automatically for $\tilde{J}=0$ or $k=0 \lor N$. 
\begin{figure}
    \includegraphics[width = 0.475\textwidth]{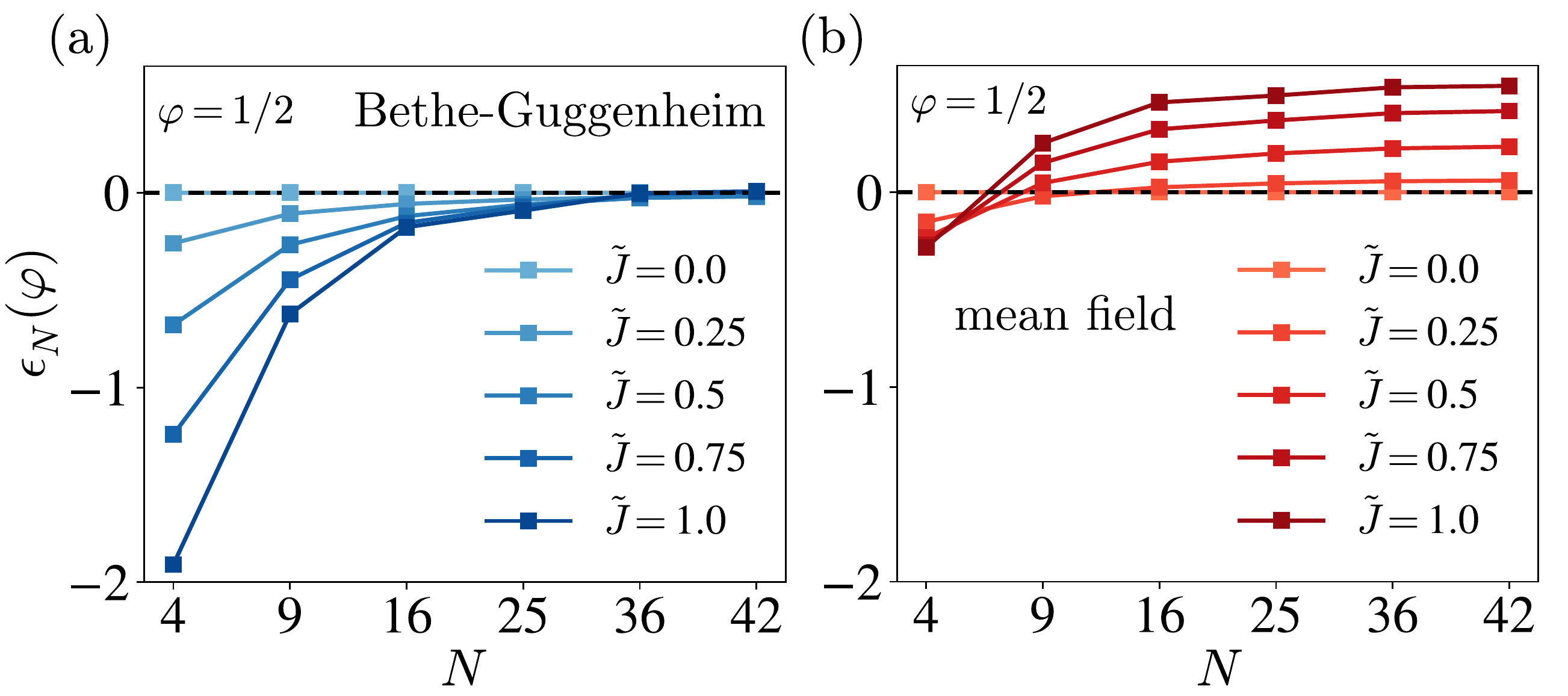}
    \caption{\textbf{Relative error of the approximate free energy density} for $\tmu=\thh=0$ obtained by the (a) Bethe-Guggenheim and (b) mean field approximation as a function of the system size. Up to $N=36$ each point corresponds to the relative error for a square lattice of size $\sqrt{N}\times\sqrt{N}$  with free boundary conditions and $\varphi=1/2$ closed bonds. For $N=42$ a rectangular lattice of size $6\times7$ with free boundary conditions is considered.}
    \label{fig A1}
\end{figure}
In Fig.~\ref{fig A1} we compare the accuracy of the
Bethe-Guggenheim (a) and mean field (b) approximations by means of the
relative error $\epsilon^{\rm BG,MF}_{N}\left(\varphi\right) \equiv 1-\tilde{f}_{N}^{\rm BG/MF}\left(\varphi\right)/\tilde{f}_{N}\left(\varphi\right)$
between the exact and approximate free energy density
for a two-dimensional lattice with $\varphi=1/2$ closed bonds as a
function of the system size $N$. We find that the Bethe-Guggenheim
approximation converges to the exact free energy density with
increasing $N$, regardless of the coupling strength
$\tilde{J}$. Conversely, the mean field approximation diverges with
both, increasing system size $N$ and coupling strength $\tj$. 
\subsection{Phase diagram}\label{Appendix B.2}
\subsubsection{Independence of the binodal and spinodal line on $\tmu$}
Let us write $\tilde{\rm f}\left(\varphi\right)=\tilde{\mu} \varphi + \tilde{\rm g}\left(\varphi\right)$, where $\tilde{\rm g}\left(\varphi\right)$ is the remainder of the free energy density after leaving out all linear terms in $\varphi$. Plugging this expression into the binodal line equations given by Eq.~\eqref{binodal} gives
\begin{eqnarray}
    \tilde{\mu}+\tilde{\rm
      g}'\left(\varphi_{l}\right)&=&\tilde{\mu}+\tilde{\rm
      g}'\left(\varphi_{g}\right),\nonumber\\
    \tilde{\mu} + \frac{\tilde{\rm g}\left(\varphi_{l}\right)-\tilde{\rm g}\left(\varphi_{g}\right)}{\varphi_{l}-\varphi_{g}}&=&\tilde{\mu}+\tilde{\rm g}'\left(\varphi_{l}\right).    
\end{eqnarray}
Clearly $\tilde{\mu}$ cancels and therefore does not affect the
binodal line. 

The spinodal line -- also known as the stability boundary -- denotes
the boundary between the metastable and unstable state, and is given
by $\rtf''(\vphi)=0$. The spinodal line is as well not affected by $\tmu$, since the second derivative of the linear
term vanishes.
\subsubsection{Binodal line for zero force}
At zero force the binodal line equations simplify to
$\rtf'(\vphi)|^{\thh=0}_{\tmu=0}=0$. Our aim is to solve this
  equation for $\Omega_{\varphi}$ (defined as Eq.~\eqref{xx} in the thermodynamic limit) in Eq.~\eqref{TD}, and then use the inverse relation
\begin{equation}
    \tilde{J}\left(\varphi,\Omega_{\varphi}\right) = \frac{1}{4}\ln{\left(\frac{\left(\varphi-\Omega_{\varphi}\right)\left(1-\varphi-\Omega_{\varphi}\right)}{\Omega_{\varphi}^{2}}\right)},
    \label{general solution y}
\end{equation}
to obtain the binodal line. Notice that it follows from Eq.~\eqref{xx}
that $\Omega_{\varphi}\geq0$ for $0\leq \varphi \leq1$, and
this constraint has to be obeyed by the implicit solution for
$\Omega_{\varphi}$. Evaluating
the total derivative with
respect to $\varphi$ of Eq.~\eqref{TD} gives
\begin{equation}
    \rtf'(\vphi,\Omega_{\vphi})|_{\tj, \tmu = 0}{=}\partial_\vphi \rtf(\vphi,\Omega_{\vphi})|_{\tj, \tmu = 0}+\partial_{\Omega_\vphi} \rtf(\vphi,\Omega_{\vphi})|_{\tj, \tmu = 0}\cdot
    \Omega'_{\varphi},
 \label{chain}   
\end{equation}
where $\Omega'_{\varphi}=\partial_{\varphi}\Omega_{\varphi}|_{\tj}$. Since $\Omega_{\varphi}$ was obtained by solving $\partial_{\Omega_\vphi}
  \rtf(\vphi,\Omega_{\vphi})|_{\tj, \tmu=0}=0$ the second term in Eq.~\eqref{chain} vanishes while the first term yields
  \begin{equation}
\frac{2\tilde{h}}{\left(1+\varphi\right)^{2}}+\left(1-\bar{z}\right)\ln{\left(\frac{\varphi}{1-\varphi}\right)}+\frac{\bar{z}}{2}\ln{\left(\frac{\varphi-\Omega_{\varphi}}{1-\varphi-\Omega_{\varphi}}\right)}.
    \label{df2}  
  \end{equation}
  Setting $\thh=0$ in Eq.~\eqref{df2}, and introducing
  $\chi_{\varphi}\equiv\varphi/(1-\varphi)$ and
  $\alpha\equiv(\bar{z}-1)/\bar{z}$ we find the following solution for
  $\Omega_{\varphi}$ evaluated on the zero-force binodal line
\begin{equation}
     \Omega_{\varphi}(\tj_{\rm b}(\varphi)|^{\thh=0}_{\tmu=0})=\frac{\chi_{\varphi}}{1+\chi_{\varphi}}\frac{1-\chi_{\varphi}^{2\alpha-1}}{1-\chi_{\varphi}^{2\alpha}}.
    \label{Implicit solution binodal line Bethe-Guggenheim}
\end{equation}
 Plugging Eq.~\eqref{Implicit solution binodal line Bethe-Guggenheim}
 into Eq.~\eqref{general solution y} yields Eq.~\eqref{bin} for
 the zero-force binodal line, which was also reported in
 \cite{de2013liquid, fowler_statistical_1939,
   guggenheim1952mixtures}. For $\tilde{h}\neq0$ the binodal line can be obtained by solving Eq.~\eqref{binodal} numerically.
\subsubsection{Spinodal line}
To determine the spinodal line we calculate the second
derivative of the Bethe-Guggenheim free energy density
\begin{widetext}
\begin{eqnarray}
   \rtf''(\vphi,\Omega_{\vphi})=
    -\frac{4\tilde{h}}{\left(1+\varphi\right)^{3}}+\frac{1-\bar{z}}{\varphi\left(1-\varphi\right)}+\frac{\bar{z}}{2}\frac{\left(1-2\Omega_{\varphi}\right)+2\left(2\varphi-1\right)\Omega_{\varphi}^{\prime}}{\left(\varphi-\Omega_{\varphi}\right)\left(1-\varphi-\Omega_{\varphi}\right)}+\frac{\bar{z}}{2}\frac{\left[2\varphi\left(1-\varphi\right)-\Omega_{\varphi}\right]\Omega_{\varphi}'^{2}}{\Omega_{\varphi}\left(\varphi-\Omega_{\varphi}\right)\left(1-\varphi-\Omega_{\varphi}\right)}.
    \label{ddf}
\end{eqnarray}
Eq.~\eqref{ddf} contains $\Omega'_{\vphi}$, which we want to express in terms of $\Omega_{\vphi}$. Therefore we use Eq.~\eqref{general solution y} and differentiate both sides with respect to $\vphi$ for fixed $\tilde{J}$, yielding
\begin{equation}
    \Omega_{\varphi}' = \frac{\Omega_{\varphi}\left(1-2\varphi\right)}{2\varphi\left(1-\varphi\right)-\Omega_{\varphi}}.
    \label{dy}
\end{equation}
Plugging Eq.~\eqref{dy} into Eq.~\eqref{ddf} gives
\begin{equation}
   \rtf''(\vphi,\Omega_{\vphi})=-\frac{4\tilde{h}}{\left(1+\varphi\right)^{3}}+\frac{1-\bar{z}}{\varphi\left(1-\varphi\right)}+\frac{\bar{z}}{2\varphi\left(1-\varphi\right)-\Omega_{\varphi}}=0.
    \label{Spinodal line equation}
\end{equation}
Solving Eq.~\eqref{Spinodal line equation} for $\Omega_{\varphi}$
yields $\Phi(\vphi,\thh)$ in Eq.~\eqref{spin}, which
  must be non-negative thus implying that Eq.~\eqref{spin} is valid for $(2-\bar{z})/8\tilde{h}\le \varphi(1-\varphi)/(1+\varphi)^3\le(1-\bar{z})/4\tilde{h}$.
\subsubsection{Statical critical point}
We derive the Bethe-Guggenheim statical critical point in the form of
a convergent Newton series \cite{Godec_2016, Hartich_2019, Hartich_2019_2}.
We determine the statical critical point from
 \begin{equation}
 \rtf'''(\vphi,\Omega_{\vphi})=\frac{12\tilde{h}}{\left(1+\varphi\right)^{4}}+\frac{\left(1-\bar{z}\right)\left(2\varphi-1\right)}{\varphi^{2}\left(1-\varphi\right)^{2}}-\frac{\bar{z}\left[2\left(1-2\varphi\right)-\Omega_{\varphi}'\right]}{\left(2\varphi\left(1-\varphi\right)-\Omega_{\varphi}\right)^{2}}=0.
 \end{equation}
Using Eq.~\eqref{dy} for $\Omega_{\varphi}'$ we get
\begin{equation}
    \frac{12\tilde{h}}{\left(1+\varphi\right)^{4}} +\frac{\left(1-\bar{z}\right)\left(2\varphi-1\right)}{\varphi^{2}\left(1-\varphi\right)^{2}}+\frac{\bar{z}\left(2\varphi-1\right)\left[4\varphi\left(1-\varphi\right)-3\Omega_{\varphi}\right]}{\left(2\varphi\left(1-\varphi\right)-\Omega_{\varphi}\right)^{3}}=0.
    \label{Critical point quasi chemical approximation}
\end{equation}
To simplify Eq.~\eqref{Critical point quasi chemical approximation}
further we introduce the auxiliary parameter $\alpha_{\tilde{h}}\equiv
12\tilde{h}$  and use the fact that the critical point lies on the
spinodal line,  which allows us to use Eq.~\eqref{spinA} for $\Omega_{\varphi}$ and leads to
\begin{equation}
    g\left(\varphi\right)\equiv\alpha_{\tilde{h}}\varphi^{2}\left(1-\varphi\right)^{2}+\left(1+\varphi\right)^{4}\left(2\varphi-1\right)\gamma_{\bar{z}}\left(\varphi\right)=0,
    \label{Critical point Bethe-Guggenheim}
\end{equation}
with
\begin{equation}
    \gamma_{\bar{z}}\left(\varphi\right) \equiv 1-\frac{1}{\bar{z}}\left(3-\frac{2}{\bar{z}}\right)+\frac{2\alpha_{\tilde{h}}}{\bar{z}}\left(1-\frac{1}{\bar{z}}\right)\Lambda\left(\varphi\right)-\frac{\alpha_{\tilde{h}}^{2}}{3\bar{z}}\left(1-\frac{2}{\bar{z}}\right)\Lambda^{2}\left(\varphi\right)-\frac{2\alpha_{\tilde{h}}^{3}}{27\bar{z}^2}\Lambda^{3}\left(\varphi\right),
    \label{gamma}
\end{equation}
\end{widetext}
and $\Lambda\left(\varphi\right) \equiv
\varphi(1-\varphi)/(1+\varphi)^{3}$. For $\tilde{h}=0$ the solution of Eq.~\eqref{Critical point
  Bethe-Guggenheim} is given by $\varphi^{\rm s}_{\rm crit,BG}=1/2$,
and the corresponding statical critical point is given by
$\tilde{J}^{\rm s}_{\rm
  crit,BG}=\ln{\left(\bar{z}/(\bar{z}-2)\right)}/\bar{2}$.

For
non-zero force we solve Eq.~\eqref{Critical point Bethe-Guggenheim} by
means of a ``quadratic'' Newton series as explained in more detail in
Appendix~\ref{Appendix D}. The
main result for the statical critical fraction obtained by the quadratic Newton series reads 
\begin{eqnarray}
 \!\!\!\!\!\!\!\!\!\varphi^{\rm s}_{\rm crit, BG} &\approx&
    \frac{1}{2}-\frac{g'\left(\frac{1}{2}\right)}{g''\left(\frac{1}{2}\right)}+\left[\frac{g'\left(\frac{1}{2}\right)^{2}}{g''\left(\frac{1}{2}\right)^2}-2\frac{g\left(\frac{1}{2}\right)}{g''\left(\frac{1}{2}\right)}\right]^{1/2}\nonumber\\
    &=&\frac{1}{2}-\frac{3}{2^4}\frac{\delta_{\bar{z}}(\tilde{h})}{\nu_{\bar{z}}(\tilde{h})}+\frac{3}{2^4}\left[\frac{\delta^2_{\bar{z}}(\tilde{h})}{\nu^2_{\bar{z}}(\tilde{h})}-\frac{2^6}{3^4}\frac{\bar{z}^2\tilde{h}}{\nu_{\bar{z}(\tilde{h})}}\right]^{1/2},
    \label{Newton series truncated2}
\end{eqnarray}
where $g'\left(\varphi\right)=\partial_{\varphi}g(\varphi)$, $g''\left(\varphi\right)=\partial^{2}_{\varphi}g(\varphi)$, and the auxiliary functions $ \delta_{\bar{z}}(\tilde{h})$ and $\nu_{\bar{z}}(\tilde{h})$ are defined as
\begin{equation}
    \delta_{\bar{z}}(\tilde{h})\equiv(\bar{z}-1-\frac{2^6}{3^5}\tilde{h}^2)(\bar{z}-2+\frac{2^4}{3^2}\tilde{h})+\frac{2^{13}}{3^9}\tilde{h}^3,
\end{equation}
and
\begin{equation}
    \nu_{\bar{z}}(\tilde{h})\equiv(\bar{z}-1+\frac{2^5}{3^5}\tilde{h}^2)(\bar{z}-2+\frac{2^2}{3^2}\tilde{h})-\frac{2}{3^2}\bar{z}^{2}\tilde{h}+\frac{2^7}{3^9}\tilde{h}^3,
\end{equation}
respectively. 

The statical critical coupling $\tj^{\rm s}_{\rm crit}$ is obtained by
inserting Eq.~\eqref{Newton series truncated2} into
Eq.~\eqref{spin}, and the result is depicted by the gradient line
Fig.~\ref{fig 5} in the main text, where the black symbols
represent the fully converged Newton's series \eqref{Newton series},
as well as in Fig.~\ref{fig A2} where the gradient line depicts the fully converged Newton's series. 
\setcounter{equation}{0}
\setcounter{figure}{0}
\renewcommand{\thefigure}{C\arabic{figure}}
\renewcommand{\theequation}{C\arabic{equation}}
\section{Mean field approximation}\label{Appendix C}
\subsection{Partition function}\label{Appendix C.1}
Within the mean field approximation the partition function reads
\begin{equation}
    Z_{k}^{\rm MF}=\binom{N}{k}\e{\tilde{J}\bar{z}(N/2-2X^{*}_{k})},
\end{equation}
with $X^{*}_{k}\equiv k\left(N-k\right)/N$, such that the corresponding free energy density in
the thermodynamic limit attains the form 
\begin{equation}
    \tilde{\rm f}^{\rm MF}(\vphi)=-\tmu\vphi + \frac{2\thh\vphi}{1+\vphi}+\frac{1}{2}\tilde{J}\bar{z}[4\varphi(1-\varphi)-1]+\Xi_\Delta(\vphi),
    \label{TD MF}
\end{equation}
where $\Xi_\Delta(x)\equiv\Xi(x)+\Xi(1-x)$. 
\subsection{Phase diagram}\label{Appendix C.2}
We now evaluate the binodal and spinodal line, as well as the statical
critical point within the mean field approximation. The corresponding
exact solutions for the Bethe-Guggenheim approximation are given by
Eqs.~\eqref{bin}-\eqref{smallforce1}. 
\subsubsection{Binodal line for zero force}
In the absence of an external field, $\thh=0$, the binodal line is given by
\begin{equation}
    \tilde{J}_{\rm b, MF}(\varphi)|_{\thh=0} = \frac{1}{2\bar{z}(2\varphi-1)}\ln{\left(\frac{\varphi}{1-\varphi}\right)}.
    \label{bin MF}
\end{equation}
This solution is well known and has been reported in the literature
extensively \cite{doi2013soft, de2013liquid, fowler_statistical_1939,
  guggenheim1952mixtures}. For $\thh\neq0$ we solve for the binodal
line numerically.
\subsubsection{Spinodal line}
The spinodal line is in turn given by the solution of $\rtf^{{\rm
    MF}''}(\vphi)=0$ and reads
\begin{equation}
    \tilde{J}_{\rm s, MF}(\varphi) = \frac{1}{4\bar{z}}\left(\frac{1}{\varphi\left(1-\varphi\right)}-\frac{4\tilde{h}}{\left(1+\varphi\right)^3}\right).
    \label{spin MF}
\end{equation}
\begin{figure*}[t!]
    \includegraphics[width = 1\textwidth]{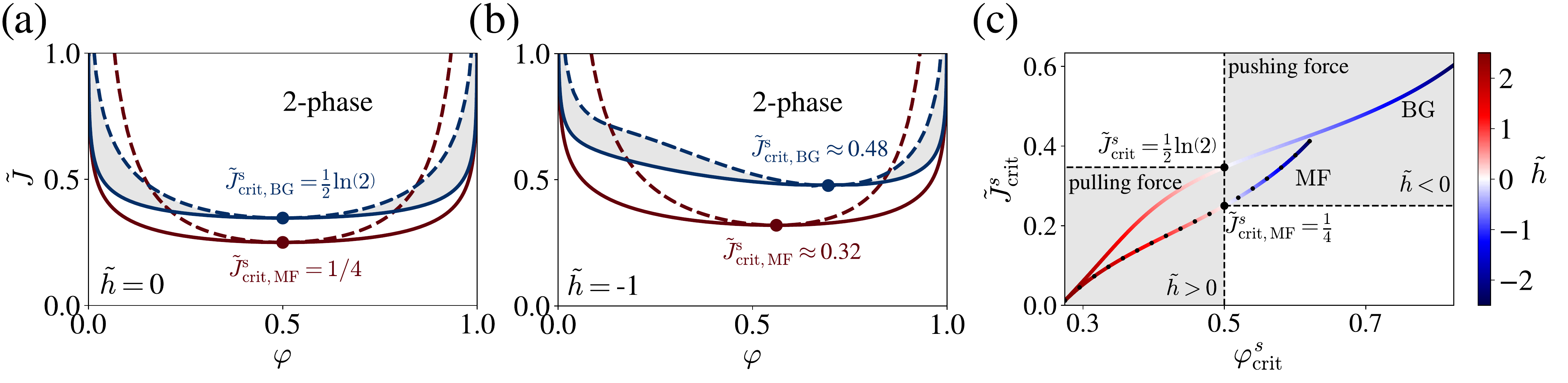}
    \caption{\textbf{Comparison of the Bethe-Guggenheim (BG) and mean field (MF) binodal line, spinodal line, and statical critical point.} (a-b) Phase diagram for (a) zero force and (b) pushing force $\thh = -1$; the full and dashes lines depict the binodal and spinodal line, respectively. Red lines correspond to the mean field approximation obtained with Eqs.~\eqref{bin MF} (for zero force) and \eqref{spin MF}. Blue lines correspond to the Bethe-Guggenheim approximation in Eqs.~\eqref{bin} (for zero force) and \eqref{spin}. The shaded area depicts the region where the system is metastable for the Bethe-Guggenheim approximation. The blue and red circle depict the statical critical point $(\varphi^{\rm s}_{\rm crit}, \tj^{\rm s}_{\rm crit})$ for the mean field and Bethe-Guggenheim approximation, respectively. (c) The statical critical point as a function of the force $\thh$. For the Bethe-Guggenheim approximation is the result obtained with a converged Newton's series (see Eq.~\eqref{Newton series}), whereas the mean field result is obtained by considering the first ten terms of Eq.~\eqref{Critical coordinate mf}.}
    \label{fig A2}
\end{figure*}
\subsubsection{Statical critical point}
The statical critical point is given by the solution of $\rtf'''^{{\rm MF}}(\vphi)=0$, which after introducing the parameter $\alpha_{\tilde{h}}\equiv 12\tilde{h}$ translates into solving the algebraic equation
\begin{equation}
    \alpha_{\tilde{h}}\varphi^{2}\left(1-\varphi\right)^{2}+\left(1+\varphi\right)^{4}\left(2\varphi-1\right)=0.
    \label{Critical point mean field rewritten}
\end{equation}
Notice that $\bar{z}$, the average coordination number, does not enter
Eq.~\eqref{Critical point mean field rewritten}. When $\tilde{h}=0$
the solution is $\varphi^{\rm s}_{\rm crit,MF}=1/2$, and the
corresponding statical critical point is given by $\tilde{J}^{\rm
  s}_{\rm crit,MF}=\bar{z}^{-1}$. To solve Eq.~\eqref{Critical point
  mean field rewritten} for non-zero force we first note that
$0\leq\varphi^{\rm s}_{\rm crit, MF}\leq 1$, and therefore we can divide Eq.~\eqref{Critical point mean field rewritten} by $\left(1+\varphi\right)^{4}$. Upon introducing the variable $w=2\varphi-1$ we get the equation
\begin{equation}
    \frac{w}{f\left(w\right)}=-\alpha_{\tilde{h}},
    \label{Critical point mean field rewritten II}
\end{equation}
with $f\left(w\right)=\left(w+1\right)^{2}\left(w-1\right)^{2}\left(w+3\right)^{-4}$. Now we recall the \emph{Lagrange inversion theorem}: Let $f(w)$ be analytic in some neighborhood of the point $w=0$
(of the complex plane) with $f(0)\ne 0$ and let it satisfy the equation
 \begin{equation}
   \frac{w}{f(w)}=\alpha.
\label{Lagrange}   
\end{equation}
Then $\exists a,b\in\mathbbm{R}^+$ such that for $|\alpha| < a$ Eq.~\eqref{Lagrange} has only a single solution in the domain $|w| <
b$. According to the Lagrange-B\"urmann formula this
unique solution is an analytical function of $\alpha$ given by
\begin{equation}
    w=\sum_{k=1}^{\infty}\frac{\alpha^k}{k!}\left[\frac{d^{k-1}}{dw^{k-1}}f(w)^k\right]_{w=0}.
    \label{Burmann}  
\end{equation}  
Notice that Eq.~\eqref{Critical point mean field rewritten II} has the form of Eq.~\eqref{Lagrange}, and therefore we can use Eq.~\eqref{Burmann} to obtain $\varphi^{\rm s}_{\rm crit, MF}$. To evaluate the derivative inside Eq.~\eqref{Burmann} it is convenient to write $f(w)^k=g(w)h(w)$, with $g(w)\equiv(w+1)^{2k}(w-1)^{2k}=(w^2-1)^{2k}$ and $h(w)\equiv(w+3)^{-4k}$. Using the fact that
\begin{equation}
    \frac{d^{n}}{dw^{n}}g(w)h(w)|_{w=0}=\sum_{k=0}^n\binom{n}{k}
    g^{(n-k)}(0)h^{(k)}(0),
\end{equation}
as well as
\begin{equation}
    \frac{d^{n}}{dw^{n}}(w^2-1)^{2k}|_{w=0}=\cos{\left(\frac{\pi n}{2}\right)}n!\binom{2k}{n/2}
\end{equation}
and
\begin{equation}
    \frac{d^{n}}{dw^{n}}\left(w+3\right)^{-4k}|_{w=0}=(-1)^{n}3^{-4k-n}\frac{(4k+n-1)!}{(4k-1)!},
\end{equation}
we find
\begin{eqnarray}
 \frac{d^{k-1}}{dw^{k-1}}f(w)^k&&|_{w=0}=\frac{\left(-1\right)^{k-1}3^{-5k+1}\left(k-1\right)!\left(2k\right)!}{\left(4k-1\right)!}\nonumber\\
\times&&  \sum_{m=0}^{\lfloor\frac{k-1}{2}\rfloor}\frac{\left(-9\right)^{m}\left(5k-2(1+m)\right)!}{m!\left(2k-m\right)!\left(k-1-2m\right)!}.
    \label{Derivative explicit}
\end{eqnarray}
Plugging Eq.~\eqref{Derivative explicit} into Eq.~\eqref{Burmann} and using the relation $\varphi=(1+w)/2$
we find $\vphi^s_{\rm
  crit,MF}=\vphi^{s,0}_{\rm
  crit,MF}-6\sum_{k=1}^{\infty}\delta\vphi_k(\thh)$ with
\begin{equation}
    \delta\vphi_k(\thh) = (4\tilde{h})^k
    \frac{\left(2k\right)!}{\left(4k\right)!}\sum\limits_{m=0}^{\lfloor\frac{k-1}{2}\rfloor}\frac{\left(-9\right)^{m-2k}\left(5k-2(1+m)\right)!}{m!\left(2k-m\right)!\left(k-1-2m\right)!}
    \label{Critical coordinate mf}.
\end{equation}
The statical critical coupling for non-zero force, $\tj^{\rm s}_{\rm
crit, MF}(\thh)$, is obtained by plugging $\vphi^{s}_{\rm crit,MF}$ into Eq.~\eqref{spin MF}.

In Fig.~\ref{fig A2}
we plot the statical critical point for the Bethe-Guggenheim and mean
field approximation as a function of the force $\thh$. For large
pulling force ($\thh \geq 2$) the statical critical point is pushed
towards lower values of $\tj$, and as a consequence the
Bethe-Guggenheim and mean field solution start to coincide. For other
values of $\thh$, however, the Bethe-Guggenheim and mean field solutions
disagree strongly, in particular for weak forces $|\thh|\to 0$. 

To compare the critical points for weak pulling/pushing forces in more
detail we inspect the perturbation series given in
Eqs.~\eqref{smallforce} and \eqref{smallforce1} and compare it with the first two terms of Eq.~\eqref{Critical coordinate mf}. Defining $\tj^s_{\rm crit,MF}= \tj^{s,0}_{\rm crit,MF}-\delta \tj^s_{\rm crit,MF}(\thh)+\mathcal{O}(\thh^3)$ and $\vphi^s_{\rm
  crit,MF}=\vphi^{s,0}_{\rm crit,MF}-\delta\vphi_{\rm crit,MF}(\thh)+\mathcal{O}(\thh^3)$, we get
\begin{equation}
    \delta \tilde{J}^{\rm s}_{\rm crit,MF}(\thh) = \frac{8}{27}\frac{1}{\tilde{z}}\left(\tilde{h}+\frac{2}{27}\tilde{h}^{2}\right),
    \label{smallforceMF}
\end{equation}
and
\begin{equation}
    \delta\varphi^{\rm s}_{\rm crit,MF}(\thh) = \frac{2}{27} \left(\thh+\frac{16}{81}\thh^{2}\right).
   \label{smallforce1MF}
\end{equation}
Interestingly, whereas the Bethe-Guggenheim critical point depends on $\bar{z}$ (see Eq.~\eqref{smallforce1}), the mean field result in Eq.~\eqref{smallforce1MF} does not. In the limit $\bar{z}\rightarrow\infty$ we find that the Bethe-Guggenheim statical critical point converges to the mean field solution, which is to be expected.
\setcounter{equation}{0}
\setcounter{figure}{0}
\renewcommand{\thefigure}{D\arabic{figure}}
\renewcommand{\theequation}{D\arabic{equation}}
\section{Equation of state in the thermodynamic limit}\label{Appendix D}
Here we derive the equation of state in the thermodynamic limit using the
saddle-point technique, i.e.
\begin{multline}
    \langle\varphi\rangle_{\rm
      TD}\equiv\lim_{N\to\infty}\frac{\displaystyle{\int_0^1 \varphi\e{-N\tilde{\rm f}(\varphi)}d\varphi} }{\displaystyle{\int_0^1
    \e{-N\tilde{\rm f}(\varphi)}d\varphi} }\simeq\\ \lim_{N\to\infty}\frac{\displaystyle{\int_0^1
    \sum\limits_{i=1}^{M}\varphi^{0}_{i}\e{-N\tilde{\rm f}''(\varphi^{0}_i)(\varphi-\varphi^{0}_i)^2}d\varphi}}{\displaystyle{\int_0^1
    \sum\limits_{i=1}^{M}\e{-N\tilde{\rm f}''(\varphi^{0}_i)(\varphi-\varphi^{0}_i)^2}d\varphi} }=\sum\limits_{i=1}^{M}c_{i}\varphi^{0}_{i},
    \label{saddle point technique}
\end{multline}
with
\begin{eqnarray}
    c_{j}&=&\lim_{N\to\infty}\frac{\displaystyle{\int_0^1
    \e{-N\tilde{\rm f}''(\varphi^{0}_j)(\varphi-\varphi^{0}_{j})^2}d\varphi}}{\displaystyle{\int_0^1
    \sum\limits_{i=1}^{M}\e{-N\tilde{\rm
        f}''(\varphi^{0}_{i})(\varphi-\varphi^{0}_{i})^2}d\varphi}}\nonumber\\
    &\simeq&
    \left(1+\sum_{i=1|i\neq j}^{M}\sqrt{\tilde{\rm f}''(\varphi^0_j)/\tilde{\rm f}''(\varphi^0_i)}\right)^{-1},
    \label{prefactor}
\end{eqnarray}
and $\varphi^0_1, \varphi^0_{2}, ..., \varphi^0_{M}$ denote the
locations of the local minima of the Bethe-Guggenheim free energy
density ${\rm \tf}(\varphi)\equiv{\rm \tf^{BG}}(\varphi)$ in
Eq.~\eqref{TD}. The idea behind Eq.~\eqref{saddle point technique} is
that in the large $N$ limit we expect the integral over $\varphi$ to
be dominated by the immediate neighborhood of the local minima of
$\tilde{\rm f}\left(\varphi\right)$. We may therefore approximate the
exponent by its Taylor expansion around these extremal points. In general
special care has to be taken when one of the global minima lies at the
boundary of the integration interval \cite{bender2013advanced}, which
turns out not to be the case here.

The locations of the local minima, maxima, and saddle points of the Bethe-Guggenheim free energy density are given by the solution of $\rtf'^{{\rm BG}}(\vphi)=0$. Notice
that here we do \emph{not} set  the intrinsic binding-affinity
$\tilde{\mu}$ to zero, since we are interested in the stationary points and not the binodal line. We solve the former equation for $\Omega_\vphi$ and substitute the solution into Eq.~\eqref{general solution y} which gives
\begin{equation}
    \tilde{J} = \frac{1}{2}\ln{\left(\frac{1-\chi_{\varphi}}{c_{\tmu,\thh}^{-1}(\chi_{\vphi})\chi_{\varphi}^{1-\alpha}-c_{\tmu,\thh}(\chi_{\vphi})\chi_{\varphi}^{\alpha}}\right)},
    \label{Implicit solution chi}
\end{equation}
where $\chi_{\varphi}\equiv\varphi/(1-\varphi)$, $\alpha\equiv
(\bar{z}-1)/\bar{z}$, and
\begin{equation}
    c_{\tmu,\thh}(\chi_{\vphi})\equiv\e{\frac{\tilde{\mu}}{\bar{z}}}\e{-\frac{2}{\bar{z}}\left(\frac{1+\chi_{\vphi}}{1+2\chi_{\vphi}}\right)^{2}\tilde{h}}.
    \label{c}
\end{equation}
Rewriting Eq.~\eqref{Implicit solution chi} gives
\begin{equation}
    \chi_{\varphi}-\e{2\tilde{J}}\left(c_{\tmu,\thh}(\chi_{\vphi})\chi^{\alpha}_{\varphi}-c^{-1}_{\tmu,\thh}(\chi_{\vphi})\chi_{\varphi}^{1-\alpha}\right)-1=0.
    \label{Implicit solution chi 2}
\end{equation}
For a
two-dimensional square lattice in the thermodynamic limit we have
$\bar{z}=4$ and so $\alpha = 3/4$. Upon introducing the
auxiliary variable $\xi\equiv\chi_{\varphi}^{1/4}$ we obtain the transcendental equation
\begin{equation}
    g(\xi)\equiv\xi^{4}-\e{2\tilde{J}}\left(c_{\tmu,\thh}(\xi)\xi- c_{\tmu,\thh}^{-1}(\xi)\xi^{-1}\right)\xi^{2}-1=0.
    \label{transcendental equation}
\end{equation}
To solve for the roots of
$g(\xi)$ we first consider the force-free scenario $\thh = 0$ and
afterwards solve for the general case. 
\subsection{Zero force}\label{Appendix D.1}
For zero force $g(\xi)$ reduces to a quartic in $\xi$. The roots of a
general quartic equation are known and are given by
\begin{subequations}
\begin{equation}
    \xi_{1,2}=\frac{\mathcal{Y}_+}{4}-S\pm\frac{1}{2}\sqrt{\frac{3}{4}\mathcal{Y}_+^2+\frac{\mathcal{Y}_--\frac{1}{8}\mathcal{Y}_+^3}{S}-4S^{2}},
    \label{solution 1-2}
\end{equation}
\begin{equation}
   \xi_{3,4}=\frac{\mathcal{Y}_+}{4}+S\pm\frac{1}{2}\sqrt{\frac{3}{4}\mathcal{Y}_+^2-\frac{\mathcal{Y}_--\frac{1}{8}\mathcal{Y}_+^3}{S}-4S^{2}},
    \label{solution 3-4}
\end{equation}
\end{subequations}
where $\mathcal{Y}_\pm=\e{2\tj\pm\tmu/4}$ and 
\begin{equation}
    S = \frac{1}{2}\sqrt{\frac{1}{4}\mathcal{Y}_+^2+\frac{1}{3}\left(W+\frac{\Delta_{0}}{W}\right)},
    \label{S}
\end{equation}
\begin{equation}
    W = \left(\frac{\Delta_{1}+\sqrt{\Delta_{1}^{2}-4\Delta_{0}^{3}}}{2}\right)^{1/3},
\end{equation}
\begin{equation}
    \Delta_{0}=3\e{4\tilde{J}}-12,\,\,\, \Delta_{1}=-54\e{4\tilde{J}}\Delta_{\tmu},\,\,\,\Delta_{\tmu}=\sinh(\tmu/2).
\end{equation}
Fig.~\ref{fig A3} depicts the four solutions in
Eqs.~\eqref{solution 1-2} and \eqref{solution 3-4} as a function of
$\tj$ for various values of $\tmu$.

To determine the local
minima we must analyze the properties of these roots starting with the sign of the discriminant, given by
\begin{equation}
    \Delta = \frac{1}{27}\left(4\Delta_{0}^{3}-\Delta_{1}^{2}\right)=4\left(\e{4\tilde{J}}-4\right)^{3}-108\e{8\tilde{J}}\Delta_{\tmu}^{2}.
    \label{Discriminant}
\end{equation}
For $\Delta<0$ there are two distinct real roots and two complex
conjugate roots, whereas for $\Delta>0$ there are either four real
roots or four imaginary roots, where the former scenario applies here. The discriminant is zero at the critical coupling value
\begin{equation}
    \tilde{J}_{\Delta = 0} = \frac{1}{4}\ln{\left(\frac{3\Delta_{\tmu}^{2}\left(9\Delta_{\tmu}^{2}+8\right)}{\Phi^{1/3}}+3\Phi^{1/3}+9\Delta_{\tmu}^{2}+4\right)},
    \label{critical coupling}
\end{equation}
with
\begin{equation}
    \Phi=27\Delta_{\tmu}^{6}+36\Delta_{\tmu}^{4}+8\Delta_{\tmu}^{2}[1+(1+\Delta_{\tmu}^{2})^{1/2}].
\end{equation}
Increasing the coupling strength above $\tilde{J}_{\Delta = 0}$ gives
rise to a local maximum in the free energy landscape, its position
being $\xi^{4}_{4}/(1+\xi^{4}_{4})$ (see purple line in Fig.~\ref{fig A3}). 
\subsubsection{Zero intrinsic binding-affinity}
For zero intrinsic binding-affinity we note that $\tj_{\Delta=0}$
coincides with the statical critical point $\tilde{J}^{s}_{\rm
  crit}=\ln{(2)}/2$. In this limit the four solutions
~\eqref{solution 1-2} and \eqref{solution 3-4} simplify substantially,
and the corresponding locations of the local minima -- which are also global minima -- are given by (see Fig.~\ref{fig A3}a)
\begin{equation}
    \varphi^{0}_{\rm 1,2} |_{\tmu=0}^{\thh=0}=
    \begin{dcases*}
        \frac{1}{2} &, $0\leq\tilde{J} \leq  \tilde{J}^{s}_{\rm crit}$\\ 
        \frac{1}{2}\left[1\pm\frac{\e{2\tilde{J}}\sqrt{\e{4 \tilde{J}}-4}}{(\e{4\tilde{J}}-2)}\right] &, $\tilde{J} \geq  \tilde{J}^{s}_{\rm crit}$
    \end{dcases*}.
    \label{bare field solution}
\end{equation}
For $\tilde{J} \leq  \tilde{J}^{s}_{\rm crit}$ there is a single
global minimum in the free energy landscape, and therefore the weight
$c_{1}$ given by Eq.~\eqref{prefactor} becomes unity. For $\tilde{J} >
\tilde{J}^{s}_{\rm crit}$ the global minima are two-fold degenerate
and located equidistantly from the local maximum at
$\varphi=1/2$. Since both global minima have the same curvature, we
find that the weights are given by $c_{1,2}=1/2$. Combining these
results we obtain the average fraction of closed bonds in the
thermodynamic limit for zero intrinsic binding-affinity in the form
\begin{equation}
    \langle \varphi \rangle_{\rm TD}|_{\tmu=0}^{\thh=0}=\frac{1}{2},
    \label{e.o.s. bare field}
\end{equation}
which was to be expected since the coupling $\tj$ does not favor bonds to be open nor closed. 
\subsubsection{Non-zero intrinsic bining-affinity}
\begin{figure*}
    \includegraphics[width = \textwidth]{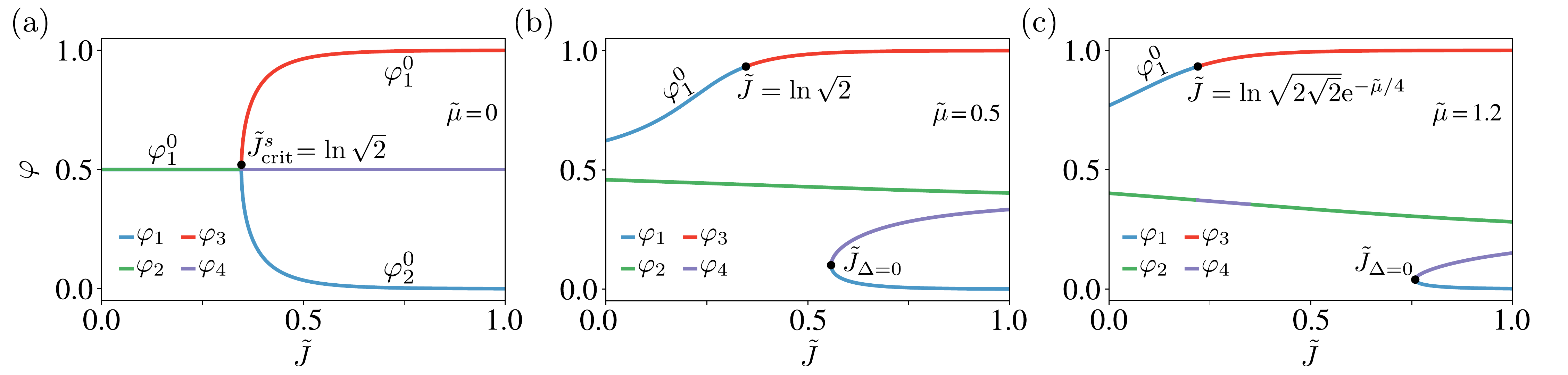}
    \caption{\textbf{Stationary points of the Bethe-Guggenheim free
        energy density for zero force}, given by Eqs.~\eqref{solution
        1-2} and \eqref{solution 3-4}, as a function of the coupling
      strength $\tj$. From left to right we consider increasing values
      of the intrinsic binding-affinity $\tmu$. The variable
      $\varphi^{0}_{1,2}$ indicates the location of the global minima,
      which is given by Eq.~\eqref{bare field solution} for zero
      intrinsic binding-affinity (a), and given by Eq.~\eqref{solution
        global minima} for non-zero intrinsic binding-affinity (b and
      c). The coupling value $\tilde{J}_{\Delta=0}$, given by
      Eq.~\eqref{critical coupling}, indicates the coupling strength
      where all four solutions become real, and denotes the first appearance
      of a local maximum given by $\varphi_{4}$.}
    \label{fig A3}
\end{figure*}

For non-zero intrinsic binding-affinity, $\tmu\ne 0$, the free energy landscape is tilted, resulting in a unique global minimum with corresponding weight $c_{1}=1$. As a result, the average fraction of closed bonds, which is dominated by this minimum $\varphi^{0}_{1}$, is given by (see Fig.~\ref{fig A3}(b) and (c))
\begin{equation}
    \langle \varphi \rangle_{\rm TD} |_{\tmu\neq0}^{\thh=0}
    \begin{dcases*}
        \frac{\xi_{1}^{4}}{1+\xi_{1}^{4}} &, $0 \leq \tj \leq \ln{\min{\{\sqrt{2},s_{0}\}}}$\\ 
        \frac{\xi_{3}^{4}}{1+\xi_{3}^{4}} &, $\tilde{J} \geq \ln{\min{\{\sqrt{2},s_{0}\}}}$,
    \end{dcases*}.
    \label{solution global minima}
\end{equation} 
where $\ln{s_{0}}\equiv\ln{\sqrt{2\sqrt{2}}\e{-\tmu/4}}$ denotes the
coupling strength that solves for the root of $S$ in Eq.~\eqref{S}.
\subsection{Non-Zero force}\label{Appendix D.2}
We determine the roots of $g(\xi)$ for a non-zero force by means of a
convergent Newton series yielding the exact result \cite{Godec_2016, Hartich_2019, Hartich_2019_2}
\begin{equation}
    \xi = \xi_{0}-\sum\limits_{k=1}^{\infty}\frac{\left[g^{\left(0\right)}\left(\xi_{0}\right)\right]^{k}}{\left[g^{\left(1\right)}\left(\xi_{0}\right)\right]^{2k-1}}\frac{\det{\mathcal{A}_{k}\left(\xi_{0}\right)}}{\left(k-1\right)!},
    \label{Newton series}
\end{equation}
where $\xi_{0}$ is an initial guess in a convex neighborhood around $\xi$,  $g^{\left(i\right)}\left(\xi_{0}\right)$ denotes the $i$th
derivative of $g(\xi)$ at the point $\xi_0$, and $\mathcal{A}_{k}\left(\xi_{0}\right)$ are almost triangular matrices of size $\left(k-1\right)\times\left(k-1\right)$ with elements
\begin{eqnarray}
    &&\mathcal{A}^{ij}_{k}\left(\xi_{0}\right)=\frac{g^{\left(i-j+2\right)}\left(\xi_{0}\right)\theta\left(i-j+1\right)}{\left(i-j+2\right)!}\times\nonumber\\&&\left(k\left[i-j+1\right]\theta\left(j-2\right)+i\theta\left(1-j\right)+j-1\right),
\end{eqnarray}
where $\theta\left(x\right)$ denotes the Heaviside step function,
i.e.\ $\theta\left(x\right)=1$ if $x\geq0$ and 0 otherwise, and we
symbolically set $\det\mathcal{A}_{1}=1$. The determinant of almost
triangular matrices, also known as upper/lower Hessenberg matrices,
can be efficiently calculated using a recursion formula
\cite{cahill2002fibonacci}, for which a numerical implementation can
be found in \cite{jia2018numerical}. If we set
$g^{\left(3\right)}=g^{\left(4\right)}=...=0$ in the almost triangular
matrices, the resulting matrix $\mathcal{\tilde{A}}_{k}$ becomes
triangular, implying that its determinant is simply given by the
product of its diagonal elements. Making the substitution
$\mathcal{A}_{k} \rightarrow \mathcal{\tilde{A}}_{k}$ in
Eq.~\eqref{Newton series} yields the so-called ``quadratic
approximation'' \cite{Hartich_2019, Hartich_2019_2}
\begin{equation}
    \xi \approx \xi_{0}-\frac{g^{\left(1\right)}\left(\xi_{0}\right)-\sqrt{g^{\left(1\right)}\left(\xi_{0}\right)^{2}-2g^{\left(0\right)}\left(\xi_{0}\right)g^{\left(2\right)}\left(\xi_{0}\right)}}{g^{\left(2\right)}\left(\xi_{0}\right)}.
    \label{Newton series truncated}
\end{equation}
that becomes exceedingly accurate when the root moves close to $\xi_0$.
For the initial point $\xi_{0}$ we use the \emph{ansatz}
\begin{equation}
  \xi_{0}= \exp\left(\frac{\tmu}{4}-\frac{2}{9}\thh+{\rm sign}\left[\frac{
      \tmu}{4}-\frac{2}{9}\thh\right] \frac{6}{5}\tj\right),
   \label{Ansatz}
\end{equation}\vspace{-0.2cm}\\
which is derived by considering an adapted form of
Eqs.~\eqref{solution 1-2} and \eqref{solution 3-4} in combination with
the implementation of the force term. The weight $2/9$ is derived from
the term $(1+\xi^{4})^2/2(1+2\xi^{4})^{2}$ in Eq.~\eqref{c} evaluated
at the point $\xi=1$ (corresponding to $\varphi = 1/2$), and
the weight $6/5$ in front of $\tj$ was selected empirically. This
choice assures that Eq.~\eqref{transcendental equation} satisfies the
Lipschitz condition between $\xi_{0}$ and the root $\xi$ and thus assures
the convergence of the Newton's series.

Plugging
Eq.~\eqref{Ansatz} into Eq.~\eqref{Newton series truncated}, and using
the relation $\varphi^0_{1}=\xi^4/(1+\xi^4)$, we obtain the location
of the global minimum - and thus $\langle \varphi \rangle_{\rm TD}$ -
for non-zero force. Notably, the \emph{ansatz} given by
Eq.~\eqref{Ansatz} also provides a numerically correct solution for a
zero force and non-zero intrinsic binding-affinity. For completeness
we write down explicitly all the terms which are used to evaluate
Eq.~\eqref{Newton series truncated} (higher order terms entering the
fully converged series in Eq.~(\ref{Newton series}) are omitted as
they are lengthy).

Let $g\left(\xi\right)$ be given by Eq.~\eqref{transcendental
  equation}. Introducing the auxiliary functions
\begin{eqnarray}
    \alpha_{\thh}\left(\xi\right)&\equiv&\frac{4\xi^{4}\left(\xi^{4}+1\right)\thh}{\left(2\xi^{4}+1\right)^{3}}
    \nonumber\\
    \beta_{\thh}\left(\xi\right)&\equiv&\frac{4\xi^4(10\xi^8+11\xi^4-3)\thh}{\left(2\xi^{4}+1\right)^{4}},
\end{eqnarray}
the first and second derivative can be written as
\begin{widetext}
\begin{equation}
    g^{\left(1\right)}\left(\xi\right)=4\xi^{3}-
    \e{2\tilde{J}}\left(c_{\tmu,\thh}(\xi)\left[3+\alpha_{\thh}\left(\xi\right)\right]\xi
    -c_{\tmu,\thh}^{-1}(\xi)\left[1-\alpha_{\thh}\left(\xi\right)\right]\xi^{-1}\right)\xi,
    \label{g1}
\end{equation}
\begin{equation}
    g^{\left(2\right)}\left(\xi\right)=12\xi^{2}-
    \e{2\tilde{J}}\left(c_{\tmu,\thh}(\xi)\left[(3+\alpha_{\thh}(\xi))^2-(3+\beta_{\thh}\left(\xi\right))\right]\xi+
    c_{\tmu,\thh}^{-1}(\xi)\left[2 \alpha_{\thh}\left(\xi\right) -3\alpha^{2}_{\thh}\left(\xi\right)+\beta_{\thh}(\xi)\right]\xi^{-1}\right),
    \label{g2}
\end{equation}
\end{widetext}
where $c_{\tmu,\thh}(\xi)$ is defined in Eq.~\eqref{c}. Eqs.~\eqref{e.o.s. bare field}, \eqref{solution global minima}, and \eqref{Newton series truncated} form our main result for the equation of state in the thermodynamic limit. In Fig.~\ref{fig 4} we show the results for various values of the force and intrinsic affinity. 
\setcounter{equation}{0}
\setcounter{figure}{0}
\renewcommand{\thefigure}{E\arabic{figure}}
\renewcommand{\theequation}{E\arabic{equation}}
\section{Kinetics of cluster formation and dissolution}\label{Appendix E}
\subsection{Exact algebraic result for small clusters}\label{Appendix E.1}
It is well known that the transition matrix for an absorbing discrete-time Markov chain with a set of
recurrent states has the canonical form \cite{iosifescu_finite_2014}
\begin{equation}
    \mathbf{P}=\begin{bmatrix} \mathbf{1} & \!\!\!\!\!\!\!\mathbf{0}
    \\ \mathbf{R} & \mathbf{T}_{\rm d,f}\end{bmatrix},
    \label{canon}
\end{equation}
where $\mathbf{1}$ is the identity matrix, $\mathbf{T}_{\rm d,f}$ is the
submatrix of transient states in dissolution/formation, and $\mathbf{R}$ the submatrix of
recurrent states. In the particular case of cluster
dissolution the $(2^N-1)\times(2^N-1)$ matrix
$\mathbf{T}_{\rm d}$ entering Eq.~(\ref{canon}) is obtained by
removing the last column and row, and the $(2^N-1)\times(2^N-1)$ matrix
$\mathbf{T}_{\rm f}$ entering Eq.~(\ref{canon}) by removing the first column and row. If we introduce the column vector
$\mathbf{\hat{e}}_k$ with components
$(\mathbf{\hat{e}}_k)_i=\delta_{ki}$ and the column vector
$\mathbf{e}$ whose elements are all equal to 1, the mean first
passage times for cluster formation and dissolution read exactly
\begin{equation}
\langle \tau_{d}\rangle=\mathbf{\hat{e}}_1^T
(\mathbf{1}-\mathbf{T}_{\rm d})^{-1}\mathbf{e},\quad\langle \tau_{f}\rangle=\mathbf{\hat{e}}_{2^N-1}^T
(\mathbf{1}-\mathbf{T}_{\rm f})^{-1}\mathbf{e}.
\label{MFPT}
\end{equation} 
In applying Eq.~(\ref{MFPT}) one must invert a $(2^N-1)\times(2^N-1)$
sparse matrix and afterwards sum over $2^N-1$ terms, which is feasible for
$N\lesssim 5\times 5$. For a system of $N=4\times5$ the exact results are shown with the blue line in Fig.~\ref{fig 7}. Larger clusters are treated within the \emph{local equilibrium
approximation}. 
\subsection{Finite-size results for a non-uniform force distribution}\label{Appendix E.1.b}
\begin{figure}
    \includegraphics[width = 0.375\textwidth]{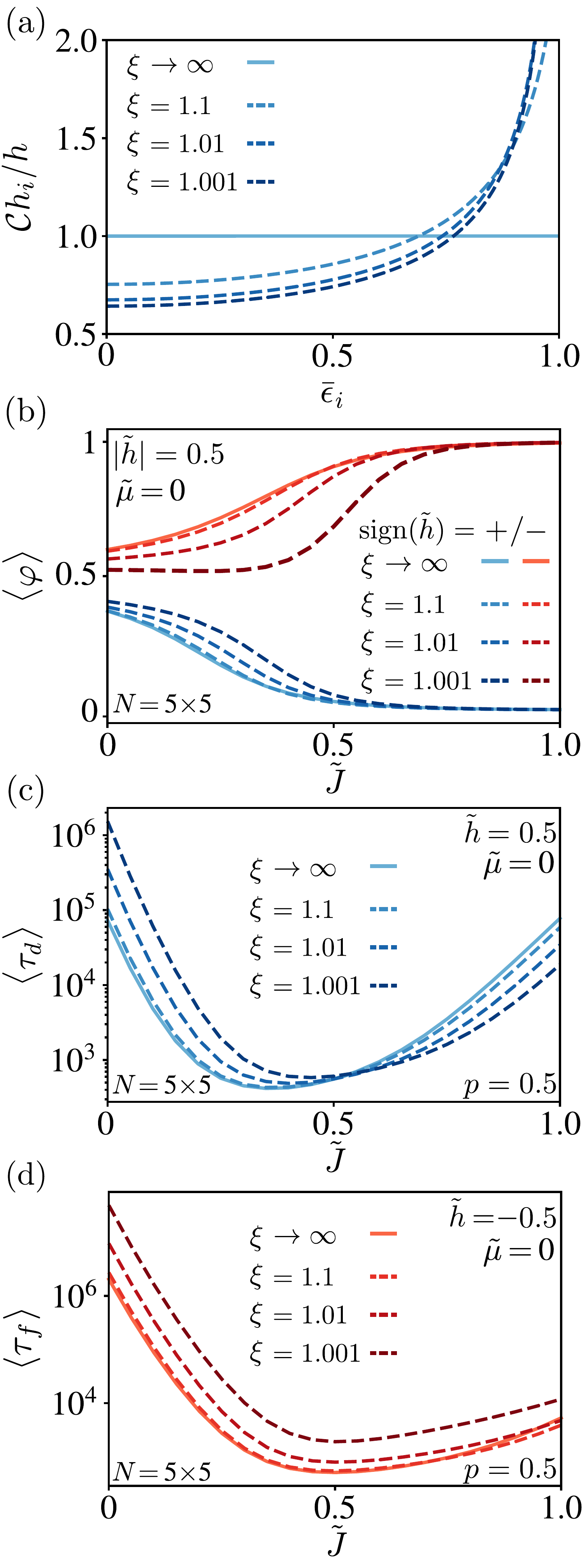}
    \caption{\textbf{Comparison between a uniform and non-uniform force load.} (a) The non-uniform force load distribution given by Eq.~\eqref{nonuniform} as a function of the normalized lattice distance $\bar{\epsilon}_{i}$. For $\xi < \infty$ adhesion bonds located at the outer edge of the lattice are subject to a larger external force than adhesion bonds located at the inner part of the lattice. (b) The equation of state for a pulling (blue) and pushing (red) force for various values of $\xi$ for a system of $N=5\times5$ adhesion bonds with zero intrinsic binding-affinity. (c-d) The mean first passage time to dissolution (c) and formation (d) for mixed Glauber-Kawasaki dynamics with constant Glauber attempt probability $p=0.5$ under a pulling (blue) and pushing (red) force for various values of $\xi$ for a system of $N=5\times5$ adhesion bonds with zero intrinsic binding-affinity.}
    \label{fig A4}
\end{figure}

Under the condition of a small combined elastic modulus, corresponding
to large values of the coupling  strength $\tilde{J}\gg 1$, the assumption of an equally shared force load is no longer valid \cite{erdmann_stability_2004,gao_probing_2011,Qian_2008}. We therefore address how a non-uniform force distribution affects the equation of state and mean first passage time to cluster dissolution/formation for finite system sizes. Based on Eq.~$(7)$ in \cite{Qian_2008} and Eq.~$(4)$ in \cite{PhysRevLett.120.268002} we introduce a non-uniform force load by making the substitution $h\rightarrow\mathcal{C}\sum_{i}h_{i}\delta_{\sigma_{i},-1}/N_{\rm c}(\{\sigma_{i}\})$ in Eq.~\eqref{force}, where $\mathcal{C}\equiv Nh/\sum_{i}h_{i}$ is a normalization constant such that initially, i.e.\ when all bonds are closed, the total force load is $h$. The load on bond $i$, denoted as $h_{i}$, is given by
\begin{equation}
    h_{i}=\frac{1}{\sqrt{\xi-\bar{\epsilon}^{2}_{i}}},
    \label{nonuniform}
\end{equation}
where $\xi \geq 1$, and $\bar{\epsilon}_{i}\equiv
(\epsilon_{i}-r)/(d-r) \in [0,1]$ is a normalized distance of bond $i$
to the center of the lattice, with $\epsilon_{i}$ defined as the
eccentricity of node $i$, which is the maximum number of edges between
node $i$ and any other node in the lattice. The radius $r\equiv
\min{\epsilon_{i}}$ and diameter $d\equiv\max{\epsilon_{i}}$ of the
lattice are defined as the minimum and maximum eccentricity,  respectively. With the force distribution given by Eq.~\eqref{nonuniform}, which is depicted in Fig.~\ref{fig A4}a, closed bonds located at the outer edge of the lattice ($\epsilon_{i}=1$) experience a larger external force than bonds located at the inner part of the lattice ($\epsilon_{i}=0$). The parameter $\xi$ is an indicator for the spread in force load among the individual bonds. For $\xi=1$, which holds when $\lim \tilde{J}\rightarrow\infty$ \cite{Qian_2008}, the force distribution at the edge of the cluster is singular and nonphysical. On the contrary, for $\lim\xi\rightarrow \infty$, which is valid for $\lim\tilde{J}\rightarrow 0$, we recover the uniform force distribution.

In Fig.~\ref{fig A4}(b-d) we depict the equation of
  state  (b) and mean first passage time to cluster dissolution (c)
  and formation (d) for mixed Glauber-Kawasaki dynamics with a
  constant Glauber attempt probability $p_{k}\rightarrow p=0.5$ and
  for various values of $\xi$ under a pulling or pushing force
  ($\tilde{h}=\pm0.5$). The results were obtained by exact
  summation/algebraic techniques. Interestingly, for $\xi \geq 1.1$
  the equation of state and mean first passage times are almost
  identical to the uniform force load solutions that correspond to
  $\xi \rightarrow \infty$. Only for $\xi < 1.1$, which is valid for
  very large coupling values corresponding to extremely floppy
  membranes, we observe deviations from the uniform force results. The
  origin of the deviations is the extreme force load on the outer
  bonds, which is $\sqrt{\xi/(\xi-1)}$ times larger than the force
  load on the inner bond. For $\xi=1.01$ this leads approximately to a
  factor of $\times 10$, and for $\xi=1.001$ this leads approximately
  to a factor of $\times 32$. Hence, for most physically meaningful  realizations of a non-uniform force distribution (i.e.\ distributions based on Eq.~\eqref{nonuniform}) the results converge to the uniform force solutions. Only under the extreme conditions where the force load on the outer bonds becomes at least an order of magnitude larger compared to the inner bonds we find large deviations from the uniform force load.

Note that the relative fraction of edge bonds in the
  limit of larger system sizes (and specifically in the thermodynamic
  limit) vanishes. Therefore we expect a non-uniform force load, which
  mainly penalizes the edge bonds for $\xi \rightarrow 1$, to have
  an even weaker effect on the equation of state and mean first
  passage times in large systems.
\subsection{Proof of detailed-balance for
local equilibrium rates}\label{Appendix E.2}
Before stating the explicit result for the mean first passage time to dissolution/formation in the local equilibrium approximation, we prove that the local equilibrium transition rates $\bar{w}_{k\to k\pm1}$ given by Eq.~\eqref{loceq} obey detailed-balance w.r.t $Q_{k}$ defined in Eq.~\eqref{QK}. The effective transition rates are obtained by mapping the full mixed Glauber-Kawasaki dynamics
onto an effective birth-death process over the number of closed bonds
(see Fig.~\ref{fig 6}), where we assume that the dynamics
reaches a local equilibrium at any number of closed bonds before any
transition. As a result, the birth-death process is a Markov
chain on the free
energy landscape for the fraction of closed bonds $\vphi$. Recall that the original Glauber rates in Eq.~\eqref{Glauber Transition rate} obey detailed-balance w.r.t.\ the Hamiltonian $\mathcal{H}(\{\sigma_{j}\})$, and therefore
\begin{multline}
    \e{-\beta\mathcal{H}(\{\sigma_{j}\})}\delta_{N_c(\{\sigma_{j}\}),k}w_i(\{\sigma_{j}\})\delta_{N_c(\{\sigma_{j}\}_i'),k\pm1}=\\
    \e{-\beta\mathcal{H}(\{\sigma_{j}\}_i')}\delta_{N_c(\{\sigma_{j}\}_i'),k\pm1}w_i(\{\sigma_{j}\}_i')\delta_{N_c(\{\sigma_{j}\}),k},
    \label{original detailed balance}
\end{multline}
where we have explicitly incorporated the constraints arising from
single-bond-flip dynamics by means of the Kronecker delta´s. Upon
summing the left hand side of Eq.~\eqref{original detailed balance} over all initial configurations with $N_c(\{\sigma_{j}\})=k$ and over all rates that jump to a configuration with $N_c(\{\sigma_{j}\}_i')=k\pm1$, we reach all possible final configurations with $N_c(\{\sigma_{j}\}_i')=k\pm1$, with a backward rate given by the sum of all rates that jump to a configuration with $N_c(\{\sigma_{j}\})=k$. Hence we find the equality
\begin{multline}    
    \sum\limits_{\{\sigma_{j}\}}\sum\limits_{i=1}^{N}\e{-\beta \mathcal{H}(\{\sigma_{j}\})}\delta_{N_c(\{\sigma_{j}\}),k}w_i(\{\sigma_{j}\})\delta_{N_c(\{\sigma_{j}\}_i'),k\pm1}=\\
    \sum\limits_{\{\sigma_{j}\}_i'}\sum\limits_{i=1}^{N}\e{-\beta\mathcal{H}(\{\sigma_{j}\}_i')}\delta_{N_c(\{\sigma_{j}\}_i'),k\pm1}w_i(\{\sigma_{j}\}_i')\delta_{N_c(\{\sigma_{j}\}),k}.
    \label{Effective detailed balance 1}
\end{multline}
Comparing Eq.~\eqref{Effective detailed balance 1} with
Eq.~\eqref{loceq}, we recognize the left and right hand side as
$\tilde{Q}_{k}\bar{w}_{k\to k\pm1}$ and $ \tilde{Q}_{k\pm1}\bar{w}_{k\pm1\to k}$,
respectively, which proves the effective detailed-balance relation
\begin{equation}
    \tilde{Q}_{k}\bar{w}_{k\to k\pm1}= \tilde{Q}_{k\pm1}\bar{w}_{k\pm1\to k}.
    \label{Effective detailed balance}
\end{equation}
\subsection{First passage time statistics within the local equilibrium approximation}\label{Appendix E.3}
The local equilibrium approximation maps the complete mixed Glauber-Kawasaki dynamics onto an effective birth-death process with a right-acting tri-diagonal transition matrix $\mathbf{P}^{\rm le}$ of size $(N+1)\times(N+1)$ with elements
\begin{multline}
    P^{\rm le}_{ij}=\Lambda_{i}\delta_{ij}+\\
    \bar{w}_{i-1\to i}\delta_{i+1j}\theta\left(N-i\right)+\bar{w}_{i-1\to i-2}\delta_{i-1j}\theta\left(i-2\right),
\end{multline}
and $\Lambda_{i}=1-\sum_{j\neq i}^{N+1}P^{\rm le}_{ij}$. To obtain the mean first passage time we use the same algebraic technique as for small clusters. Upon removing
the first/last row and column of $\mathbf{P}^{\rm le}$ we obtain the
submatrix $\mathbf{T}^{\rm le}_{\rm d,f}$ for cluster dissolution and
formation, respectively. We can invert the tri-diagonal submatrix  exactly, which leads to the following LU/UL decomposition
\begin{equation}
    (\mathbf{1}-\mathbf{T}^{\rm le}_{\rm{d,f}})^{-1}=\mathcal{A}^{\rm{d,f}}\mathcal{B}^{\rm{d,f}},
    \label{Cluster dissolution/nucleation inverse1}
\end{equation}
where $\mathcal{A}^{\rm{d}}$ and $\mathcal{B}^{\rm{d}}$ are the lower and upper triangular matrix with elements
\begin{equation}
    \mathcal{A}^{\rm d}_{ij}=\frac{\theta\left(i-j\right)}{ \tilde{Q}_{j-1}}\bar{w}_{j-1\rightarrow j}, 
    \ \
    \mathcal{B}^{\rm d}_{ij}= \tilde{Q}_{j}\theta\left(j-i\right),
\end{equation}
and $\mathcal{A}^{\rm{f}}$ and $\mathcal{B}^{\rm{f}}$ are the upper and lower triangular matrix with elements
\begin{equation}
    \mathcal{A}^{\rm f}_{ij}=\frac{\theta\left(j-i\right)}{ \tilde{Q}_{j}}\bar{w}_{j\rightarrow j-1},
    \ \
    \mathcal{B}^{\rm f}_{ij}= \tilde{Q}_{j-1}\theta\left(i-j\right).
\end{equation}
A proof that Eq.~\eqref{Cluster dissolution/nucleation inverse1} is
indeed the inverse of $\mathbf{1}-\mathbf{T}^{\rm le}_{\rm{d,f}}$ is
given in the SM. Let us denote with $\langle \tau^{\rm
 le}_{d,f} \rangle_{m}$ the mean first passage time to cluster
dissolution and formation, starting from the state with $m$ closed bonds. Using Eq.~\eqref{Cluster dissolution/nucleation inverse1} we obtain an exact expression for the first moments
\begin{equation}
    \langle \tau_{d}^{\rm le} \rangle_{0< m \leq N}=\mathbf{\hat{e}}_m^T
    \mathcal{A}^{\rm d}\mathcal{B}^{\rm d}\mathbf{e}
    =\sum_{k=0}^{m-1}\frac{1}{\bar{w}_{k\rightarrow k+1}}\sum_{l=k+1}^{N}\frac{\tilde{Q}_{l}}{ \tilde{Q}_{k}},
    \label{MFPT cluster dissolution}
\end{equation}
\begin{equation}
    \langle \tau_{f}^{\rm le} \rangle_{0\leq m < N}=\mathbf{\hat{e}}_{m+1}^T\mathcal{A}^{\rm f}\mathcal{B}^{\rm f}
    \mathbf{e}=\sum_{k=m+1}^{N}\frac{1}{\bar{w}_{k\rightarrow k-1}}\sum_{l=0}^{k-1}\frac{\tilde{Q}_{l}}{ \tilde{Q}_{k}},
    \label{MFPT cluster formation}
\end{equation}
where $\mathbf{\hat{e}}_m$ is the column vector with dimension $N$
with components $(\mathbf{\hat{e}}_m)_{i}=\delta_{mi}$, and $\mathbf{e}$
is the column vector with all components equal to 1.
  Notice that Eq.~\eqref{Cluster dissolution/nucleation
  inverse1},  and therefore Eqs.~\eqref{MFPT cluster dissolution} and
\eqref{MFPT cluster formation}, are applicable to any right-acting
tri-diagonal transition matrix with rates obeying
detailed-balance. Although we only present the mean first passage time
here,  we can easily obtain any higher order moments of the first
passage time to cluster dissolution and formation  using
Eq.~\eqref{Cluster dissolution/nucleation
  inverse1} \cite{iosifescu_finite_2014}. Notably, Eqs.~\eqref{MFPT cluster
  dissolution} and \eqref{MFPT cluster formation} appear to have a
similar structure as the largest eigenvalue of the transition matrix
in classical nucleation theory \cite{RevModPhys.62.251,
  brendel2005nucleation}.
\subsection{A bound on the effective transition rates}\label{Appendix E.4}
\begin{figure}
    \includegraphics[width = 0.45\textwidth]{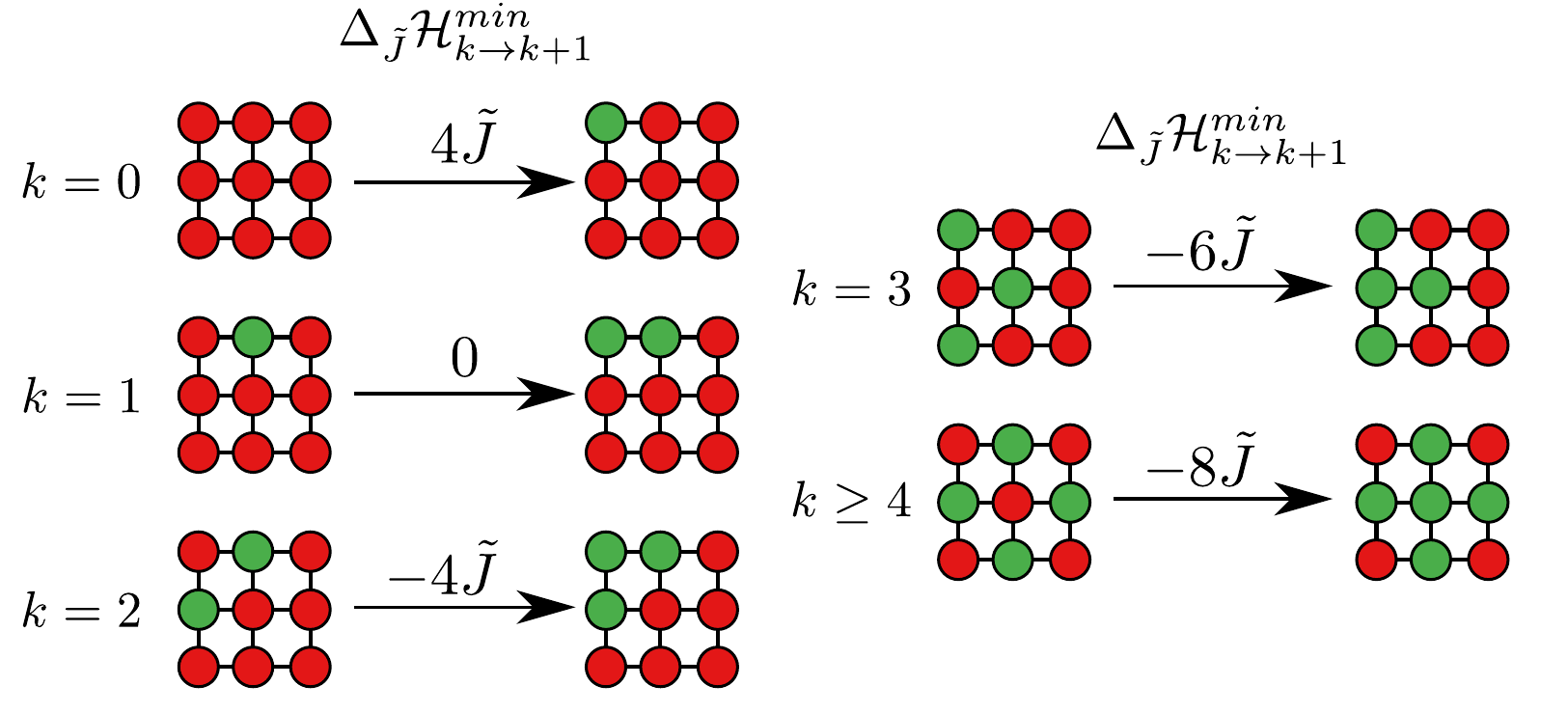}
    \caption{\textbf{Minimum energy forward transitions} between two configurations $\{\sigma_{j}\}$ and $\{\sigma_{j}\}_i'$ with $N_c(\{\sigma_{j}\})=k$ and $N_c(\{\sigma_{j}\}_i')=k+1$ respectively. Although we depict here the minimal energy differences for a lattice of size $N=3\times3$, the result holds for any two-dimensional lattice of size $N\geq3\times3$ as long as the transitions for $k=0,1,2$ are taken at the corner, and the transition for $k=3$ is taken at the edge.}
    \label{fig A5}
\end{figure}

Here we present a bound on the local equilibrium rates given by Eq.~\eqref{loceq} which proves that the transition rates are strictly sub-exponential in $N$. First, we consider a bound for the exit rates
$w_{\rm exit}^{\pm}(\{\sigma_{i}\})$ defined in Eq.~\eqref{exit}, which
contain a sum over the original Glauber rates that are defined in
Eq.~\eqref{Glauber Transition rate}. Since $1-\tanh{(x)}\geq 0 \ \forall x \in \mathds{R}$ the Glauber rates are non-negative, and therefore the exit rates obey the bound
\begin{equation}
    w^{\rm max}_{k\to k\pm 1}\leq w_{\rm exit}^{\pm}(\{\sigma_{i}\}) \leq c^{\pm}_{k}w^{\rm max}_{k\to k\pm 1},
    \label{bound1}
\end{equation}
with $c_{k}^+=N-k$ and $c_{k}^-=k$ denoting the number of terms inside the sum of Eq.~\eqref{exit}, and $w^{\rm max}_{k\to k\pm 1}$ denotes the largest transition rate to go from a state with $k$ to $k\pm1$ closed bonds. The largest transition rate can be written as
\begin{equation}
    w^{\rm max}_{k\to k\pm 1}=\frac{1}{2N}\left[1-\tanh{(\Delta \mathcal{H}^{\rm min}_{k\to k\pm1}/2)}\right],
\end{equation}
where
\begin{equation}
    \Delta \mathcal{H}^{\rm min}_{k\to k\pm 1}\equiv\inf_{\substack{N_c(\{\sigma_{j}\})=k,\\N_c(\{\sigma_{j}\}_i')=k\pm1}}\{\mathcal{H}(\{\sigma_{j}\}_i') -\mathcal{H}(\{\sigma_{j}\})\}
    \label{dHmin}
\end{equation}
denotes the smallest possible energy change between two configurations
$\{\sigma_{j}\}$ and $\{\sigma_{j}\}_i'$ with $N_c(\{\sigma_{j}\})=k$
and $N_c(\{\sigma_{j}\}_i')=k\pm1$, respectively.  To obtain a
closed-form expression for $ w^{\rm max}_{k\to k\pm 1}$ we first note
that the contribution to $\Delta \mathcal{H}^{\rm min}_{k\to k\pm 1}$
from the external force and intrinsic binding-affinity are fixed and
given by the second and third term in Eq.~\eqref{Glauber Transition rate}.
Therefore we are left to consider the smallest energy change due to
the coupling strength, which we denote as $\Delta_{\tj}
\mathcal{H}^{\rm min}_{k\to k\pm 1}$. For a square lattice with free boundary conditions the minimal energy ``forward transitions'' with energy difference $\Delta_{\tj} \mathcal{H}^{\rm
  min}_{k\to k+ 1}$ for various values of $k$ are depicted in
Fig.~\eqref{fig A5}. Similarly, the minimal energy ``downward transitions'' with
energy difference $\Delta_{\tj} \mathcal{H}^{\rm min}_{k\to k-1}$ are
obtained by interchanging the open (red) and closed (green) adhesion
pairs in Fig.~\eqref{fig A5}. Combining these two results yields
$\Delta_{\tj} \mathcal{H}^{\rm min}_{k\to k\pm 1}=2m_{k}^{\pm}\tj$,
with 
\begin{equation}
    m_{k}^{\pm} \equiv  2(c_{k}^{\mp}-1)\theta(2-c_{k}^{\mp})-\min(c_{k}^{\mp}, 4)\theta(c_{k}^{\mp}-3),
    \label{mk}
\end{equation}  
and delivers the expression for $w^{\rm max}_{k\to k\pm 1}$.
Finally, since $w^{\rm max}_{k\to k\pm 1}$ is independent of the specific configuration $\{\sigma_{i}\}$ at fixed $k$, it drops out of the sum over $\{\sigma_{i}\}$ in Eq.~\eqref{loceq} for the effective transition rates, and therefore the bound in Eq.~\eqref{bound1} can directly be applied to the effective transition rate upon multiplying both sides with the Glauber attempt probability
\begin{equation}
    p_{k}w^{\rm max}_{k\to k\pm 1}\leq \bar{w}_{k\to k\pm 1} \leq p_{k}c^{\pm}_{k}w^{\rm max}_{k\to k\pm 1},
    \label{bound 2}
\end{equation}
which yields the bound on the effective transition rate. 

The lower and upper bound for the effective transition rate are used
to determine an upper and lower bound for the mean first passage time
to cluster dissolution and formation, respectively. The specific result for a rectangular lattice of size
$N=6\times7$ for pure Glauber dynamics (i.e.\ $p_{k}=1 \ \forall k$) is shown in Fig.~\ref{fig A6}. For small values of
the coupling strength $\tj$ we find that the upper bound in
Eq.~\eqref{bound 2}, corresponding to the lower bound in
Fig.~\ref{fig A6} (since $\langle {\rm t}_{{\rm d,f}} \rangle
\propto 1/\bar{w}_{k\to k\pm 1}$), is saturated by the exact effective
transition rate. Conversely, for large values of the coupling strength it seems that the lower bound in Eq.~\eqref{bound 2} is saturated. 
\subsection{Approximate effective transition rate}\label{Appendix E.5}
For systems larger than $N\approx50$ bonds the 
combinatorics involved in the computation of $Q_k$ defined in Eq.~(\ref{QK}) and $\bar{w}_{k\to k+1}$ in Eq.~(\ref{loceq}) become prohibitive, and thus forces us to make further approximations. To get Eq.~(\ref{MFPT_le}) fully explicit we make an ``instanton'' approximation for $\bar{w}_{k\to k\pm 1}$ using the Bethe-Guggenheim approximation with the bound given by Eq.~\eqref{bound 2}, and reads
\begin{equation}
    \bar{w}_{k\to k\pm 1}\approx
    \max(1,\alpha_{k}c_k^\pm)p_{k}w^{\rm max}_{k\to k\pm 1},
    \label{le_rate2}
\end{equation}
with
$\alpha_{k}=\delta_{k0}+\delta_{kN}+2\bar{X}_{k}/N\in\left[0,1\right]$,
$c_{k}^+=N-k$, $c_{k}^-=k$, and $\bar{X}_{k}$ given by Eq.~\eqref{xx}. The prefactor $\alpha_{k}c_k^\pm$  is a measure for the number
of "favorable" adhesion bonds that are most likely to flip in a
configuration with $k$ closed bonds. For $k=0\lor N$ all bonds have an equal surrounding in the thermodynamic limit (or
for a periodic lattice), and therefore all $c_k^\pm$ open/closed bonds
are equally likely to attempt a flip. For $0<k<N$ it
becomes energetically more favorable to flip a bond which is part of an open-closed adhesion pair (see Fig.~\ref{fig A5}). To determine the number of bonds that constitute an open-closed pair, we recall 
that $\bar{z}\bar{X}_{k}$ is a measure for the number of open-closed
pairs in a lattice of size $N$ with $k$ closed bonds. Upon dividing by
the total number of pairs in the system, given by $\bar{z}N/2$, we
obtain the probability to select an open-closed pair in the lattice that
is given by
\begin{equation}
    \frac{2\bar{X}_{k}}{N}=\frac{4X^{*}_{k}/N}{\left[1+4X^{*}_{k}(\e{4\tilde{J}}-1)/N\right]^{\frac{1}{2}}+1},
    \label{probability of open-closed pair}
\end{equation}
where $4X^{*}_{k}/N=4k(N-k)/N^2\in\left[0,1\right]$. Multiplying Eq.~\eqref{probability of open-closed pair} by the total number of open/closed adhesion bonds, i.e.  $2c_k^{\pm}\bar{X}_{k}/N$, we obtain an approximate expression for the number of open/closed bonds which constitute an open-closed adhesion pair. 

To prove that the approximate effective rate given by
Eq.~\eqref{le_rate2} obeys the bound given by Eq.~\eqref{bound 2} we
apply a chain of inequalities. First we note that $0\leq
2\bar{X}_{k}/N \leq 1/2$, where the upper bound follows from
considering $\tj=0$ and $k=N/2$ in Eq.~\eqref{probability of
  open-closed pair}, and the lower bound is given for $k=0\lor N$ or
the limit $\tj \to \infty$. From this it follows that $0\leq \alpha_{k} \leq
1$, and therefore $1\leq \max(1,\alpha_{k}c_k^\pm) \leq
c^{\pm}_{k}$. Finally, since we use $p_{k}w^{\rm max}_{k\to k\pm1}$ in
Eq.~\eqref{le_rate2}, it cancels on both sides of the inequality in
Eq.~\eqref{bound 2}, which leaves to prove the
inequality we have proven above and thereby completes the proof. 

\begin{figure*}
    \includegraphics[width = 0.9\textwidth]{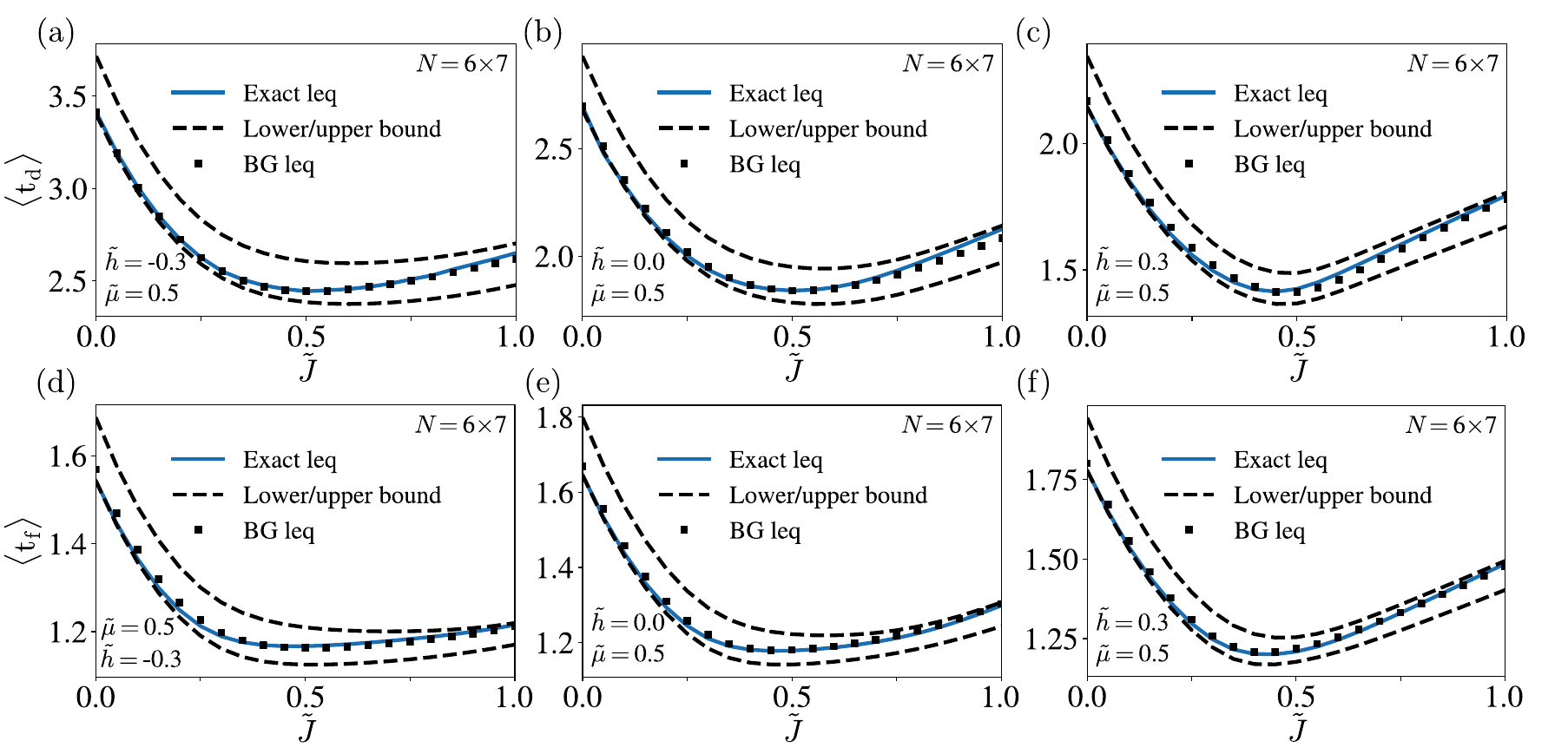}
    \caption{\textbf{Comparison of the exact local equilibrium
        effective rates and the approximate Bethe-Guggenheim local
        equilibrium effective rates for pure Glauber dynamics.} $\langle {\rm t}_{\rm d,f}
      \rangle$ for cluster dissolution (a)-(c) and cluster formation
      (d)-(f) as a function of the coupling $\tj$ for fixed intrinsic
      binding-affinity $\tmu = 0.5$ in the presence of a
      pushing force $\thh=-0.3$ (a and d), zero force
      $\thh =0$ (b and e), and a pulling force $\thh = 0.3$
      (c and f); The blue solid line is obtained with the exact local equilibrium effective transition rate and exact partition function $Q_{k}$ (Eqs.~\eqref{loceq} and \eqref{QK}, respectively) with $p_{k}=1 \ \forall k$. The black symbols are obtained with the Bethe-Guggenheim approximation to the effective rate and partition function $Q_{k}^{\rm BG}$ (Eqs.~\eqref{loceq} and \eqref{QK} in combination with Eq.~\eqref{zqc}, respectively). The black dotted line indicates the upper and lower bound to the mean dissolution/formation time, which is obtained with the upper and lower bound to the effective transition rate in combination with the exact partition function (Eqs.~\eqref{bound 2} and \eqref{QK} respectively).}
    \label{fig A6}
\end{figure*}

Fig.~\ref{fig A6} shows the mean first passage time to cluster
dissolution and formation obtained with the approximate effective rates
(\ref{le_rate2}) in combination with the Bethe-Guggenheim
approximation for $Q_{k}$ for a lattice of size $N=6\times7$ for pure Glauber dynamics (i.e.\ $p_{k}=1 \ \forall k$). The
results obtained with the approximate rates (black symbols) agree to a high degree with the results obtained by the exact effective rates (blue solid line).  
\subsection{The mean first passage time in the thermodynamic limit}\label{Appendix E.6}
Here we prove the result for the mean first passage time to dissolution/formation in the thermodynamic limit given by Eqs.~\eqref{MDT} and \eqref{ext_TD} based on the local equilibrium approximation. Intuitively $\langle\tau_{d,f}\rangle$
must scale as $\langle\tau_{d,f}\rangle\sim
\e{N\Delta\rtf^{\dagger}}$, where
$\Delta\rtf^{\dagger}\equiv\rtf_{\rm max}-\rtf_{\rm min}$ denotes the
difference in the free energy density between the minimum, $\rtf_{\rm min}=\inf_{\vphi}\rtf(\vphi)$, and the maximum, $\rtf_{\rm max}\equiv\sup_{\vphi}\rtf(\vphi)$, that for large clusters becomes independent of $N$. Indeed, according to Eq.~\eqref{MFPT_le} we have
$\tilde{Q}_l/\tilde{Q}_k=(p_{k}/p_{l})\e{N[\tilde{f}_N(k/N)-\tilde{f}_N(l/N)]}$ and recall that
$\bar{w}_{k\to k+1}$ is strictly
sub-exponential in $N$. Furthermore we make the assumption that the Glauber attempt probabilities $p_{k}$ are strictly sub-exponential in $N$. Since both series in Eq.~\eqref{MFPT_le}
are absolutely convergent, we can apply a version of the  ``squeeze'' theorem to Eq.~\eqref{MFPT_le}.

To simplify the notation we write the summands in Eq.~\eqref{MFPT_le} as
$0<(p_{k}/p_{l})a_{k,l}/\bar{w}_{k\to
  k+1}<\infty$, where
$a_{k,l}\equiv \e{N[\tilde{f}_N(k/N)-\tilde{f}_N(l/N)]}$. If 
$k^\dagger$ denotes the index of the largest $k$-dependent term
\begin{equation}
    k^{\dagger}\equiv\sup_{0\le k <N}\frac{p_{k}\exp[N\tilde{f}_N(k/N)]}{\bar{w}_{k\to
    k+1}}
    \label{kdag}
\end{equation}  
and by $l_{d,f}^\dagger$ the index of the largest l-dependent term
\begin{eqnarray}
  l_d^{\dagger}&\equiv&\sup_{k^\dagger  < l \le  N}\frac{\exp{[-N\tilde{f}_N(l/N)]}}{p_{l}}, \nonumber \\
  l_f^{\dagger}&\equiv&\sup_{0< l < k^\dagger}\frac{\exp{[-N\tilde{f}_N(l/N)]}}{p_{l}},
  \label{ldag}
\end{eqnarray}
then the following chain of inequalities holds for any $N$ 
\begin{equation}
    \frac{p_{k^\dagger}a_{k^\dagger,l^\dagger_{d,f}}}{p_{l^{\dagger}_{d,f}}\bar{w}_{k^\dagger \to k^\dagger+1}}\le \sum_{k=0}^{N-1}\!\sum_{l=m}^{M-1}\!\frac{p_{k}a_{k,l}}{p_{l}\bar{w}_{k\to k+1}} \le \frac{c_{M,m} p_{k^{\dagger}}a_{k^\dagger,l^\dagger_{d,f}}}{p_{l^{\dagger}_{d,f}}\bar{w}_{k^\dagger\to k^\dagger+1}},
    \label{ineq}  
\end{equation}
where $c_{M,m}\equiv N(M-m)$, $M=N+1$ and $m=k+1$ for dissolution, and $M=k+1$ and
$m=0$ for cluster formation. Since
$x^{1/N}$ is monotonic in $x>0$, such that $x_1<x_2$ implies
$x_1^{1/N}<x_2^{1/N}$, the inequality \eqref{ineq} is
preserved when exponentiated to $1/N$. The thermodynamic limit of Eq.~\eqref{ineq} is a
scaling limit, i.e.\ $\lim_s\equiv \lim_{N\to\infty}|_{k/N=\vphi_k}^{l/N=\vphi_l}$,
and thus the inequality \eqref{ineq} becomes
\begin{equation}
    {\rm lim}_s\!\left[\frac{p_{k^{\dagger}}a_{k^\dagger,l^\dagger_{d,f}}}{p_{l^{\dagger}_{d,f}}\bar{w}_{k^\dagger
    \to k^\dagger+1}}\right]^{\!\frac{1}{N}}\hspace{-0.25cm}\le \langle {\rm t_{d,f}}\rangle \le \mathrm{lim}_s\!\left[\frac{c_{M,m}p_{k^{\dagger}}a_{k^\dagger,l^\dagger_{d,f}}}{p_{l^{\dagger}_{d,f}}\bar{w}_{k^\dagger\to k^\dagger+1}}\right]^{\!\frac{1}{N}}\!\!\!\!.
    \label{ineq2}  
\end{equation}
Moreover, since
${\rm lim}_s\,\bar{w}^{-1/N}_{k^\dagger\to
  k^\dagger+1}=1$, ${\rm lim}_s\,[c_{M,m}]^{1/N}=1$, and ${\rm lim}_s\,[p_{k^{\dagger}}/p_{l^{\dagger}_{d,f}}]^{1/N}=1$, all four
limits in Eq.~(\ref{ineq2}) exist and thus may be taken separately,
implying the convergence of the upper bound in Eq.~(\ref{ineq2}) to
the lower bound. Thereby $\langle {\rm
  t_{d,f}}\rangle$ becomes \emph{squeezed} in-between rendering
the inequality an equality. Since $\lim_s
\tilde{f}_N(k^\dagger/N)=\rtf(\vphi_{\rm max})$ and $\lim_s
\tilde{f}_N(l_{d,f}^\dagger/N)=\rtf(\vphi^{d,f}_{\rm min})$ we finally obtain Eqs.~\eqref{MDT} and \eqref{ext_TD}, thus completing the proof.
\subsection{Evaluation of the Bethe-Guggenheim mean first passage time in the thermodynamic limit}\label{Appendix E.7}
\begin{figure*}[t!]
    \includegraphics[width = 1\textwidth]{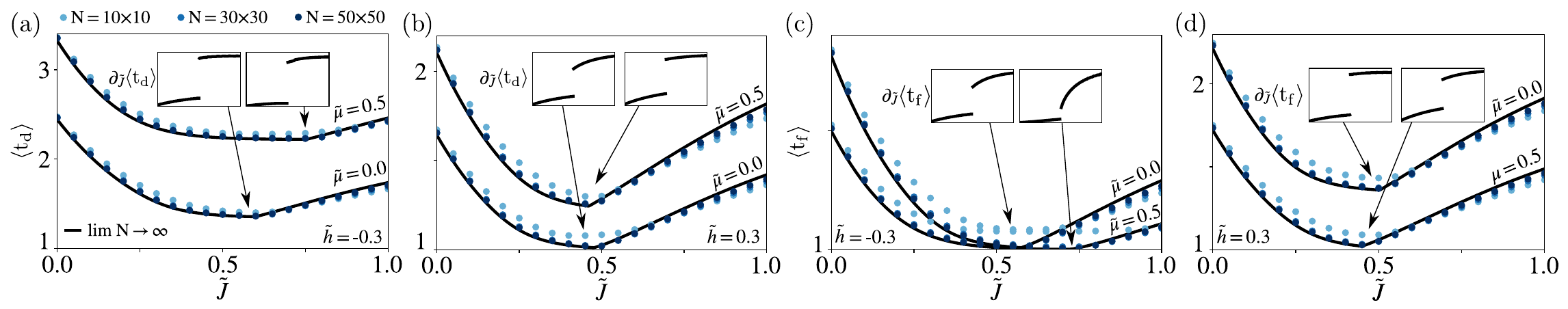}
    \caption{\textbf{Master scaling of mean dissolution and formation times per bond for finite clusters and in the thermodynamic limit.} $\langle {\rm
  t_{d,f}}\rangle$ for cluster dissolution (a-b) and formation (c-d) as a function of the
      coupling $\tj$ for a pair of intrinsic affinities $\tmu=0$ and
      $\tmu=0.5$ and various cluster sizes (symbols) as well as the
     thermodynamic limit (lines)
      in the presence of an external pushing (a,c) and pulling (b,d) force; Symbols are evaluated with local equilibrium approximation Eqs.~\eqref{MFPT_le}
      using $Q_k^{\rm BG}$ (Eqs.~\eqref{QK} and \eqref{zqc}) and $\bar{w}_{k\to
  k+1}$ from Eq.~\eqref{le_rate2} with $p_{k}=1 \ \forall k$ (i.e.\ pure Glauber dynamics)}. The discrepancy between the lines and
    symbols is due to finite-size effects.
    \label{fig A7}
\end{figure*}
In the previous section we have proven that in the thermodynamic limit
the mean first passage time to cluster dissolution/formation scales as
$\langle \tau_{d,f} \rangle \simeq \langle \rm{t}_{\rm d, f}
\rangle^{N} = \e{N\Delta \tilde{\rm f}^{\dagger}}$, where $\Delta
\tilde{\rm f}^{\dagger}$ denotes the largest left/right barrier in the
Bethe-Guggenheim free energy density Eq.~\eqref{TD}. In this section
we determine $\Delta \tilde{\rm f}^{\dagger}$ and thereby obtain a
closed-form expression for the mean first passage time per bond in the thermodynamic limit. 
\subsubsection{Case 1: $\tj \geq 0$, $\thh = \tmu = 0$}
We first consider the mean first passage time to cluster
dissolution/formation in the absence of an external force and
intrinsic binding-affinity. Due to the $\mathds{Z}_{2}$ symmetry of the coupling strength, we note that $\langle \rm t_{d} \rangle = \langle \rm t_{f} \rangle$. Our first task is to find the locations of the global maximum and minimum in the free energy landscape, denoted by $\varphi^{d,f}_{\rm max}$ and $\varphi^{d,f}_{\rm min}$, respectively. 

The position of the global minimum for zero force and intrinsic binding-affinity is given by Eq.~\eqref{bare field solution}, while the position of the global maximum is located at $\varphi^{d,f}_{\rm max}=0\land 1$ for low values of the coupling strength $\tj$, and at $\varphi^{d,f}_{\rm max}=1/2$ for large values. The coupling strength at which the global maximum changes position corresponds to the root of the equation
\begin{equation}
    \tilde{\rm f}\left(0\right)-\tilde{\rm f}\left(\frac{1}{2}\right)=\frac{\bar{z}}{2}\ln{\left(\e{2\tj}+1\right)}-\bar{z}\tj+\left(1-\frac{\bar{z}}{2}\right)\ln{\left(2\right)},
    \label{Critical value maximum}
\end{equation}
which is given by
\begin{equation}
    \tj^{\rm d}_{\rm crit, BG}=-\frac{1}{2}\ln{\left(2^{1-2/\bar{z}}-1\right)},
    \label{Dynamical critical value QC}
\end{equation}
and sets the dynamical critical coupling value for the zero-field
Ising model under the Bethe-Guggenheim approximation.

Surprisingly, for the two-dimensional square lattice with $\bar{z}=4$
we exactly recover the static critical point obtained by Onsager
\cite{PhysRev.65.117}. To check whether this is a mere coincidence we
note that for the honeycomb lattice with $\bar{z}=3$ the exact
statical critical point is given by $\tj^{\rm s}_{\rm crit} =
\frac{1}{2}\ln{\left(2+\sqrt{3}\right)}=0.65...$
\cite{RevModPhys.17.50}, whereas Eq.~\eqref{Dynamical critical value
  QC} gives $\tj^{\rm d}_{\rm
  crit}=\frac{-1}{2}\ln{\left(2^{1/3}-1\right)}=0.67...$, and so we
find direct evidence that the dynamical critical point in the
Bethe-Guggenheim approximation does not (at least not always) coincide
with the exact statical critical point.

Combining our results for the locations of the global maximum and
minimum we obtain the following result for the mean first passage time
per adhesion bond in the thermodynamic limit for the zero field Ising model on a two-dimensional square lattice 
\begin{equation}
    \ln\langle {\rm
    t_{d,f}}\rangle=
    \left\{
    \begin{array}{lc}
    \tilde{\rm f}(0)-\tilde{\rm f}(\frac{1}{2}), & 0 \le \tj\le \tj^{\rm s}_{\rm crit, BG} \\
    \tilde{\rm f}(0)-\tilde{\rm f}(\frac{1}{2}[1\pm\mathcal{C}]), & \tj^{\rm s}_{\rm crit, BG}\le\!\tj\!\le\tj^{\rm d}_{\rm crit, BG}\\
    \tilde{\rm f}(\frac{1}{2})-\tilde{\rm f}(\frac{1}{2}[1\pm\mathcal{C}]), & \tj\ge \tj^{\rm d}_{\rm crit, BG},
    \end{array}
    \right.
    \label{MFPT bare field}
\end{equation}
where
$\mathcal{C}\equiv \e{2\tj}\sqrt{\e{4\tj}-4}/(\e{4\tj}-2)$
comes from Eq.~\eqref{bare field solution}, $\tj^{\rm s}_{\rm
  crit}=\ln{(2)}/2$ denotes the statical critical point for zero
force, and $\tj^{\rm d}_{\rm crit}=\ln{(1+\sqrt{2})}/2$ the
dynamical critical point in the force free case. This ultimately leads
to Eq.~(\ref{MDT_ff}) in the main text.

In the strong coupling limit we find from Eq.~\eqref{MFPT bare field}
$\lim_{\tj \to \infty} \langle {\rm t}_{\rm d,f} \rangle = 2$, which
is identical to the result obtained for zero coupling. The physical
intuition behind this result comes from considering the average number
of steps required to change the state of a single independent adhesion
bond. For zero force and intrinsic binding-affinity the probability to
associate/dissociate is a $1/2$, and therefore the average
dissolution/formation time is given by
\begin{equation*}
1\left(\frac{1}{2}\right)+2\left(\frac{1}{2}\right)^{2}+3\left(\frac{1}{2}\right)^{3}+...=\sum\limits_{n=1}^{\infty}n\left(\frac{1}{2}\right)^{n}=2
.
\end{equation*}
For an infinite coupling strength the interaction between the bonds is
so strong that effectively the system behaves as one ``super bond'',
and therefore the average dissolution/formation time is equal to that
of a single independent adhesion bond. 
\subsubsection{Case 2: $\tj \geq 0$, $\tmu \neq 0$, $\thh = 0$}
Here we use the results obtained in Appendix~\ref{Appendix D} which  leads to
\begin{equation}
    \ln\langle {\rm t}_{\rm d} \rangle  =
    \left\{
    \begin{array}{lc}
        \tilde{\rm f}\left(0\right)-\tilde{\rm f}\left(\varphi_{1}\right), & 0 \leq \tj \leq \ln{\min{\{\sqrt{2},s_{0}\}}}\\ 
        \tilde{\rm f}\left(0\right)-\tilde{\rm f}\left(\varphi_{3}\right), & \ln{\min{\{\sqrt{2},s_{0}\}}} \leq \tj \leq \tj^{\rm d,-}_{\rm crit}\\ 
        \tilde{\rm f}\left(\varphi_{4}\right)-\tilde{\rm f}\left(\varphi_{3}\right), & \tj \geq \tj^{\rm d,-}_{\rm crit}
    \end{array},
    \right.
    \label{MFPT cluster dissolution non-zero binding-affinity}
\end{equation}
for cluster dissolution, and
\begin{equation}
    \ln\langle {\rm t}_{\rm f} \rangle =
    \left\{
    \begin{array}{lc}
        \tilde{\rm f}\left(1\right)-\tilde{\rm f}\left(\varphi_{1}\right), & 0 \leq \tj \leq \ln{\min{\{\sqrt{2},s_{0}\}}}\\ 
        \tilde{\rm f}\left(1\right)-\tilde{\rm f}\left(\varphi_{3}\right), & \ln{\min{\{\sqrt{2},s_{0}\}}} \leq \tj \leq \tj^{\rm d,+}_{\rm crit}\\ 
        \tilde{\rm f}\left(\varphi_{4}\right)-\tilde{\rm f}\left(\varphi_{1}\right), & \tj \geq \tj^{\rm d,+}_{\rm crit}
    \end{array}
    \right.
    \label{MFPT cluster nucleation non-zero binding-affinity},
\end{equation}
for cluster formation, where $\varphi_{i}=\xi_{i}^{4}/(1+\xi_{i}^{4})$ is given by
Eqs.~\eqref{solution 1-2} and \eqref{solution 3-4},
$s_{0}\equiv\sqrt{2\sqrt{2}}\e{-\tmu/4}$, and $\tj^{\rm d,-}_{\rm
  crit}$ and $\tj^{\rm d,+}_{\rm crit}$ are the dynamical critical
points for cluster dissolution and formation respectively, which are
solutions of
\begin{subequations}
\begin{equation}
    [\tilde{\rm f}\left(0\right)-\tilde{\rm f}\left(\varphi_{4}\right)]|_{\tj^{\rm d,-}_{\rm crit}}\stackrel{!}{=}0,
    \label{dynamical critical coupling dissolution}
\end{equation}
\begin{equation}
  [  \tilde{\rm f}\left(1\right)-\tilde{\rm f}\left(\varphi_{3}\right)-\tilde{\rm f}\left(\varphi_{4}\right)+\tilde{\rm f}\left(\varphi_{1}\right)]|_{\tj^{\rm d,+}_{\rm crit}}\stackrel{!}{=}0.
    \label{dynamical critical coupling nucleation}
\end{equation}
\end{subequations}
\subsubsection{Case 3: $\tj \geq 0$, $\tmu \neq 0$, $\thh \neq 0$}
Using a quadratic Newton series (which is defined in
Appendix~\ref{Appendix D}), Eqs.~\eqref{MFPT cluster dissolution non-zero binding-affinity} and \eqref{MFPT cluster nucleation non-zero binding-affinity} are directly applicable to the non-zero force scenario upon applying the transformation $\varphi_{i} \to \varphi^{*}_{i}$, where $\varphi^{*}_{i}=\xi^{*4}_{i}/(1+\xi^{*4}_{i})$ and
\begin{equation}
    \xi^{*}_{i}=\xi_{i}-\frac{g^{\left(1\right)}\left(\xi_{i}\right)\pm\sqrt{g^{\left(1\right)}\left(\xi_{i}\right)^{2}-2g^{\left(0\right)}\left(\xi_{i}\right)g^{\left(2\right)}\left(\xi_{i}\right)}}{g^{\left(2\right)}\left(\xi_{i}\right)},
\end{equation}
with a minus sign for the global minimum $\xi^{*}_{1,3}$, and a plus sign for the global maximum $\xi^{*}_{4}$. The function $g^{(0)}(\xi)$ and its first and second derivative $g^{(1,2)}(\xi)$ are given in Eqs.~\eqref{transcendental equation},\eqref{g1}, and \eqref{g2}, respectively. 

Our analytical results for the mean first passage time to cluster
dissolution and formation per adhesion bond are depicted in
Fig.~\ref{fig 8} and Fig~\ref{fig A7} for zero and nonzero external force respectively; note the remarkable agreement between the black
solid line depicting the thermodynamic limit and the results for finite system sizes on the order of $N\geq 10\times10$.
\subsection{Evaluation of the mean first passage time in the thermodynamic limit in the mean field approximation}\label{Appendix E.8}
\begin{figure}
    \includegraphics[width = 0.40\textwidth]{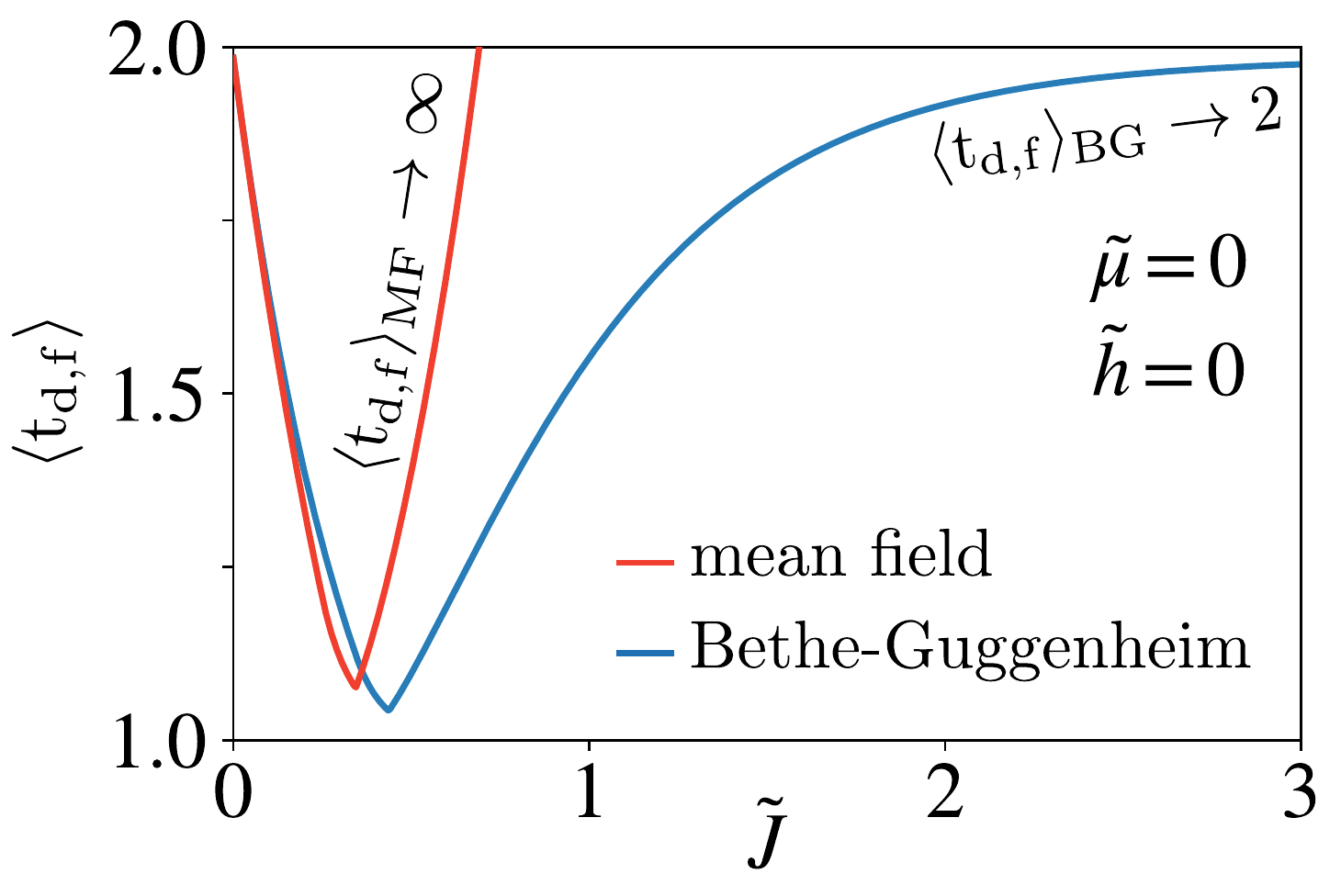}
    \caption{\textbf{Mean dissolution/formation time in the
        thermodynamic limit: Bethe-Guggenheim versus mean field
        approximation.} The Bethe-Guggenheim approximation is given by
      Eq.~\eqref{MFPT bare field}, and the mean field approximation is given by Eq.~\eqref{MFPT bare field MF explicit}, where we solve for Eq.~\eqref{MF equation} numerically.}
    \label{fig A8}
\end{figure}
Similarly to our previous analysis we must determine the global minimum and maximum, respectively, of the mean field free energy density given by Eq.~\eqref{TD MF}. For convenience we here consider only zero force and zero binding-affinity.

Below the statical critical coupling $\tj^{\rm
    s}_{\rm crit,MF}=1/\bar{z}$ there is a unique global minimum
  located at $\varphi^{d,f}_{\rm min}=1/2$. Above the statical
  critical coupling there exist two global minima, which are given by the non-zero solutions of the transcendental \emph{mean field equation}
\begin{equation}
    s=\tanh{\left(\bar{z}\tilde{J}s\right)},
    \label{MF equation}
\end{equation}
with $s=2\varphi-1$. Eq.~\eqref{MF equation} is obtained directly from Eq.~\eqref{bin MF} by using the relation $\ln{|\frac{1+x}{1-x}|}=2\tanh^{-1}{x}$.

Similarly to the Bethe-Guggenheim free energy density, the position of the global maximum is located at $\varphi^{d,f}_{\rm max}=0 \land 1$ for small values of the coupling strength $\tj$, and at $\varphi^{d,f}_{\rm max}=1/2$ for large values. The transition at which the global maximum changes location is given by the root of the equation
\begin{equation}
    \tilde{\rm f^{MF}}\left(0\right)-\tilde{\rm f^{MF}}\left(\frac{1}{2}\right)=-\frac{1}{2}\tj\bar{z}+\ln{(2)},
\end{equation}
which is given by
\begin{equation}
    \tj^{\rm d}_{\rm crit, MF}=\frac{2}{\bar{z}}\ln{(2)},
\end{equation}
and sets the mean field dynamical critical coupling. For $\bar{z}=4$ we coincidentally recover the Bethe-Guggenheim statical critical coupling.

Combining our results for the locations of the global minimum and maximum we obtain the following results for the mean first passage time to cluster dissolution and formation per bond
\begin{equation}
    \ln\langle {\rm
    t_{d,f}}\rangle_{\rm MF}=
    \left\{
    \begin{array}{lc}
    \tilde{\rm f^{MF}}(0)-\tilde{\rm f^{MF}}(\frac{1}{2}), & 0 \le \tj\le \frac{1}{\bar{z}} \\
    \tilde{\rm f^{MF}}(0)-\tilde{\rm f^{MF}}(\frac{1}{2}[1\pm s]), & \frac{1}{\bar{z}}\le\!\tj\!\le\frac{2}{\bar{z}}\ln{(2)}\\
    \tilde{\rm f^{MF}}(\frac{1}{2})-\tilde{\rm f^{MF}}(\frac{1}{2}[1\pm s]), & \tj \!\ge\frac{2}{\bar{z}}\ln{(2)},
    \end{array},
    \right.
    \label{MFPT bare field MF}
\end{equation}
where $s$ is the non-zero solution of Eq.~\eqref{MF equation}. Evaluating Eq.~\eqref{MFPT bare field MF} explicitly using Eq.~\eqref{TD MF} we find
\begin{equation}
    \langle {\rm
    t_{d,f}}\rangle_{\rm MF}=
    \left\{
    \begin{array}{lc}
    2\e{-\bar{z}\tilde{J}/2}, & 0 \le \tj\le \frac{1}{\bar{z}} \\
    2\e{-\bar{z}\tilde{J}(1-s^{2})/2}\lambda(s), & \frac{1}{\bar{z}}\le\!\tj\!\le\frac{2}{\bar{z}}\ln{2}\\
    \e{\bar{z}\tilde{J}s^{2}/2}\lambda(s), & \tj \!\ge\frac{2}{\bar{z}}\ln{(2)},
    \end{array}
    \right.
    \label{MFPT bare field MF explicit}
\end{equation}
where 
\begin{equation}
    \lambda(s)\equiv\frac{\exp(-s\arctanh{(s)})}{\sqrt{1-s^{2}}}=\frac{\exp(-\bar{z}\tj s^{2})}{\sqrt{1-s^{2}}},
\end{equation}
and in the second equality we used Eq.~\eqref{TD MF} to make the substitution $\arctanh{(s)}=\bar{z}\tj s$.

As with the Bethe-Guggenheim analysis, we find a global minimum of $\langle {\rm t}_{\rm d,f} \rangle_{\rm MF}$ at the dynamical critical coupling given by
\begin{equation}
    \langle {\rm t}_{\rm d,f} \rangle_{\rm MF}|_{\tj=\frac{2}{\bar{z}}\ln{(2)}}=\frac{\exp{(-\ln{(2)}s_{\rm crit}^{2})}}{\sqrt{1-s^{2}_{\rm crit}}}=1.0785(1)
    \label{min time MF},
\end{equation}
where $s_{\rm crit}$ is the solution of the transcendental equation 
$s_{\rm crit}=\tanh{(2\ln{(2)}s_{\rm crit})}$. Remarkably, the minimum for $\langle {\rm t}_{\rm d,f} \rangle_{\rm MF}$ is independent of the coordination number $\bar{z}$, and therefore the mean field approximation predicts a universal lower bound on the mean first passage time to cluster dissolution and formation per bond.

At the dynamical critical coupling there is a first-order discontinuity w.r.t. the coupling $\tj$ given by
\begin{eqnarray}
    \lim_{\tj \nearrow \frac{2}{\bar{z}}\ln(2)}\partial_{\tj}\langle {\rm
    t_{d,f}}\rangle&=&-\frac{\bar{z}}{2}(1-s^{2}_{\rm crit})\langle {\rm t}_{\rm d,f} \rangle_{\rm MF}|_{\tj=\frac{2}{\bar{z}}\ln{(2)}}\nonumber\\ 
    \lim_{\tj \searrow \frac{2}{\bar{z}}\ln(2)}\partial_{\tj}\langle {\rm
    t_{d,f}}\rangle&=&\frac{\bar{z}}{2}s^{2}_{\rm crit}\langle {\rm t}_{\rm d,f} \rangle_{\rm MF}|_{\tj=\frac{2}{\bar{z}}\ln{(2)}},
\end{eqnarray}
where we use Eq.~\eqref{min time MF}. The derivatives have a trivial dependence on $\bar{z}$ while the ratios of the two derivatives are independent of $\bar{z}$.

In the range $1/\bar{z}<\tilde{J}<\infty$ we need
to solve Eq.~\eqref{MF equation} numerically to get Eq.~\eqref{MFPT bare field MF explicit} fully explicit. In the limit $\tilde{J} \rightarrow \infty$ Eq.~\eqref{MF equation} translates to
\begin{equation}
    s=\theta\left(s\right)-\theta\left(-s\right),
    \label{MF equation strong coupling limit}
\end{equation}
where $\theta\left(s\right)=1$ for $s\geq0$ and zero otherwise. The
two non-zero solutions of Eq.~\eqref{MF equation strong coupling limit} are given by $s=\pm1$, corresponding to $\varphi=0,1$. Finally we use that $\lim_{s\to\pm1}\lambda(s)=1/2$ to find 
\begin{equation}
    \lim\limits_{\tilde{J}\rightarrow\infty}\langle {\rm t}_{\rm d,f} \rangle_{\rm MF}=\lim\limits_{\tilde{J}\rightarrow\infty}\frac{1}{2}\exp{\left(\frac{\bar{z}\tilde{J}}{2}\right)}=\infty \ \forall \bar{z} > 0.
    \label{MF infinite coupling}
\end{equation}
The mean field approximation thus predicts unphysical dynamics in the
strong coupling limit, which is also depicted in Fig.~\ref{fig A8}. The consideration of correlations is required in order to
arrive at a physically correct and consistent result. It is therefore paramount to go beyond the mean field
approximation and consider correlations explicitly.
\bibliography{mybibliography.bib}

\begin{thebibliography}{149}%
\makeatletter
\providecommand \@ifxundefined [1]{%
 \@ifx{#1\undefined}
}%
\providecommand \@ifnum [1]{%
 \ifnum #1\expandafter \@firstoftwo
 \else \expandafter \@secondoftwo
 \fi
}%
\providecommand \@ifx [1]{%
 \ifx #1\expandafter \@firstoftwo
 \else \expandafter \@secondoftwo
 \fi
}%
\providecommand \natexlab [1]{#1}%
\providecommand \enquote  [1]{``#1''}%
\providecommand \bibnamefont  [1]{#1}%
\providecommand \bibfnamefont [1]{#1}%
\providecommand \citenamefont [1]{#1}%
\providecommand \href@noop [0]{\@secondoftwo}%
\providecommand \href [0]{\begingroup \@sanitize@url \@href}%
\providecommand \@href[1]{\@@startlink{#1}\@@href}%
\providecommand \@@href[1]{\endgroup#1\@@endlink}%
\providecommand \@sanitize@url [0]{\catcode `\\12\catcode `\$12\catcode
  `\&12\catcode `\#12\catcode `\^12\catcode `\_12\catcode `\%12\relax}%
\providecommand \@@startlink[1]{}%
\providecommand \@@endlink[0]{}%
\providecommand \url  [0]{\begingroup\@sanitize@url \@url }%
\providecommand \@url [1]{\endgroup\@href {#1}{\urlprefix }}%
\providecommand \urlprefix  [0]{URL }%
\providecommand \Eprint [0]{\href }%
\providecommand \doibase [0]{https://doi.org/}%
\providecommand \selectlanguage [0]{\@gobble}%
\providecommand \bibinfo  [0]{\@secondoftwo}%
\providecommand \bibfield  [0]{\@secondoftwo}%
\providecommand \translation [1]{[#1]}%
\providecommand \BibitemOpen [0]{}%
\providecommand \bibitemStop [0]{}%
\providecommand \bibitemNoStop [0]{.\EOS\space}%
\providecommand \EOS [0]{\spacefactor3000\relax}%
\providecommand \BibitemShut  [1]{\csname bibitem#1\endcsname}%
\let\auto@bib@innerbib\@empty
\bibitem [{\citenamefont {Huang}\ and\ \citenamefont
  {Ingber}(1999)}]{regulation}%
  \BibitemOpen
  \bibfield  {author} {\bibinfo {author} {\bibfnamefont {S.}~\bibnamefont
  {Huang}}\ and\ \bibinfo {author} {\bibfnamefont {D.~E.}\ \bibnamefont
  {Ingber}},\ }\bibfield  {title} {\bibinfo {title} {The structural and
  mechanical complexity of cell-growth control},\ }\href
  {https://doi.org/10.1038/13043} {\bibfield  {journal} {\bibinfo  {journal}
  {Nat. Cell Biol.}\ }\textbf {\bibinfo {volume} {1}},\ \bibinfo {pages}
  {E131–E138} (\bibinfo {year} {1999})}\BibitemShut {NoStop}%
\bibitem [{\citenamefont {Gille}\ and\ \citenamefont
  {Swerlick}(1996)}]{commun}%
  \BibitemOpen
  \bibfield  {author} {\bibinfo {author} {\bibfnamefont {J.}~\bibnamefont
  {Gille}}\ and\ \bibinfo {author} {\bibfnamefont {R.~A.}\ \bibnamefont
  {Swerlick}},\ }\bibfield  {title} {\bibinfo {title} {Integrins: Role in cell
  adhesion and communication},\ }\href
  {https://doi.org/10.1111/j.1749-6632.1996.tb52952.x} {\bibfield  {journal}
  {\bibinfo  {journal} {Ann. N. Y. Acad. Sci.}\ }\textbf {\bibinfo {volume}
  {797}},\ \bibinfo {pages} {93–106} (\bibinfo {year} {1996})}\BibitemShut
  {NoStop}%
\bibitem [{\citenamefont {Zarnitsyna}\ and\ \citenamefont
  {Zhu}(2012)}]{zarnitsyna_t_2012}%
  \BibitemOpen
  \bibfield  {author} {\bibinfo {author} {\bibfnamefont {V.}~\bibnamefont
  {Zarnitsyna}}\ and\ \bibinfo {author} {\bibfnamefont {C.}~\bibnamefont
  {Zhu}},\ }\bibfield  {title} {\bibinfo {title} {T cell triggering: insights
  from {2D} kinetics analysis of molecular interactions},\ }\href
  {https://doi.org/10.1088/1478-3975/9/4/045005} {\bibfield  {journal}
  {\bibinfo  {journal} {Phys. Biol.}\ }\textbf {\bibinfo {volume} {9}},\
  \bibinfo {pages} {045005} (\bibinfo {year} {2012})}\BibitemShut {NoStop}%
\bibitem [{\citenamefont {Brugu{\'e}s}\ \emph {et~al.}(2014)\citenamefont
  {Brugu{\'e}s}, \citenamefont {Anon}, \citenamefont {Conte}, \citenamefont
  {Veldhuis}, \citenamefont {Gupta}, \citenamefont {Colombelli}, \citenamefont
  {Mu\~noz}, \citenamefont {Brodland}, \citenamefont {Ladoux},\ and\
  \citenamefont {Trepat}}]{wound}%
  \BibitemOpen
  \bibfield  {author} {\bibinfo {author} {\bibfnamefont {A.}~\bibnamefont
  {Brugu{\'e}s}}, \bibinfo {author} {\bibfnamefont {E.}~\bibnamefont {Anon}},
  \bibinfo {author} {\bibfnamefont {V.}~\bibnamefont {Conte}}, \bibinfo
  {author} {\bibfnamefont {J.~H.}\ \bibnamefont {Veldhuis}}, \bibinfo {author}
  {\bibfnamefont {M.}~\bibnamefont {Gupta}}, \bibinfo {author} {\bibfnamefont
  {J.}~\bibnamefont {Colombelli}}, \bibinfo {author} {\bibfnamefont {J.~J.}\
  \bibnamefont {Mu\~noz}}, \bibinfo {author} {\bibfnamefont {G.~W.}\
  \bibnamefont {Brodland}}, \bibinfo {author} {\bibfnamefont {B.}~\bibnamefont
  {Ladoux}},\ and\ \bibinfo {author} {\bibfnamefont {X.}~\bibnamefont
  {Trepat}},\ }\bibfield  {title} {\bibinfo {title} {Forces driving epithelial
  wound healing},\ }\href {https://doi.org/10.1038/nphys3040} {\bibfield
  {journal} {\bibinfo  {journal} {Nat. Phys.}\ }\textbf {\bibinfo {volume}
  {10}},\ \bibinfo {pages} {683–690} (\bibinfo {year} {2014})}\BibitemShut
  {NoStop}%
\bibitem [{\citenamefont {Borghi}\ and\ \citenamefont {Nelson}(2009)}]{morpho}%
  \BibitemOpen
  \bibfield  {author} {\bibinfo {author} {\bibfnamefont {N.}~\bibnamefont
  {Borghi}}\ and\ \bibinfo {author} {\bibfnamefont {W.~J.}\ \bibnamefont
  {Nelson}},\ }\bibfield  {title} {\bibinfo {title} {Chapter 1 intercellular
  adhesion in morphogenesis: Molecular and biophysical considerations},\ }in\
  \href {https://doi.org/https://doi.org/10.1016/S0070-2153(09)89001-7} {\emph
  {\bibinfo {booktitle} {Current Topics in Developmental Biology}}},\ \bibinfo
  {series} {Current Topics in Developmental Biology}, Vol.~\bibinfo {volume}
  {89}\ (\bibinfo  {publisher} {Academic Press},\ \bibinfo {year}
  {2009})\BibitemShut {NoStop}%
\bibitem [{\citenamefont {Hynes}(1999)}]{hynes_cell_1999}%
  \BibitemOpen
  \bibfield  {author} {\bibinfo {author} {\bibfnamefont {R.~O.}\ \bibnamefont
  {Hynes}},\ }\bibfield  {title} {\bibinfo {title} {Cell adhesion: old and new
  questions},\ }\href@noop {} {\bibfield  {journal} {\bibinfo  {journal}
  {Trends Cell Biol.}\ }\textbf {\bibinfo {volume} {9}},\ \bibinfo {pages}
  {M33} (\bibinfo {year} {1999})}\BibitemShut {NoStop}%
\bibitem [{\citenamefont {Cavallaro}\ and\ \citenamefont
  {Christofori}(2004)}]{cancer}%
  \BibitemOpen
  \bibfield  {author} {\bibinfo {author} {\bibfnamefont {U.}~\bibnamefont
  {Cavallaro}}\ and\ \bibinfo {author} {\bibfnamefont {G.}~\bibnamefont
  {Christofori}},\ }\bibfield  {title} {\bibinfo {title} {Cell adhesion and
  signalling by cadherins and ig-cams in cancer},\ }\href
  {https://doi.org/10.1038/nrc1276} {\bibfield  {journal} {\bibinfo  {journal}
  {Nat. Rev. Cancer.}\ }\textbf {\bibinfo {volume} {4}},\ \bibinfo {pages}
  {118–132} (\bibinfo {year} {2004})}\BibitemShut {NoStop}%
\bibitem [{\citenamefont {Andl}(2010)}]{andl_misregulation_2010}%
  \BibitemOpen
  \bibfield  {author} {\bibinfo {author} {\bibfnamefont {C.~D.}\ \bibnamefont
  {Andl}},\ }\bibfield  {title} {\bibinfo {title} {The misregulation of cell
  adhesion components during tumorigenesis: overview and commentary},\
  }\href@noop {} {\bibfield  {journal} {\bibinfo  {journal} {J. Oncol.}\
  }\textbf {\bibinfo {volume} {2010}} (\bibinfo {year} {2010})}\BibitemShut
  {NoStop}%
\bibitem [{\citenamefont {Morone}\ \emph {et~al.}(2008)\citenamefont {Morone},
  \citenamefont {Nakada}, \citenamefont {Umemura}, \citenamefont {Usukura},\
  and\ \citenamefont {Kusumi}}]{anchor}%
  \BibitemOpen
  \bibfield  {author} {\bibinfo {author} {\bibfnamefont {N.}~\bibnamefont
  {Morone}}, \bibinfo {author} {\bibfnamefont {C.}~\bibnamefont {Nakada}},
  \bibinfo {author} {\bibfnamefont {Y.}~\bibnamefont {Umemura}}, \bibinfo
  {author} {\bibfnamefont {J.}~\bibnamefont {Usukura}},\ and\ \bibinfo {author}
  {\bibfnamefont {A.}~\bibnamefont {Kusumi}},\ }\bibfield  {title} {\bibinfo
  {title} {Chapter 12 three-dimensional molecular architecture of the
  plasma-membrane-associated cytoskeleton as reconstructed by freeze-etch
  electron tomography},\ }in\ \href
  {https://doi.org/https://doi.org/10.1016/S0091-679X(08)00412-3} {\emph
  {\bibinfo {booktitle} {Introduction to Electron Microscopy for
  Biologists}}},\ \bibinfo {series} {Methods in Cell Biology}, Vol.~\bibinfo
  {volume} {88}\ (\bibinfo  {publisher} {Academic Press},\ \bibinfo {year}
  {2008})\ pp.\ \bibinfo {pages} {207 -- 236}\BibitemShut {NoStop}%
\bibitem [{\citenamefont {DeMond}\ \emph {et~al.}(2008)\citenamefont {DeMond},
  \citenamefont {Mossman}, \citenamefont {Starr}, \citenamefont {Dustin},\ and\
  \citenamefont {Groves}}]{DEMOND20083286}%
  \BibitemOpen
  \bibfield  {author} {\bibinfo {author} {\bibfnamefont {A.~L.}\ \bibnamefont
  {DeMond}}, \bibinfo {author} {\bibfnamefont {K.~D.}\ \bibnamefont {Mossman}},
  \bibinfo {author} {\bibfnamefont {T.}~\bibnamefont {Starr}}, \bibinfo
  {author} {\bibfnamefont {M.~L.}\ \bibnamefont {Dustin}},\ and\ \bibinfo
  {author} {\bibfnamefont {J.~T.}\ \bibnamefont {Groves}},\ }\bibfield  {title}
  {\bibinfo {title} {T cell receptor microcluster transport through molecular
  mazes reveals mechanism of translocation},\ }\href
  {https://doi.org/https://doi.org/10.1529/biophysj.107.119099} {\bibfield
  {journal} {\bibinfo  {journal} {Biophys. J.}\ }\textbf {\bibinfo {volume}
  {94}},\ \bibinfo {pages} {3286} (\bibinfo {year} {2008})}\BibitemShut
  {NoStop}%
\bibitem [{\citenamefont {Bell}(1978)}]{bell_models_1978}%
  \BibitemOpen
  \bibfield  {author} {\bibinfo {author} {\bibfnamefont {G.}~\bibnamefont
  {Bell}},\ }\bibfield  {title} {\bibinfo {title} {Models for the specific
  adhesion of cells to cells},\ }\href {https://doi.org/10.1126/science.347575}
  {\bibfield  {journal} {\bibinfo  {journal} {Science}\ }\textbf {\bibinfo
  {volume} {200}},\ \bibinfo {pages} {618} (\bibinfo {year}
  {1978})}\BibitemShut {NoStop}%
\bibitem [{\citenamefont {Schwarz}\ and\ \citenamefont
  {Safran}(2013)}]{schwarz_physics_2013}%
  \BibitemOpen
  \bibfield  {author} {\bibinfo {author} {\bibfnamefont {U.~S.}\ \bibnamefont
  {Schwarz}}\ and\ \bibinfo {author} {\bibfnamefont {S.~A.}\ \bibnamefont
  {Safran}},\ }\bibfield  {title} {\bibinfo {title} {Physics of adherent
  cells},\ }\href {https://doi.org/10.1103/RevModPhys.85.1327} {\bibfield
  {journal} {\bibinfo  {journal} {Rev. Mod. Phys.}\ }\textbf {\bibinfo {volume}
  {85}},\ \bibinfo {pages} {1327} (\bibinfo {year} {2013})}\BibitemShut
  {NoStop}%
\bibitem [{\citenamefont {Evans}\ and\ \citenamefont
  {Ritchie}(1997)}]{evans_dynamic_1997}%
  \BibitemOpen
  \bibfield  {author} {\bibinfo {author} {\bibfnamefont {E.}~\bibnamefont
  {Evans}}\ and\ \bibinfo {author} {\bibfnamefont {K.}~\bibnamefont
  {Ritchie}},\ }\bibfield  {title} {\bibinfo {title} {Dynamic strength of
  molecular adhesion bonds},\ }\href
  {https://doi.org/10.1016/S0006-3495(97)78802-7} {\bibfield  {journal}
  {\bibinfo  {journal} {Biophys. J.}\ }\textbf {\bibinfo {volume} {72}},\
  \bibinfo {pages} {1541} (\bibinfo {year} {1997})}\BibitemShut {NoStop}%
\bibitem [{\citenamefont {Schmitz}\ and\ \citenamefont
  {Gottschalk}(2008)}]{Schmitz_2008}%
  \BibitemOpen
  \bibfield  {author} {\bibinfo {author} {\bibfnamefont {J.}~\bibnamefont
  {Schmitz}}\ and\ \bibinfo {author} {\bibfnamefont {K.-E.}\ \bibnamefont
  {Gottschalk}},\ }\bibfield  {title} {\bibinfo {title} {Mechanical regulation
  of cell adhesion},\ }\href {https://doi.org/10.1039/b716805p} {\bibfield
  {journal} {\bibinfo  {journal} {Soft Matter}\ }\textbf {\bibinfo {volume}
  {4}},\ \bibinfo {pages} {1373} (\bibinfo {year} {2008})}\BibitemShut
  {NoStop}%
\bibitem [{\citenamefont {Zuckerman}\ and\ \citenamefont
  {Bruinsma}(1995)}]{Bruinsma}%
  \BibitemOpen
  \bibfield  {author} {\bibinfo {author} {\bibfnamefont {D.}~\bibnamefont
  {Zuckerman}}\ and\ \bibinfo {author} {\bibfnamefont {R.}~\bibnamefont
  {Bruinsma}},\ }\bibfield  {title} {\bibinfo {title} {Statistical mechanics of
  membrane adhesion by reversible molecular bonds},\ }\href
  {https://doi.org/10.1103/PhysRevLett.74.3900} {\bibfield  {journal} {\bibinfo
   {journal} {Phys. Rev. Lett.}\ }\textbf {\bibinfo {volume} {74}},\ \bibinfo
  {pages} {3900} (\bibinfo {year} {1995})}\BibitemShut {NoStop}%
\bibitem [{\citenamefont {Speck}(2011)}]{Speck}%
  \BibitemOpen
  \bibfield  {author} {\bibinfo {author} {\bibfnamefont {T.}~\bibnamefont
  {Speck}},\ }\bibfield  {title} {\bibinfo {title} {Effective free energy for
  pinned membranes},\ }\href {https://doi.org/10.1103/PhysRevE.83.050901}
  {\bibfield  {journal} {\bibinfo  {journal} {Phys. Rev. E}\ }\textbf {\bibinfo
  {volume} {83}},\ \bibinfo {pages} {050901} (\bibinfo {year}
  {2011})}\BibitemShut {NoStop}%
\bibitem [{\citenamefont {Speck}\ \emph {et~al.}(2010)\citenamefont {Speck},
  \citenamefont {Reister},\ and\ \citenamefont
  {Seifert}}]{speck_specific_2010}%
  \BibitemOpen
  \bibfield  {author} {\bibinfo {author} {\bibfnamefont {T.}~\bibnamefont
  {Speck}}, \bibinfo {author} {\bibfnamefont {E.}~\bibnamefont {Reister}},\
  and\ \bibinfo {author} {\bibfnamefont {U.}~\bibnamefont {Seifert}},\
  }\bibfield  {title} {\bibinfo {title} {Specific adhesion of membranes:
  Mapping to an effective bond lattice gas},\ }\href
  {https://doi.org/10.1103/PhysRevE.82.021923} {\bibfield  {journal} {\bibinfo
  {journal} {Phys. Rev. E}\ }\textbf {\bibinfo {volume} {82}},\ \bibinfo
  {pages} {021923} (\bibinfo {year} {2010})}\BibitemShut {NoStop}%
\bibitem [{\citenamefont {Bihr}\ \emph {et~al.}(2012)\citenamefont {Bihr},
  \citenamefont {Seifert},\ and\ \citenamefont {Smith}}]{bihr_nucleation_2012}%
  \BibitemOpen
  \bibfield  {author} {\bibinfo {author} {\bibfnamefont {T.}~\bibnamefont
  {Bihr}}, \bibinfo {author} {\bibfnamefont {U.}~\bibnamefont {Seifert}},\ and\
  \bibinfo {author} {\bibfnamefont {A.-S.}\ \bibnamefont {Smith}},\ }\bibfield
  {title} {\bibinfo {title} {Nucleation of {Ligand}-{Receptor} {Domains} in
  {Membrane} {Adhesion}},\ }\href
  {https://doi.org/10.1103/PhysRevLett.109.258101} {\bibfield  {journal}
  {\bibinfo  {journal} {Phys. Rev. Lett.}\ }\textbf {\bibinfo {volume} {109}},\
  \bibinfo {pages} {258101} (\bibinfo {year} {2012})}\BibitemShut {NoStop}%
\bibitem [{\citenamefont {Farago}(2010)}]{farago_fluctuation-induced_2010}%
  \BibitemOpen
  \bibfield  {author} {\bibinfo {author} {\bibfnamefont {O.}~\bibnamefont
  {Farago}},\ }\bibfield  {title} {\bibinfo {title} {Fluctuation-induced
  attraction between adhesion sites of supported membranes},\ }\href
  {https://doi.org/10.1103/PhysRevE.81.050902} {\bibfield  {journal} {\bibinfo
  {journal} {Phys. Rev. E}\ }\textbf {\bibinfo {volume} {81}},\ \bibinfo
  {pages} {050902} (\bibinfo {year} {2010})}\BibitemShut {NoStop}%
\bibitem [{\citenamefont {Tolentino}\ \emph {et~al.}(2008)\citenamefont
  {Tolentino}, \citenamefont {Wu}, \citenamefont {Zarnitsyna}, \citenamefont
  {Fang}, \citenamefont {Dustin},\ and\ \citenamefont
  {Zhu}}]{tolentino_measuring_2008}%
  \BibitemOpen
  \bibfield  {author} {\bibinfo {author} {\bibfnamefont {T.~P.}\ \bibnamefont
  {Tolentino}}, \bibinfo {author} {\bibfnamefont {J.}~\bibnamefont {Wu}},
  \bibinfo {author} {\bibfnamefont {V.~I.}\ \bibnamefont {Zarnitsyna}},
  \bibinfo {author} {\bibfnamefont {Y.}~\bibnamefont {Fang}}, \bibinfo {author}
  {\bibfnamefont {M.~L.}\ \bibnamefont {Dustin}},\ and\ \bibinfo {author}
  {\bibfnamefont {C.}~\bibnamefont {Zhu}},\ }\bibfield  {title} {\bibinfo
  {title} {Measuring {Diffusion} and {Binding} {Kinetics} by {Contact} {Area}
  {FRAP}},\ }\href {https://doi.org/10.1529/biophysj.107.114447} {\bibfield
  {journal} {\bibinfo  {journal} {Biophys. J.}\ }\textbf {\bibinfo {volume}
  {95}},\ \bibinfo {pages} {920} (\bibinfo {year} {2008})}\BibitemShut
  {NoStop}%
\bibitem [{\citenamefont {Huppa}\ \emph {et~al.}(2010)\citenamefont {Huppa},
  \citenamefont {Axmann}, \citenamefont {Mörtelmaier}, \citenamefont
  {Lillemeier}, \citenamefont {Newell}, \citenamefont {Brameshuber},
  \citenamefont {Klein}, \citenamefont {Schütz},\ and\ \citenamefont
  {Davis}}]{huppa_tcrpeptidemhc_2010}%
  \BibitemOpen
  \bibfield  {author} {\bibinfo {author} {\bibfnamefont {J.~B.}\ \bibnamefont
  {Huppa}}, \bibinfo {author} {\bibfnamefont {M.}~\bibnamefont {Axmann}},
  \bibinfo {author} {\bibfnamefont {M.~A.}\ \bibnamefont {Mörtelmaier}},
  \bibinfo {author} {\bibfnamefont {B.~F.}\ \bibnamefont {Lillemeier}},
  \bibinfo {author} {\bibfnamefont {E.~W.}\ \bibnamefont {Newell}}, \bibinfo
  {author} {\bibfnamefont {M.}~\bibnamefont {Brameshuber}}, \bibinfo {author}
  {\bibfnamefont {L.~O.}\ \bibnamefont {Klein}}, \bibinfo {author}
  {\bibfnamefont {G.~J.}\ \bibnamefont {Schütz}},\ and\ \bibinfo {author}
  {\bibfnamefont {M.~M.}\ \bibnamefont {Davis}},\ }\bibfield  {title} {\bibinfo
  {title} {{TCR}–peptide–{MHC} interactions in situ show accelerated
  kinetics and increased affinity},\ }\href
  {https://doi.org/10.1038/nature08746} {\bibfield  {journal} {\bibinfo
  {journal} {Nature}\ }\textbf {\bibinfo {volume} {463}},\ \bibinfo {pages}
  {963} (\bibinfo {year} {2010})}\BibitemShut {NoStop}%
\bibitem [{\citenamefont {Baronsky}\ \emph {et~al.}(2017)\citenamefont
  {Baronsky}, \citenamefont {Ruhlandt}, \citenamefont {Brückner},
  \citenamefont {Schäfer}, \citenamefont {Karedla}, \citenamefont {Isbaner},
  \citenamefont {Hähnel}, \citenamefont {Gregor}, \citenamefont {Enderlein},
  \citenamefont {Janshoff},\ and\ \citenamefont
  {Chizhik}}]{baronsky_cellsubstrate_2017}%
  \BibitemOpen
  \bibfield  {author} {\bibinfo {author} {\bibfnamefont {T.}~\bibnamefont
  {Baronsky}}, \bibinfo {author} {\bibfnamefont {D.}~\bibnamefont {Ruhlandt}},
  \bibinfo {author} {\bibfnamefont {B.~R.}\ \bibnamefont {Brückner}}, \bibinfo
  {author} {\bibfnamefont {J.}~\bibnamefont {Schäfer}}, \bibinfo {author}
  {\bibfnamefont {N.}~\bibnamefont {Karedla}}, \bibinfo {author} {\bibfnamefont
  {S.}~\bibnamefont {Isbaner}}, \bibinfo {author} {\bibfnamefont
  {D.}~\bibnamefont {Hähnel}}, \bibinfo {author} {\bibfnamefont
  {I.}~\bibnamefont {Gregor}}, \bibinfo {author} {\bibfnamefont
  {J.}~\bibnamefont {Enderlein}}, \bibinfo {author} {\bibfnamefont
  {A.}~\bibnamefont {Janshoff}},\ and\ \bibinfo {author} {\bibfnamefont
  {A.~I.}\ \bibnamefont {Chizhik}},\ }\bibfield  {title} {\bibinfo {title}
  {Cell–{Substrate} {Dynamics} of the {Epithelial}-to-{Mesenchymal}
  {Transition}},\ }\href {https://doi.org/10.1021/acs.nanolett.7b01558}
  {\bibfield  {journal} {\bibinfo  {journal} {Nano Lett.}\ }\textbf {\bibinfo
  {volume} {17}},\ \bibinfo {pages} {3320} (\bibinfo {year}
  {2017})}\BibitemShut {NoStop}%
\bibitem [{\citenamefont {Limozin}\ and\ \citenamefont
  {Sengupta}(2009)}]{limozin_quantitative_2009}%
  \BibitemOpen
  \bibfield  {author} {\bibinfo {author} {\bibfnamefont {L.}~\bibnamefont
  {Limozin}}\ and\ \bibinfo {author} {\bibfnamefont {K.}~\bibnamefont
  {Sengupta}},\ }\bibfield  {title} {\bibinfo {title} {Quantitative
  {Reflection} {Interference} {Contrast} {Microscopy} ({RICM}) in {Soft}
  {Matter} and {Cell} {Adhesion}},\ }\href
  {https://doi.org/10.1002/cphc.200900601} {\bibfield  {journal} {\bibinfo
  {journal} {ChemPhysChem}\ }\textbf {\bibinfo {volume} {10}},\ \bibinfo
  {pages} {2752} (\bibinfo {year} {2009})}\BibitemShut {NoStop}%
\bibitem [{\citenamefont {Fällman}\ \emph {et~al.}(2004)\citenamefont
  {Fällman}, \citenamefont {Schedin}, \citenamefont {Jass}, \citenamefont
  {Andersson}, \citenamefont {Uhlin},\ and\ \citenamefont
  {Axner}}]{fallman_optical_2004}%
  \BibitemOpen
  \bibfield  {author} {\bibinfo {author} {\bibfnamefont {E.}~\bibnamefont
  {Fällman}}, \bibinfo {author} {\bibfnamefont {S.}~\bibnamefont {Schedin}},
  \bibinfo {author} {\bibfnamefont {J.}~\bibnamefont {Jass}}, \bibinfo {author}
  {\bibfnamefont {M.}~\bibnamefont {Andersson}}, \bibinfo {author}
  {\bibfnamefont {B.~E.}\ \bibnamefont {Uhlin}},\ and\ \bibinfo {author}
  {\bibfnamefont {O.}~\bibnamefont {Axner}},\ }\bibfield  {title} {\bibinfo
  {title} {Optical tweezers based force measurement system for quantitating
  binding interactions: system design and application for the study of
  bacterial adhesion},\ }\href {https://doi.org/10.1016/j.bios.2003.12.029}
  {\bibfield  {journal} {\bibinfo  {journal} {Biosens. Bioelectron.}\ }\bibinfo
  {series} {Micro and {Nano} {Bioengineering}},\ \textbf {\bibinfo {volume}
  {19}},\ \bibinfo {pages} {1429} (\bibinfo {year} {2004})}\BibitemShut
  {NoStop}%
\bibitem [{\citenamefont {Alon}\ \emph {et~al.}(1995)\citenamefont {Alon},
  \citenamefont {Hammer},\ and\ \citenamefont {Springer}}]{alon_lifetime_1995}%
  \BibitemOpen
  \bibfield  {author} {\bibinfo {author} {\bibfnamefont {R.}~\bibnamefont
  {Alon}}, \bibinfo {author} {\bibfnamefont {D.~A.}\ \bibnamefont {Hammer}},\
  and\ \bibinfo {author} {\bibfnamefont {T.~A.}\ \bibnamefont {Springer}},\
  }\bibfield  {title} {\bibinfo {title} {Lifetime of the
  {P}-selectin-carbohydrate bond and its response to tensile force in
  hydrodynamic flow},\ }\href {https://doi.org/10.1038/374539a0} {\bibfield
  {journal} {\bibinfo  {journal} {Nature}\ }\textbf {\bibinfo {volume} {374}},\
  \bibinfo {pages} {539} (\bibinfo {year} {1995})}\BibitemShut {NoStop}%
\bibitem [{\citenamefont {Juliano}\ and\ \citenamefont
  {Gagalang}(1977)}]{juliano_adhesion_1977}%
  \BibitemOpen
  \bibfield  {author} {\bibinfo {author} {\bibfnamefont {R.~L.}\ \bibnamefont
  {Juliano}}\ and\ \bibinfo {author} {\bibfnamefont {E.}~\bibnamefont
  {Gagalang}},\ }\bibfield  {title} {\bibinfo {title} {The adhesion of
  {Chinese} hamster cells. {I}. {Effects} of temperature, metabolic inhibitors
  and proteolytic dissection of cell surface macromolecules},\ }\href
  {https://doi.org/10.1002/jcp.1040920209} {\bibfield  {journal} {\bibinfo
  {journal} {J. Cell}\ }\textbf {\bibinfo {volume} {92}},\ \bibinfo {pages}
  {209} (\bibinfo {year} {1977})}\BibitemShut {NoStop}%
\bibitem [{\citenamefont {Piper}\ \emph {et~al.}(1998)\citenamefont {Piper},
  \citenamefont {Swerlick},\ and\ \citenamefont
  {Zhu}}]{piper_determining_1998}%
  \BibitemOpen
  \bibfield  {author} {\bibinfo {author} {\bibfnamefont {J.~W.}\ \bibnamefont
  {Piper}}, \bibinfo {author} {\bibfnamefont {R.~A.}\ \bibnamefont
  {Swerlick}},\ and\ \bibinfo {author} {\bibfnamefont {C.}~\bibnamefont
  {Zhu}},\ }\bibfield  {title} {\bibinfo {title} {Determining force dependence
  of two-dimensional receptor-ligand binding affinity by centrifugation.},\
  }\href {https://www.ncbi.nlm.nih.gov/pmc/articles/PMC1299402/} {\bibfield
  {journal} {\bibinfo  {journal} {Biophys. J.}\ }\textbf {\bibinfo {volume}
  {74}},\ \bibinfo {pages} {492} (\bibinfo {year} {1998})}\BibitemShut
  {NoStop}%
\bibitem [{\citenamefont {Marlin}\ and\ \citenamefont
  {Springer}(1987)}]{marlin_purified_1987}%
  \BibitemOpen
  \bibfield  {author} {\bibinfo {author} {\bibfnamefont {S.~D.}\ \bibnamefont
  {Marlin}}\ and\ \bibinfo {author} {\bibfnamefont {T.~A.}\ \bibnamefont
  {Springer}},\ }\bibfield  {title} {\bibinfo {title} {Purified intercellular
  adhesion molecule-1 ({ICAM}-1) is a ligand for lymphocyte function-associated
  antigen 1 ({LFA}-1)},\ }\href {https://doi.org/10.1016/0092-8674(87)90104-8}
  {\bibfield  {journal} {\bibinfo  {journal} {Cell}\ }\textbf {\bibinfo
  {volume} {51}},\ \bibinfo {pages} {813} (\bibinfo {year} {1987})}\BibitemShut
  {NoStop}%
\bibitem [{\citenamefont {Chen}\ \emph {et~al.}(2008)\citenamefont {Chen},
  \citenamefont {Evans}, \citenamefont {McEver},\ and\ \citenamefont
  {Zhu}}]{chen_monitoring_2008}%
  \BibitemOpen
  \bibfield  {author} {\bibinfo {author} {\bibfnamefont {W.}~\bibnamefont
  {Chen}}, \bibinfo {author} {\bibfnamefont {E.~A.}\ \bibnamefont {Evans}},
  \bibinfo {author} {\bibfnamefont {R.~P.}\ \bibnamefont {McEver}},\ and\
  \bibinfo {author} {\bibfnamefont {C.}~\bibnamefont {Zhu}},\ }\bibfield
  {title} {\bibinfo {title} {Monitoring {Receptor}-{Ligand} {Interactions}
  between {Surfaces} by {Thermal} {Fluctuations}},\ }\href
  {https://doi.org/10.1529/biophysj.107.117895} {\bibfield  {journal} {\bibinfo
   {journal} {Biophys. J.}\ }\textbf {\bibinfo {volume} {94}},\ \bibinfo
  {pages} {694} (\bibinfo {year} {2008})}\BibitemShut {NoStop}%
\bibitem [{\citenamefont {Evans}\ \emph {et~al.}(1995)\citenamefont {Evans},
  \citenamefont {Ritchie},\ and\ \citenamefont
  {Merkel}}]{evans_sensitive_1995}%
  \BibitemOpen
  \bibfield  {author} {\bibinfo {author} {\bibfnamefont {E.}~\bibnamefont
  {Evans}}, \bibinfo {author} {\bibfnamefont {K.}~\bibnamefont {Ritchie}},\
  and\ \bibinfo {author} {\bibfnamefont {R.}~\bibnamefont {Merkel}},\
  }\bibfield  {title} {\bibinfo {title} {Sensitive force technique to probe
  molecular adhesion and structural linkages at biological interfaces.},\
  }\href {https://doi.org/10.1016/S0006-3495(95)80441-8} {\bibfield  {journal}
  {\bibinfo  {journal} {Biophys. J.}\ }\textbf {\bibinfo {volume} {68}},\
  \bibinfo {pages} {2580} (\bibinfo {year} {1995})}\BibitemShut {NoStop}%
\bibitem [{\citenamefont {Prechtel}\ \emph {et~al.}(2002)\citenamefont
  {Prechtel}, \citenamefont {Bausch}, \citenamefont {Marchi-Artzner},
  \citenamefont {Kantlehner}, \citenamefont {Kessler},\ and\ \citenamefont
  {Merkel}}]{prechtel_dynamic_2002}%
  \BibitemOpen
  \bibfield  {author} {\bibinfo {author} {\bibfnamefont {K.}~\bibnamefont
  {Prechtel}}, \bibinfo {author} {\bibfnamefont {A.~R.}\ \bibnamefont
  {Bausch}}, \bibinfo {author} {\bibfnamefont {V.}~\bibnamefont
  {Marchi-Artzner}}, \bibinfo {author} {\bibfnamefont {M.}~\bibnamefont
  {Kantlehner}}, \bibinfo {author} {\bibfnamefont {H.}~\bibnamefont
  {Kessler}},\ and\ \bibinfo {author} {\bibfnamefont {R.}~\bibnamefont
  {Merkel}},\ }\bibfield  {title} {\bibinfo {title} {Dynamic {Force}
  {Spectroscopy} to {Probe} {Adhesion} {Strength} of {Living} {Cells}},\ }\href
  {https://doi.org/10.1103/PhysRevLett.89.028101} {\bibfield  {journal}
  {\bibinfo  {journal} {Phys. Rev. Lett.}\ }\textbf {\bibinfo {volume} {89}},\
  \bibinfo {pages} {028101} (\bibinfo {year} {2002})}\BibitemShut {NoStop}%
\bibitem [{\citenamefont {Lomakina}\ and\ \citenamefont
  {Waugh}(2004)}]{lomakina_micromechanical_2004}%
  \BibitemOpen
  \bibfield  {author} {\bibinfo {author} {\bibfnamefont {E.~B.}\ \bibnamefont
  {Lomakina}}\ and\ \bibinfo {author} {\bibfnamefont {R.~E.}\ \bibnamefont
  {Waugh}},\ }\bibfield  {title} {\bibinfo {title} {Micromechanical {Tests} of
  {Adhesion} {Dynamics} between {Neutrophils} and {Immobilized} {ICAM}-1},\
  }\href {https://doi.org/10.1016/S0006-3495(04)74196-X} {\bibfield  {journal}
  {\bibinfo  {journal} {Biophys. J.}\ }\textbf {\bibinfo {volume} {86}},\
  \bibinfo {pages} {1223} (\bibinfo {year} {2004})}\BibitemShut {NoStop}%
\bibitem [{\citenamefont {Heymann}\ and\ \citenamefont
  {Grubmüller}(2000)}]{heymann_dynamic_2000}%
  \BibitemOpen
  \bibfield  {author} {\bibinfo {author} {\bibfnamefont {B.}~\bibnamefont
  {Heymann}}\ and\ \bibinfo {author} {\bibfnamefont {H.}~\bibnamefont
  {Grubmüller}},\ }\bibfield  {title} {\bibinfo {title} {Dynamic {Force}
  {Spectroscopy} of {Molecular} {Adhesion} {Bonds}},\ }\href
  {https://doi.org/10.1103/PhysRevLett.84.6126} {\bibfield  {journal} {\bibinfo
   {journal} {Phys. Rev. Lett.}\ }\textbf {\bibinfo {volume} {84}},\ \bibinfo
  {pages} {6126} (\bibinfo {year} {2000})}\BibitemShut {NoStop}%
\bibitem [{\citenamefont {Sanyour}\ \emph {et~al.}(2018)\citenamefont
  {Sanyour}, \citenamefont {Childs}, \citenamefont {Meininger},\ and\
  \citenamefont {Hong}}]{sanyour_spontaneous_2018}%
  \BibitemOpen
  \bibfield  {author} {\bibinfo {author} {\bibfnamefont {H.}~\bibnamefont
  {Sanyour}}, \bibinfo {author} {\bibfnamefont {J.}~\bibnamefont {Childs}},
  \bibinfo {author} {\bibfnamefont {G.~A.}\ \bibnamefont {Meininger}},\ and\
  \bibinfo {author} {\bibfnamefont {Z.}~\bibnamefont {Hong}},\ }\bibfield
  {title} {\bibinfo {title} {Spontaneous oscillation in cell adhesion and
  stiffness measured using atomic force microscopy},\ }\href
  {https://doi.org/10.1038/s41598-018-21253-9} {\bibfield  {journal} {\bibinfo
  {journal} {Sci. Rep.}\ }\textbf {\bibinfo {volume} {8}},\ \bibinfo {pages}
  {1} (\bibinfo {year} {2018})}\BibitemShut {NoStop}%
\bibitem [{\citenamefont {Rico}\ \emph {et~al.}(2010)\citenamefont {Rico},
  \citenamefont {Chu}, \citenamefont {Abdulreda}, \citenamefont {Qin},\ and\
  \citenamefont {Moy}}]{rico_temperature_2010}%
  \BibitemOpen
  \bibfield  {author} {\bibinfo {author} {\bibfnamefont {F.}~\bibnamefont
  {Rico}}, \bibinfo {author} {\bibfnamefont {C.}~\bibnamefont {Chu}}, \bibinfo
  {author} {\bibfnamefont {M.~H.}\ \bibnamefont {Abdulreda}}, \bibinfo {author}
  {\bibfnamefont {Y.}~\bibnamefont {Qin}},\ and\ \bibinfo {author}
  {\bibfnamefont {V.~T.}\ \bibnamefont {Moy}},\ }\bibfield  {title} {\bibinfo
  {title} {Temperature {Modulation} of {Integrin}-{Mediated} {Cell}
  {Adhesion}},\ }\href {https://doi.org/10.1016/j.bpj.2010.06.037} {\bibfield
  {journal} {\bibinfo  {journal} {Biophys. J.}\ }\textbf {\bibinfo {volume}
  {99}},\ \bibinfo {pages} {1387} (\bibinfo {year} {2010})}\BibitemShut
  {NoStop}%
\bibitem [{\citenamefont {Sagvolden}\ \emph {et~al.}(1999)\citenamefont
  {Sagvolden}, \citenamefont {Giaever}, \citenamefont {Pettersen},\ and\
  \citenamefont {Feder}}]{sagvolden_cell_1999}%
  \BibitemOpen
  \bibfield  {author} {\bibinfo {author} {\bibfnamefont {G.}~\bibnamefont
  {Sagvolden}}, \bibinfo {author} {\bibfnamefont {I.}~\bibnamefont {Giaever}},
  \bibinfo {author} {\bibfnamefont {E.~O.}\ \bibnamefont {Pettersen}},\ and\
  \bibinfo {author} {\bibfnamefont {J.}~\bibnamefont {Feder}},\ }\bibfield
  {title} {\bibinfo {title} {Cell adhesion force microscopy},\ }\href
  {https://doi.org/10.1073/pnas.96.2.471} {\bibfield  {journal} {\bibinfo
  {journal} {Proc. Natl. Acad. Sci. U.S.A.}\ }\textbf {\bibinfo {volume}
  {96}},\ \bibinfo {pages} {471} (\bibinfo {year} {1999})}\BibitemShut
  {NoStop}%
\bibitem [{\citenamefont {Makarov}(2009)}]{Makarov}%
  \BibitemOpen
  \bibfield  {author} {\bibinfo {author} {\bibfnamefont {D.~E.}\ \bibnamefont
  {Makarov}},\ }\bibfield  {title} {\bibinfo {title} {A theoretical model for
  the mechanical unfolding of repeat proteins},\ }\href
  {https://doi.org/https://doi.org/10.1016/j.bpj.2008.12.3899} {\bibfield
  {journal} {\bibinfo  {journal} {Biophys. J.}\ }\textbf {\bibinfo {volume}
  {96}},\ \bibinfo {pages} {2160 } (\bibinfo {year} {2009})}\BibitemShut
  {NoStop}%
\bibitem [{\citenamefont {Makarov}\ \emph {et~al.}(2002)\citenamefont
  {Makarov}, \citenamefont {Wang}, \citenamefont {Thompson},\ and\
  \citenamefont {Hansma}}]{Makarov_2}%
  \BibitemOpen
  \bibfield  {author} {\bibinfo {author} {\bibfnamefont {D.~E.}\ \bibnamefont
  {Makarov}}, \bibinfo {author} {\bibfnamefont {Z.}~\bibnamefont {Wang}},
  \bibinfo {author} {\bibfnamefont {J.~B.}\ \bibnamefont {Thompson}},\ and\
  \bibinfo {author} {\bibfnamefont {H.~G.}\ \bibnamefont {Hansma}},\ }\bibfield
   {title} {\bibinfo {title} {On the interpretation of force extension curves
  of single protein molecules},\ }\href {https://doi.org/10.1063/1.1466835}
  {\bibfield  {journal} {\bibinfo  {journal} {J. Chem. Phys.}\ }\textbf
  {\bibinfo {volume} {116}},\ \bibinfo {pages} {7760} (\bibinfo {year}
  {2002})}\BibitemShut {NoStop}%
\bibitem [{\citenamefont {Williams}\ \emph {et~al.}(2001)\citenamefont
  {Williams}, \citenamefont {Nagarajan}, \citenamefont {Selvaraj},\ and\
  \citenamefont {Zhu}}]{williams_quantifying_2001}%
  \BibitemOpen
  \bibfield  {author} {\bibinfo {author} {\bibfnamefont {T.~E.}\ \bibnamefont
  {Williams}}, \bibinfo {author} {\bibfnamefont {S.}~\bibnamefont {Nagarajan}},
  \bibinfo {author} {\bibfnamefont {P.}~\bibnamefont {Selvaraj}},\ and\
  \bibinfo {author} {\bibfnamefont {C.}~\bibnamefont {Zhu}},\ }\bibfield
  {title} {\bibinfo {title} {Quantifying the {Impact} of {Membrane}
  {Microtopology} on {Effective} {Two}-dimensional {Affinity}},\ }\href
  {https://doi.org/10.1074/jbc.M010427200} {\bibfield  {journal} {\bibinfo
  {journal} {J. Biol. Chem.}\ }\textbf {\bibinfo {volume} {276}},\ \bibinfo
  {pages} {13283} (\bibinfo {year} {2001})}\BibitemShut {NoStop}%
\bibitem [{\citenamefont {Wu}\ \emph {et~al.}(2011)\citenamefont {Wu},
  \citenamefont {Vendome}, \citenamefont {Shapiro}, \citenamefont {Ben-Shaul},\
  and\ \citenamefont {Honig}}]{wu_transforming_2011}%
  \BibitemOpen
  \bibfield  {author} {\bibinfo {author} {\bibfnamefont {Y.}~\bibnamefont
  {Wu}}, \bibinfo {author} {\bibfnamefont {J.}~\bibnamefont {Vendome}},
  \bibinfo {author} {\bibfnamefont {L.}~\bibnamefont {Shapiro}}, \bibinfo
  {author} {\bibfnamefont {A.}~\bibnamefont {Ben-Shaul}},\ and\ \bibinfo
  {author} {\bibfnamefont {B.}~\bibnamefont {Honig}},\ }\bibfield  {title}
  {\bibinfo {title} {Transforming binding affinities from {3D} to {2D} with
  application to cadherin clustering},\ }\href
  {https://doi.org/10.1038/nature10183} {\bibfield  {journal} {\bibinfo
  {journal} {Nature}\ }\textbf {\bibinfo {volume} {475}},\ \bibinfo {pages}
  {510} (\bibinfo {year} {2011})}\BibitemShut {NoStop}%
\bibitem [{\citenamefont {Fenz}\ \emph {et~al.}(2017)\citenamefont {Fenz},
  \citenamefont {Bihr}, \citenamefont {Schmidt}, \citenamefont {Merkel},
  \citenamefont {Seifert}, \citenamefont {Sengupta},\ and\ \citenamefont
  {Smith}}]{fenz_membrane_2017}%
  \BibitemOpen
  \bibfield  {author} {\bibinfo {author} {\bibfnamefont {S.~F.}\ \bibnamefont
  {Fenz}}, \bibinfo {author} {\bibfnamefont {T.}~\bibnamefont {Bihr}}, \bibinfo
  {author} {\bibfnamefont {D.}~\bibnamefont {Schmidt}}, \bibinfo {author}
  {\bibfnamefont {R.}~\bibnamefont {Merkel}}, \bibinfo {author} {\bibfnamefont
  {U.}~\bibnamefont {Seifert}}, \bibinfo {author} {\bibfnamefont
  {K.}~\bibnamefont {Sengupta}},\ and\ \bibinfo {author} {\bibfnamefont
  {A.-S.}\ \bibnamefont {Smith}},\ }\bibfield  {title} {\bibinfo {title}
  {Membrane fluctuations mediate lateral interaction between cadherin bonds},\
  }\href {https://doi.org/10.1038/nphys4138} {\bibfield  {journal} {\bibinfo
  {journal} {Nat. Phys.}\ }\textbf {\bibinfo {volume} {13}},\ \bibinfo {pages}
  {906} (\bibinfo {year} {2017})}\BibitemShut {NoStop}%
\bibitem [{\citenamefont {Merkel}(2001)}]{merkel_force_2001}%
  \BibitemOpen
  \bibfield  {author} {\bibinfo {author} {\bibfnamefont {R.}~\bibnamefont
  {Merkel}},\ }\bibfield  {title} {\bibinfo {title} {Force spectroscopy on
  single passive biomolecules and single biomolecular bonds},\ }\href
  {https://doi.org/10.1016/S0370-1573(00)00103-4} {\bibfield  {journal}
  {\bibinfo  {journal} {Phys. Rep.}\ }\textbf {\bibinfo {volume} {346}},\
  \bibinfo {pages} {343} (\bibinfo {year} {2001})}\BibitemShut {NoStop}%
\bibitem [{\citenamefont {Gao}\ \emph {et~al.}(2011)\citenamefont {Gao},
  \citenamefont {Qian},\ and\ \citenamefont {Chen}}]{gao_probing_2011}%
  \BibitemOpen
  \bibfield  {author} {\bibinfo {author} {\bibfnamefont {H.}~\bibnamefont
  {Gao}}, \bibinfo {author} {\bibfnamefont {J.}~\bibnamefont {Qian}},\ and\
  \bibinfo {author} {\bibfnamefont {B.}~\bibnamefont {Chen}},\ }\bibfield
  {title} {\bibinfo {title} {Probing mechanical principles of focal contacts in
  cell–matrix adhesion with a coupled stochastic–elastic modelling
  framework},\ }\href {https://doi.org/10.1098/rsif.2011.0157} {\bibfield
  {journal} {\bibinfo  {journal} {J. R. Soc. Interface}\ }\textbf {\bibinfo
  {volume} {8}},\ \bibinfo {pages} {1217} (\bibinfo {year} {2011})}\BibitemShut
  {NoStop}%
\bibitem [{\citenamefont {Hu}\ \emph {et~al.}(2013)\citenamefont {Hu},
  \citenamefont {Lipowsky},\ and\ \citenamefont {Weikl}}]{hu_binding_2013}%
  \BibitemOpen
  \bibfield  {author} {\bibinfo {author} {\bibfnamefont {J.}~\bibnamefont
  {Hu}}, \bibinfo {author} {\bibfnamefont {R.}~\bibnamefont {Lipowsky}},\ and\
  \bibinfo {author} {\bibfnamefont {T.~R.}\ \bibnamefont {Weikl}},\ }\bibfield
  {title} {\bibinfo {title} {Binding constants of membrane-anchored receptors
  and ligands depend strongly on the nanoscale roughness of membranes},\ }\href
  {https://doi.org/10.1073/pnas.1305766110} {\bibfield  {journal} {\bibinfo
  {journal} {Proc. Natl. Acad. Sci. U.S.A.}\ }\textbf {\bibinfo {volume}
  {110}},\ \bibinfo {pages} {15283} (\bibinfo {year} {2013})}\BibitemShut
  {NoStop}%
\bibitem [{\citenamefont {Steink{\"u}hler}\ \emph {et~al.}(2019)\citenamefont
  {Steink{\"u}hler}, \citenamefont {R{\'o}{\.z}ycki}, \citenamefont {Alvey},
  \citenamefont {Lipowsky}, \citenamefont {Weikl}, \citenamefont {Dimova},\
  and\ \citenamefont {Discher}}]{steinkuhler2019membrane}%
  \BibitemOpen
  \bibfield  {author} {\bibinfo {author} {\bibfnamefont {J.}~\bibnamefont
  {Steink{\"u}hler}}, \bibinfo {author} {\bibfnamefont {B.}~\bibnamefont
  {R{\'o}{\.z}ycki}}, \bibinfo {author} {\bibfnamefont {C.}~\bibnamefont
  {Alvey}}, \bibinfo {author} {\bibfnamefont {R.}~\bibnamefont {Lipowsky}},
  \bibinfo {author} {\bibfnamefont {T.~R.}\ \bibnamefont {Weikl}}, \bibinfo
  {author} {\bibfnamefont {R.}~\bibnamefont {Dimova}},\ and\ \bibinfo {author}
  {\bibfnamefont {D.~E.}\ \bibnamefont {Discher}},\ }\bibfield  {title}
  {\bibinfo {title} {Membrane fluctuations and acidosis regulate cooperative
  binding of ‘marker of self’protein cd47 with the macrophage checkpoint
  receptor sirp$\alpha$},\ }\href@noop {} {\bibfield  {journal} {\bibinfo
  {journal} {J. Cell Sci.}\ }\textbf {\bibinfo {volume} {132}} (\bibinfo {year}
  {2019})}\BibitemShut {NoStop}%
\bibitem [{\citenamefont {Hu}\ \emph {et~al.}(2015)\citenamefont {Hu},
  \citenamefont {Xu}, \citenamefont {Lipowsky},\ and\ \citenamefont
  {Weikl}}]{hu_binding_2015}%
  \BibitemOpen
  \bibfield  {author} {\bibinfo {author} {\bibfnamefont {J.}~\bibnamefont
  {Hu}}, \bibinfo {author} {\bibfnamefont {G.-K.}\ \bibnamefont {Xu}}, \bibinfo
  {author} {\bibfnamefont {R.}~\bibnamefont {Lipowsky}},\ and\ \bibinfo
  {author} {\bibfnamefont {T.~R.}\ \bibnamefont {Weikl}},\ }\bibfield  {title}
  {\bibinfo {title} {Binding kinetics of membrane-anchored receptors and
  ligands: {Molecular} dynamics simulations and theory},\ }\href
  {https://doi.org/10.1063/1.4936135} {\bibfield  {journal} {\bibinfo
  {journal} {J. Chem. Phys.}\ }\textbf {\bibinfo {volume} {143}},\ \bibinfo
  {pages} {243137} (\bibinfo {year} {2015})}\BibitemShut {NoStop}%
\bibitem [{\citenamefont {Krobath}\ \emph {et~al.}(2009)\citenamefont
  {Krobath}, \citenamefont {Rózycki}, \citenamefont {Lipowsky},\ and\
  \citenamefont {Weikl}}]{krobath_binding_2009}%
  \BibitemOpen
  \bibfield  {author} {\bibinfo {author} {\bibfnamefont {H.}~\bibnamefont
  {Krobath}}, \bibinfo {author} {\bibfnamefont {B.}~\bibnamefont {Rózycki}},
  \bibinfo {author} {\bibfnamefont {R.}~\bibnamefont {Lipowsky}},\ and\
  \bibinfo {author} {\bibfnamefont {T.~R.}\ \bibnamefont {Weikl}},\ }\bibfield
  {title} {\bibinfo {title} {Binding cooperativity of membrane adhesion
  receptors},\ }\href {https://doi.org/10.1039/B902036E} {\bibfield  {journal}
  {\bibinfo  {journal} {Soft Matter}\ }\textbf {\bibinfo {volume} {5}},\
  \bibinfo {pages} {3354} (\bibinfo {year} {2009})}\BibitemShut {NoStop}%
\bibitem [{\citenamefont {Reister-Gottfried}\ \emph {et~al.}(2008)\citenamefont
  {Reister-Gottfried}, \citenamefont {Sengupta}, \citenamefont {Lorz},
  \citenamefont {Sackmann}, \citenamefont {Seifert},\ and\ \citenamefont
  {Smith}}]{reister-gottfried_dynamics_2008}%
  \BibitemOpen
  \bibfield  {author} {\bibinfo {author} {\bibfnamefont {E.}~\bibnamefont
  {Reister-Gottfried}}, \bibinfo {author} {\bibfnamefont {K.}~\bibnamefont
  {Sengupta}}, \bibinfo {author} {\bibfnamefont {B.}~\bibnamefont {Lorz}},
  \bibinfo {author} {\bibfnamefont {E.}~\bibnamefont {Sackmann}}, \bibinfo
  {author} {\bibfnamefont {U.}~\bibnamefont {Seifert}},\ and\ \bibinfo {author}
  {\bibfnamefont {A.-S.}\ \bibnamefont {Smith}},\ }\bibfield  {title} {\bibinfo
  {title} {Dynamics of specific vesicle-substrate adhesion: from local events
  to global dynamics},\ }\href {https://doi.org/10.1103/PhysRevLett.101.208103}
  {\bibfield  {journal} {\bibinfo  {journal} {Phys. Rev. Lett.}\ }\textbf
  {\bibinfo {volume} {101}},\ \bibinfo {pages} {208103} (\bibinfo {year}
  {2008})}\BibitemShut {NoStop}%
\bibitem [{\citenamefont {Schmidt}\ \emph {et~al.}(2012)\citenamefont
  {Schmidt}, \citenamefont {Bihr}, \citenamefont {Seifert},\ and\ \citenamefont
  {Smith}}]{schmidt_coexistence_2012}%
  \BibitemOpen
  \bibfield  {author} {\bibinfo {author} {\bibfnamefont {D.}~\bibnamefont
  {Schmidt}}, \bibinfo {author} {\bibfnamefont {T.}~\bibnamefont {Bihr}},
  \bibinfo {author} {\bibfnamefont {U.}~\bibnamefont {Seifert}},\ and\ \bibinfo
  {author} {\bibfnamefont {A.-S.}\ \bibnamefont {Smith}},\ }\bibfield  {title}
  {\bibinfo {title} {Coexistence of dilute and densely packed domains of
  ligand-receptor bonds in membrane adhesion},\ }\href
  {https://doi.org/10.1209/0295-5075/99/38003} {\bibfield  {journal} {\bibinfo
  {journal} {EPL}\ }\textbf {\bibinfo {volume} {99}},\ \bibinfo {pages} {38003}
  (\bibinfo {year} {2012})}\BibitemShut {NoStop}%
\bibitem [{\citenamefont {Xu}\ \emph {et~al.}(2015)\citenamefont {Xu},
  \citenamefont {Hu}, \citenamefont {Lipowsky},\ and\ \citenamefont
  {Weikl}}]{xu_binding_2015}%
  \BibitemOpen
  \bibfield  {author} {\bibinfo {author} {\bibfnamefont {G.-K.}\ \bibnamefont
  {Xu}}, \bibinfo {author} {\bibfnamefont {J.}~\bibnamefont {Hu}}, \bibinfo
  {author} {\bibfnamefont {R.}~\bibnamefont {Lipowsky}},\ and\ \bibinfo
  {author} {\bibfnamefont {T.~R.}\ \bibnamefont {Weikl}},\ }\bibfield  {title}
  {\bibinfo {title} {Binding constants of membrane-anchored receptors and
  ligands: {A} general theory corroborated by {Monte} {Carlo} simulations},\
  }\href {https://doi.org/10.1063/1.4936134} {\bibfield  {journal} {\bibinfo
  {journal} {J. Chem. Phys.}\ }\textbf {\bibinfo {volume} {143}},\ \bibinfo
  {pages} {243136} (\bibinfo {year} {2015})}\BibitemShut {NoStop}%
\bibitem [{\citenamefont {Monzel}\ and\ \citenamefont
  {Sengupta}(2016)}]{monzel2016measuring}%
  \BibitemOpen
  \bibfield  {author} {\bibinfo {author} {\bibfnamefont {C.}~\bibnamefont
  {Monzel}}\ and\ \bibinfo {author} {\bibfnamefont {K.}~\bibnamefont
  {Sengupta}},\ }\bibfield  {title} {\bibinfo {title} {Measuring shape
  fluctuations in biological membranes},\ }\href@noop {} {\bibfield  {journal}
  {\bibinfo  {journal} {Journal of Physics D: Applied Physics}\ }\textbf
  {\bibinfo {volume} {49}},\ \bibinfo {pages} {243002} (\bibinfo {year}
  {2016})}\BibitemShut {NoStop}%
\bibitem [{\citenamefont {Gov}\ and\ \citenamefont
  {Safran}(2005)}]{gov2005red}%
  \BibitemOpen
  \bibfield  {author} {\bibinfo {author} {\bibfnamefont {N.}~\bibnamefont
  {Gov}}\ and\ \bibinfo {author} {\bibfnamefont {S.}~\bibnamefont {Safran}},\
  }\bibfield  {title} {\bibinfo {title} {Red blood cell membrane fluctuations
  and shape controlled by atp-induced cytoskeletal defects},\ }\href@noop {}
  {\bibfield  {journal} {\bibinfo  {journal} {Biophysical journal}\ }\textbf
  {\bibinfo {volume} {88}},\ \bibinfo {pages} {1859} (\bibinfo {year}
  {2005})}\BibitemShut {NoStop}%
\bibitem [{\citenamefont {Tuvia}\ \emph {et~al.}(1997)\citenamefont {Tuvia},
  \citenamefont {Almagor}, \citenamefont {Bitler}, \citenamefont {Levin},
  \citenamefont {Korenstein},\ and\ \citenamefont {Yedgar}}]{tuvia1997cell}%
  \BibitemOpen
  \bibfield  {author} {\bibinfo {author} {\bibfnamefont {S.}~\bibnamefont
  {Tuvia}}, \bibinfo {author} {\bibfnamefont {A.}~\bibnamefont {Almagor}},
  \bibinfo {author} {\bibfnamefont {A.}~\bibnamefont {Bitler}}, \bibinfo
  {author} {\bibfnamefont {S.}~\bibnamefont {Levin}}, \bibinfo {author}
  {\bibfnamefont {R.}~\bibnamefont {Korenstein}},\ and\ \bibinfo {author}
  {\bibfnamefont {S.}~\bibnamefont {Yedgar}},\ }\bibfield  {title} {\bibinfo
  {title} {Cell membrane fluctuations are regulated by medium macroviscosity:
  evidence for a metabolic driving force},\ }\href@noop {} {\bibfield
  {journal} {\bibinfo  {journal} {Proceedings of the National Academy of
  Sciences}\ }\textbf {\bibinfo {volume} {94}},\ \bibinfo {pages} {5045}
  (\bibinfo {year} {1997})}\BibitemShut {NoStop}%
\bibitem [{\citenamefont {Biswas}\ \emph {et~al.}(2017)\citenamefont {Biswas},
  \citenamefont {Alex},\ and\ \citenamefont {Sinha}}]{biswas_mapping_2017}%
  \BibitemOpen
  \bibfield  {author} {\bibinfo {author} {\bibfnamefont {A.}~\bibnamefont
  {Biswas}}, \bibinfo {author} {\bibfnamefont {A.}~\bibnamefont {Alex}},\ and\
  \bibinfo {author} {\bibfnamefont {B.}~\bibnamefont {Sinha}},\ }\bibfield
  {title} {\bibinfo {title} {Mapping {Cell} {Membrane} {Fluctuations} {Reveals}
  {Their} {Active} {Regulation} and {Transient} {Heterogeneities}},\ }\href
  {https://doi.org/10.1016/j.bpj.2017.08.041} {\bibfield  {journal} {\bibinfo
  {journal} {Biophys. J.}\ }\textbf {\bibinfo {volume} {113}},\ \bibinfo
  {pages} {1768} (\bibinfo {year} {2017})}\BibitemShut {NoStop}%
\bibitem [{\citenamefont {Simson}\ \emph {et~al.}(1998)\citenamefont {Simson},
  \citenamefont {Wallraff}, \citenamefont {Faix}, \citenamefont
  {Niew{\"o}hner}, \citenamefont {Gerisch},\ and\ \citenamefont
  {Sackmann}}]{simson1998membrane}%
  \BibitemOpen
  \bibfield  {author} {\bibinfo {author} {\bibfnamefont {R.}~\bibnamefont
  {Simson}}, \bibinfo {author} {\bibfnamefont {E.}~\bibnamefont {Wallraff}},
  \bibinfo {author} {\bibfnamefont {J.}~\bibnamefont {Faix}}, \bibinfo {author}
  {\bibfnamefont {J.}~\bibnamefont {Niew{\"o}hner}}, \bibinfo {author}
  {\bibfnamefont {G.}~\bibnamefont {Gerisch}},\ and\ \bibinfo {author}
  {\bibfnamefont {E.}~\bibnamefont {Sackmann}},\ }\bibfield  {title} {\bibinfo
  {title} {Membrane bending modulus and adhesion energy of wild-type and mutant
  cells of dictyostelium lacking talin or cortexillins},\ }\href@noop {}
  {\bibfield  {journal} {\bibinfo  {journal} {Biophys. J.}\ }\textbf {\bibinfo
  {volume} {74}},\ \bibinfo {pages} {514} (\bibinfo {year} {1998})}\BibitemShut
  {NoStop}%
\bibitem [{\citenamefont {Alsteens}\ \emph {et~al.}(2010)\citenamefont
  {Alsteens}, \citenamefont {Garcia}, \citenamefont {Lipke},\ and\
  \citenamefont {Dufrene}}]{Fungal}%
  \BibitemOpen
  \bibfield  {author} {\bibinfo {author} {\bibfnamefont {D.}~\bibnamefont
  {Alsteens}}, \bibinfo {author} {\bibfnamefont {M.~C.}\ \bibnamefont
  {Garcia}}, \bibinfo {author} {\bibfnamefont {P.~N.}\ \bibnamefont {Lipke}},\
  and\ \bibinfo {author} {\bibfnamefont {Y.~F.}\ \bibnamefont {Dufrene}},\
  }\bibfield  {title} {\bibinfo {title} {Force-induced formation and
  propagation of adhesion nanodomains in living fungal cells},\ }\href
  {https://doi.org/10.1073/pnas.1013893107} {\bibfield  {journal} {\bibinfo
  {journal} {Proc. Natl. Acad. Sci. U.S.A.}\ }\textbf {\bibinfo {volume}
  {107}},\ \bibinfo {pages} {20744–20749} (\bibinfo {year}
  {2010})}\BibitemShut {NoStop}%
\bibitem [{\citenamefont {Hong}\ \emph {et~al.}(2014)\citenamefont {Hong},
  \citenamefont {Sun}, \citenamefont {Li}, \citenamefont {Li}, \citenamefont
  {Bunyak}, \citenamefont {Ersoy}, \citenamefont {Trzeciakowski}, \citenamefont
  {Staiculescu}, \citenamefont {Jin}, \citenamefont {Martinez‐Lemus},
  \citenamefont {Hill}, \citenamefont {Palaniappan},\ and\ \citenamefont
  {Meininger}}]{hong_vasoactive_2014}%
  \BibitemOpen
  \bibfield  {author} {\bibinfo {author} {\bibfnamefont {Z.}~\bibnamefont
  {Hong}}, \bibinfo {author} {\bibfnamefont {Z.}~\bibnamefont {Sun}}, \bibinfo
  {author} {\bibfnamefont {M.}~\bibnamefont {Li}}, \bibinfo {author}
  {\bibfnamefont {Z.}~\bibnamefont {Li}}, \bibinfo {author} {\bibfnamefont
  {F.}~\bibnamefont {Bunyak}}, \bibinfo {author} {\bibfnamefont
  {I.}~\bibnamefont {Ersoy}}, \bibinfo {author} {\bibfnamefont {J.~P.}\
  \bibnamefont {Trzeciakowski}}, \bibinfo {author} {\bibfnamefont {M.~C.}\
  \bibnamefont {Staiculescu}}, \bibinfo {author} {\bibfnamefont
  {M.}~\bibnamefont {Jin}}, \bibinfo {author} {\bibfnamefont {L.}~\bibnamefont
  {Martinez‐Lemus}}, \bibinfo {author} {\bibfnamefont {M.~A.}\ \bibnamefont
  {Hill}}, \bibinfo {author} {\bibfnamefont {K.}~\bibnamefont {Palaniappan}},\
  and\ \bibinfo {author} {\bibfnamefont {G.~A.}\ \bibnamefont {Meininger}},\
  }\bibfield  {title} {\bibinfo {title} {Vasoactive agonists exert dynamic and
  coordinated effects on vascular smooth muscle cell elasticity, cytoskeletal
  remodelling and adhesion},\ }\href
  {https://doi.org/10.1113/jphysiol.2013.264929} {\bibfield  {journal}
  {\bibinfo  {journal} {J. Physiol.}\ }\textbf {\bibinfo {volume} {592}},\
  \bibinfo {pages} {1249} (\bibinfo {year} {2014})}\BibitemShut {NoStop}%
\bibitem [{\citenamefont {Zhu}\ \emph {et~al.}(2012)\citenamefont {Zhu},
  \citenamefont {Qiu}, \citenamefont {Trzeciakowski}, \citenamefont {Sun},
  \citenamefont {Li}, \citenamefont {Hong}, \citenamefont {Hill}, \citenamefont
  {Hunter}, \citenamefont {Vatner}, \citenamefont {Vatner},\ and\ \citenamefont
  {Meininger}}]{zhu_temporal_2012}%
  \BibitemOpen
  \bibfield  {author} {\bibinfo {author} {\bibfnamefont {Y.}~\bibnamefont
  {Zhu}}, \bibinfo {author} {\bibfnamefont {H.}~\bibnamefont {Qiu}}, \bibinfo
  {author} {\bibfnamefont {J.~P.}\ \bibnamefont {Trzeciakowski}}, \bibinfo
  {author} {\bibfnamefont {Z.}~\bibnamefont {Sun}}, \bibinfo {author}
  {\bibfnamefont {Z.}~\bibnamefont {Li}}, \bibinfo {author} {\bibfnamefont
  {Z.}~\bibnamefont {Hong}}, \bibinfo {author} {\bibfnamefont {M.~A.}\
  \bibnamefont {Hill}}, \bibinfo {author} {\bibfnamefont {W.~C.}\ \bibnamefont
  {Hunter}}, \bibinfo {author} {\bibfnamefont {D.~E.}\ \bibnamefont {Vatner}},
  \bibinfo {author} {\bibfnamefont {S.~F.}\ \bibnamefont {Vatner}},\ and\
  \bibinfo {author} {\bibfnamefont {G.~A.}\ \bibnamefont {Meininger}},\
  }\bibfield  {title} {\bibinfo {title} {Temporal analysis of vascular smooth
  muscle cell elasticity and adhesion reveals oscillation waveforms that differ
  with aging},\ }\href {https://doi.org/10.1111/j.1474-9726.2012.00840.x}
  {\bibfield  {journal} {\bibinfo  {journal} {Aging Cell}\ }\textbf {\bibinfo
  {volume} {11}},\ \bibinfo {pages} {741} (\bibinfo {year} {2012})}\BibitemShut
  {NoStop}%
\bibitem [{\citenamefont {Hong}\ \emph {et~al.}(2012)\citenamefont {Hong},
  \citenamefont {Sun}, \citenamefont {Li}, \citenamefont {Mesquitta},
  \citenamefont {Trzeciakowski},\ and\ \citenamefont
  {Meininger}}]{hong_coordination_2012}%
  \BibitemOpen
  \bibfield  {author} {\bibinfo {author} {\bibfnamefont {Z.}~\bibnamefont
  {Hong}}, \bibinfo {author} {\bibfnamefont {Z.}~\bibnamefont {Sun}}, \bibinfo
  {author} {\bibfnamefont {Z.}~\bibnamefont {Li}}, \bibinfo {author}
  {\bibfnamefont {W.-T.}\ \bibnamefont {Mesquitta}}, \bibinfo {author}
  {\bibfnamefont {J.~P.}\ \bibnamefont {Trzeciakowski}},\ and\ \bibinfo
  {author} {\bibfnamefont {G.~A.}\ \bibnamefont {Meininger}},\ }\bibfield
  {title} {\bibinfo {title} {Coordination of fibronectin adhesion with
  contraction and relaxation in microvascular smooth muscle},\ }\href
  {https://doi.org/10.1093/cvr/cvs239} {\bibfield  {journal} {\bibinfo
  {journal} {Cardiovasc. Res.}\ }\textbf {\bibinfo {volume} {96}},\ \bibinfo
  {pages} {73} (\bibinfo {year} {2012})}\BibitemShut {NoStop}%
\bibitem [{\citenamefont {Swaminathan}\ \emph {et~al.}(2011)\citenamefont
  {Swaminathan}, \citenamefont {Mythreye}, \citenamefont {O'Brien},
  \citenamefont {Berchuck}, \citenamefont {Blobe},\ and\ \citenamefont
  {Superfine}}]{swaminathan_mechanical_2011}%
  \BibitemOpen
  \bibfield  {author} {\bibinfo {author} {\bibfnamefont {V.}~\bibnamefont
  {Swaminathan}}, \bibinfo {author} {\bibfnamefont {K.}~\bibnamefont
  {Mythreye}}, \bibinfo {author} {\bibfnamefont {E.~T.}\ \bibnamefont
  {O'Brien}}, \bibinfo {author} {\bibfnamefont {A.}~\bibnamefont {Berchuck}},
  \bibinfo {author} {\bibfnamefont {G.~C.}\ \bibnamefont {Blobe}},\ and\
  \bibinfo {author} {\bibfnamefont {R.}~\bibnamefont {Superfine}},\ }\bibfield
  {title} {\bibinfo {title} {Mechanical {Stiffness} {Grades} {Metastatic}
  {Potential} in {Patient} {Tumor} {Cells} and in {Cancer} {Cell} {Lines}},\
  }\href {https://doi.org/10.1158/0008-5472.CAN-11-0247} {\bibfield  {journal}
  {\bibinfo  {journal} {Cancer Res.}\ }\textbf {\bibinfo {volume} {71}},\
  \bibinfo {pages} {5075} (\bibinfo {year} {2011})}\BibitemShut {NoStop}%
\bibitem [{\citenamefont {Wang}\ \emph {et~al.}(2017)\citenamefont {Wang},
  \citenamefont {Jacobi}, \citenamefont {Waschke}, \citenamefont {Hartmann},
  \citenamefont {L{\"o}wen},\ and\ \citenamefont {Schmidt}}]{wang2017elastic}%
  \BibitemOpen
  \bibfield  {author} {\bibinfo {author} {\bibfnamefont {H.}~\bibnamefont
  {Wang}}, \bibinfo {author} {\bibfnamefont {F.}~\bibnamefont {Jacobi}},
  \bibinfo {author} {\bibfnamefont {J.}~\bibnamefont {Waschke}}, \bibinfo
  {author} {\bibfnamefont {L.}~\bibnamefont {Hartmann}}, \bibinfo {author}
  {\bibfnamefont {H.}~\bibnamefont {L{\"o}wen}},\ and\ \bibinfo {author}
  {\bibfnamefont {S.}~\bibnamefont {Schmidt}},\ }\bibfield  {title} {\bibinfo
  {title} {Elastic modulus dependence on the specific adhesion of hydrogels},\
  }\href@noop {} {\bibfield  {journal} {\bibinfo  {journal} {Adv. Funct.
  Mater.}\ }\textbf {\bibinfo {volume} {27}},\ \bibinfo {pages} {1702040}
  (\bibinfo {year} {2017})}\BibitemShut {NoStop}%
\bibitem [{\citenamefont {Seifert}(2000)}]{seifert_rupture_2000}%
  \BibitemOpen
  \bibfield  {author} {\bibinfo {author} {\bibfnamefont {U.}~\bibnamefont
  {Seifert}},\ }\bibfield  {title} {\bibinfo {title} {Rupture of {Multiple}
  {Parallel} {Molecular} {Bonds} under {Dynamic} {Loading}},\ }\href
  {https://doi.org/10.1103/PhysRevLett.84.2750} {\bibfield  {journal} {\bibinfo
   {journal} {Phys. Rev. Lett.}\ }\textbf {\bibinfo {volume} {84}},\ \bibinfo
  {pages} {2750} (\bibinfo {year} {2000})}\BibitemShut {NoStop}%
\bibitem [{\citenamefont {Erdmann}\ and\ \citenamefont
  {Schwarz}(2004{\natexlab{a}})}]{erdmann_stochastic_2004}%
  \BibitemOpen
  \bibfield  {author} {\bibinfo {author} {\bibfnamefont {T.}~\bibnamefont
  {Erdmann}}\ and\ \bibinfo {author} {\bibfnamefont {U.~S.}\ \bibnamefont
  {Schwarz}},\ }\bibfield  {title} {\bibinfo {title} {Stochastic dynamics of
  adhesion clusters under shared constant force and with rebinding},\ }\href
  {https://doi.org/10.1063/1.1805496} {\bibfield  {journal} {\bibinfo
  {journal} {J. Chem. Phys.}\ }\textbf {\bibinfo {volume} {121}},\ \bibinfo
  {pages} {8997} (\bibinfo {year} {2004}{\natexlab{a}})}\BibitemShut {NoStop}%
\bibitem [{\citenamefont {Dasanna}\ \emph {et~al.}(2020)\citenamefont
  {Dasanna}, \citenamefont {Gompper},\ and\ \citenamefont {Fedosov}}]{Dasanna}%
  \BibitemOpen
  \bibfield  {author} {\bibinfo {author} {\bibfnamefont {A.~K.}\ \bibnamefont
  {Dasanna}}, \bibinfo {author} {\bibfnamefont {G.}~\bibnamefont {Gompper}},\
  and\ \bibinfo {author} {\bibfnamefont {D.~A.}\ \bibnamefont {Fedosov}},\
  }\bibfield  {title} {\bibinfo {title} {Stability of heterogeneous
  parallel-bond adhesion clusters under load},\ }\href
  {https://doi.org/10.1103/PhysRevResearch.2.043063} {\bibfield  {journal}
  {\bibinfo  {journal} {Phys. Rev. Res.}\ }\textbf {\bibinfo {volume} {2}},\
  \bibinfo {pages} {043063} (\bibinfo {year} {2020})}\BibitemShut {NoStop}%
\bibitem [{\citenamefont {Smith}\ and\ \citenamefont
  {Sackmann}(2009)}]{smith_progress_2009}%
  \BibitemOpen
  \bibfield  {author} {\bibinfo {author} {\bibfnamefont {A.-S.}\ \bibnamefont
  {Smith}}\ and\ \bibinfo {author} {\bibfnamefont {E.}~\bibnamefont
  {Sackmann}},\ }\bibfield  {title} {\bibinfo {title} {Progress in {Mimetic}
  {Studies} of {Cell} {Adhesion} and the {Mechanosensing}},\ }\href
  {https://doi.org/10.1002/cphc.200800683} {\bibfield  {journal} {\bibinfo
  {journal} {ChemPhysChem}\ }\textbf {\bibinfo {volume} {10}},\ \bibinfo
  {pages} {66} (\bibinfo {year} {2009})}\BibitemShut {NoStop}%
\bibitem [{\citenamefont {Smith}\ \emph {et~al.}(2008)\citenamefont {Smith},
  \citenamefont {Sengupta}, \citenamefont {Goennenwein}, \citenamefont
  {Seifert},\ and\ \citenamefont {Sackmann}}]{smith_force-induced_2008}%
  \BibitemOpen
  \bibfield  {author} {\bibinfo {author} {\bibfnamefont {A.-S.}\ \bibnamefont
  {Smith}}, \bibinfo {author} {\bibfnamefont {K.}~\bibnamefont {Sengupta}},
  \bibinfo {author} {\bibfnamefont {S.}~\bibnamefont {Goennenwein}}, \bibinfo
  {author} {\bibfnamefont {U.}~\bibnamefont {Seifert}},\ and\ \bibinfo {author}
  {\bibfnamefont {E.}~\bibnamefont {Sackmann}},\ }\bibfield  {title} {\bibinfo
  {title} {Force-induced growth of adhesion domains is controlled by receptor
  mobility},\ }\href {https://doi.org/10.1073/pnas.0801706105} {\bibfield
  {journal} {\bibinfo  {journal} {Proc. Natl. Acad. Sci. U.S.A.}\ }\textbf
  {\bibinfo {volume} {105}},\ \bibinfo {pages} {6906} (\bibinfo {year}
  {2008})}\BibitemShut {NoStop}%
\bibitem [{\citenamefont {Boudjemaa}\ \emph {et~al.}(2019)\citenamefont
  {Boudjemaa}, \citenamefont {Steenkeste}, \citenamefont {Canette},
  \citenamefont {Briandet}, \citenamefont {Fontaine-Aupart},\ and\
  \citenamefont {Marlière}}]{boudjemaa_direct_2019}%
  \BibitemOpen
  \bibfield  {author} {\bibinfo {author} {\bibfnamefont {R.}~\bibnamefont
  {Boudjemaa}}, \bibinfo {author} {\bibfnamefont {K.}~\bibnamefont
  {Steenkeste}}, \bibinfo {author} {\bibfnamefont {A.}~\bibnamefont {Canette}},
  \bibinfo {author} {\bibfnamefont {R.}~\bibnamefont {Briandet}}, \bibinfo
  {author} {\bibfnamefont {M.-P.}\ \bibnamefont {Fontaine-Aupart}},\ and\
  \bibinfo {author} {\bibfnamefont {C.}~\bibnamefont {Marlière}},\ }\bibfield
  {title} {\bibinfo {title} {Direct observation of the cell-wall remodeling in
  adhering {Staphylococcus} aureus 27217: {An} {AFM} study supported by {SEM}
  and {TEM}},\ }\href {https://doi.org/10.1016/j.tcsw.2019.100018} {\bibfield
  {journal} {\bibinfo  {journal} {The Cell Surface}\ }\textbf {\bibinfo
  {volume} {5}},\ \bibinfo {pages} {100018} (\bibinfo {year}
  {2019})}\BibitemShut {NoStop}%
\bibitem [{\citenamefont {Huang}\ \emph {et~al.}(2010)\citenamefont {Huang},
  \citenamefont {Zarnitsyna}, \citenamefont {Liu}, \citenamefont {Edwards},
  \citenamefont {Jiang}, \citenamefont {Evavold},\ and\ \citenamefont
  {Zhu}}]{huang_kinetics_2010}%
  \BibitemOpen
  \bibfield  {author} {\bibinfo {author} {\bibfnamefont {J.}~\bibnamefont
  {Huang}}, \bibinfo {author} {\bibfnamefont {V.~I.}\ \bibnamefont
  {Zarnitsyna}}, \bibinfo {author} {\bibfnamefont {B.}~\bibnamefont {Liu}},
  \bibinfo {author} {\bibfnamefont {L.~J.}\ \bibnamefont {Edwards}}, \bibinfo
  {author} {\bibfnamefont {N.}~\bibnamefont {Jiang}}, \bibinfo {author}
  {\bibfnamefont {B.~D.}\ \bibnamefont {Evavold}},\ and\ \bibinfo {author}
  {\bibfnamefont {C.}~\bibnamefont {Zhu}},\ }\bibfield  {title} {\bibinfo
  {title} {The kinetics of two-dimensional {TCR} and {pMHC} interactions
  determine {T}-cell responsiveness},\ }\href
  {https://doi.org/10.1038/nature08944} {\bibfield  {journal} {\bibinfo
  {journal} {Nature}\ }\textbf {\bibinfo {volume} {464}},\ \bibinfo {pages}
  {932} (\bibinfo {year} {2010})}\BibitemShut {NoStop}%
\bibitem [{\citenamefont {Katsamba}\ \emph {et~al.}(2009)\citenamefont
  {Katsamba}, \citenamefont {Carroll}, \citenamefont {Ahlsen}, \citenamefont
  {Bahna}, \citenamefont {Vendome}, \citenamefont {Posy}, \citenamefont
  {Rajebhosale}, \citenamefont {Price}, \citenamefont {Jessell}, \citenamefont
  {Ben-Shaul}, \citenamefont {Shapiro},\ and\ \citenamefont
  {Honig}}]{affinity}%
  \BibitemOpen
  \bibfield  {author} {\bibinfo {author} {\bibfnamefont {P.}~\bibnamefont
  {Katsamba}}, \bibinfo {author} {\bibfnamefont {K.}~\bibnamefont {Carroll}},
  \bibinfo {author} {\bibfnamefont {G.}~\bibnamefont {Ahlsen}}, \bibinfo
  {author} {\bibfnamefont {F.}~\bibnamefont {Bahna}}, \bibinfo {author}
  {\bibfnamefont {J.}~\bibnamefont {Vendome}}, \bibinfo {author} {\bibfnamefont
  {S.}~\bibnamefont {Posy}}, \bibinfo {author} {\bibfnamefont {M.}~\bibnamefont
  {Rajebhosale}}, \bibinfo {author} {\bibfnamefont {S.}~\bibnamefont {Price}},
  \bibinfo {author} {\bibfnamefont {T.~M.}\ \bibnamefont {Jessell}}, \bibinfo
  {author} {\bibfnamefont {A.}~\bibnamefont {Ben-Shaul}}, \bibinfo {author}
  {\bibfnamefont {L.}~\bibnamefont {Shapiro}},\ and\ \bibinfo {author}
  {\bibfnamefont {B.~H.}\ \bibnamefont {Honig}},\ }\bibfield  {title} {\bibinfo
  {title} {Linking molecular affinity and cellular specificity in
  cadherin-mediated adhesion},\ }\href
  {https://doi.org/10.1073/pnas.0905349106} {\bibfield  {journal} {\bibinfo
  {journal} {Proc. Natl. Acad. Sci. U.S.A.}\ }\textbf {\bibinfo {volume}
  {106}},\ \bibinfo {pages} {11594} (\bibinfo {year} {2009})}\BibitemShut
  {NoStop}%
\bibitem [{\citenamefont {Buda}\ and\ \citenamefont
  {Pignatelli}(2011)}]{buda_e-cadherin_2011}%
  \BibitemOpen
  \bibfield  {author} {\bibinfo {author} {\bibfnamefont {A.}~\bibnamefont
  {Buda}}\ and\ \bibinfo {author} {\bibfnamefont {M.}~\bibnamefont
  {Pignatelli}},\ }\bibfield  {title} {\bibinfo {title} {E-cadherin and the
  cytoskeletal network in colorectal cancer development and metastasis},\
  }\href {https://doi.org/10.3109/15419061.2011.636465} {\bibfield  {journal}
  {\bibinfo  {journal} {Cell Commun Adhes.}\ }\textbf {\bibinfo {volume}
  {18}},\ \bibinfo {pages} {133} (\bibinfo {year} {2011})}\BibitemShut
  {NoStop}%
\bibitem [{\citenamefont {Korb}\ \emph {et~al.}(2004)\citenamefont {Korb},
  \citenamefont {Schlüter}, \citenamefont {Enns}, \citenamefont {Spiegel},
  \citenamefont {Senninger}, \citenamefont {Nicolson},\ and\ \citenamefont
  {Haier}}]{korb_integrity_2004}%
  \BibitemOpen
  \bibfield  {author} {\bibinfo {author} {\bibfnamefont {T.}~\bibnamefont
  {Korb}}, \bibinfo {author} {\bibfnamefont {K.}~\bibnamefont {Schlüter}},
  \bibinfo {author} {\bibfnamefont {A.}~\bibnamefont {Enns}}, \bibinfo {author}
  {\bibfnamefont {H.-U.}\ \bibnamefont {Spiegel}}, \bibinfo {author}
  {\bibfnamefont {N.}~\bibnamefont {Senninger}}, \bibinfo {author}
  {\bibfnamefont {G.~L.}\ \bibnamefont {Nicolson}},\ and\ \bibinfo {author}
  {\bibfnamefont {J.}~\bibnamefont {Haier}},\ }\bibfield  {title} {\bibinfo
  {title} {Integrity of actin fibers and microtubules influences metastatic
  tumor cell adhesion},\ }\href {https://doi.org/10.1016/j.yexcr.2004.06.001}
  {\bibfield  {journal} {\bibinfo  {journal} {Exp.}\ }\textbf {\bibinfo
  {volume} {299}},\ \bibinfo {pages} {236} (\bibinfo {year}
  {2004})}\BibitemShut {NoStop}%
\bibitem [{\citenamefont {Zeng}\ \emph {et~al.}(2019)\citenamefont {Zeng},
  \citenamefont {Cao}, \citenamefont {Liu}, \citenamefont {Zhao}, \citenamefont
  {Zhang}, \citenamefont {Xiao}, \citenamefont {Jia}, \citenamefont {Tian},
  \citenamefont {Yu}, \citenamefont {Chen},\ and\ \citenamefont
  {Cai}}]{zeng_sept9_i1_2019}%
  \BibitemOpen
  \bibfield  {author} {\bibinfo {author} {\bibfnamefont {Y.}~\bibnamefont
  {Zeng}}, \bibinfo {author} {\bibfnamefont {Y.}~\bibnamefont {Cao}}, \bibinfo
  {author} {\bibfnamefont {L.}~\bibnamefont {Liu}}, \bibinfo {author}
  {\bibfnamefont {J.}~\bibnamefont {Zhao}}, \bibinfo {author} {\bibfnamefont
  {T.}~\bibnamefont {Zhang}}, \bibinfo {author} {\bibfnamefont
  {L.}~\bibnamefont {Xiao}}, \bibinfo {author} {\bibfnamefont {M.}~\bibnamefont
  {Jia}}, \bibinfo {author} {\bibfnamefont {Q.}~\bibnamefont {Tian}}, \bibinfo
  {author} {\bibfnamefont {H.}~\bibnamefont {Yu}}, \bibinfo {author}
  {\bibfnamefont {S.}~\bibnamefont {Chen}},\ and\ \bibinfo {author}
  {\bibfnamefont {Y.}~\bibnamefont {Cai}},\ }\bibfield  {title} {\bibinfo
  {title} {{SEPT9}\_i1 regulates human breast cancer cell motility through
  cytoskeletal and {RhoA}/{FAK} signaling pathway regulation},\ }\href
  {https://doi.org/10.1038/s41419-019-1947-9} {\bibfield  {journal} {\bibinfo
  {journal} {Cell Death Dis.}\ }\textbf {\bibinfo {volume} {10}},\ \bibinfo
  {pages} {1} (\bibinfo {year} {2019})}\BibitemShut {NoStop}%
\bibitem [{\citenamefont {Ising}(1925)}]{ising_beitrag_1925}%
  \BibitemOpen
  \bibfield  {author} {\bibinfo {author} {\bibfnamefont {E.}~\bibnamefont
  {Ising}},\ }\bibfield  {title} {\bibinfo {title} {Beitrag zur {Theorie} des
  {Ferromagnetismus}},\ }\href {https://doi.org/10.1007/BF02980577} {\bibfield
  {journal} {\bibinfo  {journal} {Zeitschrift für Physik}\ }\textbf {\bibinfo
  {volume} {31}},\ \bibinfo {pages} {253} (\bibinfo {year} {1925})}\BibitemShut
  {NoStop}%
\bibitem [{\citenamefont {Erdmann}\ and\ \citenamefont
  {Schwarz}(2004{\natexlab{b}})}]{erdmann_stability_2004}%
  \BibitemOpen
  \bibfield  {author} {\bibinfo {author} {\bibfnamefont {T.}~\bibnamefont
  {Erdmann}}\ and\ \bibinfo {author} {\bibfnamefont {U.~S.}\ \bibnamefont
  {Schwarz}},\ }\bibfield  {title} {\bibinfo {title} {Stability of adhesion
  clusters under constant force},\ }\href
  {https://doi.org/10.1103/PhysRevLett.92.108102} {\bibfield  {journal}
  {\bibinfo  {journal} {Phys. Rev. Lett.}\ }\textbf {\bibinfo {volume} {92}},\
  \bibinfo {pages} {108102} (\bibinfo {year} {2004}{\natexlab{b}})}\BibitemShut
  {NoStop}%
\bibitem [{\citenamefont {Qian}\ \emph {et~al.}(2008)\citenamefont {Qian},
  \citenamefont {Wang},\ and\ \citenamefont {Gao}}]{Qian_2008}%
  \BibitemOpen
  \bibfield  {author} {\bibinfo {author} {\bibfnamefont {J.}~\bibnamefont
  {Qian}}, \bibinfo {author} {\bibfnamefont {J.}~\bibnamefont {Wang}},\ and\
  \bibinfo {author} {\bibfnamefont {H.}~\bibnamefont {Gao}},\ }\bibfield
  {title} {\bibinfo {title} {Lifetime and strength of adhesive molecular bond
  clusters between elastic media†},\ }\href
  {https://doi.org/10.1021/la702401b} {\bibfield  {journal} {\bibinfo
  {journal} {Langmuir}\ }\textbf {\bibinfo {volume} {24}},\ \bibinfo {pages}
  {1262–1270} (\bibinfo {year} {2008})}\BibitemShut {NoStop}%
\bibitem [{\citenamefont {Glauber}(1963)}]{glauber_timedependent_1963}%
  \BibitemOpen
  \bibfield  {author} {\bibinfo {author} {\bibfnamefont {R.~J.}\ \bibnamefont
  {Glauber}},\ }\bibfield  {title} {\bibinfo {title} {Time‐{Dependent}
  {Statistics} of the {Ising} {Model}},\ }\href
  {https://doi.org/10.1063/1.1703954} {\bibfield  {journal} {\bibinfo
  {journal} {J. Math. Phys.}\ }\textbf {\bibinfo {volume} {4}},\ \bibinfo
  {pages} {294} (\bibinfo {year} {1963})}\BibitemShut {NoStop}%
\bibitem [{\citenamefont {Kawasaki}(1966)}]{PhysRev.145.224}%
  \BibitemOpen
  \bibfield  {author} {\bibinfo {author} {\bibfnamefont {K.}~\bibnamefont
  {Kawasaki}},\ }\bibfield  {title} {\bibinfo {title} {Diffusion constants near
  the critical point for time-dependent ising models. i},\ }\href
  {https://doi.org/10.1103/PhysRev.145.224} {\bibfield  {journal} {\bibinfo
  {journal} {Phys. Rev.}\ }\textbf {\bibinfo {volume} {145}},\ \bibinfo {pages}
  {224} (\bibinfo {year} {1966})}\BibitemShut {NoStop}%
\bibitem [{\citenamefont {Iosifescu}(2014)}]{iosifescu_finite_2014}%
  \BibitemOpen
  \bibfield  {author} {\bibinfo {author} {\bibfnamefont {M.}~\bibnamefont
  {Iosifescu}},\ }\href@noop {} {\emph {\bibinfo {title} {Finite {Markov}
  {Processes} and {Their} {Applications}}}}\ (\bibinfo  {publisher} {Courier
  Corporation},\ \bibinfo {year} {2014})\BibitemShut {NoStop}%
\bibitem [{\citenamefont {Fowler}(1939)}]{fowler_statistical_1939}%
  \BibitemOpen
  \bibfield  {author} {\bibinfo {author} {\bibfnamefont {R.~H.}\ \bibnamefont
  {Fowler}},\ }\href@noop {} {\emph {\bibinfo {title} {Statistical
  {Thermodynamics}}}}\ (\bibinfo  {publisher} {CUP Archive},\ \bibinfo {year}
  {1939})\BibitemShut {NoStop}%
\bibitem [{\citenamefont {Godec}\ and\ \citenamefont
  {Metzler}(2016)}]{Godec_2016}%
  \BibitemOpen
  \bibfield  {author} {\bibinfo {author} {\bibfnamefont {A.}~\bibnamefont
  {Godec}}\ and\ \bibinfo {author} {\bibfnamefont {R.}~\bibnamefont
  {Metzler}},\ }\bibfield  {title} {\bibinfo {title} {Universal proximity
  effect in target search kinetics in the few-encounter limit},\ }\href
  {https://doi.org/10.1103/PhysRevX.6.041037} {\bibfield  {journal} {\bibinfo
  {journal} {Phys. Rev. X}\ }\textbf {\bibinfo {volume} {6}},\ \bibinfo {pages}
  {041037} (\bibinfo {year} {2016})}\BibitemShut {NoStop}%
\bibitem [{\citenamefont {Hartich}\ and\ \citenamefont
  {Godec}(2018)}]{Hartich_2018}%
  \BibitemOpen
  \bibfield  {author} {\bibinfo {author} {\bibfnamefont {D.}~\bibnamefont
  {Hartich}}\ and\ \bibinfo {author} {\bibfnamefont {A.}~\bibnamefont
  {Godec}},\ }\bibfield  {title} {\bibinfo {title} {Duality between relaxation
  and first passage in reversible markov dynamics: rugged energy landscapes
  disentangled},\ }\href {https://doi.org/10.1088/1367-2630/aaf038} {\bibfield
  {journal} {\bibinfo  {journal} {New J. Phys.}\ }\textbf {\bibinfo {volume}
  {20}},\ \bibinfo {pages} {112002} (\bibinfo {year} {2018})}\BibitemShut
  {NoStop}%
\bibitem [{\citenamefont {Hartich}\ and\ \citenamefont
  {Godec}(2019{\natexlab{a}})}]{Hartich_2019}%
  \BibitemOpen
  \bibfield  {author} {\bibinfo {author} {\bibfnamefont {D.}~\bibnamefont
  {Hartich}}\ and\ \bibinfo {author} {\bibfnamefont {A.}~\bibnamefont
  {Godec}},\ }\bibfield  {title} {\bibinfo {title} {Interlacing relaxation and
  first-passage phenomena in reversible discrete and continuous space markovian
  dynamics},\ }\href {https://doi.org/10.1088/1742-5468/ab00df} {\bibfield
  {journal} {\bibinfo  {journal} {J. Stat. Mech. Theory Exp.}\ }\textbf
  {\bibinfo {volume} {2019}},\ \bibinfo {pages} {024002} (\bibinfo {year}
  {2019}{\natexlab{a}})}\BibitemShut {NoStop}%
\bibitem [{\citenamefont {Onsager}(1944)}]{PhysRev.65.117}%
  \BibitemOpen
  \bibfield  {author} {\bibinfo {author} {\bibfnamefont {L.}~\bibnamefont
  {Onsager}},\ }\bibfield  {title} {\bibinfo {title} {Crystal statistics. i. a
  two-dimensional model with an order-disorder transition},\ }\href
  {https://doi.org/10.1103/PhysRev.65.117} {\bibfield  {journal} {\bibinfo
  {journal} {Phys. Rev.}\ }\textbf {\bibinfo {volume} {65}},\ \bibinfo {pages}
  {117} (\bibinfo {year} {1944})}\BibitemShut {NoStop}%
\bibitem [{\citenamefont {Paszek}\ \emph {et~al.}(2009)\citenamefont {Paszek},
  \citenamefont {Boettiger}, \citenamefont {Weaver},\ and\ \citenamefont
  {Hammer}}]{paszek2009integrin}%
  \BibitemOpen
  \bibfield  {author} {\bibinfo {author} {\bibfnamefont {M.~J.}\ \bibnamefont
  {Paszek}}, \bibinfo {author} {\bibfnamefont {D.}~\bibnamefont {Boettiger}},
  \bibinfo {author} {\bibfnamefont {V.~M.}\ \bibnamefont {Weaver}},\ and\
  \bibinfo {author} {\bibfnamefont {D.~A.}\ \bibnamefont {Hammer}},\ }\bibfield
   {title} {\bibinfo {title} {Integrin clustering is driven by mechanical
  resistance from the glycocalyx and the substrate},\ }\href@noop {} {\bibfield
   {journal} {\bibinfo  {journal} {PLoS Comput Biol}\ }\textbf {\bibinfo
  {volume} {5}},\ \bibinfo {pages} {e1000604} (\bibinfo {year}
  {2009})}\BibitemShut {NoStop}%
\bibitem [{\citenamefont {Caputo}\ and\ \citenamefont
  {Hammer}(2005)}]{caputo2005effect}%
  \BibitemOpen
  \bibfield  {author} {\bibinfo {author} {\bibfnamefont {K.~E.}\ \bibnamefont
  {Caputo}}\ and\ \bibinfo {author} {\bibfnamefont {D.~A.}\ \bibnamefont
  {Hammer}},\ }\bibfield  {title} {\bibinfo {title} {Effect of microvillus
  deformability on leukocyte adhesion explored using adhesive dynamics
  simulations},\ }\href@noop {} {\bibfield  {journal} {\bibinfo  {journal}
  {Biophys. J.}\ }\textbf {\bibinfo {volume} {89}},\ \bibinfo {pages} {187}
  (\bibinfo {year} {2005})}\BibitemShut {NoStop}%
\bibitem [{\citenamefont {Nermut}\ \emph {et~al.}(1988)\citenamefont {Nermut},
  \citenamefont {Green}, \citenamefont {Eason}, \citenamefont {Yamada},\ and\
  \citenamefont {Yamada}}]{nermut1988electron}%
  \BibitemOpen
  \bibfield  {author} {\bibinfo {author} {\bibfnamefont {M.}~\bibnamefont
  {Nermut}}, \bibinfo {author} {\bibfnamefont {N.}~\bibnamefont {Green}},
  \bibinfo {author} {\bibfnamefont {P.}~\bibnamefont {Eason}}, \bibinfo
  {author} {\bibfnamefont {S.~S.}\ \bibnamefont {Yamada}},\ and\ \bibinfo
  {author} {\bibfnamefont {K.}~\bibnamefont {Yamada}},\ }\bibfield  {title}
  {\bibinfo {title} {Electron microscopy and structural model of human
  fibronectin receptor.},\ }\href@noop {} {\bibfield  {journal} {\bibinfo
  {journal} {The EMBO journal}\ }\textbf {\bibinfo {volume} {7}},\ \bibinfo
  {pages} {4093} (\bibinfo {year} {1988})}\BibitemShut {NoStop}%
\bibitem [{\citenamefont {Pelta}\ \emph {et~al.}(2000)\citenamefont {Pelta},
  \citenamefont {Berry}, \citenamefont {Fadda}, \citenamefont {Pauthe},\ and\
  \citenamefont {Lairez}}]{pelta2000statistical}%
  \BibitemOpen
  \bibfield  {author} {\bibinfo {author} {\bibfnamefont {J.}~\bibnamefont
  {Pelta}}, \bibinfo {author} {\bibfnamefont {H.}~\bibnamefont {Berry}},
  \bibinfo {author} {\bibfnamefont {G.}~\bibnamefont {Fadda}}, \bibinfo
  {author} {\bibfnamefont {E.}~\bibnamefont {Pauthe}},\ and\ \bibinfo {author}
  {\bibfnamefont {D.}~\bibnamefont {Lairez}},\ }\bibfield  {title} {\bibinfo
  {title} {Statistical conformation of human plasma fibronectin},\ }\href@noop
  {} {\bibfield  {journal} {\bibinfo  {journal} {Biochemistry}\ }\textbf
  {\bibinfo {volume} {39}},\ \bibinfo {pages} {5146} (\bibinfo {year}
  {2000})}\BibitemShut {NoStop}%
\bibitem [{\citenamefont {Dimova}(2014)}]{dimova2014recent}%
  \BibitemOpen
  \bibfield  {author} {\bibinfo {author} {\bibfnamefont {R.}~\bibnamefont
  {Dimova}},\ }\bibfield  {title} {\bibinfo {title} {Recent developments in the
  field of bending rigidity measurements on membranes},\ }\href@noop {}
  {\bibfield  {journal} {\bibinfo  {journal} {Adv. Colloid Interface Sci.}\
  }\textbf {\bibinfo {volume} {208}},\ \bibinfo {pages} {225} (\bibinfo {year}
  {2014})}\BibitemShut {NoStop}%
\bibitem [{\citenamefont {Faizi}\ \emph {et~al.}(2019)\citenamefont {Faizi},
  \citenamefont {Frey}, \citenamefont {Steink{\"u}hler}, \citenamefont
  {Dimova},\ and\ \citenamefont {Vlahovska}}]{faizi2019bending}%
  \BibitemOpen
  \bibfield  {author} {\bibinfo {author} {\bibfnamefont {H.~A.}\ \bibnamefont
  {Faizi}}, \bibinfo {author} {\bibfnamefont {S.~L.}\ \bibnamefont {Frey}},
  \bibinfo {author} {\bibfnamefont {J.}~\bibnamefont {Steink{\"u}hler}},
  \bibinfo {author} {\bibfnamefont {R.}~\bibnamefont {Dimova}},\ and\ \bibinfo
  {author} {\bibfnamefont {P.~M.}\ \bibnamefont {Vlahovska}},\ }\bibfield
  {title} {\bibinfo {title} {Bending rigidity of charged lipid bilayer
  membranes},\ }\href@noop {} {\bibfield  {journal} {\bibinfo  {journal} {Soft
  Matter}\ }\textbf {\bibinfo {volume} {15}},\ \bibinfo {pages} {6006}
  (\bibinfo {year} {2019})}\BibitemShut {NoStop}%
\bibitem [{\citenamefont {Braig}\ \emph {et~al.}(2015)\citenamefont {Braig},
  \citenamefont {Schmidt}, \citenamefont {Stoiber}, \citenamefont {H{\"a}ndel},
  \citenamefont {M{\"o}hn}, \citenamefont {Werz}, \citenamefont {M{\"u}ller},
  \citenamefont {Zahler}, \citenamefont {Koeberle}, \citenamefont {K{\"a}s}
  \emph {et~al.}}]{braig2015pharmacological}%
  \BibitemOpen
  \bibfield  {author} {\bibinfo {author} {\bibfnamefont {S.}~\bibnamefont
  {Braig}}, \bibinfo {author} {\bibfnamefont {B.~S.}\ \bibnamefont {Schmidt}},
  \bibinfo {author} {\bibfnamefont {K.}~\bibnamefont {Stoiber}}, \bibinfo
  {author} {\bibfnamefont {C.}~\bibnamefont {H{\"a}ndel}}, \bibinfo {author}
  {\bibfnamefont {T.}~\bibnamefont {M{\"o}hn}}, \bibinfo {author}
  {\bibfnamefont {O.}~\bibnamefont {Werz}}, \bibinfo {author} {\bibfnamefont
  {R.}~\bibnamefont {M{\"u}ller}}, \bibinfo {author} {\bibfnamefont
  {S.}~\bibnamefont {Zahler}}, \bibinfo {author} {\bibfnamefont
  {A.}~\bibnamefont {Koeberle}}, \bibinfo {author} {\bibfnamefont {J.~A.}\
  \bibnamefont {K{\"a}s}}, \emph {et~al.},\ }\bibfield  {title} {\bibinfo
  {title} {Pharmacological targeting of membrane rigidity: implications on
  cancer cell migration and invasion},\ }\href@noop {} {\bibfield  {journal}
  {\bibinfo  {journal} {New J. Phys.}\ }\textbf {\bibinfo {volume} {17}},\
  \bibinfo {pages} {083007} (\bibinfo {year} {2015})}\BibitemShut {NoStop}%
\bibitem [{\citenamefont {Ayee}\ \emph
  {et~al.}(2018{\natexlab{a}})\citenamefont {Ayee}, \citenamefont {LeMaster},
  \citenamefont {Teng}, \citenamefont {Lee},\ and\ \citenamefont
  {Levitan}}]{ayee_hypotonic_2018}%
  \BibitemOpen
  \bibfield  {author} {\bibinfo {author} {\bibfnamefont {M.~A.~A.}\
  \bibnamefont {Ayee}}, \bibinfo {author} {\bibfnamefont {E.}~\bibnamefont
  {LeMaster}}, \bibinfo {author} {\bibfnamefont {T.}~\bibnamefont {Teng}},
  \bibinfo {author} {\bibfnamefont {J.}~\bibnamefont {Lee}},\ and\ \bibinfo
  {author} {\bibfnamefont {I.}~\bibnamefont {Levitan}},\ }\bibfield  {title}
  {\bibinfo {title} {Hypotonic {Challenge} of {Endothelial} {Cells} {Increases}
  {Membrane} {Stiffness} with {No} {Effect} on {Tether} {Force}},\ }\href
  {https://doi.org/10.1016/j.bpj.2017.12.032} {\bibfield  {journal} {\bibinfo
  {journal} {Biophys. J.}\ }\textbf {\bibinfo {volume} {114}},\ \bibinfo
  {pages} {929} (\bibinfo {year} {2018}{\natexlab{a}})}\BibitemShut {NoStop}%
\bibitem [{\citenamefont {Callies}\ \emph {et~al.}(2011)\citenamefont
  {Callies}, \citenamefont {Fels}, \citenamefont {Liashkovich}, \citenamefont
  {Kliche}, \citenamefont {Jeggle}, \citenamefont {Kusche-Vihrog},\ and\
  \citenamefont {Oberleithner}}]{callies_membrane_2011}%
  \BibitemOpen
  \bibfield  {author} {\bibinfo {author} {\bibfnamefont {C.}~\bibnamefont
  {Callies}}, \bibinfo {author} {\bibfnamefont {J.}~\bibnamefont {Fels}},
  \bibinfo {author} {\bibfnamefont {I.}~\bibnamefont {Liashkovich}}, \bibinfo
  {author} {\bibfnamefont {K.}~\bibnamefont {Kliche}}, \bibinfo {author}
  {\bibfnamefont {P.}~\bibnamefont {Jeggle}}, \bibinfo {author} {\bibfnamefont
  {K.}~\bibnamefont {Kusche-Vihrog}},\ and\ \bibinfo {author} {\bibfnamefont
  {H.}~\bibnamefont {Oberleithner}},\ }\bibfield  {title} {\bibinfo {title}
  {Membrane potential depolarization decreases the stiffness of vascular
  endothelial cells},\ }\href {https://doi.org/10.1242/jcs.084657} {\bibfield
  {journal} {\bibinfo  {journal} {J. Cell Sci.}\ }\textbf {\bibinfo {volume}
  {124}},\ \bibinfo {pages} {1936} (\bibinfo {year} {2011})}\BibitemShut
  {NoStop}%
\bibitem [{\citenamefont {Sliogeryte}\ \emph {et~al.}(2016)\citenamefont
  {Sliogeryte}, \citenamefont {Botto}, \citenamefont {Lee},\ and\ \citenamefont
  {Knight}}]{sliogeryte_chondrocyte_2016}%
  \BibitemOpen
  \bibfield  {author} {\bibinfo {author} {\bibfnamefont {K.}~\bibnamefont
  {Sliogeryte}}, \bibinfo {author} {\bibfnamefont {L.}~\bibnamefont {Botto}},
  \bibinfo {author} {\bibfnamefont {D.~A.}\ \bibnamefont {Lee}},\ and\ \bibinfo
  {author} {\bibfnamefont {M.~M.}\ \bibnamefont {Knight}},\ }\bibfield  {title}
  {\bibinfo {title} {Chondrocyte dedifferentiation increases cell stiffness by
  strengthening membrane-actin adhesion},\ }\href
  {https://doi.org/10.1016/j.joca.2015.12.007} {\bibfield  {journal} {\bibinfo
  {journal} {Osteoarthritis and Cartilage}\ }\textbf {\bibinfo {volume} {24}},\
  \bibinfo {pages} {912} (\bibinfo {year} {2016})}\BibitemShut {NoStop}%
\bibitem [{\citenamefont {Sanyour}\ \emph {et~al.}(2019)\citenamefont
  {Sanyour}, \citenamefont {Li}, \citenamefont {Rickel}, \citenamefont
  {Childs}, \citenamefont {Kinser},\ and\ \citenamefont
  {Hong}}]{sanyour_membrane_2019}%
  \BibitemOpen
  \bibfield  {author} {\bibinfo {author} {\bibfnamefont {H.~J.}\ \bibnamefont
  {Sanyour}}, \bibinfo {author} {\bibfnamefont {N.}~\bibnamefont {Li}},
  \bibinfo {author} {\bibfnamefont {A.~P.}\ \bibnamefont {Rickel}}, \bibinfo
  {author} {\bibfnamefont {J.~D.}\ \bibnamefont {Childs}}, \bibinfo {author}
  {\bibfnamefont {C.~N.}\ \bibnamefont {Kinser}},\ and\ \bibinfo {author}
  {\bibfnamefont {Z.}~\bibnamefont {Hong}},\ }\bibfield  {title} {\bibinfo
  {title} {Membrane cholesterol and substrate stiffness co-ordinate to induce
  the remodelling of the cytoskeleton and the alteration in the biomechanics of
  vascular smooth muscle cells},\ }\href {https://doi.org/10.1093/cvr/cvy276}
  {\bibfield  {journal} {\bibinfo  {journal} {Cardiovasc. Res.}\ }\textbf
  {\bibinfo {volume} {115}},\ \bibinfo {pages} {1369} (\bibinfo {year}
  {2019})}\BibitemShut {NoStop}%
\bibitem [{\citenamefont {Fowler}\ \emph {et~al.}(2016)\citenamefont {Fowler},
  \citenamefont {Hélie}, \citenamefont {Duncan}, \citenamefont {Chavent},
  \citenamefont {Koldsø},\ and\ \citenamefont
  {Sansom}}]{fowler_membrane_2016}%
  \BibitemOpen
  \bibfield  {author} {\bibinfo {author} {\bibfnamefont {P.~W.}\ \bibnamefont
  {Fowler}}, \bibinfo {author} {\bibfnamefont {J.}~\bibnamefont {Hélie}},
  \bibinfo {author} {\bibfnamefont {A.}~\bibnamefont {Duncan}}, \bibinfo
  {author} {\bibfnamefont {M.}~\bibnamefont {Chavent}}, \bibinfo {author}
  {\bibfnamefont {H.}~\bibnamefont {Koldsø}},\ and\ \bibinfo {author}
  {\bibfnamefont {M.~S.~P.}\ \bibnamefont {Sansom}},\ }\bibfield  {title}
  {\bibinfo {title} {Membrane stiffness is modified by integral membrane
  proteins},\ }\href {https://doi.org/10.1039/C6SM01186A} {\bibfield  {journal}
  {\bibinfo  {journal} {Soft Matter}\ }\textbf {\bibinfo {volume} {12}},\
  \bibinfo {pages} {7792} (\bibinfo {year} {2016})}\BibitemShut {NoStop}%
\bibitem [{\citenamefont {Faris}\ \emph {et~al.}(2009)\citenamefont {Faris},
  \citenamefont {Lacoste}, \citenamefont {Pécréaux}, \citenamefont {Joanny},
  \citenamefont {Prost},\ and\ \citenamefont
  {Bassereau}}]{faris_membrane_2009}%
  \BibitemOpen
  \bibfield  {author} {\bibinfo {author} {\bibfnamefont {M.~D. E.~A.}\
  \bibnamefont {Faris}}, \bibinfo {author} {\bibfnamefont {D.}~\bibnamefont
  {Lacoste}}, \bibinfo {author} {\bibfnamefont {J.}~\bibnamefont {Pécréaux}},
  \bibinfo {author} {\bibfnamefont {J.-F.}\ \bibnamefont {Joanny}}, \bibinfo
  {author} {\bibfnamefont {J.}~\bibnamefont {Prost}},\ and\ \bibinfo {author}
  {\bibfnamefont {P.}~\bibnamefont {Bassereau}},\ }\bibfield  {title} {\bibinfo
  {title} {Membrane {Tension} {Lowering} {Induced} by {Protein} {Activity}},\
  }\href {https://doi.org/10.1103/PhysRevLett.102.038102} {\bibfield  {journal}
  {\bibinfo  {journal} {Phys. Rev. Lett.}\ }\textbf {\bibinfo {volume} {102}},\
  \bibinfo {pages} {038102} (\bibinfo {year} {2009})}\BibitemShut {NoStop}%
\bibitem [{\citenamefont {Roli}\ \emph {et~al.}(2018)\citenamefont {Roli},
  \citenamefont {Villani}, \citenamefont {Filisetti},\ and\ \citenamefont
  {Serra}}]{roli2018dynamical}%
  \BibitemOpen
  \bibfield  {author} {\bibinfo {author} {\bibfnamefont {A.}~\bibnamefont
  {Roli}}, \bibinfo {author} {\bibfnamefont {M.}~\bibnamefont {Villani}},
  \bibinfo {author} {\bibfnamefont {A.}~\bibnamefont {Filisetti}},\ and\
  \bibinfo {author} {\bibfnamefont {R.}~\bibnamefont {Serra}},\ }\bibfield
  {title} {\bibinfo {title} {Dynamical criticality: overview and open
  questions},\ }\href@noop {} {\bibfield  {journal} {\bibinfo  {journal}
  {Journal of Systems Science and Complexity}\ }\textbf {\bibinfo {volume}
  {31}},\ \bibinfo {pages} {647} (\bibinfo {year} {2018})}\BibitemShut
  {NoStop}%
\bibitem [{\citenamefont {Doroudian}\ \emph {et~al.}(2013)\citenamefont
  {Doroudian}, \citenamefont {Curtis}, \citenamefont {Gang},\ and\
  \citenamefont {Russell}}]{Russel}%
  \BibitemOpen
  \bibfield  {author} {\bibinfo {author} {\bibfnamefont {G.}~\bibnamefont
  {Doroudian}}, \bibinfo {author} {\bibfnamefont {M.~W.}\ \bibnamefont
  {Curtis}}, \bibinfo {author} {\bibfnamefont {A.}~\bibnamefont {Gang}},\ and\
  \bibinfo {author} {\bibfnamefont {B.}~\bibnamefont {Russell}},\ }\bibfield
  {title} {\bibinfo {title} {Cyclic strain dominates over microtopography in
  regulating cytoskeletal and focal adhesion remodeling of human mesenchymal
  stem cells},\ }\href
  {https://doi.org/https://doi.org/10.1016/j.bbrc.2012.11.120} {\bibfield
  {journal} {\bibinfo  {journal} {Biochem. Biophys. Res. Commun}\ }\textbf
  {\bibinfo {volume} {430}},\ \bibinfo {pages} {1040 } (\bibinfo {year}
  {2013})}\BibitemShut {NoStop}%
\bibitem [{\citenamefont {Greiner}\ \emph {et~al.}(2013)\citenamefont
  {Greiner}, \citenamefont {Chen}, \citenamefont {Spatz},\ and\ \citenamefont
  {Kemkemer}}]{Spatz_cyc}%
  \BibitemOpen
  \bibfield  {author} {\bibinfo {author} {\bibfnamefont {A.~M.}\ \bibnamefont
  {Greiner}}, \bibinfo {author} {\bibfnamefont {H.}~\bibnamefont {Chen}},
  \bibinfo {author} {\bibfnamefont {J.~P.}\ \bibnamefont {Spatz}},\ and\
  \bibinfo {author} {\bibfnamefont {R.}~\bibnamefont {Kemkemer}},\ }\bibfield
  {title} {\bibinfo {title} {Cyclic tensile strain controls cell shape and
  directs actin stress fiber formation and focal adhesion alignment in
  spreading cells},\ }\href {https://doi.org/10.1371/journal.pone.0077328}
  {\bibfield  {journal} {\bibinfo  {journal} {PLOS ONE}\ }\textbf {\bibinfo
  {volume} {8}},\ \bibinfo {pages} {1} (\bibinfo {year} {2013})}\BibitemShut
  {NoStop}%
\bibitem [{\citenamefont {Parsons}\ \emph {et~al.}(2010)\citenamefont
  {Parsons}, \citenamefont {Horwitz},\ and\ \citenamefont
  {Schwartz}}]{Parsons}%
  \BibitemOpen
  \bibfield  {author} {\bibinfo {author} {\bibfnamefont {J.~T.}\ \bibnamefont
  {Parsons}}, \bibinfo {author} {\bibfnamefont {A.~R.}\ \bibnamefont
  {Horwitz}},\ and\ \bibinfo {author} {\bibfnamefont {M.~A.}\ \bibnamefont
  {Schwartz}},\ }\bibfield  {title} {\bibinfo {title} {Cell adhesion:
  integrating cytoskeletal dynamics and cellular tension},\ }\href
  {https://doi.org/10.1038/nrm2957} {\bibfield  {journal} {\bibinfo  {journal}
  {Nat. Rev. Mol. Cell Biol.}\ }\textbf {\bibinfo {volume} {11}},\ \bibinfo
  {pages} {633–643} (\bibinfo {year} {2010})}\BibitemShut {NoStop}%
\bibitem [{\citenamefont {Chesla}\ \emph {et~al.}(1998)\citenamefont {Chesla},
  \citenamefont {Selvaraj},\ and\ \citenamefont {Zhu}}]{chesla1998measuring}%
  \BibitemOpen
  \bibfield  {author} {\bibinfo {author} {\bibfnamefont {S.~E.}\ \bibnamefont
  {Chesla}}, \bibinfo {author} {\bibfnamefont {P.}~\bibnamefont {Selvaraj}},\
  and\ \bibinfo {author} {\bibfnamefont {C.}~\bibnamefont {Zhu}},\ }\bibfield
  {title} {\bibinfo {title} {Measuring two-dimensional receptor-ligand binding
  kinetics by micropipette},\ }\href@noop {} {\bibfield  {journal} {\bibinfo
  {journal} {Biophys. J.}\ }\textbf {\bibinfo {volume} {75}},\ \bibinfo {pages}
  {1553} (\bibinfo {year} {1998})}\BibitemShut {NoStop}%
\bibitem [{\citenamefont {Zhelev}\ \emph {et~al.}(1994)\citenamefont {Zhelev},
  \citenamefont {Needham},\ and\ \citenamefont {Hochmuth}}]{zhelev1994role}%
  \BibitemOpen
  \bibfield  {author} {\bibinfo {author} {\bibfnamefont {D.~V.}\ \bibnamefont
  {Zhelev}}, \bibinfo {author} {\bibfnamefont {D.}~\bibnamefont {Needham}},\
  and\ \bibinfo {author} {\bibfnamefont {R.~M.}\ \bibnamefont {Hochmuth}},\
  }\bibfield  {title} {\bibinfo {title} {Role of the membrane cortex in
  neutrophil deformation in small pipets},\ }\href@noop {} {\bibfield
  {journal} {\bibinfo  {journal} {Biophys. J.}\ }\textbf {\bibinfo {volume}
  {67}},\ \bibinfo {pages} {696} (\bibinfo {year} {1994})}\BibitemShut
  {NoStop}%
\bibitem [{\citenamefont {Sharma}(1993)}]{sharma1993cellular}%
  \BibitemOpen
  \bibfield  {author} {\bibinfo {author} {\bibfnamefont {K.}~\bibnamefont
  {Sharma}},\ }\bibfield  {title} {\bibinfo {title} {Cellular deformability
  studies in leukemia.},\ }\href@noop {} {\bibfield  {journal} {\bibinfo
  {journal} {Physiol. Chem. Phys. Med. NMR.}\ }\textbf {\bibinfo {volume}
  {25}},\ \bibinfo {pages} {293} (\bibinfo {year} {1993})}\BibitemShut
  {NoStop}%
\bibitem [{\citenamefont {Hasan}\ \emph {et~al.}(1998)\citenamefont {Hasan},
  \citenamefont {Adams}, \citenamefont {Joiner}, \citenamefont {Marshall},\
  and\ \citenamefont {Hart}}]{Detach}%
  \BibitemOpen
  \bibfield  {author} {\bibinfo {author} {\bibfnamefont {N.}~\bibnamefont
  {Hasan}}, \bibinfo {author} {\bibfnamefont {G.}~\bibnamefont {Adams}},
  \bibinfo {author} {\bibfnamefont {M.}~\bibnamefont {Joiner}}, \bibinfo
  {author} {\bibfnamefont {J.}~\bibnamefont {Marshall}},\ and\ \bibinfo
  {author} {\bibfnamefont {I.}~\bibnamefont {Hart}},\ }\bibfield  {title}
  {\bibinfo {title} {Hypoxia facilitates tumour cell detachment by reducing
  expression of surface adhesion molecules and adhesion to extracellular
  matrices without loss of cell viability},\ }\href
  {https://doi.org/10.1038/bjc.1998.299} {\bibfield  {journal} {\bibinfo
  {journal} {Br. J. Cancer}\ }\textbf {\bibinfo {volume} {77}},\ \bibinfo
  {pages} {1799–1805} (\bibinfo {year} {1998})}\BibitemShut {NoStop}%
\bibitem [{\citenamefont {Yamada}\ and\ \citenamefont {Sixt}(2019)}]{Yamada}%
  \BibitemOpen
  \bibfield  {author} {\bibinfo {author} {\bibfnamefont {K.~M.}\ \bibnamefont
  {Yamada}}\ and\ \bibinfo {author} {\bibfnamefont {M.}~\bibnamefont {Sixt}},\
  }\bibfield  {title} {\bibinfo {title} {Mechanisms of 3d cell migration},\
  }\href {https://doi.org/10.1038/s41580-019-0172-9} {\bibfield  {journal}
  {\bibinfo  {journal} {Nat. Rev. Mol. Cell Biol.}\ }\textbf {\bibinfo {volume}
  {20}},\ \bibinfo {pages} {738–752} (\bibinfo {year} {2019})}\BibitemShut
  {NoStop}%
\bibitem [{\citenamefont {De~Pascalis}\ and\ \citenamefont
  {Etienne-Manneville}(2017)}]{De_Pascalis}%
  \BibitemOpen
  \bibfield  {author} {\bibinfo {author} {\bibfnamefont {C.}~\bibnamefont
  {De~Pascalis}}\ and\ \bibinfo {author} {\bibfnamefont {S.}~\bibnamefont
  {Etienne-Manneville}},\ }\bibfield  {title} {\bibinfo {title} {Single and
  collective cell migration: the mechanics of adhesions},\ }\href
  {https://doi.org/10.1091/mbc.e17-03-0134} {\bibfield  {journal} {\bibinfo
  {journal} {Mol. Biol. Cell}\ }\textbf {\bibinfo {volume} {28}},\ \bibinfo
  {pages} {1833–1846} (\bibinfo {year} {2017})}\BibitemShut {NoStop}%
\bibitem [{\citenamefont {Kirfel}(2004)}]{Kirfel}%
  \BibitemOpen
  \bibfield  {author} {\bibinfo {author} {\bibfnamefont {G.}~\bibnamefont
  {Kirfel}},\ }\bibfield  {title} {\bibinfo {title} {Cell migration: mechanisms
  of rear detachment and the formation of migration tracks},\ }\href
  {https://doi.org/10.1078/0171-9335-00421} {\bibfield  {journal} {\bibinfo
  {journal} {Eur. J. Cell Biol}\ }\textbf {\bibinfo {volume} {83}},\ \bibinfo
  {pages} {717–724} (\bibinfo {year} {2004})}\BibitemShut {NoStop}%
\bibitem [{\citenamefont {Dwir}\ \emph {et~al.}(2003)\citenamefont {Dwir},
  \citenamefont {Solomon}, \citenamefont {Mangan}, \citenamefont {Kansas},
  \citenamefont {Schwarz},\ and\ \citenamefont {Alon}}]{Dwir}%
  \BibitemOpen
  \bibfield  {author} {\bibinfo {author} {\bibfnamefont {O.}~\bibnamefont
  {Dwir}}, \bibinfo {author} {\bibfnamefont {A.}~\bibnamefont {Solomon}},
  \bibinfo {author} {\bibfnamefont {S.}~\bibnamefont {Mangan}}, \bibinfo
  {author} {\bibfnamefont {G.~S.}\ \bibnamefont {Kansas}}, \bibinfo {author}
  {\bibfnamefont {U.~S.}\ \bibnamefont {Schwarz}},\ and\ \bibinfo {author}
  {\bibfnamefont {R.}~\bibnamefont {Alon}},\ }\bibfield  {title} {\bibinfo
  {title} {Avidity enhancement of l-selectin bonds by flow},\ }\href
  {https://doi.org/10.1083/jcb.200303134} {\bibfield  {journal} {\bibinfo
  {journal} {J. Cell Biol.}\ }\textbf {\bibinfo {volume} {163}},\ \bibinfo
  {pages} {649–659} (\bibinfo {year} {2003})}\BibitemShut {NoStop}%
\bibitem [{\citenamefont {Erbeldinger}\ \emph {et~al.}(2017)\citenamefont
  {Erbeldinger}, \citenamefont {Rapp}, \citenamefont {Ktitareva}, \citenamefont
  {Wendel}, \citenamefont {Bothe}, \citenamefont {Dettmering}, \citenamefont
  {Durante}, \citenamefont {Friedrich}, \citenamefont {Bertulat}, \citenamefont
  {Meyer}, \citenamefont {Cardoso}, \citenamefont {Hehlgans}, \citenamefont
  {Rödel},\ and\ \citenamefont {Fournier}}]{Erbeldinger}%
  \BibitemOpen
  \bibfield  {author} {\bibinfo {author} {\bibfnamefont {N.}~\bibnamefont
  {Erbeldinger}}, \bibinfo {author} {\bibfnamefont {F.}~\bibnamefont {Rapp}},
  \bibinfo {author} {\bibfnamefont {S.}~\bibnamefont {Ktitareva}}, \bibinfo
  {author} {\bibfnamefont {P.}~\bibnamefont {Wendel}}, \bibinfo {author}
  {\bibfnamefont {A.~S.}\ \bibnamefont {Bothe}}, \bibinfo {author}
  {\bibfnamefont {T.}~\bibnamefont {Dettmering}}, \bibinfo {author}
  {\bibfnamefont {M.}~\bibnamefont {Durante}}, \bibinfo {author} {\bibfnamefont
  {T.}~\bibnamefont {Friedrich}}, \bibinfo {author} {\bibfnamefont
  {B.}~\bibnamefont {Bertulat}}, \bibinfo {author} {\bibfnamefont
  {S.}~\bibnamefont {Meyer}}, \bibinfo {author} {\bibfnamefont {M.~C.}\
  \bibnamefont {Cardoso}}, \bibinfo {author} {\bibfnamefont {S.}~\bibnamefont
  {Hehlgans}}, \bibinfo {author} {\bibfnamefont {F.}~\bibnamefont {Rödel}},\
  and\ \bibinfo {author} {\bibfnamefont {C.}~\bibnamefont {Fournier}},\
  }\bibfield  {title} {\bibinfo {title} {Measuring leukocyte adhesion to
  (primary) endothelial cells after photon and charged particle exposure with a
  dedicated laminar flow chamber},\ }\href
  {https://doi.org/10.3389/fimmu.2017.00627} {\bibfield  {journal} {\bibinfo
  {journal} {Front. Immunol.}\ }\textbf {\bibinfo {volume} {8}},\ \bibinfo
  {pages} {627} (\bibinfo {year} {2017})}\BibitemShut {NoStop}%
\bibitem [{\citenamefont {Zhou}\ \emph {et~al.}(2014)\citenamefont {Zhou},
  \citenamefont {Kucik}, \citenamefont {Szalai},\ and\ \citenamefont
  {Edberg}}]{Zhou}%
  \BibitemOpen
  \bibfield  {author} {\bibinfo {author} {\bibfnamefont {Y.}~\bibnamefont
  {Zhou}}, \bibinfo {author} {\bibfnamefont {D.~F.}\ \bibnamefont {Kucik}},
  \bibinfo {author} {\bibfnamefont {A.~J.}\ \bibnamefont {Szalai}},\ and\
  \bibinfo {author} {\bibfnamefont {J.~C.}\ \bibnamefont {Edberg}},\ }\bibfield
   {title} {\bibinfo {title} {Human neutrophil flow chamber adhesion assay},\
  }\href@noop {} {\bibfield  {journal} {\bibinfo  {journal} {JoVE}\ ,\ \bibinfo
  {pages} {e51410}} (\bibinfo {year} {2014})}\BibitemShut {NoStop}%
\bibitem [{\citenamefont {Cross}\ \emph {et~al.}(2007)\citenamefont {Cross},
  \citenamefont {Jin}, \citenamefont {Rao},\ and\ \citenamefont
  {Gimzewski}}]{compliant}%
  \BibitemOpen
  \bibfield  {author} {\bibinfo {author} {\bibfnamefont {S.~E.}\ \bibnamefont
  {Cross}}, \bibinfo {author} {\bibfnamefont {Y.-S.}\ \bibnamefont {Jin}},
  \bibinfo {author} {\bibfnamefont {J.}~\bibnamefont {Rao}},\ and\ \bibinfo
  {author} {\bibfnamefont {J.~K.}\ \bibnamefont {Gimzewski}},\ }\bibfield
  {title} {\bibinfo {title} {Nanomechanical analysis of cells from cancer
  patients},\ }\href {https://doi.org/10.1038/nnano.2007.388} {\bibfield
  {journal} {\bibinfo  {journal} {Nat. Nanotechnol.}\ }\textbf {\bibinfo
  {volume} {2}},\ \bibinfo {pages} {780–783} (\bibinfo {year}
  {2007})}\BibitemShut {NoStop}%
\bibitem [{\citenamefont {Plischke}\ and\ \citenamefont
  {Mattis}(1970)}]{PhysRevB.2.2660}%
  \BibitemOpen
  \bibfield  {author} {\bibinfo {author} {\bibfnamefont {M.}~\bibnamefont
  {Plischke}}\ and\ \bibinfo {author} {\bibfnamefont {D.}~\bibnamefont
  {Mattis}},\ }\bibfield  {title} {\bibinfo {title} {Two-dimensional ising
  model in a finite magnetic field},\ }\href
  {https://doi.org/10.1103/PhysRevB.2.2660} {\bibfield  {journal} {\bibinfo
  {journal} {Phys. Rev. B}\ }\textbf {\bibinfo {volume} {2}},\ \bibinfo {pages}
  {2660} (\bibinfo {year} {1970})}\BibitemShut {NoStop}%
\bibitem [{\citenamefont {Morgenstern}\ \emph {et~al.}(1981)\citenamefont
  {Morgenstern}, \citenamefont {Binder},\ and\ \citenamefont
  {Hornreich}}]{PhysRevB.23.287}%
  \BibitemOpen
  \bibfield  {author} {\bibinfo {author} {\bibfnamefont {I.}~\bibnamefont
  {Morgenstern}}, \bibinfo {author} {\bibfnamefont {K.}~\bibnamefont
  {Binder}},\ and\ \bibinfo {author} {\bibfnamefont {R.~M.}\ \bibnamefont
  {Hornreich}},\ }\bibfield  {title} {\bibinfo {title} {Two-dimensional ising
  model in random magnetic fields},\ }\href
  {https://doi.org/10.1103/PhysRevB.23.287} {\bibfield  {journal} {\bibinfo
  {journal} {Phys. Rev. B}\ }\textbf {\bibinfo {volume} {23}},\ \bibinfo
  {pages} {287} (\bibinfo {year} {1981})}\BibitemShut {NoStop}%
\bibitem [{\citenamefont {Stump}(1987)}]{PhysRevA.36.4439}%
  \BibitemOpen
  \bibfield  {author} {\bibinfo {author} {\bibfnamefont {D.~R.}\ \bibnamefont
  {Stump}},\ }\bibfield  {title} {\bibinfo {title} {Entropy of the
  two-dimensional ising model},\ }\href
  {https://doi.org/10.1103/PhysRevA.36.4439} {\bibfield  {journal} {\bibinfo
  {journal} {Phys. Rev. A}\ }\textbf {\bibinfo {volume} {36}},\ \bibinfo
  {pages} {4439} (\bibinfo {year} {1987})}\BibitemShut {NoStop}%
\bibitem [{\citenamefont {Kaufman}(1987)}]{PhysRevB.36.3697}%
  \BibitemOpen
  \bibfield  {author} {\bibinfo {author} {\bibfnamefont {M.}~\bibnamefont
  {Kaufman}},\ }\bibfield  {title} {\bibinfo {title} {Square-lattice ising
  model in a weak uniform magnetic field: Renormalization-group analysis},\
  }\href {https://doi.org/10.1103/PhysRevB.36.3697} {\bibfield  {journal}
  {\bibinfo  {journal} {Phys. Rev. B}\ }\textbf {\bibinfo {volume} {36}},\
  \bibinfo {pages} {3697} (\bibinfo {year} {1987})}\BibitemShut {NoStop}%
\bibitem [{\citenamefont {de~With~(G.)}(2013)}]{de2013liquid}%
  \BibitemOpen
  \bibfield  {author} {\bibinfo {author} {\bibnamefont {de~With~(G.)}},\
  }\href@noop {} {\emph {\bibinfo {title} {Liquid-state Physical Chemistry:
  Fundamentals, Modeling, and Applications}}}\ (\bibinfo  {publisher}
  {Wiley-VCH Verlag},\ \bibinfo {year} {2013})\BibitemShut {NoStop}%
\bibitem [{\citenamefont {Datta}\ \emph {et~al.}(2018)\citenamefont {Datta},
  \citenamefont {Acharyya},\ and\ \citenamefont
  {Dhar}}]{datta2018magnetisation}%
  \BibitemOpen
  \bibfield  {author} {\bibinfo {author} {\bibfnamefont {R.}~\bibnamefont
  {Datta}}, \bibinfo {author} {\bibfnamefont {M.}~\bibnamefont {Acharyya}},\
  and\ \bibinfo {author} {\bibfnamefont {A.}~\bibnamefont {Dhar}},\ }\bibfield
  {title} {\bibinfo {title} {Magnetisation reversal in ising ferromagnet by
  thermal and field gradients},\ }\href@noop {} {\bibfield  {journal} {\bibinfo
   {journal} {Heliyon}\ }\textbf {\bibinfo {volume} {4}},\ \bibinfo {pages}
  {e00892} (\bibinfo {year} {2018})}\BibitemShut {NoStop}%
\bibitem [{\citenamefont {Brendel}\ \emph {et~al.}(2003)\citenamefont
  {Brendel}, \citenamefont {Barkema},\ and\ \citenamefont {van
  Beijeren}}]{brendel2003magnetization}%
  \BibitemOpen
  \bibfield  {author} {\bibinfo {author} {\bibfnamefont {K.}~\bibnamefont
  {Brendel}}, \bibinfo {author} {\bibfnamefont {G.}~\bibnamefont {Barkema}},\
  and\ \bibinfo {author} {\bibfnamefont {H.}~\bibnamefont {van Beijeren}},\
  }\bibfield  {title} {\bibinfo {title} {Magnetization reversal times in the
  two-dimensional ising model},\ }\href@noop {} {\bibfield  {journal} {\bibinfo
   {journal} {Phys. Rev. E}\ }\textbf {\bibinfo {volume} {67}},\ \bibinfo
  {pages} {026119} (\bibinfo {year} {2003})}\BibitemShut {NoStop}%
\bibitem [{\citenamefont {Garc{\'\i}a-Pablos}\ \emph
  {et~al.}(1996)\citenamefont {Garc{\'\i}a-Pablos}, \citenamefont
  {Garc{\'\i}a-Mochales}, \citenamefont {Garcia},\ and\ \citenamefont
  {Serena}}]{garcia1996nonhomogeneous}%
  \BibitemOpen
  \bibfield  {author} {\bibinfo {author} {\bibfnamefont {D.}~\bibnamefont
  {Garc{\'\i}a-Pablos}}, \bibinfo {author} {\bibfnamefont {P.}~\bibnamefont
  {Garc{\'\i}a-Mochales}}, \bibinfo {author} {\bibfnamefont {N.}~\bibnamefont
  {Garcia}},\ and\ \bibinfo {author} {\bibfnamefont {P.}~\bibnamefont
  {Serena}},\ }\bibfield  {title} {\bibinfo {title} {Nonhomogeneous
  magnetization reversal in 2d ising clusters},\ }\href@noop {} {\bibfield
  {journal} {\bibinfo  {journal} {J. Appl. Phys.}\ }\textbf {\bibinfo {volume}
  {79}},\ \bibinfo {pages} {6019} (\bibinfo {year} {1996})}\BibitemShut
  {NoStop}%
\bibitem [{\citenamefont {Brendel}\ \emph {et~al.}(2005)\citenamefont
  {Brendel}, \citenamefont {Barkema},\ and\ \citenamefont {van
  Beijeren}}]{brendel2005nucleation}%
  \BibitemOpen
  \bibfield  {author} {\bibinfo {author} {\bibfnamefont {K.}~\bibnamefont
  {Brendel}}, \bibinfo {author} {\bibfnamefont {G.}~\bibnamefont {Barkema}},\
  and\ \bibinfo {author} {\bibfnamefont {H.}~\bibnamefont {van Beijeren}},\
  }\bibfield  {title} {\bibinfo {title} {Nucleation time distribution in the
  two-dimensional ising model with spin-flip dynamics},\ }in\ \href@noop {}
  {\emph {\bibinfo {booktitle} {AIP Conference Proceedings}}},\ Vol.\ \bibinfo
  {volume} {800}\ (\bibinfo {organization} {American Institute of Physics},\
  \bibinfo {year} {2005})\ pp.\ \bibinfo {pages} {39--49}\BibitemShut {NoStop}%
\bibitem [{\citenamefont {Sear}(2006)}]{sear2006heterogeneous}%
  \BibitemOpen
  \bibfield  {author} {\bibinfo {author} {\bibfnamefont {R.~P.}\ \bibnamefont
  {Sear}},\ }\bibfield  {title} {\bibinfo {title} {Heterogeneous and
  homogeneous nucleation compared: Rapid nucleation on microscopic
  impurities},\ }\href@noop {} {\bibfield  {journal} {\bibinfo  {journal} {J.
  Phys. Chem. B}\ }\textbf {\bibinfo {volume} {110}},\ \bibinfo {pages} {4985}
  (\bibinfo {year} {2006})}\BibitemShut {NoStop}%
\bibitem [{\citenamefont {Hohenberg}\ and\ \citenamefont
  {Halperin}(1977)}]{RevModPhys.49.435}%
  \BibitemOpen
  \bibfield  {author} {\bibinfo {author} {\bibfnamefont {P.~C.}\ \bibnamefont
  {Hohenberg}}\ and\ \bibinfo {author} {\bibfnamefont {B.~I.}\ \bibnamefont
  {Halperin}},\ }\bibfield  {title} {\bibinfo {title} {Theory of dynamic
  critical phenomena},\ }\href {https://doi.org/10.1103/RevModPhys.49.435}
  {\bibfield  {journal} {\bibinfo  {journal} {Rev. Mod. Phys.}\ }\textbf
  {\bibinfo {volume} {49}},\ \bibinfo {pages} {435} (\bibinfo {year}
  {1977})}\BibitemShut {NoStop}%
\bibitem [{\citenamefont {Binder}\ and\ \citenamefont
  {Young}(1986)}]{RevModPhys.58.801}%
  \BibitemOpen
  \bibfield  {author} {\bibinfo {author} {\bibfnamefont {K.}~\bibnamefont
  {Binder}}\ and\ \bibinfo {author} {\bibfnamefont {A.~P.}\ \bibnamefont
  {Young}},\ }\bibfield  {title} {\bibinfo {title} {Spin glasses: Experimental
  facts, theoretical concepts, and open questions},\ }\href
  {https://doi.org/10.1103/RevModPhys.58.801} {\bibfield  {journal} {\bibinfo
  {journal} {Rev. Mod. Phys.}\ }\textbf {\bibinfo {volume} {58}},\ \bibinfo
  {pages} {801} (\bibinfo {year} {1986})}\BibitemShut {NoStop}%
\bibitem [{\citenamefont {Krobath}\ \emph {et~al.}(2011)\citenamefont
  {Krobath}, \citenamefont {Rózycki}, \citenamefont {Lipowsky},\ and\
  \citenamefont {Weikl}}]{krobath_line_2011}%
  \BibitemOpen
  \bibfield  {author} {\bibinfo {author} {\bibfnamefont {H.}~\bibnamefont
  {Krobath}}, \bibinfo {author} {\bibfnamefont {B.}~\bibnamefont {Rózycki}},
  \bibinfo {author} {\bibfnamefont {R.}~\bibnamefont {Lipowsky}},\ and\
  \bibinfo {author} {\bibfnamefont {T.~R.}\ \bibnamefont {Weikl}},\ }\bibfield
  {title} {\bibinfo {title} {Line {Tension} and {Stability} of {Domains} in
  {Cell}-{Adhesion} {Zones} {Mediated} by {Long} and {Short}
  {Receptor}-{Ligand} {Complexes}},\ }\href
  {https://dx.plos.org/10.1371/journal.pone.0023284} {\bibfield  {journal}
  {\bibinfo  {journal} {PLoS ONE}\ }\textbf {\bibinfo {volume} {6}},\ \bibinfo
  {pages} {e23284} (\bibinfo {year} {2011})}\BibitemShut {NoStop}%
\bibitem [{\citenamefont {Erdmann}\ and\ \citenamefont
  {Schwarz}(2007)}]{erdmann_impact_2007}%
  \BibitemOpen
  \bibfield  {author} {\bibinfo {author} {\bibfnamefont {T.}~\bibnamefont
  {Erdmann}}\ and\ \bibinfo {author} {\bibfnamefont {U.~S.}\ \bibnamefont
  {Schwarz}},\ }\bibfield  {title} {\bibinfo {title} {Impact of receptor-ligand
  distance on adhesion cluster stability},\ }\href
  {https://doi.org/10.1140/epje/e2007-00019-8} {\bibfield  {journal} {\bibinfo
  {journal} {Eur. Phys. J. E}\ }\textbf {\bibinfo {volume} {22}},\ \bibinfo
  {pages} {123} (\bibinfo {year} {2007})}\BibitemShut {NoStop}%
\bibitem [{\citenamefont {Ayee}\ \emph
  {et~al.}(2018{\natexlab{b}})\citenamefont {Ayee}, \citenamefont {LeMaster},
  \citenamefont {Teng}, \citenamefont {Lee},\ and\ \citenamefont
  {Levitan}}]{Ayee_2018}%
  \BibitemOpen
  \bibfield  {author} {\bibinfo {author} {\bibfnamefont {M.~A.~A.}\
  \bibnamefont {Ayee}}, \bibinfo {author} {\bibfnamefont {E.}~\bibnamefont
  {LeMaster}}, \bibinfo {author} {\bibfnamefont {T.}~\bibnamefont {Teng}},
  \bibinfo {author} {\bibfnamefont {J.}~\bibnamefont {Lee}},\ and\ \bibinfo
  {author} {\bibfnamefont {I.}~\bibnamefont {Levitan}},\ }\bibfield  {title}
  {\bibinfo {title} {Hypotonic challenge of endothelial cells increases
  membrane stiffness with no effect on tether force},\ }\href
  {https://doi.org/10.1016/j.bpj.2017.12.032} {\bibfield  {journal} {\bibinfo
  {journal} {Biophysical Journal}\ }\textbf {\bibinfo {volume} {114}},\
  \bibinfo {pages} {929–938} (\bibinfo {year}
  {2018}{\natexlab{b}})}\BibitemShut {NoStop}%
\bibitem [{\citenamefont {Maan}\ \emph {et~al.}(2018)\citenamefont {Maan},
  \citenamefont {Loiseau},\ and\ \citenamefont {Bausch}}]{Bausch}%
  \BibitemOpen
  \bibfield  {author} {\bibinfo {author} {\bibfnamefont {R.}~\bibnamefont
  {Maan}}, \bibinfo {author} {\bibfnamefont {E.}~\bibnamefont {Loiseau}},\ and\
  \bibinfo {author} {\bibfnamefont {A.~R.}\ \bibnamefont {Bausch}},\ }\bibfield
   {title} {\bibinfo {title} {Adhesion of active cytoskeletal vesicles},\
  }\href {https://doi.org/10.1016/j.bpj.2018.10.013} {\bibfield  {journal}
  {\bibinfo  {journal} {Biophysical Journal}\ }\textbf {\bibinfo {volume}
  {115}},\ \bibinfo {pages} {2395–2402} (\bibinfo {year} {2018})}\BibitemShut
  {NoStop}%
\bibitem [{\citenamefont {Mu\~noz}(2018)}]{Munoz}%
  \BibitemOpen
  \bibfield  {author} {\bibinfo {author} {\bibfnamefont {M.~A.}\ \bibnamefont
  {Mu\~noz}},\ }\bibfield  {title} {\bibinfo {title} {Colloquium: Criticality
  and dynamical scaling in living systems},\ }\href
  {https://doi.org/10.1103/RevModPhys.90.031001} {\bibfield  {journal}
  {\bibinfo  {journal} {Rev. Mod. Phys.}\ }\textbf {\bibinfo {volume} {90}},\
  \bibinfo {pages} {031001} (\bibinfo {year} {2018})}\BibitemShut {NoStop}%
\bibitem [{\citenamefont {Balaban}\ \emph {et~al.}(2001)\citenamefont
  {Balaban}, \citenamefont {Schwarz}, \citenamefont {Riveline}, \citenamefont
  {Goichberg}, \citenamefont {Tzur}, \citenamefont {Sabanay}, \citenamefont
  {Mahalu}, \citenamefont {Safran}, \citenamefont {Bershadsky}, \citenamefont
  {Addadi},\ and\ \citenamefont {et~al.}}]{constant_1}%
  \BibitemOpen
  \bibfield  {author} {\bibinfo {author} {\bibfnamefont {N.~Q.}\ \bibnamefont
  {Balaban}}, \bibinfo {author} {\bibfnamefont {U.~S.}\ \bibnamefont
  {Schwarz}}, \bibinfo {author} {\bibfnamefont {D.}~\bibnamefont {Riveline}},
  \bibinfo {author} {\bibfnamefont {P.}~\bibnamefont {Goichberg}}, \bibinfo
  {author} {\bibfnamefont {G.}~\bibnamefont {Tzur}}, \bibinfo {author}
  {\bibfnamefont {I.}~\bibnamefont {Sabanay}}, \bibinfo {author} {\bibfnamefont
  {D.}~\bibnamefont {Mahalu}}, \bibinfo {author} {\bibfnamefont
  {S.}~\bibnamefont {Safran}}, \bibinfo {author} {\bibfnamefont
  {A.}~\bibnamefont {Bershadsky}}, \bibinfo {author} {\bibfnamefont
  {L.}~\bibnamefont {Addadi}},\ and\ \bibinfo {author} {\bibnamefont
  {et~al.}},\ }\bibfield  {title} {\bibinfo {title} {Force and focal adhesion
  assembly: a close relationship studied using elastic micropatterned
  substrates},\ }\href {https://doi.org/10.1038/35074532} {\bibfield  {journal}
  {\bibinfo  {journal} {Nat. Cell Biol.}\ }\textbf {\bibinfo {volume} {3}},\
  \bibinfo {pages} {466–472} (\bibinfo {year} {2001})}\BibitemShut {NoStop}%
\bibitem [{\citenamefont {Thomas}\ \emph {et~al.}(2008)\citenamefont {Thomas},
  \citenamefont {Vogel},\ and\ \citenamefont {Sokurenko}}]{catch_rev}%
  \BibitemOpen
  \bibfield  {author} {\bibinfo {author} {\bibfnamefont {W.~E.}\ \bibnamefont
  {Thomas}}, \bibinfo {author} {\bibfnamefont {V.}~\bibnamefont {Vogel}},\ and\
  \bibinfo {author} {\bibfnamefont {E.}~\bibnamefont {Sokurenko}},\ }\bibfield
  {title} {\bibinfo {title} {Biophysics of catch bonds},\ }\href
  {https://doi.org/10.1146/annurev.biophys.37.032807.125804} {\bibfield
  {journal} {\bibinfo  {journal} {Annu. Rev. Biophys.}\ }\textbf {\bibinfo
  {volume} {37}},\ \bibinfo {pages} {399} (\bibinfo {year} {2008})}\BibitemShut
  {NoStop}%
\bibitem [{\citenamefont {Chakrabarti}\ \emph {et~al.}(2014)\citenamefont
  {Chakrabarti}, \citenamefont {Hinczewski},\ and\ \citenamefont
  {Thirumalai}}]{Hin}%
  \BibitemOpen
  \bibfield  {author} {\bibinfo {author} {\bibfnamefont {S.}~\bibnamefont
  {Chakrabarti}}, \bibinfo {author} {\bibfnamefont {M.}~\bibnamefont
  {Hinczewski}},\ and\ \bibinfo {author} {\bibfnamefont {D.}~\bibnamefont
  {Thirumalai}},\ }\bibfield  {title} {\bibinfo {title} {Plasticity of hydrogen
  bond networks regulates mechanochemistry of cell adhesion complexes},\ }\href
  {https://doi.org/10.1073/pnas.1405384111} {\bibfield  {journal} {\bibinfo
  {journal} {Proc. Natl. Acad. Sci. U.S.A.}\ }\textbf {\bibinfo {volume}
  {111}},\ \bibinfo {pages} {9048} (\bibinfo {year} {2014})}\BibitemShut
  {NoStop}%
\bibitem [{\citenamefont {Chakrabarti}\ \emph {et~al.}(2017)\citenamefont
  {Chakrabarti}, \citenamefont {Hinczewski},\ and\ \citenamefont
  {Thirumalai}}]{Hin1}%
  \BibitemOpen
  \bibfield  {author} {\bibinfo {author} {\bibfnamefont {S.}~\bibnamefont
  {Chakrabarti}}, \bibinfo {author} {\bibfnamefont {M.}~\bibnamefont
  {Hinczewski}},\ and\ \bibinfo {author} {\bibfnamefont {D.}~\bibnamefont
  {Thirumalai}},\ }\bibfield  {title} {\bibinfo {title} {Phenomenological and
  microscopic theories for catch bonds},\ }\href@noop {} {\bibfield  {journal}
  {\bibinfo  {journal} {J. Struct. Biol.}\ }\textbf {\bibinfo {volume} {197}},\
  \bibinfo {pages} {50} (\bibinfo {year} {2017})}\BibitemShut {NoStop}%
\bibitem [{\citenamefont {Adhikari}\ \emph {et~al.}(2018)\citenamefont
  {Adhikari}, \citenamefont {Moran}, \citenamefont {Weddle},\ and\
  \citenamefont {Hinczewski}}]{Hin2}%
  \BibitemOpen
  \bibfield  {author} {\bibinfo {author} {\bibfnamefont {S.}~\bibnamefont
  {Adhikari}}, \bibinfo {author} {\bibfnamefont {J.}~\bibnamefont {Moran}},
  \bibinfo {author} {\bibfnamefont {C.}~\bibnamefont {Weddle}},\ and\ \bibinfo
  {author} {\bibfnamefont {M.}~\bibnamefont {Hinczewski}},\ }\bibfield  {title}
  {\bibinfo {title} {Unraveling the mechanism of the cadherin-catenin-actin
  catch bond},\ }\href {https://doi.org/10.1371/journal.pcbi.1006399}
  {\bibfield  {journal} {\bibinfo  {journal} {PLOS Comput. Biol.}\ }\textbf
  {\bibinfo {volume} {14}},\ \bibinfo {pages} {1} (\bibinfo {year}
  {2018})}\BibitemShut {NoStop}%
\bibitem [{\citenamefont {Zhuravlev}\ \emph {et~al.}(2016)\citenamefont
  {Zhuravlev}, \citenamefont {Hinczewski}, \citenamefont {Chakrabarti},
  \citenamefont {Marqusee},\ and\ \citenamefont {Thirumalai}}]{Hin3}%
  \BibitemOpen
  \bibfield  {author} {\bibinfo {author} {\bibfnamefont {P.~I.}\ \bibnamefont
  {Zhuravlev}}, \bibinfo {author} {\bibfnamefont {M.}~\bibnamefont
  {Hinczewski}}, \bibinfo {author} {\bibfnamefont {S.}~\bibnamefont
  {Chakrabarti}}, \bibinfo {author} {\bibfnamefont {S.}~\bibnamefont
  {Marqusee}},\ and\ \bibinfo {author} {\bibfnamefont {D.}~\bibnamefont
  {Thirumalai}},\ }\bibfield  {title} {\bibinfo {title} {Force-dependent switch
  in protein unfolding pathways and transition-state movements},\ }\href
  {https://doi.org/10.1073/pnas.1515730113} {\bibfield  {journal} {\bibinfo
  {journal} {Proc. Natl. Acad. Sci. U.S.A.}\ }\textbf {\bibinfo {volume}
  {113}},\ \bibinfo {pages} {E715} (\bibinfo {year} {2016})}\BibitemShut
  {NoStop}%
\bibitem [{\citenamefont {Blom}\ and\ \citenamefont {Godec}(2021)}]{Godec2021}%
  \BibitemOpen
  \bibfield  {author} {\bibinfo {author} {\bibfnamefont {K.}~\bibnamefont
  {Blom}}\ and\ \bibinfo {author} {\bibfnamefont {A.}~\bibnamefont {Godec}},\
  }\href@noop {} {}\bibinfo {howpublished}
  {\url{https://gitlab.gwdg.de/kblom/criticality-in-cell-adhesion}} (\bibinfo
  {year} {2021})\BibitemShut {NoStop}%
\bibitem [{\citenamefont {Retter}(1987)}]{retter1987adsorption}%
  \BibitemOpen
  \bibfield  {author} {\bibinfo {author} {\bibfnamefont {U.}~\bibnamefont
  {Retter}},\ }\bibfield  {title} {\bibinfo {title} {On adsorption according to
  the lattice gas model (ising model)},\ }\href@noop {} {\bibfield  {journal}
  {\bibinfo  {journal} {J. electroanal. chem. interfacial electrochem.}\
  }\textbf {\bibinfo {volume} {236}},\ \bibinfo {pages} {21} (\bibinfo {year}
  {1987})}\BibitemShut {NoStop}%
\bibitem [{\citenamefont {Guggenheim}(1935)}]{guggenheim1935statistical}%
  \BibitemOpen
  \bibfield  {author} {\bibinfo {author} {\bibfnamefont {E.~A.}\ \bibnamefont
  {Guggenheim}},\ }\bibfield  {title} {\bibinfo {title} {The statistical
  mechanics of regular solutions},\ }\href@noop {} {\bibfield  {journal}
  {\bibinfo  {journal} {Proc. Math. Phys. Eng. Sci.}\ }\textbf {\bibinfo
  {volume} {148}},\ \bibinfo {pages} {304} (\bibinfo {year}
  {1935})}\BibitemShut {NoStop}%
\bibitem [{\citenamefont {Bethe}(1935)}]{bethe1935statistical}%
  \BibitemOpen
  \bibfield  {author} {\bibinfo {author} {\bibfnamefont {H.~A.}\ \bibnamefont
  {Bethe}},\ }\bibfield  {title} {\bibinfo {title} {Statistical theory of
  superlattices},\ }\href@noop {} {\bibfield  {journal} {\bibinfo  {journal}
  {Proc. Math. Phys. Eng. Sci.}\ }\textbf {\bibinfo {volume} {150}},\ \bibinfo
  {pages} {552} (\bibinfo {year} {1935})}\BibitemShut {NoStop}%
\bibitem [{\citenamefont {Guggenheim}(1952)}]{guggenheim1952mixtures}%
  \BibitemOpen
  \bibfield  {author} {\bibinfo {author} {\bibfnamefont {E.~A.}\ \bibnamefont
  {Guggenheim}},\ }\href@noop {} {\emph {\bibinfo {title} {Mixtures: the theory
  of the equilibrium properties of some simple classes of mixtures solutions
  and alloys}}}\ (\bibinfo  {publisher} {Clarendon Press},\ \bibinfo {year}
  {1952})\BibitemShut {NoStop}%
\bibitem [{\citenamefont {{McQuarrie}}(1976)}]{mcquarrie1965statistical}%
  \BibitemOpen
  \bibfield  {author} {\bibinfo {author} {\bibfnamefont {D.~A.}\ \bibnamefont
  {{McQuarrie}}},\ }\href@noop {} {\emph {\bibinfo {title} {Statistical
  Mechanics}}},\ Harper's Chemistry Series\ (\bibinfo  {publisher}
  {HarperCollins Publishing, Inc.},\ \bibinfo {address} {New York},\ \bibinfo
  {year} {1976})\BibitemShut {NoStop}%
\bibitem [{\citenamefont {Hill}(1986)}]{hill1986introduction}%
  \BibitemOpen
  \bibfield  {author} {\bibinfo {author} {\bibfnamefont {T.~L.}\ \bibnamefont
  {Hill}},\ }\href@noop {} {\emph {\bibinfo {title} {An introduction to
  statistical thermodynamics}}}\ (\bibinfo  {publisher} {Courier Corporation},\
  \bibinfo {year} {1986})\BibitemShut {NoStop}%
\bibitem [{\citenamefont {Hartich}\ and\ \citenamefont
  {Godec}(2019{\natexlab{b}})}]{Hartich_2019_2}%
  \BibitemOpen
  \bibfield  {author} {\bibinfo {author} {\bibfnamefont {D.}~\bibnamefont
  {Hartich}}\ and\ \bibinfo {author} {\bibfnamefont {A.}~\bibnamefont
  {Godec}},\ }\bibfield  {title} {\bibinfo {title} {Extreme value statistics of
  ergodic markov processes from first passage times in the large deviation
  limit},\ }\href {https://doi.org/10.1088/1751-8121/ab1eca} {\bibfield
  {journal} {\bibinfo  {journal} {J. Phys. A: Math. Theor.}\ }\textbf {\bibinfo
  {volume} {52}},\ \bibinfo {pages} {244001} (\bibinfo {year}
  {2019}{\natexlab{b}})}\BibitemShut {NoStop}%
\bibitem [{\citenamefont {Doi}(2013)}]{doi2013soft}%
  \BibitemOpen
  \bibfield  {author} {\bibinfo {author} {\bibfnamefont {M.}~\bibnamefont
  {Doi}},\ }\href@noop {} {\emph {\bibinfo {title} {Soft matter physics}}}\
  (\bibinfo  {publisher} {Oxford University Press},\ \bibinfo {year}
  {2013})\BibitemShut {NoStop}%
\bibitem [{\citenamefont {Bender}\ and\ \citenamefont
  {Orszag}(2013)}]{bender2013advanced}%
  \BibitemOpen
  \bibfield  {author} {\bibinfo {author} {\bibfnamefont {C.~M.}\ \bibnamefont
  {Bender}}\ and\ \bibinfo {author} {\bibfnamefont {S.~A.}\ \bibnamefont
  {Orszag}},\ }\href@noop {} {\emph {\bibinfo {title} {Advanced mathematical
  methods for scientists and engineers I: Asymptotic methods and perturbation
  theory}}}\ (\bibinfo  {publisher} {Springer Science \& Business Media},\
  \bibinfo {year} {2013})\BibitemShut {NoStop}%
\bibitem [{\citenamefont {Cahill}\ \emph {et~al.}(2002)\citenamefont {Cahill},
  \citenamefont {D'Errico}, \citenamefont {Narayan},\ and\ \citenamefont
  {Narayan}}]{cahill2002fibonacci}%
  \BibitemOpen
  \bibfield  {author} {\bibinfo {author} {\bibfnamefont {N.~D.}\ \bibnamefont
  {Cahill}}, \bibinfo {author} {\bibfnamefont {J.~R.}\ \bibnamefont
  {D'Errico}}, \bibinfo {author} {\bibfnamefont {D.~A.}\ \bibnamefont
  {Narayan}},\ and\ \bibinfo {author} {\bibfnamefont {J.~Y.}\ \bibnamefont
  {Narayan}},\ }\bibfield  {title} {\bibinfo {title} {Fibonacci determinants},\
  }\href@noop {} {\bibfield  {journal} {\bibinfo  {journal} {The College
  Mathematics Journal}\ }\textbf {\bibinfo {volume} {33}},\ \bibinfo {pages}
  {221} (\bibinfo {year} {2002})}\BibitemShut {NoStop}%
\bibitem [{\citenamefont {Jia}(2018)}]{jia2018numerical}%
  \BibitemOpen
  \bibfield  {author} {\bibinfo {author} {\bibfnamefont {J.-T.}\ \bibnamefont
  {Jia}},\ }\bibfield  {title} {\bibinfo {title} {Numerical algorithms for the
  determinant evaluation of general hessenberg matrices},\ }\href@noop {}
  {\bibfield  {journal} {\bibinfo  {journal} {J. Math. Chem.}\ }\textbf
  {\bibinfo {volume} {56}},\ \bibinfo {pages} {247} (\bibinfo {year}
  {2018})}\BibitemShut {NoStop}%
\bibitem [{\citenamefont {Mulla}\ \emph {et~al.}(2018)\citenamefont {Mulla},
  \citenamefont {Oliveri}, \citenamefont {Overvelde},\ and\ \citenamefont
  {Koenderink}}]{PhysRevLett.120.268002}%
  \BibitemOpen
  \bibfield  {author} {\bibinfo {author} {\bibfnamefont {Y.}~\bibnamefont
  {Mulla}}, \bibinfo {author} {\bibfnamefont {G.}~\bibnamefont {Oliveri}},
  \bibinfo {author} {\bibfnamefont {J.~T.~B.}\ \bibnamefont {Overvelde}},\ and\
  \bibinfo {author} {\bibfnamefont {G.~H.}\ \bibnamefont {Koenderink}},\
  }\bibfield  {title} {\bibinfo {title} {Crack initiation in viscoelastic
  materials},\ }\href {https://doi.org/10.1103/PhysRevLett.120.268002}
  {\bibfield  {journal} {\bibinfo  {journal} {Phys. Rev. Lett.}\ }\textbf
  {\bibinfo {volume} {120}},\ \bibinfo {pages} {268002} (\bibinfo {year}
  {2018})}\BibitemShut {NoStop}%
\bibitem [{\citenamefont {H\"anggi}\ \emph {et~al.}(1990)\citenamefont
  {H\"anggi}, \citenamefont {Talkner},\ and\ \citenamefont
  {Borkovec}}]{RevModPhys.62.251}%
  \BibitemOpen
  \bibfield  {author} {\bibinfo {author} {\bibfnamefont {P.}~\bibnamefont
  {H\"anggi}}, \bibinfo {author} {\bibfnamefont {P.}~\bibnamefont {Talkner}},\
  and\ \bibinfo {author} {\bibfnamefont {M.}~\bibnamefont {Borkovec}},\
  }\bibfield  {title} {\bibinfo {title} {Reaction-rate theory: fifty years
  after kramers},\ }\href {https://doi.org/10.1103/RevModPhys.62.251}
  {\bibfield  {journal} {\bibinfo  {journal} {Rev. Mod. Phys.}\ }\textbf
  {\bibinfo {volume} {62}},\ \bibinfo {pages} {251} (\bibinfo {year}
  {1990})}\BibitemShut {NoStop}%
\bibitem [{\citenamefont {Wannier}(1945)}]{RevModPhys.17.50}%
  \BibitemOpen
  \bibfield  {author} {\bibinfo {author} {\bibfnamefont {G.~H.}\ \bibnamefont
  {Wannier}},\ }\bibfield  {title} {\bibinfo {title} {The statistical problem
  in cooperative phenomena},\ }\href {https://doi.org/10.1103/RevModPhys.17.50}
  {\bibfield  {journal} {\bibinfo  {journal} {Rev. Mod. Phys.}\ }\textbf
  {\bibinfo {volume} {17}},\ \bibinfo {pages} {50} (\bibinfo {year}
  {1945})}\BibitemShut {NoStop}%
\end{thebibliography}%
\clearpage
\newpage
\onecolumngrid
\renewcommand{\thefigure}{S\arabic{figure}}
\renewcommand{\theequation}{S\arabic{equation}}
\setcounter{equation}{0}
\setcounter{figure}{0}
\setcounter{page}{1}
\setcounter{section}{0}
\begin{center}
  \textbf{Supplemental Material for:\\Criticality in Cell Adhesion}\\[0.2cm]
Kristian Blom and Alja\v{z} Godec\\
\emph{Mathematical bioPhysics Group, Max Planck Institute for Biophysical Chemistry, 37077 Göttingen, Germany}
\\[0.6cm]  
\end{center}

In this Supplemental Material we prove the expression for the inverse
of the matrix $\mathbf{1}-\mathbf{T}^{\rm le}_{\rm d,f}$ used in Appendix~E4, where $\mathbf{T}^{\rm le}_{\rm d,f}$ is the sub-matrix for cluster dissolution/formation,
The elements of $\mathbf{1}-\mathbf{T}^{\rm le}_{\rm d,f}$ are given by
\begin{equation}
    \left[1-T^{\rm le}_{\rm d}\right]_{ij}=\Lambda^{\rm d}_{i}\delta_{ij}-
    \bar{w}_{i\to i+1}\delta_{i+1j}\theta\left(N-1-i\right)-\bar{w}_{i\to i-1}\delta_{i-1j}\theta\left(i-2\right),
\end{equation}
\begin{equation}
    \left[1-T^{\rm le}_{\rm f}\right]_{ij}=\Lambda^{\rm f}_{i}\delta_{ij}-
    \bar{w}_{i-1\to i}\delta_{i+1j}\theta\left(N-1-i\right)-\bar{w}_{i-1\to i-2}\delta_{i-1j}\theta\left(i-2\right),
\end{equation}
where $\Lambda^{\rm d}_{i}=-\sum_{j\neq i}^{N}\left[1-T^{\rm le}_{\rm d}\right]_{ij}+\bar{w}_{1\to0}\delta_{i1}$ and $\Lambda^{\rm f}_{i}=-\sum_{j\neq i}^{N}\left[1-T^{\rm le}_{\rm f}\right]_{ij}+\bar{w}_{N-1\to N}\delta_{iN}$. The inverse can be expressed as
\begin{equation}
    \mathcal{N}_{\rm d,f}\equiv(\mathbf{1}-\mathbf{T}^{\rm le}_{\rm{d,f}})^{-1}=\mathcal{A}^{\rm{d,f}}\mathcal{B}^{\rm{d,f}},
    \label{Cluster dissolution/nucleation inverse}
\end{equation}
where $\mathcal{A}^{\rm{d}}$ and $\mathcal{B}^{\rm{d}}$ are the lower
and upper triangular matrix, respectively, with elements
\begin{equation}
     \mathcal{A}^{\rm d}_{ij}=\frac{\theta\left(i-j\right)}{\textcolor{black}{\tilde{Q}_{j-1}}\bar{w}_{j-1\rightarrow j}}, 
    \ \
    \mathcal{B}^{\rm d}_{ij}=\textcolor{black}{\tilde{Q}_{j}}\theta\left(j-i\right),
\end{equation}
and $\mathcal{A}^{\rm{f}}$ and $\mathcal{B}^{\rm{f}}$ are the upper
and lower triangular matrix, respectively, with elements
\begin{equation}
    \mathcal{A}^{\rm f}_{ij}=\frac{\theta\left(j-i\right)}{\textcolor{black}{\tilde{Q}_{j}}\bar{w}_{j\rightarrow j-1}},
    \ \
    \mathcal{B}^{\rm f}_{ij}=\textcolor{black}{\tilde{Q}_{j-1}}\theta\left(i-j\right),
\end{equation}
\textcolor{black}{and $\tilde{Q}_{j}\equiv Q_{j}/p_{j}$ where $p_{j}$
  is the Glauber attempt probability and $Q_{j}$ the partition
  function constrained to $j$, the number of closed bonds defined in Eq.~\eqref{QK}.} 
To prove that Eq.~\eqref{Cluster dissolution/nucleation inverse} is
the inverse of $\mathbf{1}-\mathbf{T}_{\rm d,f}$ we evaluate  the
corresponding matrix product. We first write the elements of the
fundamental matrices as 
\begin{equation}
    \left[\mathcal{N}_{\rm d}\right]_{ij}=\sum\limits_{k=1}^{N}\mathcal{A}^{\rm d}_{ik}\mathcal{B}^{\rm d}_{kj}=\sum\limits_{k=1}^{N}\frac{\textcolor{black}{\tilde{Q}_{j}}\theta\left(i-k\right)\theta\left(j-k\right)}{\textcolor{black}{\tilde{Q}_{k-1}}\bar{w}_{k-1\rightarrow k}}=\sum\limits_{k=1}^{\min{\left(i,j\right)}}\frac{\textcolor{black}{\tilde{Q}_{j}}}{\textcolor{black}{\tilde{Q}_{k-1}}\bar{w}_{k-1\rightarrow k}},
    \label{Nd rewritten}
\end{equation}
and
\begin{equation}
    \left[\mathcal{N}_{\rm f}\right]_{ij}=\sum\limits_{k=1}^{N}\mathcal{A}^{\rm f}_{ik}\mathcal{B}^{\rm f}_{kj}=\sum\limits_{k=1}^{N}\frac{\textcolor{black}{\tilde{Q}_{j-1}}\theta\left(k-i\right)\theta\left(k-j\right)}{\textcolor{black}{\tilde{Q}_{k}}\bar{w}_{k\rightarrow k-1}}=\sum\limits_{k=\max{\left(i,j\right)}}^{N-1}\frac{\textcolor{black}{\tilde{Q}_{j-1}}}{\textcolor{black}{\tilde{Q}_{k}}\bar{w}_{k\rightarrow k-1}}.
    \label{Nf rewritten}
\end{equation}
For cluster dissolution, the matrix product is given by
\begin{multline}
    \sum\limits_{j=1}^{N}\left[1-T^{\rm le}_{\rm d}\right]_{ij}\left[\mathcal{N}_{\rm d}\right]_{jk}=\left[1-T^{\rm le}_{\rm d}\right]_{ii-1}\left[\mathcal{N}_{\rm d}\right]_{i-1k}\theta\left(i-2\right)+\left[1-T^{\rm le}_{\rm d}\right]_{ii}\left[\mathcal{N}_{\rm d}\right]_{ik}+\left[1-T^{\rm le}_{\rm d}\right]_{ii+1}\left[\mathcal{N}_{\rm d}\right]_{i+1k}\theta\left(N-1-i\right)=\\
    \bar{w}_{i\rightarrow i-1}\left(\left[\mathcal{N}_{\rm d}\right]_{ik}-\left[\mathcal{N}_{\rm d}\right]_{i-1k}\theta\left(i-2\right)\right)+\bar{w}_{i\rightarrow i+1}\left(\left[\mathcal{N}_{\rm d}\right]_{ik}-\left[\mathcal{N}_{\rm d}\right]_{i+1k}\right)\theta\left(N-1-i\right).
    \label{matrix product dissolution}
\end{multline}
The difference between two consecutive elements of the fundamental matrix can be obtained from Eq.~\eqref{Nd rewritten}, and reads
\begin{equation}
    \left[\mathcal{N}_{\rm d}\right]_{ik}-\left[\mathcal{N}_{\rm d}\right]_{i-1k}\theta\left(i-2\right)=
    \begin{dcases*}
        0 &, $k\leq i-1$\\ 
        \frac{\textcolor{black}{\tilde{Q}_{k}}}{\textcolor{black}{\tilde{Q}_{i-1}}\bar{w}_{i-1\rightarrow i}} &, $ k\geq i$
    \end{dcases*},
    \label{diff 1}
\end{equation}
and similarly
\begin{equation}
    \left(\left[\mathcal{N}_{\rm d}\right]_{ik}-\left[\mathcal{N}_{\rm d}\right]_{i+1k}\right)\theta\left(N-1-i\right)=
    \begin{dcases*}
        0 &, $k\leq i$\\ 
        -\frac{\textcolor{black}{\tilde{Q}_{k}}}{\textcolor{black}{\tilde{Q}_{i}}\bar{w}_{i\rightarrow i+1}} &, $k\geq i+1$\\
    \end{dcases*}.
    \label{diff 2}
\end{equation}
Plugging Eqs.~\eqref{diff 1} and \eqref{diff 2} into Eq.~\eqref{matrix product dissolution} we get
\begin{equation}
    \sum\limits_{j=1}^{N}\left[1-T^{\rm le}_{\rm d}\right]_{ij}\left[\mathcal{N}_{\rm d}\right]_{jk}=
    \begin{dcases*}
        0 &, $k \leq i-1$\\
        \frac{\textcolor{black}{\tilde{Q}_{i}}\bar{w}_{i\rightarrow i-1}}{\textcolor{black}{\tilde{Q}_{i-1}}\bar{w}_{i-1\rightarrow i}}=1&, $k = i$\\
        \frac{\textcolor{black}{\tilde{Q}_{k}}\bar{w}_{i\rightarrow i-1}}{\textcolor{black}{\tilde{Q}_{i-1}}\bar{w}_{i-1\rightarrow i}}-\frac{\textcolor{black}{\tilde{Q}_{k}}}{\textcolor{black}{\tilde{Q}_{i}}}=0&, $k \geq i+1$\\
    \end{dcases*},
    \label{final1}
\end{equation}
where we have used the detailed-balance relation $\textcolor{black}{\tilde{Q}_{i}}\bar{w}_{i\to i-1}=\textcolor{black}{\tilde{Q}_{i-1}}\bar{w}_{i-1\to i}$ \textcolor{black}{(see proof in Appendix E.3)} to obtain the final result. The matrix product for cluster formation becomes
\begin{multline}
    \sum\limits_{j=1}^{N}\left[1-T^{\rm le}_{\rm f}\right]_{ij}\left[\mathcal{N}_{\rm f}\right]_{jk}=\left[1-T_{\rm f}\right]_{ii-1}\left[\mathcal{N}_{\rm f}\right]_{i-1k}\theta\left(i-2\right)+\left[1-T_{\rm f}\right]_{ii}\left[\mathcal{N}_{\rm f}\right]_{ik}+\left[1-T_{\rm f}\right]_{ii+1}\left[\mathcal{N}_{\rm f}\right]_{i+1k}\theta\left(N-1-i\right)=\\
    \bar{w}_{i-1\rightarrow i-2}\left(\left[\mathcal{N}_{\rm f}\right]_{ik}-\left[\mathcal{N}_{\rm f}\right]_{i-1k}\right)\theta\left(i-2\right)+\bar{w}_{i-1\rightarrow i}\left(\left[\mathcal{N}_{\rm f}\right]_{ik}-\left[\mathcal{N}_{\rm f}\right]_{i+1k}\theta\left(N-1-i\right)\right).
    \label{matrix product formation}
\end{multline}
We again obtain the difference between two consecutive elements of the fundamental matrix for cluster formation using Eq.~\eqref{Nf rewritten}
\begin{equation}
    \left(\left[\mathcal{N}_{\rm f}\right]_{ik}-\left[\mathcal{N}_{\rm f}\right]_{i-1k}\right)\theta\left(i-2\right)=
    \begin{dcases*}
        -\frac{\textcolor{black}{\tilde{Q}_{k-1}}}{\textcolor{black}{\tilde{Q}_{i-1}}\bar{w}_{i-1\rightarrow i-2}} &, $k\leq i-1$\\ 
        0 &, $ k\geq i$
    \end{dcases*},
    \label{diff 3}
\end{equation}
and similarly
\begin{equation}
    \left(\left[\mathcal{N}_{\rm f}\right]_{ik}-\left[\mathcal{N}_{\rm f}\right]_{i+1k}\theta\left(N-1-i\right)\right)=
    \begin{dcases*}
        \frac{\textcolor{black}{\tilde{Q}_{k-1}}}{\textcolor{black}{\tilde{Q}_{i}}\bar{w}_{i\rightarrow i-1}} &, $k\leq i$\\ 
        0 &, $k\geq i+1$\\
    \end{dcases*}.
    \label{diff 4}
\end{equation}
Plugging Eqs.~\eqref{diff 3} and \eqref{diff 4} into Eq.~\eqref{matrix
  product formation} we arrive at
\begin{equation}
    \sum\limits_{j=1}^{N}\left[1-T^{\rm le}_{\rm f}\right]_{ij}\left[\mathcal{N}_{\rm f}\right]_{jk}=
    \begin{dcases*}
        \frac{\textcolor{black}{\tilde{Q}_{k-1}}\bar{w}_{i-1\rightarrow i}}{\textcolor{black}{\tilde{Q}_{i}}\bar{w}_{i\rightarrow i-1}}-\frac{\textcolor{black}{\tilde{Q}_{k-1}}}{\textcolor{black}{\tilde{Q}_{i-1}}}=0 &, $k \leq i-1$\\
        \frac{\textcolor{black}{\tilde{Q}_{i-1}}\bar{w}_{i-1\rightarrow
            i}}{\textcolor{black}{\tilde{Q}_{i}}\bar{w}_{i\rightarrow i-1}}=1&, $k = i$\\
        0 &, $k \geq i+1$\\
    \end{dcases*},
    \label{final2}
\end{equation}
where  we have again used the detailed-balance relation. Finally, Eqs.~\eqref{final1} and \eqref{final2} show that 
\begin{equation}
    \sum_{j=1}^{N}\left[1-T^{\rm le}_{\rm d,f}\right]_{ij}\left[\mathcal{N}_{\rm d,f}\right]_{jk}=\delta_{ik},
\end{equation} 
which completes the proof.
\end{document}